\documentclass[11pt]{article}

\usepackage[margin=1in]{geometry} 
\usepackage{amsmath} 
\usepackage{amsthm} 
\usepackage{amsfonts, amssymb} 
\usepackage{graphicx} 
\usepackage{hyperref} 
\usepackage{color} 
\usepackage[dvipsnames]{xcolor} 
\usepackage{bm} 
\usepackage{bbm} 
\usepackage{natbib}
\usepackage{multirow}
\usepackage{algorithmic} 
\usepackage{algorithm} 
\usepackage{tikz}
\usetikzlibrary{arrows, chains, positioning, quotes, shapes.geometric}
\usepackage{cases}
\usepackage{setspace}
\usepackage{enumitem}
\usepackage{etoolbox}
\newrobustcmd*{\mytriangle}[1]{\tikz{\filldraw[draw=#1,fill=#1] (0,0) --
(0.2cm,0) -- (0.1cm,0.2cm);}}

\newtheorem{theorem}{Theorem}[section]
\newtheorem{corollary}{Corollary}[section]
\newtheorem{lemma}{Lemma}[section]

\theoremstyle{definition}

\newtheorem{assumption}{Assumption}[section]

\theoremstyle{remark}

\def\T{^\mathrm{\scriptscriptstyle T}}

\newcommand{\indep}{\rotatebox[origin=c]{90}{$\models$} \, }

\newcommand{\R}{\mathbb{R}}
\newcommand{\cond}{\mid }
\newcommand{\con}{\, ; \,}

\newcommand{\INDEX}{\mathcal{I}}

\newcommand{\bY}{ {{Y}} }

\newcommand{\bZ}{ {{Z}} }
\newcommand{\bX}{ {{X}} }
\newcommand{\bW}{ {{W}} }
\newcommand{\bA}{ {{A}} }
\newcommand{\bO}{ {{O}} }
\newcommand{\ATE}{ {\rm ATE} }

\newcommand{\bv}{ {{v}} }
\newcommand{\bp}{ {{p}} }
\newcommand{\bT}{ {\theta} }
\newcommand{\uT}{ {\theta} }
\newcommand{\oT}{ {\tau} }

\newcommand{\by}{ {{y}} }

\newcommand{\bz}{ {{z}} }
\newcommand{\bx}{ {{x}} }
\newcommand{\bw}{ {{w}} }
\newcommand{\ba}{ {{a}} }

\newcommand{\ind}{\mathbbm{1}}
\newcommand{\balpha}{{\alpha}}
\newcommand{\Paraset}{ \mathrm{T} }
\newcommand{\Parasetk}{ \Theta }

\newcommand{\type}{L}
\newcommand{\NC}{N}
\newcommand{\NT}{K}
\newcommand{\NI}{M}
\newcommand{\zosets}{\mathcal{A}}

\newcommand{\EXP}{{E}}
\newcommand{\VAR}{{\rm var}}
\newcommand{\ee}{\Psi}
\newcommand{\bep}{ {{\epsilon}} }

\newcommand{\DE}{ {\rm DE} }
\newcommand{\IE}{ {\rm IE} }

\newcommand{\tDE}{ \oT^\DE }

\newcommand{\tIE}{ \oT^\IE }
\newcommand{\OR}{ {{g}} }
\newcommand{\POR}{ \OR }
\newcommand{\ePOR}{ \widehat{\OR} }
\newcommand{\Pe}{ e}
\newcommand{\ePe}{ \widehat{e} }

\newcommand{\indOR}{ g }
\newcommand{\indPOR}{ \indOR}
\newcommand{\indePOR}{ \widehat{\indOR}^{{\rm Par}} }
\newcommand{\indIOR}{ \indOR^{\rm NoInt} }
\newcommand{\indPIOR}{ \indOR^{\rm Par,NoInt} }

\newcommand{ \paraOR }{ {\beta}_\indOR }
\newcommand{ \paraORtype }[1]{ {\beta}_{\indOR,#1} }
\newcommand{ \estparaOR }[1]{ \widehat{{\beta}}_{\indOR #1} }
\newcommand{ \paraPS }{ {\beta}_e }
\newcommand{ \paraPStype }[1]{ {\beta}_{e,#1} }
\newcommand{ \estparaPS }[1]{ \widehat{{\beta}}_{e #1} }

\newcommand{ \paraT }{ {\beta} }

\newcommand{ \nparaPS }{ {\zeta}_e }
\newcommand{ \nparaPStype }[1]{ {\zeta}_{e, #1} }
\newcommand{ \nparaOR }{ {\zeta}_\indOR }
\newcommand{ \nparaORtype }[1]{ {\zeta}_{\indOR, #1} }

\newcommand{\eij}{ {i(-j) } }

\newcommand{\px}{ P_{X} }
\newcommand{\py}{ P_{Y} }
\newcommand{\pyi}{ P_{Y,{\rm NoInt}}}

\newcommand{\modeliv}{ \mathcal{M}_{{\rm NP}} }
\newcommand{\modelive}{ \mathcal{M}_{{\rm NP},e^*} }
\newcommand{\modelii}{ \mathcal{M}_{{\rm NoInt}} }
\newcommand{\modelie}{ \mathcal{M}_{e} }
\newcommand{\modeliig}{ \mathcal{M}_{\indOR,{\rm NoInt}} }
\newcommand{\modelig}{ \mathcal{M}_{\indOR} }

\newcommand{\eifv}{  \varphi }
\newcommand{\eifuv}{  \varphi }

\hypersetup{
colorlinks=true,
linkcolor=blue,
filecolor=blue,
urlcolor=blue,
citecolor=blue
}

\newcommand{\reals}{\mathbb{R}}

\definecolor{red1}{RGB}{255,129,129}
\definecolor{blue1}{RGB}{143,199,255}

\graphicspath{ {plot/} }

\allowdisplaybreaks

\begin{document}

\setlength{\abovedisplayskip}{8pt}
\setlength{\belowdisplayskip}{8pt}
\setlength{\abovedisplayshortskip}{8pt}
\setlength{\belowdisplayshortskip}{8pt}

\title{Efficient Semiparametric Estimation of Network Treatment Effects Under Partial Interference}
 \author{
  Chan Park and Hyunseung Kang
  \\[0.2cm]
  {\small Department of Statistics, University of Wisconsin--Madison}}
\date{ }
\maketitle
\begin{abstract}
Recently, many estimators for network treatment effects have been proposed. But, their optimality properties in terms of semiparametric efficiency have yet to be resolved. We present a simple, yet flexible asymptotic framework to derive the efficient influence function and the semiparametric efficiency lower bound for a family of network causal effects under partial interference. An important corollary of our results is that one of the existing estimators by \citet{Liu2019} is locally efficient. We also present other estimators that are efficient and discuss results on adaptive estimation. We conclude by using the efficient estimators to study the direct and spillover effects of conditional cash transfer programs in Colombia. 
\end{abstract}
\noindent%

\section{Introduction}	
\subsection{Motivation: Efficient Estimators Under Partial Interference}	
Recently, there has been growing interest in studying causal effects under interference \citep{Cox1958, Rubin1986} where the potential outcome of a study unit is affected by the treatment assignment of other study units. 
The most well-studied type of interference is partial interference \citep{Sobel2006} where study units are partitioned into non-overlapping clusters and interference only arises within units in the same cluster. \citet{HH2008} defined various network causal effects under partial interference, notably the direct and indirect causal effects, and proposed an experimental design to estimate them. Since then, many works have proposed innovative identification and estimation strategies for various causal estimands in network settings. However, an unresolved question in this literature is determining which of the several proposed estimators is optimal in terms of semiparametric efficiency. For example, several works \citep{PH2014,Liu2016,Liu2019,Barkley2020} have examined the statistical properties of the developed estimators, but none have shown whether they achieve the semiparametric efficiency bound because the efficient influence function for the network effects has not been established yet. In contrast, without interference, it is well-established that the augmented inverse probability-weighted estimator is adaptive, locally efficient, and doubly robust for the average treatment effect; see \citet{Robins1994},  \citet{Hahn1998}, \citet{Scharfstein1999}, 
\citet{vvLaan2003}, \citet{Hirano2003}, 
and many other works on efficient estimation under no interference. 

\subsection{Our Contribution}						\label{sec:Intro2}
The goal of the paper is to study optimal, semiparametric estimation of network effects under partial interference. 
Unfortunately, we cannot directly use traditional semiparametric theory because it assumes independent and identically distributed data \citep{BKRW1998}, which is not compatible with network data. Instead, our main contribution is to re-purpose what \citet{Bickel2001} calls a ``nonparametric model for Markov chains'' which embeds non-independent and non-identically distributed data into locally independent, linear sums so that typical semiparametric theory can be applied in a local sense; see \citet{McNeney2000}, \citet{Bickel2001}, \citet{Sofrygin2016}, and Section \ref{sec:appendix1-NPmodel} of the supplementary materials for additional discussions on locally linear, asymptotic embedding. 
 
Formally, if $\bO_{i}$ represents all observed data from cluster $i=1,\ldots,N$, the approach supposes $\bO_i$ are independent of each other and each $\bO_i$ is generated from one of $\NT <\infty$ densities labeled by $\type_i = 1,\ldots,\NT$, i.e.
\begin{subequations} \label{eq:NPmodel}
\begin{align}
&P(\bO_1, \type_1,\ldots,\bO_\NC,\type_\NC) = \prod_{i=1}^\NC P(\bO_i,\type_i) \label{eq:NPmodel0} \\
&P(\bO_i,\type_i) =  \prod_{k=1}^\NT  \big\{ P(L_i = k) P (\bO_i \cond \type_i = k) \big\}^{\ind(\type_i=k)}, \quad{} {\rm dim}(\bO_i) < \infty, \quad \NT < \infty \ . \label{eq:NPmodel1}
\end{align}
\end{subequations}
In words, model \eqref{eq:NPmodel} makes the following assumptions: (a)  data from each cluster $i$, $\bO_i $, are independent of each other, (b) each $\bO_i$ follows some, potentially different, nonparametric distribution labeled by cluster types $\type_i$, and (c) the asymptotics increase the number of clusters $\NC$ while keeping the cluster size ${\rm dim}(O_i)$ bounded. Property (a) is a common assumption in partial interference \citep{VWTTH2014, Liu2016, Yang2018, Liu2019, Barkley2020, Smith2020, Kilpatrick2021}. Property (b) is our approach to deal with $\bO_i$'s having varying dimensions for each $i$ due to differences in cluster size. For example, without any covariates, if household $i=1$ has 2 individuals, $\bO_1$ is 4-dimensional. But, if household $i = 2$ has 5 individuals, $\bO_2$ is $10$ dimensional, and thus the density of $\bO_1$ is different from the density of $\bO_2$. Critically, it is likely that the interference pattern in a two-person household is different from that in a five-person household, and we use $\type_i$ to allow for different interference patterns; see below. For property (c), to the best of our knowledge, there is no established semiparametric theory that allows ${\rm dim}(\bO_i)$ to grow to infinity while the elements of $\bO_i$ remain dependent arbitrarily and asymptotically, i.e. the dependence does not vanish to zero as sample size increases. Instead, (c) bounds the cluster size to allow for arbitrary dependence between units in a cluster, critically between the treatment of an individual and the outcome of his/her peer in the same cluster, and the effective sample size increases with the number of clusters $\NC$.

As mentioned earlier, the variable $\type_i$ is a key technical device to deal with a situation where two clusters $i$ and $i'$ have different numbers of units, and under a nonparametric framework, two different nonparametric densities, labeled by $\type_i$, are needed to model $\bO_i$ and $\bO_{i'}$. An alternative to using $\type_i$ would be to assume a fixed, known, dimension-reducing model on $\bO_i$ so that clusters of varying size and critically, the dependence between units within each cluster are comparable with each other; the most popular dimension-reducing model is based on a scalar function of peers' data \citep{vvLaan2014,PH2014, Liu2016, Sofrygin2016,Ogburn2017arxiv,Liu2019, Barkley2020}. Instead, our setup 
allows a very general, nonparametric factorization of $P(\bO_i \cond L_i)$, say $P(\bO_i \cond L_i) = P(O_{i1} \cond O_{i2},\ldots,O_{i{\rm dim}(O_i)}, L_i) 
\times \cdots \times P(O_{i{\rm dim}(O_i)} \cond L_i)$ 

Also, while \eqref{eq:NPmodel} resembles a mixture model, the goal of the paper is not to identify or estimate unknown mixture labels $\type_i$ typical in mixture modeling. Instead, $\type_i$ is a tool to embed/approximate studies under partial interference into \eqref{eq:NPmodel},  and hence $\type_i$ is known by construction. For example, in a study on student absenteeism in Philadelphia with households as clusters, \citet{Basse2018} proposed stratifying households by their size. Thus, a natural embedding with $\type_i$ is by household size where a household of size 2 belongs to one cluster type and a household of size 4 belongs to another cluster type. Or, in a twins study, $\type_i$ could be defined by different types or twins such as identical and fraternal twins. Section \ref{sec:disc} contains additional discussions of model \eqref{eq:NPmodel}.

Finally, our setting differs from existing semiparametric settings under independent, but (non-)identically distributed multivariate data where the dimension of the multivariate data is often identical \citep{Robins1995, Rotnitzky1998, Chen2006, Vanstteelandt2007}. For example, in existing theoretical work on locally efficient estimators of causal effects with multivariate or repeated outcomes,  it is common to make a simplifying assumption that everyone's data have identical dimensions, say ${\rm dim}(\bO_i) = T$ for all individual $i=1,\ldots,\NC$, and $T$ is typically defined by the number of repeated response from individual $i$. Instead, the setting in the paper is closer to a conditionally independent and identically distributed setting in example 1 of \citet{Bickel2001} where we embed different parts of the observed data using $\type_i$ so that conditional on $\type_i$, $\bO_i$ becomes independent and identically distributed, and we can apply the usual semiparametric theory locally in $\type_i$.

Under model \eqref{eq:NPmodel}, we derive the globally and locally efficient influence functions and the semiparametric efficiency lower bounds for a family of network causal effects under partial interference. We remark that the target estimand is still defined as the contrasts of individual $ij$'s potential outcomes, not contrasts of cluster-level potential outcomes, and our identification, estimation, and inference strategies use individual-level observed data $\bO_{ij}$ instead of cluster-level summaries of individual-level observed data. An important corollary of our results is that the bias-corrected doubly robust estimator from \citet{Liu2019} is locally efficient for estimating the direct and indirect effects in \citet{HH2008} and \citet{TTV2012}; in short, \citet{Liu2019}'s estimator is the partial interference equivalent to the aforementioned augmented inverse probability-weighted estimator of the average treatment effect under no interference. We also present other estimators that can achieve the efficiency bound, notably a simple variant of a cross-fitting estimator  \citep{Victor2018} under partial interference that achieves the global efficiency bound. Additionally, we briefly discuss adaptive estimation, which mirrors the adaptation properties of the augmented inverse probability-weighted estimator under no interference \citep{Scharfstein1999_Rejoinder}. We conclude by using our efficient estimators to study the direct and spillover effects of conditional cash transfer programs on student attendance in Colombia. 

\section{Setup}									\label{sec:setup0}
\subsection{Notation}						\label{sec:setup}

We lay out the notations for the observed data.  We denote the cluster type by $L_i \in \{ 1, \ldots, K \}$ where $L_i$ are cluster-level variables and $K< \infty$ is the number of cluster types. For mathematical convenience, we use cluster size to define cluster types $L_i$ hereafter, but we emphasize that any $L_i$ satisfying model \eqref{eq:NPmodel} is valid for the results below.  
Let $\NC$, $\NC_k$, and $\NI_k$ be the number of clusters, the number of clusters from cluster type $k$, and the cluster size of type $k$, respectively. For each cluster $i=1,\ldots,\NC$, let $Y_{ij} \in \R$ be unit $j$'s univariate outcome, $A_{ij} \in \{0 , 1\} $ be unit $j$'s treatment indicator where $A_{ij} = 1$ indicates unit $j$ is assigned to treatment and $A_{ij} = 0$ indicates unit $j$ is assigned to control, and $\bX_{ij}$ be unit $j$'s vector of pre-treatment covariates. Let $\bY_i = ( Y_{i1} , \ldots , Y_{i \NI_k}) \T \in \R^{\NI_k} $, $\bA_i = ( A_{i1} , \ldots , A_{ i \NI_k}) \T \in \zosets(\NI_k)$, and $\bX_i = ( \bX_{i1} \T , \ldots , \bX_{i \NI_k} \T) \T \in \mathcal{X}(k)$ be the vectorized outcome, treatment assignment, and pre-treatment covariates, respectively, for each cluster $i$ in cluster type $k$; 
here, $\zosets(t)$ is a collection of $t$-dimensional binary vectors 
and $\mathcal{X}(k)$ is the finite dimensional support of $\bX_i$ for cluster type $k$. Let $\bO_i = (\bY_i, \bA_i, \bX_i)$ be all the observed data from cluster $i$ and let $\type_{i}$ indicate which type cluster $i$ belongs to. 

We use potential outcomes 
 to define causal effects. Let $\bA_{\eij} \in \zosets(\NI_k-1) $ be the vector of treatment indicators for all units in cluster $i$ except unit $j$. Let  $a_{ij}$, $\ba_i$, and $\ba_\eij$ be the realized values of $A_{ij}$, $\bA_i$, and $\bA_{\eij}$, respectively. 
Let $Y_{ij} ( \ba_i )$ be the potential outcome of unit $j$ in cluster $i$ under treatment vector $\ba_i$ and let $\bY_i ( \ba_i ) = \big( Y_{i1}(\ba_i ) , \ldots , Y_{i \NI_k}(\ba_i ) \big) \T $ be the potential outcomes of all units in cluster $i$. Following \citet{HH2008} and \citet{TTV2012}, we define the unit average potential outcome for $a \in \{0,1\}$ under a treatment allocation strategy $\alpha \in (0,1)$ as 
  \begin{align*}
	& \overline{Y}_{ij} (a \con \alpha) 
	= \sum_{ \ba_i \in \zosets(\NI_k)  } Y_{ij} ( \ba_i )   \ind (a_{ij} = a)  \pi( \ba_{\eij} \con \alpha )
	\ , \
	\pi(\ba_{\eij} \con \alpha) =  \prod_{j' \neq j} \alpha^{a_{ij' }} (1-\alpha)^{1 - a_{ij'}} 
	\ .
\end{align*}
In words, $ \overline{Y}_{ij} (a \con \alpha)$ is the average of unit $ij$'s potential outcomes when the unit's treatment is fixed at $a$ and the unit's peers in a cluster are assigned to treatment independently with probability $\alpha \in (0,1)$. We also define the cluster average potential outcome as $\overline{Y}_{i}(a \con \alpha) =  \NI_k^{-1} \sum_{j=1}^{\NI_k} \overline{Y}_{ij} (a \con \alpha)$.

Finally, 
for a vector $V_N$, let $V_\NC = O_P(1)$ and $V_\NC = o_P(1)$ be the usual big-O and little-O notations, respectively. Let $\| \cdot \|_2$ be the 2-norm of a vector and a matrix. 
Let $ V \indep W $ 
denote independence between two random variables $V$ and $W$. 

\subsection{Family of Causal Estimands Under Partial Interference} 
				\label{sec:potout}
Consider the set of causal estimands in cluster type $k$, denoted by $\Parasetk_k$, and the set of causal estimands across all clusters, denoted by $\Paraset$:
\begin{align*}		
	\hspace*{0cm} \Parasetk_k = \Big\{
		\uT_k \in \R \, \Big| \, &
		\uT_k = \uT_k(\alpha_k , \alpha_k') = \sum_{\ba_i \in \zosets(\NI_k) } \EXP \big\{  \bw_k\T(\ba_i , \bX_i \con \alpha_k, \alpha_k') \bY_i(\ba_i) \cond \type_i = k \big\}, \\ 
		\nonumber
		&  \alpha_k , \alpha_k' \in (0,1), \ \bw_k ( \ba_i , \bx_i \con \alpha_k, \alpha_k') \in \R^{\NI_k} 
		,
		\big\| \bw_k ( \ba_i , \bx_i \con \alpha_k , \alpha_k') \big\|_2 < \infty \
	\Big\} \ ,
	\\
	\hspace*{-1.0cm} \Paraset
	= \Big\{
		\oT  \in \R 
		\, \Big| \, &  
		\oT =  \bv\T(\bp)  \bT, \ \bv(\bp) = \big( v_1(p_1), \ldots, v_\NT(p_\NT) \big)\T, \ \bT = (\uT_1,\ldots,\uT_\NT)\T, \\
		\nonumber & \ v_k(\cdot)  \text{ is continuously differentiable}, \ p_k = {\rm pr}(L_i = k), \ \uT_k \in \Parasetk_k 
	\Big\} \ .
\end{align*}
In words, for each cluster type $k$, the set $\Theta_k$ consists of parameters $\theta_k(\alpha_k, \alpha_k')$, which are linear, weighted sums of expectation of potential outcomes in cluster $i$.	
The weights $\bw_k(\ba_i, \bX_i \con \alpha_k, \alpha_k')$ are determined by the causal estimand of interest and the sum is over all possible values of the treatment vector $\ba_i \in \zosets(\NI_k)$. Second, the set  ${\rm T}$ consists of estimands $\tau$, which are weighted sums of $\theta_k(\alpha_k, \alpha_k')$ with weights $v_k(p_k)$. The weights $v_k(\cdot)$ are also determined by the causal estimand of interest. 

At a high level, both sets $\Theta_{k}$ and ${\rm T}$ are abstractions of familiar causal parameters under partial interference.
For example, suppose the target estimand is the direct effect $\tDE (\alpha) = \EXP \big\{ \overline{Y}_i ( 1 \con \alpha)   - \overline{Y}_i ( 0 \con \alpha) \big\}$, which is counterpart of the direct effect defined in \citet{HH2008} under an infinite population framework. If we choose the weights $\bw_k$ and $v_k$ as 
	\begin{align}						\label{example-DE}
		\hspace*{-0.1cm}
		\bw_k( \ba_i \con \alpha) 
		=
		\frac{1}{\NI_k} 
		\begin{bmatrix}
			\big\{ \ind \big( a_{i1}=1 ) - \ind \big( a_{i1}=0 ) \big\} 
			\pi (\ba_{i(-1)} \con \alpha)
			\\[-0.1cm] \vdots \\[-0.1cm]
			\big\{ \ind \big( a_{ij}=1 ) - \ind \big( a_{ij}=0 ) \big\} 
			\pi (\ba_{i(-j)} \con \alpha)
			\\[-0.1cm] \vdots \\[-0.1cm]
			\big\{ \ind \big( a_{i\NI_k}=1 ) - \ind \big( a_{i\NI_k}=0 ) \big\}
			\pi (\ba_{i(-\NI_k)} \con \alpha)
		\end{bmatrix} \, , \,
		v_k(p_k) = p_k = P(\type_i = k) \, , 
		\hspace*{-0.1cm}
	\end{align}
this leads $\theta_k = \uT_k^\DE(\alpha) = \EXP \big\{ \overline{Y}_i ( 1 \con \alpha)   - \overline{Y}_i ( 0 \con \alpha) \cond \type_i = k \big\} \in \Theta_k$, the direct effect in cluster type $k$, and $\tau = \tDE(\alpha) =
		\EXP \big\{ \overline{Y}_i ( 1 \con \alpha)   - \overline{Y}_i ( 0 \con \alpha) \big\} \in \Paraset $.
	Similarly, by taking each entry of the weights $w_k$ as $\ind(a_{ij} = 0 )  \{	\pi( \ba_\eij \con \alpha) - \pi( \ba_\eij \con \alpha') \} / \NI_k $, we arrive at $\uT_k = \uT_k^\IE(\alpha,\alpha') = \EXP \big\{ \overline{Y}_i ( 0 \con \alpha)   - \overline{Y}_i ( 0 \con \alpha') \cond \type_i = k \big\} \in \Theta_k$, the indirect effect in cluster type $k$, and $\oT = \tIE (\alpha, \alpha') = \EXP \big\{ \overline{Y}_i ( 0 \con \alpha)   - \overline{Y}_i ( 0 \con \alpha') \big\} \in \Paraset$, the indirect effect. Section \ref{sec:AE} of the supplementary material shows that under certain growth conditions, existing finite sample causal effects in partial interference, say total effects, overall effects, or spillover effects among subgroups, can be asymptotically embedded into $\Parasetk_k$ and $\Paraset$.

To identify the causal estimands in $\Paraset$, let $\OR(\ba , \bx, k) = \EXP ( \bY_i \cond \bA_i = \ba , \bX_i = \bx , \type_i=k )$ be the vector of conditional expected outcomes in cluster $i$ 
and 
$\Sigma(\ba , \bx, k) = \VAR ( \bY_i \cond \bA_i = \ba , \bX_i = \bx, \type_i = k )$ be its conditional covariance. 
The outcome model $\OR(\ba , \bx, k)$ is a generalization of the usual outcome model to partial interference settings where now, the outcomes in a cluster are jointly modeled. Also, 
let $e(\ba \cond \bx, k) = {\rm pr}(\bA_i = \ba \cond \bX_i  = \bx, \type_i = k) $ be the probability of observing treatment vector $\ba \in \zosets(\NI_k)$ given covariates $\bx$ in cluster type $k$. The propensity score model $e(\ba \cond \bx, k)$ is a generalization of the propensity score \citep{Rosenbaum1983} to partial interference settings where all treatment assignments in a cluster are jointly modeled. Assumption \ref{Assump:VC} lays out the identifying assumptions for a parameter in $\Paraset$; see \citet{Liu2019} for similar conditions.
\begin{assumption} \label{Assump:VC} For all $\ba \in \zosets(\NI_k)$, $\bx \in \mathcal{X}(k)$, and $k = 1,\ldots,\NT$, we have the following conditions: (A1) \textit{Consistency}:  $\bY_i =  \sum_{k=1}^\NT \ind (\type_i = k) \sum_{\ba_i \in \zosets(\NI_k)} \ind(\bA_i = \ba_i) \bY_i(\ba_i ) $; (A2) \textit{Conditional Ignorability}: $\bY_i ( \ba ) \indep \bA_i \cond  (\bX_i = \bx , \type_i = k)$; 	(A3) \textit{Positivity/Overlap}: There exists a positive constant $c$ so that $ c < e(\ba \cond \bx , k) < 1-c$; (A4) \textit{Moments}: $\EXP\big\{ \OR(\ba, \bX_i, k) \cond \type_i = k \big\}$ and $\EXP\big\{ \Sigma(\ba, \bX_i, k) \cond \type_i = k \big\}$ exist and are finite. Also, $\Sigma(\ba, \bx, k)$ is positive definite.

\end{assumption}
\noindent Conditions (A1)--(A3) are natural extensions of consistency, 
conditional ignorability and overlap to partial interference settings; see \citet{ImbensRubin2015} and \citet{HR2020} for textbook discussions. Condition (A4) ensures that the expectations and covariances are well-defined. 
Under Assumption \ref{Assump:VC}, 
a causal estimand $\oT \in \Paraset$ can be identified from $(\bO_i, \type_i)$ as 
\begin{align}							\label{eq-31003-1}
	\oT = \bv\T(\bp)  \bT 
	= \sum_{k=1}^\NT v_k(p_k)
	\bigg[ \sum_{\ba_i \in \zosets(\NI_k)}
	\EXP \big\{ \bw_k \T (\ba_i, \bX_i) \OR(\ba_i , \bX_i,  k) \cond \type_i = k \big\} \bigg] \ .
\end{align}
Here, $ \bw_k (\ba_i, \bX_i) $ is shorthand for $ \bw_k (\ba_i, \bX_i \con \alpha_k, \alpha_k')$. 
The rest of the paper will focus on efficient estimation of the functional based on the observed data in \eqref{eq-31003-1} under model \eqref{eq:NPmodel}.

\section{Semiparametric Efficiency Under Partial Interference} \label{sec:EffNonpar}
\subsection{Global Efficiency}							\label{sec:GloEff}
Let $\bp^*$, $\bT^*$, $\oT^*$, $e^*$, $\OR^*$, and $\Sigma^*$ denote the true values of  $\bp$, $\bT$, $\oT$, $e$, $\OR$, and $\Sigma$, respectively. Theorem \ref{thm:EIFVC} presents our first main result where we derive the globally efficient influence function and the semiparametric efficiency bound of $\oT^* \in \Paraset$ in model $\modeliv 
	=
	\big\{  P_{\bO,\type} = P({\bO_1,\type_1,\ldots,\bO_N,\type_N}) \, \big| \, 
	 P(\cdot) \text{ satisfies \eqref{eq:NPmodel}} \big\}$. 
\begin{theorem}[Global Efficiency] 		\label{thm:EIFVC}
	Let $\oT^* \in \Paraset$ be the parameter defined in \eqref{eq-31003-1} and suppose Assumption \ref{Assump:VC} holds. 
	If $p_k^*$s are unknown,
	 the efficient influence function of $\oT^*$ under model $\modeliv$ is 
	\begin{align*}
		 \eifuv ( \oT^* )
		 &
		 =  \sum_{k=1}^\NT  v_k(p_k^*) \varphi_k ( \uT_k^* )
		 + \sum_{k=1}^\NT \Big\{ \ind(\type_i=k) -p_k^* \Big\} \frac{\partial v_k(p_k^*)}{\partial p_k }   \uT_k^*
	\end{align*}
	where $\varphi_k(\uT_k^*) = \ind (\type_i = k) \big\{ \phi_k(\bO_i, e^*, \OR^*) - \uT_k^* \big\} / p_k^*$ with 
	\begin{align}									\label{eq-34001}
	&
	\hspace*{-0.2cm}
	 \phi_k(\bO_i, e^*, \OR^*)
	= 
	\ind \big( \type_i = k \big)
	\bigg[
		 \frac{ \bw_k \T (\bA_i,\bX_i ) \big\{  \bY_i - \OR^*(\bA_i, \bX_i, k) \big\} }{e^*(\bA_i \cond \bX_i, k)}   
		 +
	\hspace*{-0.3cm}
	\sum_{\ba_i \in \zosets( \NI_k ) } 
	\hspace*{-0.3cm}
	\bw_k \T (\ba_i,\bX_i ) 
		  \OR^*(\ba_i, \bX_i, k) \bigg]  \ .
\end{align}
Moreover, the semiparametric efficiency bound of $\oT^*$ in model $\modeliv$ is $\VAR \big\{  \eifuv (\oT^*) \big\} $.
\end{theorem}
See the supplementary materials for the result when $p_k^*$s are known.
We make some remarks about Theorem \ref{thm:EIFVC}. First, if $\NT = 1$ and $\NI_k =1$, our result would reduce to Theorem 1 in \citet{Hahn1998}. Second, some of the usual components from the efficient influence function of the average treatment effect without interference are still present in equation \eqref{eq-34001}, most notably the residual-weighting term by the propensity score, i.e. $ \big\{ \bY_i - \OR^*(\bA_i, \bX_i, k) \big\} / e^*(\bA_i \cond \bX_i, k)$, and the outcome regression term, i.e. $\sum_{\ba_i \in \zosets(\NI_k)} \bw_k\T(\ba_i,\bX_i) \OR^*(\ba_i, \bX_i, k) - \uT_k^*$. However, there are new terms to account for under dependence between units, specifically (i) a weighing term $\bw_k$ that weighs peers' influence on one's own outcome, (ii) a non-diagonal covariance matrix $\Sigma^*$, (iii) a multivariate outcome regression, which leads to multivariate residuals, and (iv) a propensity score that depends on a vector of treatment assignments instead of one's own treatment assignment. 
Third, in Section \ref{sec:GloEffsupp} of the supplementary materials, we discuss the efficient influence function in \eqref{eq-34001} can be used to construct a doubly robust estimator, resolving a conjecture discussed in Section 7 of \citet{Liu2019} about the property of doubly robust estimators under partial interference.

\subsection{Local Efficiency} \label{sec:Mestimation-general}
Often in practice, investigators posit parametric or semiparametric models to estimate the outcome model $\OR$ or the propensity score model $e$ \citep{PH2014, Liu2016, Liu2019, Barkley2020}. To this end, this section presents our second main result where we derive locally efficient semiparametric estimators for parameters in $\Paraset$. 
To begin, we define model spaces $\modelig \subseteq \modeliv$ and $\modelie \subseteq \modeliv$ that restrict the outcome and propensity score models to those specified by the investigator; i.e.
\begin{align*}
	\modelig = \big\{ P_{\bO,\type} \in \modeliv \, \big| \, & 
	 \text{there is a unique } \paraOR^* \text{ such that }  
	\OR^*(\ba, \bx, k) = \POR(\ba, \bx,k \con \paraOR^* ) 	\big\}	
	\ ,
	\\
	\modelie = \big\{ P_{\bO,\type} \in \modeliv \, \big| \, & 
	\text{there is a unique } \paraPS^* \text{ such that }  
	e^*(\ba \cond \bx, k) = \Pe(\ba \cond \bx,k \con \paraPS^* ) 	\big\}	\ .
\end{align*}
Some commonly used models for $\modelig$ and $\modelie$ include generalized mixed effects models, score equations, quasi-likelihoods, or generalized estimating equations. 

The outcome regression model $\modelig$ can encode information about a known exposure mapping. For example, consider a two-person household where individual 1's outcome depends on individual 2's treatment status, but individual 2's outcome does not depend on individual 1's treatment status. In short, there is asymmetric interference where there is no interference from individual 1 to individual 2, but there is interference from individual 2 to individual 1. Also, practically speaking, this type of asymmetric interference may be plausible in some vaccine studies where, depending on the vaccine, vaccinated individuals' outcomes are unlikely to be affected by their peers' vaccination status, i.e. treatment, but the unvaccinated individuals' outcomes may be affected by their peers' vaccination status. Then, one way to encode this exposure map is through a simple linear model for $\OR = (g_1, g_2)\T$, i.e.
	\begin{align*}
		\OR (\bA_i, \bX_i, k \con \beta_g) 
		= 
		\begin{bmatrix}
			g_1 (\bA_i, \bX_i, k \con \beta_g) \\
			g_2 (\bA_i, \bX_i, k \con \beta_g)
		\end{bmatrix}
		= 
		\begin{bmatrix}
			\beta_{g10} + \beta_{g11} A_{i1} + \beta_{g12} A_{i2} + \beta_{g13}\T \bX_{i1}  + \beta_{g14}\T \bX_{i2}   \\
			\beta_{g20} + \beta_{g21} A_{i2} + \beta_{g22}\T \bX_{i1}  + \beta_{g23}\T \bX_{i2}
		\end{bmatrix}
		\ .
	\end{align*}
As another example, consider the setting by \citet{Sofrygin2016} and \citet{Ogburn2017arxiv} where the exposure mapping is restricted to a map where (i) the individual-level outcome regression $g_j (\bA_i, \bX_i , L_i) $ is identical across $j$, i.e. symmetric interference, and (ii) the dependence on $(\bA_\eij, \bX_\eij)$ occurs only through finite, fixed dimensional summary statistics, say $(\bA_\eij^s, \bX_\eij^s)$. Then, one way to encode this type of exposure map is through a linear model for $\OR = (g_1,\ldots,g_{\NI_k})\T$, i.e.
	\begin{align*}
		&
		\OR (\bA_i, \bX_i, k \con \beta_g) 
		= 
		\begin{bmatrix}
			g_1 (\bA_i, \bX_i, k \con \beta_g) \\[-0.1cm]
			\vdots \\
			g_{\NI_k} (\bA_i, \bX_i, k \con \beta_g)
		\end{bmatrix}
		\ , \
		g_j (\bA_i, \bX_i, k \con \beta_g)
		=
		\beta_{g0} + 
		\beta_{g1}\T
		\begin{bmatrix}
			A_{ij} \\ \bA_\eij^s
		\end{bmatrix}
		+ \beta_{g2}\T 
		\begin{bmatrix}
			\bX_{ij} \\ \bX_\eij^s
		\end{bmatrix} \ .
	\end{align*}

Let  $\widehat{\paraT}= \big( \estparaPS{}\T , \estparaOR{}\T\big)\T$ be an estimate of $\paraT^* = (\paraPS^{*,\mathrm{\scriptscriptstyle T}},\paraOR^{*,\mathrm{\scriptscriptstyle T}})\T$ in $\modelig$ and $\modelie$.
For $\oT^* \in \Paraset$, let $\widehat{\oT} = \bv \T (\widehat{\bp}) \widehat{\bT}$ be the estimator of $\oT^*$ where $\widehat{\bp} = (\widehat{p}_1,\ldots, \widehat{p}_\NT)\T$, $ \widehat{p}_k = \NC_k / \NC$,
and 
$\widehat{\bT}$ 
is the solution to the equation $0 = \sum_{i=1}^\NC \ee_\uT ( \widehat{\bT} , \widehat{\paraT})$. Here, $\ee_\uT(\bT , \widehat{\paraT} ) = \big( \ee_{\uT,1} ( \uT_1, \widehat{\paraT} ) , \ldots , \ee_{\uT,\NT} ( \uT_\NT, \widehat{\paraT} ) \big) \T$ and $\ee_{\uT,k} (\uT_k, \widehat{\paraT} ) 
= \ind(\type_i = k) \big\{ \phi_k \big( \bO_i, e(\cdot \con \estparaPS{}), \OR(\cdot \con \estparaOR{}) \big) - \uT_k \big\}
$ where $\phi_k \big( \bO_i, e(\cdot \con \estparaPS{}), \OR(\cdot \con \estparaOR{}) \big)$ is obtained by plugging in parametrically estimated $e$ and $\OR$ in \eqref{eq-34001}, i.e. 
\begin{align*}
	& \phi_k \big( \bO_i, e(\cdot \con \estparaPS{}), \OR(\cdot \con \estparaOR{}) \big)
	\\
	& = 
	\ind( \type_i = k) 
		\bigg[ \frac{ \bw_k \T ( \bA_i ,\bX_i) \big\{ \bY_i - \POR(\bA_i , \bX_i , k \con \estparaOR{}) \big\} }{ \Pe (\bA_i \cond \bX_i , k \con \estparaPS{} )} 
		+ 
		\hspace*{-0.3cm}
		\sum_{\ba_i \in \zosets(\NI_k) } 
		\hspace*{-0.3cm}
		\bw_k \T ( \ba_i ,\bX_i) \POR(\ba_i , \bX_i , k \con \estparaOR{}) \bigg]
		\ .
\end{align*}
Theorem \ref{thm:ParaPopEst} presents asymptotic properties of $\widehat{\oT}$ under mild regularity conditions on the estimated model parameters $\widehat{\paraT}$; these are typical for semiparametric estimators of the propensity score or the outcome model \citep{Vaart1998}. 
\begin{theorem}[Local Efficiency] 						\label{thm:ParaPopEst}
	Suppose Assumption \ref{Assump:VC} and conditions (R1)--(R4) in the supplementary material hold. 
	Let $\paraT^\dagger$ be the probability limit of $\widehat{\paraT}$. 	Then, under model $\modelie \cup \modelig$, we have  $
	\NC^{1/2} \big( \widehat{\oT} - \oT^* \big) 
	= \NC^{-1/2} \sum_{i=1}^\NC \varphi( \oT^* , \paraT^\dagger)  + o_P(1) $ where
		\begin{align*}
			\varphi( \oT^* , \paraT^\dagger) 
			&= \sum_{k=1}^\NT \bigg[ v_k(p_k^*) \varphi_k( \uT_k^* , \paraT^\dagger)
			+ \Big\{ \ind(\type_i=k) -p_k^* \Big\} \frac{\partial v_k(p_k^*) }{\partial p_k }  \uT_k^* \bigg]  \ .
			\end{align*}
			Here, $\varphi_k( \uT_k^* , \paraT^\dagger) = \ind (\type_i = k) \big\{ \phi_k \big(\bO_i, e(\cdot \con \paraPS^\dagger), \OR(\cdot \con \paraOR^\dagger) \big) - \uT_k^* \big\} / p_k^*$ where $\phi_k \big(\bO_i, e(\cdot \con \paraPS^\dagger), \OR(\cdot \con \paraOR^\dagger) \big)$ is obtained from equation \eqref{eq-34001} by plugging in $e (\cdot \con \paraPS^\dagger) \in \modelie$ and $\OR (\cdot \con \paraOR^\dagger) \in \modelig$. Also, $\widehat{\oT}$ is locally efficient under $\modelie \cap \modelig$. 
\end{theorem}
Theorem \ref{thm:ParaPopEst} states that $\widehat{\oT}$ 
is a consistent estimator of $\oT^*$ so long as either the propensity score or the outcome regression is correctly modeled by the investigator. For example, consider the following efficient estimators for the direct and indirect effects:
\begin{align}																	\label{eq-32005}
	&
	\hspace*{-0.1cm}
	\widehat{\oT}^{\DE} (\alpha)
\hspace*{-0.05cm}
	=
	\hspace*{-0.05cm}
	\frac{1}{\NC} \sum_{k=1}^\NT  \sum_{i: \type_i = k}  
	\hspace*{-0.1cm}
	 \big\{ \widehat{\psi}_k ( 1 ,  \alpha ) - \widehat{\psi}_k (0 ,\alpha )  \big\}
	\ , \
	\widehat{\oT}^{\IE} (\alpha, \alpha')
\hspace*{-0.05cm}
	=\hspace*{-0.05cm}
	\frac{1}{\NC} \sum_{k=1}^\NT \sum_{i: \type_i = k} 
	\hspace*{-0.1cm}
	\big\{ \widehat{\psi}_k ( 0 ,  \alpha )  - \widehat{\psi}_k ( 0 ,  \alpha' )  \big\}
\end{align} 
where
	\begin{align*}
	& \widehat{\psi}_k ( a , \alpha ) 
	= \frac{1}{\NI_k} \sum_{j=1}^{\NI_k} 
		\sum_{ \substack{ \ba_i \in \zosets(\NI_k) \\ {\rm s.t. }a_{ij} = a} } \Bigg[ \frac{ \ind ( \bA_i = \ba_i ) \big\{ Y_{ij} - \indOR_j( \ba_i , \bX_i , k \con \estparaOR{} ) \big\}  }{ e(\ba_i \cond \bX_i , k \con \estparaPS{} )  } 
		+ 
	 \indOR_j ( \ba_i , \bX_i, k\con \estparaOR{} ) \Bigg] \pi ( \ba_\eij \con \alpha )
	 \ .	
	 \nonumber
\end{align*}
Here, $\indOR_j$ is $j$th component of $\OR$, i.e. the outcome regression of individual $j$. If the propensity score $e^*$ is known, say because the data was generated from a network randomized experiment \citep{HH2008}, the estimators in \eqref{eq-32005} will be consistent. If, in addition, the outcome regression $g^*$ is correctly specified, the estimators will be locally efficient. Critically, the existing bias-corrected doubly robust estimator of \citet{Liu2019} for the direct and indirect effects is asymptotically equivalent to \eqref{eq-32005} and hence efficient; this resolves a long-standing question on optimal semiparametric estimation direct and spillover effects under partial interference.
\begin{corollary}[Efficiency of the Estimator of \citet{Liu2019}]								\label{thm:Liu}
	Suppose that 
	(i) estimators of $g$ and $e$ in Section \ref{sec:Mestimation-general} are the same as those used in \citet{Liu2019} and (ii) both the propensity score and the outcome model are correctly specified, i.e. $\modelie \cap \modelig$. Then, the bias-corrected doubly robust estimator of \citet{Liu2019} is locally efficient under $\modelie \cap \modelig$.
\end{corollary}
We remark that our results can also be used to derive efficient estimators of overall and total effects in \citet{Liu2019}. Also, Section \ref{sec:LocEffsupp} of the supplementary material numerically illustrates Theorem \ref{thm:ParaPopEst}. Broadly speaking, Theorem \ref{thm:ParaPopEst} and the bias-corrected estimator of \citet{Liu2019} can be seen as the partial interference analog of the well-known result on the efficiency and double-robustness of the augmented inverse probability-weighted estimator under no interference \citep{Robins1994,Scharfstein1999}. 

We end the section by briefly summarizing two properties related to adaptive estimation under partial interference; see Section \ref{sec:detail5} of the supplementary material for details. First, the doubly robust estimators in Theorem \ref{thm:ParaPopEst} can still achieve the best possible variance regardless of the knowledge of the propensity score. Second, if the investigator uses estimators that account for interference, but the true data generating model has no interference, the doubly robust estimators in Theorem \ref{thm:ParaPopEst} are consistent, but generally inefficient. In other words, the estimators do not adapt to the knowledge about exposure mappings. This suggests that potential side-information about exposure mappings may play a critical role, both in terms of consistency and efficiency of estimators. Also, these adaptation properties are similar to those without interference where the augmented inverse probability-weighted estimator adapts to the knowledge of the propensity score, but does not adapt to the knowledge of the outcome model \citep{Scharfstein1999_Rejoinder}; under partial interference, the outcome model encodes knowledge about the exposure mapping.

\section{Some Examples of Efficient Estimators In Practice} 						\label{sec:Estimator}
\subsection{Parametric Case: Generalized Mixed Effect Models With Linear Summary of Peers' Covariates}								\label{sec:GLMM}
A popular class of estimators used in studies under partial interference is based on generalized mixed effect models where the peers' data are summarized with a linear statistic \citep{PH2014, Liu2016, Liu2019, Barkley2020}. Specifically, consider the following models for $g$ and $e$ to estimate the direct and indirect effects.
\begin{align}
\label{eq-50003}		
	& \text{logit} \big\{ {\rm pr}( A_{ij} = 1 \cond \bX_i = \bx_i , \type_i = k , b_i \con \paraPS ) \big\} 
	= \bigg[ 1, \bx_{ij} \T , \sum_{\ell \neq j} \bx_{i\ell} \T  \bigg] \paraPStype{k} + b_i , 
	\end{align}
	and
	\begin{align}	
	& Y_{ij} 
	= \sum_{k=1}^\NT \ind(\type_i = k) \hspace*{-0.3cm} \sum_{\ba_i \in \zosets(\NI_k)} \hspace*{-0.3cm} \ind (\bA_i = \ba_i) 
	\bigg[ 1 , a_{ij} , \sum_{\ell \neq j} a_{j\ell} , \bx_{ij}\T ,  \sum_{\ell \neq j} \bx_{i\ell} \T \bigg] \paraORtype{k}
	+  \xi_i + \epsilon_{ij} \ .
	\label{eq-50002}
	\end{align}
	 where $b_i \cond ( \bX_i , \type_i = k ) \sim N(0, \lambda_k^{-1}) $, $\xi_i \cond ( \bA_i , \bX_i , \type_i =k ) \sim N(0, \rho_k)$, and $	\epsilon_{ij} \cond ( \bA_i, \bX_i, \type_i=k ) \sim N(0, \eta_k^{-1})$. 
	Here, $\paraORtype{k}$ is the $k$th block entry of $\paraOR = \big( \paraORtype{1}\T, \ldots, \paraORtype{\NT}\T \big)\T$ and parametrizes the outcome regression for cluster type $k$. Similarly, $\paraPStype{k}$ is the $k$th block entry of $\paraPS = \big( \paraPStype{1} \T, \ldots, \paraPStype{\NT} \T \big)\T$ and parametrizes the propensity score for cluster type $k$. The terms $\xi_i$ and $b_i$ are random effect terms and introduce dependence between observations within cluster $i$. The term $\epsilon_{ij}$ is the unit-level error term. $\xi_i$ and $\epsilon_{ij}$ are assumed to be conditionally independent given $(\bA_i,\bX_i,\type_i)$. 
Overall, model \eqref{eq-50003} and \eqref{eq-50002} roughly state that the treatment and the outcome of unit $j$ depend on the total number of peers treated as well as peers' covariates.  

Despite its popularity, to the best of our knowledge, prior works have not formally laid out the exact conditions demonstrating that they are efficient. In particular, the prior works \citep{PH2014, Liu2016, Liu2019, Barkley2020} have shown that they are asymptotically normal, but they did not show that the asymptotic variance achieves the local semiparametric efficiency bound. The following theorem rectifies this by show that these estimators can be locally efficient under mild and interpretable assumptions. 
\begin{corollary}[Local Efficiency of Mixed Effects Models]								\label{thm:ParaRand}
Suppose Assumption \ref{Assump:VC} holds. Furthermore, for each $k$, suppose that (a) $(\paraPStype{k},\lambda_k)$ is globally identifiable; (b) $\EXP\big\{ \| \bX_i \|_2^3 \cond \type_i = k \big\}$ is finite; (c) the Fisher information of \eqref{eq-50003} is positive definite and $\bX_i$ is non-degenerate. Then, $\widehat{\oT}$ is asymptotically normal so long as $\Pe$ or $\POR$, but not necessarily both, is correctly specified. Also, $\widehat{\oT}$ is locally efficient if both models are correctly specified.
\end{corollary}
For the interested reader, while the proof generally follows from the theory of maximum likelihood, some under-appreciated technical difficulties arise, especially dealing with a non-separable logistic mixed effects model where the distribution of the unobserved random effect is spherical.

\subsection{Nonparametric Case: Cross-Fitting Estimators Under Partial Interference} \label{sec:NPOR}
In this section, we propose an extension of cross-fitting under no interference to partial interference. Briefly, cross-fitting was originally developed by \citet{Victor2018} under no interference as a way to utilize off-the-shelf machine learning methods to estimate treatment effects and avoid Donsker conditions on nuisance parameters, say the outcome or the propensity score model. In the exposition below, we discuss a simple extension of cross-fitting for dependent data following model \eqref{eq:NPmodel}. 

A key step in the extension is to let $\INDEX_1$ and $\INDEX_2$ be the disjoint partitions of the sample where both partitions contain all cluster types and the proportion of each cluster type in both partitions are nearly identical; see Section \ref{sec:DetailSec4} of the supplementary materials for a simple algorithm to achieve such partitions. For each partition $\ell = 1,2$, let $\widetilde{\OR}_{(-\ell)}$ and $\widetilde{e}_{(-\ell)}$ be the nonparametrically estimated outcome regression model and the propensity score, respectively, using subsample $\INDEX_\ell^c = \INDEX_{3-\ell}$. We evaluate $\widetilde{\OR}_{(-\ell)}$ and  $\widetilde{e}_{(-\ell)}$ on the samples in $\INDEX_\ell$. Then, similar to the original cross-fitting estimator, we change the role of the partitions to fully use the observed data. We plug in the evaluated outcome and the propensity score into \eqref{eq-34001} and obtain the estimator of $\uT_k^*$ denoted by $\widetilde{\uT}_k = \sum_{\ell = 1}^2 \sum_{ i \in \INDEX_\ell} \phi_k( \bO_i, \widetilde{e}_{(-\ell)}, \widetilde{\OR}_{(-\ell)}) / \NC_k$. We then obtain the corresponding estimator of $\oT^*$ denoted by $\widetilde{\oT} = \bv\T (\widehat{\bp}) \widetilde{\bT}$ where $\widetilde{\bT} = (\widetilde{\uT}_1,\ldots,\widetilde{\uT}_\NT)\T$.  Corollary \ref{thm:MLOR} and \ref{thm:MLOR_KnownPS} describe the properties of $\widetilde{\oT}$ under different assumptions about the propensity score and the outcome model.

\begin{corollary}[Global Efficiency of Cross-Fitting Estimator $\widetilde{\oT}$ Under Partial Interference]						\label{thm:MLOR}
	Let $P_k$ be the probability law of $\bX_i \cond \type_i = k$ and suppose Assumption 1 holds. Additionally, suppose we have the following conditions for any $\ell \in \{ 1, 2 \}$, $\ba \in \zosets(\NI_k)$, and $k \in \{1,\ldots,\NT\}$: 
	\begin{itemize}
		\item[(a)] (Moments and boundedness of nuisance functions): For all $\bx \in \mathcal{X}(k)$, there exist constants $\widetilde{C} \in (0,\infty)$ and $\widetilde{c} \in (0,1)$ satisfying $\| g^*(\ba, \bx, k) \|_2 \leq \widetilde{C}$, $\| \widetilde{g}_{(-\ell)}(\ba, \bx, k) \|_2 \leq \widetilde{C}$,  $\| \Sigma^*(\ba, \bx, k) \|_2 \leq \widetilde{C}$, and $\widetilde{e}_{(-\ell)}(\ba \cond \bx, k) \in [ \widetilde{c}, 1-\widetilde{c} ]$.
		\item[(b)] (Convergence rate of estimated nuisance functions): $\widetilde{\OR}_{(-\ell)}$ and $\widetilde{e}_{(-\ell)}$ satisfy
			\begin{align}						
		&
		\int \big\| \OR^* (\ba, x,k) - \widetilde{\OR}_{(-\ell)} (\ba, x,k) \big\|_2^2 \, dP_{k}(x) = O_P(r_{g,\NC}^2) \ ,
		\label{assp:MLOR}
		\\
		&
		\int \big| e^* (\ba \cond x,k) - \widetilde{e}_{(-\ell)} (\ba \cond x,k) \big|^2 \, dP_{k}(x) = O_P(r_{e,\NC}^2)
		\label{assp:MLOR2}
	\end{align}
	where $r_{g,\NC} = o(1)$, $r_{e,\NC} = o(1)$, and $r_{g,\NC} r_{e,\NC} = o(\NC^{-1/2})$, respectively, as $\NC \rightarrow \infty$. 
	\end{itemize}
	Then, $\NC^{1/2}  \big( \widetilde{\oT} - \oT^* \big)$ weakly converges to $N \big( 0, \VAR \big\{  \eifuv (\oT^*) \big\} \big)$ as $\NC \rightarrow \infty$ where $\VAR \big\{  \eifuv (\oT^*) \big\} $ is the global efficiency bound presented in Theorem \ref{thm:EIFVC}. Moreover, let $\widetilde{\sigma}^2$ be 
	\begin{align*}
		\widetilde{\sigma}^2
		=
		\frac{1}{\NC} \sum_{\ell=1}^2 \sum_{i \in \INDEX_\ell} \Bigg[ \sum_{k=1}^\NT
		\bigg[
		\frac{\ind ( \type_i = k) v_k(\widehat{p}_k) }{\widehat{p}_k} \Big\{  \phi_k(\bO_i, \widetilde{e}_{(-\ell)} ,\widetilde{\OR}_{(-\ell)} ) - \widetilde{\uT}_k \Big\} + \Big\{  \ind  (\type_i = k) - \widehat{p}_k \Big\} \frac{\partial v_k (\widehat{p}_k)}{\partial p_k} \widetilde{\uT}_k
		\bigg]
		 \Bigg]^2 \ .
	\end{align*}	
	Then, $\widetilde{\sigma}^2$ is consistent for $\VAR \big\{  \eifuv (\oT^*) \big\}$, i.e. $\widetilde{\sigma}^2$ converges to $ \VAR \big\{  \eifuv (\oT^*) \big\} $ in probability as $\NC \rightarrow \infty$.
\end{corollary}
\begin{corollary}[$\NC^{1/2}$-Consistency of $\widetilde{\oT}$ Under Randomized Experiments]			\label{thm:MLOR_KnownPS}
Furthermore, if $\widetilde{e}_{(-\ell)} = e^*$ and equation \eqref{assp:MLOR} holds with $\OR'$ instead of $\OR^*$ where $\|\OR' (\ba, \bx, k) \|_2 \leq \widetilde{C}$, then $\NC^{1/2}  \big( \widetilde{\tau} - \tau^* \big)$ weakly converges to $N \big( 0, \sigma^2 \big)$ and $\sigma^2$ can be consistently estimated with $\widetilde{\sigma}^2$.
\end{corollary}

In words, Corollary \ref{thm:MLOR} means that $\widetilde{\oT}$ is asymptotically normal and globally efficient for $\oT^*$ so long as the outcome regression model and the propensity score satisfy some regularity conditions. Specifically, condition (a) in Corollary \ref{thm:MLOR} states the true/estimated outcome regressions and the true conditional variance of the outcome are uniformly bounded and the estimated propensity score satisfies the positivity/overlap condition. Condition (b) in Corollary \ref{thm:MLOR} states that both the estimated propensity score $\widetilde{e}_{(-\ell)}$ and the estimate outcome regression $\widetilde{g}_{(-\ell)}$ are consistently estimators where the product of their convergence rates is $o_P(\NC^{-1/2})$.  In a randomized experiment such as the cash transfer program study in Section \ref{sec:application}, condition (b) is automatically satisfied if the investigator uses the propensity score from the study design and the estimated outcome regression model converges to the true outcome model at any rate. In an observational study, condition (b) is satisfied if both the estimated outcome regression model and the estimated propensity score are converging to their true counterparts with rates faster than $o_P(\NC^{-1/4})$. Corollary \ref{thm:MLOR_KnownPS} is a special case of Corollary \ref{thm:MLOR} where $\widetilde{\oT}$ is asymptotically normal so long as the propensity score is known, say in a randomized experiment, and $\widetilde{\OR}_{(-\ell)}$ can be inconsistent, i.e. $\OR'\neq\OR^*$. In particular, if $\widetilde{\OR}$ is inconsistent, the standard error of $\widetilde{\oT}$ will be larger than the efficiency bound.

We remark that for some estimands, notably the direct and the indirect effects, the variance expression simplifies. Specifically, consider the estimators $\widetilde{\oT}^\DE (\alpha)$ and $\widetilde{\oT}^\IE (\alpha, \alpha')$, which have the same form as $\widehat{\oT}^\DE(\alpha)$ and $\widehat{\oT}^\IE(\alpha, \alpha')$ in equation \eqref{eq-32005} except $\widehat{\psi}$ in that equation is replaced by 
\begin{align*}
	& \widetilde{\psi}_k ( a , \alpha ) 
	= \frac{1}{\NI_k} \sum_{j=1}^{\NI_k} 
		\sum_{ \substack{ \ba_i \in \zosets(\NI_k) \\ {\rm s.t. }a_{ij} = a} } \Bigg[ \frac{ \ind ( \bA_i = \ba_i ) \big\{ Y_{ij} - \widetilde{\indOR}_{(-\ell)}( \ba_i , \bX_i , k ) \big\}  }{ \widetilde{e}_{(-\ell)} (\ba_i \cond \bX_i , k )  } 
		+ 
	 \widetilde{ \indOR}_{j,(-\ell)} ( \ba_i , \bX_i, k ) \Bigg] \pi ( \ba_\eij \con \alpha )
	 \ .	
	 \nonumber
\end{align*}
Then, 
the consistent variance estimator $\widetilde{\sigma}^2$ reduces to the usual mean squared deviation, i.e. $\widetilde{\sigma}^{\DE,2} = \sum_{k=1}^\NT \sum_{\ell=1}^2 \sum_{ i \in \INDEX_\ell} \ind (\type_i = k) 
	\big\{
		\widetilde{\psi}_k(1, \alpha) - \widetilde{\psi}_k(0, \alpha)  - \widetilde{\oT}^\DE(\alpha)
	\big\}^2 / \NC$ and $\widetilde{\sigma}^{\IE,2} = \sum_{k=1}^\NT \sum_{\ell=1}^2 \sum_{ i \in \INDEX_\ell} \ind (\type_i = k) 
	\big\{
		\widetilde{\psi}_k(0, \alpha) - \widetilde{\psi}_k(0, \alpha')  - \widetilde{\oT}^\IE(\alpha, \alpha')
	\big\}^2 / \NC$.

Finally, equations \eqref{assp:MLOR} and \eqref{assp:MLOR2} are often stated in the literature on cross-fitting and impose properties that a nonparametric estimator must satisfy. While such estimators do exist under no interference settings, a natural question arises on whether they can be satisfied for dependent data. Here, we show one way to satisfy the conditions under partial interference settings by modifying existing nonparametric regression estimators initially designed for independent data. The discussion focuses on the outcome regression model, but a similar principle could be used to train the propensity score.

Formally, for models in $\modeliv$, suppose the conditional moments of $Y_{ij}$ given $(\bA_i, \bX_i, \type_i)$ satisfy
\begin{align} \label{eq:equalMeanVar}
\EXP\big(Y_{ij} \cond \bA_i , \bX_i , \type_i = k \big) &= \mu_k \big( A_{ij}, \bA_\eij, \bX_{ij}, \bX_\eij \big), \\
\VAR\big( Y_{ij} \cond \bA_i , \bX_i , \type_i = k \big) &= \sigma_k^2 \big( A_{ij}, \bA_\eij, \bX_{ij}, \bX_\eij \big). \nonumber
\end{align}
In words, equation \eqref{eq:equalMeanVar} states that the mean and variance of each unit's outcome $Y_{ij}$ depend on her treatment and covariates $(A_{ij},X_{ij})$ as well as her peers' treatment and covariates $(\bA_\eij, \bX_\eij)$. A notable violation of \eqref{eq:equalMeanVar} is when a unit's conditional mean and variance vary across $j$, say in an autoregressive model based on lags of $Y_{ij}$. But, the assumption still allows for heteroskedastic variance as defined in \eqref{eq:equalMeanVar}. Then, for each element of $ \widetilde{\OR}_{(-\ell)} (\ba, x,k)$, denoted as $ \widetilde{\OR}_{j,(-\ell)} (\ba, x,k) $, consider a smoothed kernel regression \citep[Chapter 4.4]{LiRacine2007} under mixed data with a mixed kernel $\mathcal{K}_{h_c,h_d}$ and bandwidths $h_c,h_d>0$ trained in subsample $\mathcal{I}_\ell^c$.
\begin{align*}
&\widetilde{\OR}_{j,(-\ell)}^{\rm NW}(a_{ij}, \ba_{\eij}, x_{ij},\bx_{\eij},k) 
\\
&
= \frac{\sum_{i=1}^\NC \sum_{j=1}^{\NI_k} Y_{ij} \mathcal{K}_{h_c,h_d}(a_{i(-j)} - \bA_\eij, x_{ij}-X_{ij}, \bx_{\eij} - X_{i(-j)}) \ind(A_{ij} = a_{ij},\type_i = k)} { \sum_{i=1}^\NC \sum_{j=1}^{\NI_k}  \mathcal{K}_{h_c,h_d}(a_{i(-j)} - \bA_\eij, x_{ij}-X_{ij}, \bx_{\eij} - X_{i(-j)}) \ind(A_{ij} = a_{ij},\type_i = k)}.
\end{align*}
At a high level, the bandwidth $h_d$ deals with smoothing over the discrete variables, i.e. $A_{\eij}$ and discrete $\bX_i$, and the bandwidth $h_c$ deals with the continuous variables, i.e. continuous $\bX_i$; see Section \ref{proof:NWkernel} of the supplementary materials for the exact construction of $\mathcal{K}$. 
Corollary \ref{thm:Kernel} shows that under the usual regularity conditions for kernel regression  \citep{Stone1982,Hall1984}, the estimator $\widetilde{g}_{(-\ell)}^{\rm NW}$ 
satisfies \eqref{assp:MLOR}.
\begin{corollary} 					\label{thm:Kernel}
Suppose equation \eqref{eq:equalMeanVar} and regularity assumptions in Section \ref{proof:NWkernel} of the supplementary materials hold. 
If $h_c = O(\NC^{-1/(4+p)})$ and $h_d = O(\NC^{-2/(4+p)})$ where $p$ is the number of continuous components, 
the estimator $\widetilde{g}_{(-\ell)}^{\rm NW}$ using $\widetilde{\OR}_{j,(-\ell)}^{\rm NW}$ satisfies equation \eqref{assp:MLOR} with rate $r_{g,\NC} = \NC^{-2/(4+p)}$.
\end{corollary}
Similar to Section \ref{sec:GLMM}, if the investigator believes that the peers' treatment vector can be collapsed into a scalar value, say the sum of the treated number of peers, it may improve the estimation performance of the kernel estimator for small samples. More generally, if information about the network structure is available, it can inform the choice of the kernel, say using an ordered discrete kernel of \citet{AA1976} for the peers' treatment vector; see \citet{LiRacine2007} for a textbook discussion on choosing kernels. Finally, using similar proof techniques in Corollary \ref{thm:Kernel}, under \eqref{eq:equalMeanVar}, other nonparametric estimators initially designed for independent and identically distributed data could be adapted to satisfy \eqref{assp:MLOR}.


\section{Application: Network Effects of Cash Transfer Programs in Colombia}						\label{sec:application}

We apply our method in Section \ref{sec:NPOR} to a randomized experiment to study the effect of conditional cash transfer program, i.e. treatment, on students' attendance rate \citep{BO2011}. Briefly, the experiment was conducted in two regions of Bogota, Columbia: San Cristobal and Suba. For each region, investigators recruited households with school children, ranging from 1 to 5 school children, and within each household, school children were randomized to enroll into the cash transfer program via stratified randomization. Specifically,  investigators defined strata using locality (San Cristobal/Suba), type of school (public/private), gender, and grade level, and students in each stratum were randomly chosen to be enrolled into the program; see Section \ref{sec:detailapplication} of the supplementary materials for additional details on the distribution of treatment by household size. It is possible to have more than one children be treated in a household and within each stratum, each child was equally likely to be treated. Also, because the treatment assignment is known and the outcome is bounded in the unit interval, conditions (a) and (b) in Corollaries \ref{thm:MLOR} and \ref{thm:MLOR_KnownPS} are satisfied and $\NC^{1/2}$-inference of the proposed estimator remains valid. That is, if our nonparametrically estimated outcome model is correct, Corollary \ref{thm:MLOR} states that our estimate's standard error is the smallest possible among all regular estimators, i.e. semiparametrically efficient. Otherwise, Corollary \ref{thm:MLOR_KnownPS} states that the estimate is still asymptotically normal and the Wald test based on it has the asymptotically correct size.

 \citet{BO2011} was interested in both the direct effect as well as the spillover effect on the enrolled student's siblings in the same household. In particular, enrolled students received cash subsidies if they attended school at least 80\% of the time in a month. Additionally, due to peer pressure to attend school, enrolling one student in the program could increase, or decrease, the attendance rate of his/her sibling in the same household. 
Our goal is to use the proposed estimators to analyze these direct and spillover effects in the two regions of Bogota, Columbia.

Formally, for student $j$ in household $i$, let $Y_{ij} \in [0,1]$ be the self-reported attendance rate, $A_{ij}$ be equal to 1 if student $j$ was enrolled into the program and 0 otherwise, and $\bX_{ij}$ be the following pre-treatment covariates: student's age, student's grade, student's gender, household head's age, indicator of single parent household, household size, household's poverty score, household's income status, locality, and number of students in household participated in the lottery. We restrict the sample to households with complete data and to households having more than one child. We define three cluster types based on the two regions and the size of the households: households in Suba with two students, households in San Cristobal with two students, and households in San Cristobal with three or more students. We omit the two three-person households from Suba because there are only two such households in the dataset. In total, we analyze 1,010 households containing 2,129 students. 

We denote the treatment allocation strategy for San Cristobal and Suba by $\alpha_{\rm SC}$ and $\alpha_{\rm Su}$, respectively, and let $\balpha=(\alpha_{\rm SC}, \alpha_{\rm Su})$. We also use the original treatment randomization probability as the treatment allocation strategy and denote it as $\alpha^* = (\alpha_{\rm SC}^*, \alpha_{\rm Su}^*) = (0.628,0.449)$. The target network estimands are the direct effect $\oT^\DE(\alpha)$ and the spillover effect $\oT^\IE(\alpha, \alpha^*)$. 
Here, the spillover effect measures the difference between the attendance rate of an untreated student when his/her sibling is treated under $\alpha$  and that under the original experiment $\alpha^*$. 

We estimate the effects using our nonparametric, efficient estimators in Section \ref{sec:NPOR}  and the inverse probability-weighted estimators of $\oT^\DE(\alpha)$ and $\oT^\IE(\alpha, \alpha^*)$ by \citet{Liu2016}. For our estimators, we use the study design's treatment randomization as the propensity score and we nonparametrically model the outcome regression model by using ensembles of multiple machine learning methods via the super learner algorithm \citep{SL2007, Polley2010}. 
As explanatory variables in the outcome regression, we use the student's treatment status and covariates, the treated proportion of his/her peers and the average of his peers' covariate, i.e. $(A_{ij}, \bX_{ij}, \overline{\bA}_\eij$, $\overline{\bX}_\eij)$. We remark that given the small range of the cluster size, the peers' treated proportion, i.e. $\overline{A}_\eij$, is more close to a discrete random variable rather than a continuous random variable; in our analysis, the treated proportion $\overline{A}_{\eij}$ takes on six values $\{0, 1/3,1/2, 2/3, 3/4,1\}$. To preserve this discrete nature in our analysis, we code the peers' treated proportion as discrete, dummy variables in the outcome regression model. Specifically, $\overline{A}_{\eij}$ is stratified into four strata of $\{\overline{A}_\eij = 0\}$, $\{ 0 < \overline{A}_\eij \leq 0.5 \}$, $\{ 0.5 < \overline{A}_\eij < 1 \}$, and $\{ \overline{A}_\eij=1\}$. This stratification pools 2 individuals with $\overline{A}_{\eij} = 1/3$ and 124 individuals with $\overline{A}_{\eij} = 1/2$ into the same stratum. It also pools 4 individuals with $\overline{A}_{\eij} = 3/4$ and 17 individuals with $\overline{A}_{\eij} = 2/3$ in the same stratum. 
Because the treatment was randomized, our estimators and the inverse probability-weighted estimators are consistent for the network treatment effects, but may have different variances. Section \ref{sec:detailapplication} of the supplementary materials contains additional discussions on estimating nuisance functions for this data.

\begin{figure}[!htb]
	\centering
	\vspace*{-0.2cm}
	\includegraphics[width=1\textwidth]{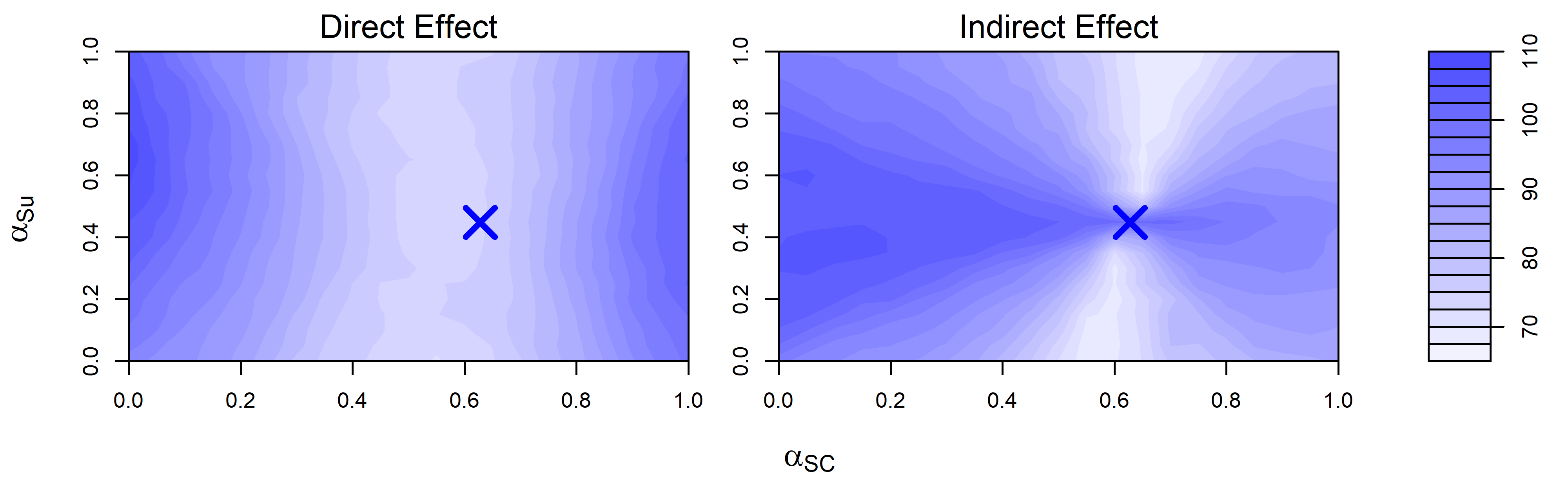}
	 \vspace*{-0.8cm}
   \caption{Relative efficiency between the proposed efficient estimator and the inverse probability-weighted estimator in \citet{Liu2016}. 
	The colors on the plot show the ratio of the variance of the efficient estimator to the variance of the inverse probability-weighted estimator, with darker colors indicating higher values of the ratio, or more improvements in efficiency for the proposed estimator. The left and right plots show the relative efficiencies for the direct and spillover effects, respectively. The $x$-axis varies the treatment allocation strategy $\alpha_{\rm SC}$ for San Cristobal and the $y$-axis varies the treatment allocation strategy $\alpha_{\rm Su}$ for Suba. The blue cross ({\color{blue}$\times$}) is the original treatment probability from the experiment, $\alpha^*=(0.628,0.449)$.}
   \label{Fig:NPOR2}
\end{figure}

Figure \ref{Fig:NPOR2} shows the relative efficiency of our estimators and the inverse probability-weighted estimators. Across all treatment allocation strategies $\alpha = (\alpha_{\rm SC},\alpha_{\rm Su})$, our estimator $\widetilde{\tau}$ is more efficient than the inverse probability-weighted estimator, with our estimator showing 68.46 to 109.26 times improvements in efficiency. This empirical result corroborates our theoretical result in Corollary \ref{thm:MLOR} on the semiparametric efficiency of our estimator. We also notice that for the direct effect, efficiency gain does not change when $\alpha_{\rm Su}$ varies. In contrast, for the spillover effect, the efficiency gain does change with both $\alpha_{\rm SU}$ and $\alpha_{\rm SC}$. 

\begin{figure}[!htb]
	\centering
	\includegraphics[width=1\textwidth]{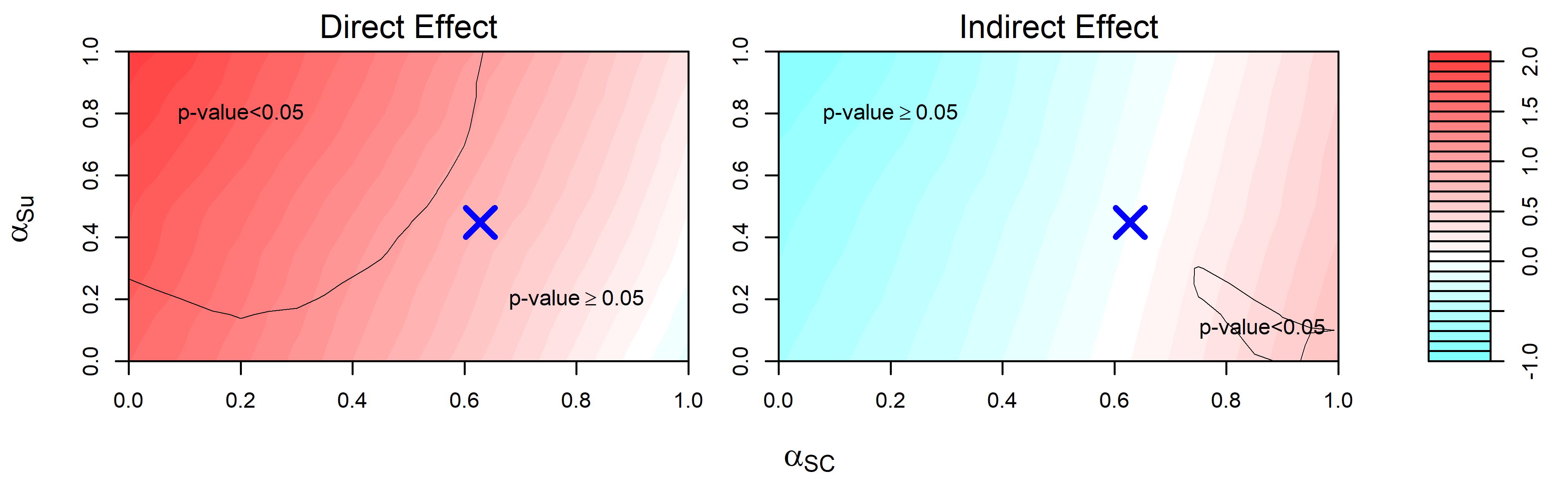}
	 \vspace*{-0.8cm}
   \caption{Effect estimates in percentage points using the proposed efficient estimator. The colors on the left and right plots show the magnitude and sign of the estimated effects, respectively. The $x$-axis varies the treatment allocation strategy $\alpha_{\rm SC}$ for San Cristobal and the $y$-axis varies the treatment allocation strategy $\alpha_{\rm Su}$ for Suba.  We also show the regions where the effects are significant at level 0.05. The blue cross ({\color{blue}$\times$}) is the original treatment probability from the experiment, $\alpha^*=(0.628,0.449)$.}
   \label{Fig:NPOR1}
\end{figure}

Figure \ref{Fig:NPOR1} shows the direct and spillover effect estimates based on the proposed efficient estimator. In general, the direct effect tends to be positive and significant for large $\alpha_{\rm Su}$. But, the spillover effect has a phase-transition behavior along $\alpha_{\rm SC}$ where the effect is negative for small $\alpha_{\rm SC}$ and positive
for large $\alpha_{\rm SC}$. Practically speaking, the analysis suggests that enrolling more students in San Cristobal induce a more stronger spillover effect towards their siblings compared to doing it in Suba. But, enrolling more students in Suba can yield a more positive direct effect compared to doing it in San Cristobal. Combined, San Cristobal may have stronger sibling spillover effects compared to Suba, potentially providing information about structuring conditional cash transfer programs to take advantage of the differences in how the treatment affects attendance rates in the two regions.

\section{Discussion}													\label{sec:disc}

This paper proposes a framework to derive globally and locally efficient influence functions of network treatment effects in $\Paraset$. Our results complement the rich set of results on efficient estimation of treatment effects without interference and establish that one of the estimator in \citet{Liu2019} is locally efficient. We also discuss results on adaptivity, notably that the efficiency is not affected by the knowledge of the propensity score, but can be affected by the knowledge of the interference pattern. Finally, we show other parametric and nonparametric estimation methods that, under some assumptions, can achieve the efficiency bound. Our empirical application corroborates our theoretical results, showing that our estimators have smaller standard errors than those by \citet{Liu2016}.

We take a moment to discuss the strengths and limitations of our framework using model \eqref{eq:NPmodel}, 
notably the restrictions on cluster size. While our framework is useful to characterize optimality of estimators in small cluster settings, say twins, households, classrooms, small villages or clinics, or certain settings in neuroimaging \citep{Luo2012}, it is not  appropriate in settings where the cluster size is large compared to the number of individuals per cluster, say states, i.e. 50 states/clusters, with many individuals per cluster, or hospital systems. In fact, if $\bO_i$ grows in dimension, \citet{Liu2014} showed that the limiting distributions of popular estimators under partial interference are no longer asymptotically normal. In this setting, we likely need a dimension-reducing assumption to make the dependence among units theoretically manageable, say by assuming the dependence is characterized by known summary functions of peers' data; see \citet{vvLaan2014}, \citet{Sofrygin2016}, and \citet{Ogburn2017arxiv} for examples.
A limitation of such works is that, as suggested in our result concerning adaptivity with respect to the interference pattern, efficiency depends on this summarizing function. That is, an estimator that was efficient under one summarizing function may no longer be efficient, and potentially inconsistent, under a different summarizing function. Ultimately, these limitations can be thought of as a cost for obtaining efficient estimators in large networks. In contrast, if the number of study units in a cluster is not large and thus model \eqref{eq:NPmodel} is plausible, we do not have to make assumptions about how peers influence each other in order to construct our estimators. 

Overall, no asymptotic framework for networks is uniformly superior and investigators should use estimators based on the data at hand. In particular, with the current theory, if investigators are working with large online networks and they know how peers might affect each other, the estimators by \citet{Sofrygin2016} and \citet{Ogburn2017arxiv} show promise. On the other hand, if investigators are working with studies involving twins, households, small villages, or neuroimaging, and they have insufficient knowledge about how units affect each other, our framework and the proposed optimal estimators show promise.

\newpage

\appendix

\section*{Supplementary Material}

This document contains supplementary materials for ``Efficient Semiparametric Estimation of Network Treatment Effects Under Partial Interference.'' 
	Section \ref{sec:appendix1} presents additional results related to the main paper.
	Section \ref{sec:appendix2} proves the theorems stated in the main paper. 
	Section \ref{sec:appendix3} proves the theorems and lemmas stated in Section \ref{sec:appendix1}.

	\section{Additional Results}										\label{sec:appendix1}

In this section, we introduce additional results which are related to the main paper.

\subsection{Comments about Model \eqref{eq:NPmodel}}						\label{sec:appendix1-NPmodel}

In relation to others works, model \eqref{eq:NPmodel} is a generalization of Example 1 of \citet{Bickel2001} where $P^{(k)}$ in their notation is equivalent to our $P(\bO_i \cond \type_i = k)$ and Example 2 of \citet{McNeney2000} where we allow for different densities. Also, \eqref{eq:NPmodel} complements \citet{Sofrygin2016} who worked under general interference and as such, had to assume (a) a single, known summary function of peers' data and (b) conditional on the summary function, a study unit's data $O_{ij}$ are independent and identically distributed; see their assumptions (A2), (A3), and (B1). In our setup, we leverage the partial interference structure where we have independence across clusters and as such, allow for different non-parametric functionals to model dependencies within a cluster.
Additionally, while \eqref{eq:NPmodel} is a type of mixture model, the goal in our paper is not to identify or estimate unknown mixture labels $\type_i$ typical in mixture modeling. Instead, we use $\type_i$ as a technical device to embed studies under partial interference into \eqref{eq:NPmodel} and hence $\type_i$ is known by construction; see the next paragraph for some examples. Finally, when $\NT=1$ and there is only one study unit in a cluster, \eqref{eq:NPmodel} reduces to the usual independent and identically distributed setting without interference; in fact, as $\NT > 1$ and there are at least two study unit per cluster, our asymptotic framework has both independent and identically distributed and non-independent and non-identically distributed components. Specifically, we observe independent and identically distributed copies of $\bO_i$ within each cluster type, but elements of the vector $\bO_i$ may exhibit arbitrary dependence structure and each cluster type may have different densities. 
As we'll see below, this blend of independent and identically distributed and non-independent and identically distributed structure leads to efficient influence functions that look similar, but not identical, to efficient influence functions under independence.
  
As mentioned in the main manuscript, many, but not all, studies under partial interference can be modeled by \eqref{eq:NPmodel}. For example, when the number of study units within each cluster is fixed, such as studies involving dyads  \citep{Rosenbaum2007, Elwert2008, Nickerson2008, Lu2015, Baird2018}, or when the number of study units is bounded, as in households surveys or in neuroimaging studies \citep{HongKong2009, Luo2012}, we can define $\type_i$ by the size of each cluster and model these studies as data generated from \eqref{eq:NPmodel}; see  \citet{Barkley2020} and \citet{Basse2018} for other examples of organizing clusters by size. More broadly, so long as clusters are organized by finite number of $\type_i$s, studies under partial interference can be reasonably modeled by \eqref{eq:NPmodel}. However, if the number of study units in a cluster is growing and there is no restrictions on $P(\bO_i \cond \type_i)$, our setting does not apply. In such settings, one likely requires assumptions on $P(\bO_i \cond \type_i)$ to reduce its dimension, say by assuming the aforementioned summary function to represent peers' data. If we also make such assumptions, \eqref{eq:NPmodel} can be used. In general, in settings where the size of the cluster grows, assumptions are likely necessary to deal with the curse of dimensionality and, perhaps more importantly, to define a reasonable efficiency or variance-based criterion.

\subsection{Details of Section \ref{sec:setup0}}						\label{sec:AE}

We show that the direct effect $ \tDE(\alpha)= \EXP \big\{ \overline{Y}_i ( 1 \con \alpha)   - \overline{Y}_i ( 0 \con \alpha) \big\}$ and the indirect effect $  \tIE(\alpha, \alpha')= \EXP \big\{ \overline{Y}_i ( 1 \con \alpha)   - \overline{Y}_i ( 0 \con \alpha) \big\}$, which are counterpart of the direct and indirect effects defined in \citet{HH2008} under an infinite population framework, belong to ${\rm T}$. If we choose the weights $\bw_k$ as
	\begin{align}						\label{example-DE2}
		&
		\bw_k( \ba_i \con \alpha) 
		=
		\frac{1}{\NI_k} 
		\begin{bmatrix}
			\big\{ \ind \big( a_{i1}=1 ) - \ind \big( a_{i1}=0 ) \big\} 
			\pi (\ba_{i(-1)} \con \alpha)
			\\ 
			\vdots
			\\
			\big\{ \ind \big( a_{ij}=1 ) - \ind \big( a_{ij}=0 ) \big\} 
			\pi (\ba_{i(-j)} \con \alpha)
			\\ 
			\vdots		
			\\
			\big\{ \ind \big( a_{i\NI_k}=1 ) - \ind \big( a_{i\NI_k}=0 ) \big\}
			\pi (\ba_{i(-\NI_k)} \con \alpha)
		\end{bmatrix} \ ,
	\end{align}
this leads to the following for $\theta_k$:
	\begin{align*}
		\theta_k
		&
		= \frac{1}{\NI_k}
		\bigg[
		 \sum_{\ba_{i(-1)}}
		\EXP \Big[ \big\{ Y_{i1} (a_{i1} = 1, \ba_{i(-1)} ) - Y_{i1} (a_{i1} = 0, \ba_{i(-1)} ) \big\}  \pi (\ba_{i(-1)} \con \alpha) \, \Big| \, \type_i = k \Big] + \cdots
		\\
		& \hspace*{1cm} +
		\sum_{\ba_{i(-j)}}
		\EXP \Big[ \big\{ Y_{ij} (a_{ij} = 1, \ba_{i(-j)} ) - Y_{ij} (a_{ij} = 0, \ba_{i(-j)} ) \big\}  \pi (\ba_{i(-j)} \con \alpha) \, \Big| \, \type_i = k \Big] + \cdots
		\\
		& \hspace*{1cm} +
		\sum_{\ba_{i(-\NI_k)}}
		\EXP \Big[ \big\{ Y_{i\NI_k} (a_{i\NI_k} = 1, \ba_{i(-\NI_k)} ) - Y_{i\NI_k} (a_{i\NI_k} = 0, \ba_{i(-\NI_k)} ) \big\}  \pi (\ba_{i(-\NI_k)} \con \alpha) \, \Big| \, \type_i = k \Big]
		\bigg]
		\\
		&
		= \frac{1}{\NI_k}
		\Big[
		\EXP \big\{
		\overline{Y}_{i1}(1; \alpha) + \cdots + \overline{Y}_{i\NI_k}(1; \alpha)  \cond \type_i=k \big\}
		- \EXP \big\{ \overline{Y}_{i1}(0 \con \alpha) + \cdots + \overline{Y}_{i\NI_k}(0 \con \alpha)  \cond \type_i=k \big\}
		\Big] \\
		& 
		= \EXP \big\{ \overline{Y}_i ( 1 \con \alpha)   - \overline{Y}_i ( 0 \con \alpha) \cond \type_i = k \big\}
	\end{align*}
	where the second identity is from the definition of $\overline{Y}_{ij}(a \con \alpha)$ and the last identity is from the definition of $\overline{Y}_i(a \con \alpha)$. Thus, the weight in \eqref{example-DE2} makes the parameter $\theta_k$ equal to the direct effect in cluster type $k$. If we also take $v_k(p_k) = p_k = P(\type_i = k)$, then $\tau$ is the same as the direct effect $ \EXP \big\{ \overline{Y}_i ( 1 \con \alpha)   - \overline{Y}_i ( 0 \con \alpha) \big\}$, i.e.
	\begin{align*}
		\tau 
		=
		\sum_{k=1}^K p_k \theta_k
		=
		\sum_{k=1}^K P ( L_i = k)  \EXP \big\{ \overline{Y}_i ( 1 \con \alpha)   - \overline{Y}_i ( 0 \con \alpha) \cond \type_i = k \big\}
		=
		\EXP \big\{ \overline{Y}_i ( 1 \con \alpha)   - \overline{Y}_i ( 0 \con \alpha) \big\} \ .
	\end{align*}

 If we choose the weights $\bw_k$ as
	\begin{align}						\label{example-IE}
		&
		\bw_k( \ba_i \con \alpha) 
		=
		\frac{1}{\NI_k} 
		\begin{bmatrix}
			\ind(a_{i1} = 0 )  \{	\pi( \ba_{i(-1)} \con \alpha) - \pi( \ba_{i(-1)}  \con \alpha') \} 
			\\ 
			\vdots
			\\
			\ind(a_{ij} = 0 )  \{	\pi( \ba_\eij \con \alpha) - \pi( \ba_\eij \con \alpha') \} 
			\\ 
			\vdots		
			\\
			\ind(a_{i\NI_k} = 0 )  \{	\pi( \ba_{i(-\NI_k)}  \con \alpha) - \pi( \ba_{i(-\NI_k)} \con \alpha') \} 
		\end{bmatrix} \ ,
	\end{align}
this leads to the following for $\theta_k$:
	\begin{align*}
		\theta_k 
		&
		= \frac{1}{\NI_k}
		\bigg[
		 \sum_{\ba_{i(-1)}}
		\EXP \Big[ Y_{i1} (a_{i1} = 0, \ba_{i(-1)} ) \big\{ \pi (\ba_{i(-1)} \con \alpha) - \pi (\ba_{i(-1)} \con \alpha') \big\} \, \Big| \, \type_i = k \Big] + \cdots
		\\
		& \hspace*{1cm} +
		\sum_{\ba_{i(-\NI_k)}}
		\EXP \Big[ Y_{i\NI_k} (a_{i\NI_k} = 0, \ba_{i(-\NI_k)} ) \big\{ \pi (\ba_{i(-\NI_k)} \con \alpha) - \pi (\ba_{i(-\NI_k)} \con \alpha') \big\}  \, \Big| \, \type_i = k \Big]
		\bigg]
		\\
		&
		= \frac{1}{\NI_k}
		\Big[
		\EXP \big\{
		\overline{Y}_{i1}(0; \alpha) + \cdots + \overline{Y}_{i\NI_k}(0; \alpha)  \cond \type_i=k \big\}
		- \EXP \big\{ \overline{Y}_{i1}(0 \con \alpha') + \cdots + \overline{Y}_{i\NI_k}(0 \con \alpha')  \cond \type_i=k \big\}
		\Big] \\
		& 
		= \EXP \big\{ \overline{Y}_i ( 0 \con \alpha)   - \overline{Y}_i ( 0 \con \alpha') \cond \type_i = k \big\}
	\end{align*}
	Thus, the weight in \eqref{example-IE} makes the parameter $\theta_k$ equal to the indirect effect in cluster type $k$. If we also take $v_k(p_k) = p_k = P(\type_i = k)$, then $\tau$ is the same as the indirect effect $ \EXP \big\{ \overline{Y}_i ( 0 \con \alpha)   - \overline{Y}_i ( 0 \con \alpha') \big\}$, i.e.
	\begin{align*}
		\tau
		=
		\sum_{k=1}^K p_k \theta_k
		=
		\sum_{k=1}^K P ( L_i = k)  \EXP \big\{ \overline{Y}_i ( 0 \con \alpha)   - \overline{Y}_i ( 0 \con \alpha') \cond \type_i = k \big\}
		=
		\EXP \big\{ \overline{Y}_i ( 0 \con \alpha)   - \overline{Y}_i ( 0 \con \alpha') \big\} \ .
	\end{align*}

Next, we introduce the details of the asymptotic embedding of network treatment effects.

\begin{lemma}[Asymptotic Embedding of Network Treatment Effects]						\label{lem-AE}
Suppose that the potential outcome $\bY_i(\ba_i)$ and the type variable $\type_i$ of cluster $i$ are a random sample from a super-population satisfying condition  \eqref{eq:NPmodel} in the main paper where $\bO_i$ is replaced with $\bY_i(\ba_i)$. Consider a network causal estimand $\oT^F(\balpha, \balpha')$ which is a linear combination of individual average potential outcome $\overline{Y}_{ij} ( a \con \alpha_k)$ defined in Section \ref{sec:setup} of the main paper, i.e., 
\begin{align*}
	&
	\oT^F(\balpha, \balpha') 
	=
	\frac{1}{\NC} \sum_{i=1}^\NC \sum_{k=1}^\NT \ind(\type_i = k)  
	\sum_{j=1}^{\NI_k} 
	\left\{
		\begin{matrix}
		C_{1jk}  \overline{Y}_{ij} (1 \con \alpha_k) + C_{2jk} \overline{Y}_{ij} (0 \con \alpha_k) 
		\hspace*{2cm}
		\\
		\hspace*{2cm}
		+ C_{3jk}  \overline{Y}_{ij} (1 \con \alpha'_k)  + C_{4jk} \overline{Y}_{ij} (0 \con \alpha'_k) 
		\end{matrix}
	\right\}
	\ .
\end{align*}
Here, $C_{\ell jk}$s $(\ell=1,2,3,4,j=1,\ldots,\NI_k, k=1,\ldots,\NT)$ are fixed constants. If (i) $\NT$ and $\NI_k$ are bounded by a constant for all $k=1,\ldots,\NT$ and (ii) the conditional expectation $\EXP\big\{ \bY_i(\ba_i) \cond \type_i = k\big\}$ is finite for all $\ba_i \in \zosets(\NI_k)$ and $k=1,\ldots,\NT$, then there exists $\oT(\balpha, \balpha') \in \Paraset$ for all $\balpha$ and $\balpha'$ where $\oT^F(\balpha, \balpha')$ converges to $\oT(\balpha, \balpha')$ in probability as $\NC \rightarrow \infty$. 
\end{lemma}
The proof of Lemma \ref{lem-AE} is in Section \ref{lem:embedding}. Lemma \ref{lem-AE} generalizes this observation and shows that under certain growth conditions, finite sample causal estimands in partial interference can be asymptotically embedded into our framework. We remark that while Lemma \ref{lem-AE} was restricted to estimands with $\alpha$-policy weights, we can pick any causal estimand where the weights in $\Paraset$ are not based on $\alpha$-policies. So long as these weights are pre-specified a-priori and satisfy the constraints of $\Paraset$, the results below will still hold.

\subsection{Details of Section \ref{sec:GloEff} in the Main Paper}							\label{sec:GloEffsupp}
	
We introduce Lemma \ref{lmm:EIFbasic}, which is a key step to proving Theorem \ref{sec:EffNonpar}. 

\begin{lemma}[Semiparametric Efficiency Bound of $\bT^*$] 		\label{lmm:EIFbasic}
Let $\bT^* =  ( \uT_1^*,\ldots,\uT_\NT^*)$ where $\uT_k^* \in \Parasetk_k$. Under the conditions in Assumption \ref{Assump:VC} in the main paper, the efficient influence function of $\bT^* = (\uT_1^*,\ldots,\uT_\NT^*)\T$ in model $\modeliv$, denoted by $\varphi(\bT^*) = \big( \varphi_1(\uT_1^*) , \ldots , \varphi_\NT(\uT_\NT^*) \big) \T $, is
\begin{align*}	
	\varphi_k ( \uT_k^*  ) 
	=
	\frac{\ind (\type_i = k)}{p_k^*}  \Bigg[ \sum_{\ba_i \in \zosets( \NI_k ) } \hspace*{-0.25cm} \bw_k \T (\ba_i,\bX_i ) \bigg[
		 \frac{\ind(\bA_i = \ba_i)}{e^*(\ba_i \cond \bX_i, k)} \Big\{ \bY_i - \OR^*(\ba_i, \bX_i, k) \Big\}  
		+
		  \OR^*(\ba_i, \bX_i, k) \bigg]- \uT_k^* \Bigg]
		 \nonumber
\end{align*}
and the corresponding semiparametric efficiency bound of $\bT^*$ in model $\modeliv$, denoted by $\VAR \big\{ \varphi(\bT^*) \big\}$, is $\VAR \big\{ \varphi(\bT^*) \big\} 
=  \text{diag} \big[ \text{SEB}_1 \big( \uT_1^* \big) , \ldots , \text{SEB}_\NT \big( \uT_\NT^* \big) \big] $ with
\begin{align*}
	& \text{SEB}_k \big( \uT_k^* \big)
	= 
		\frac{1}{p_k^*} \EXP \bigg[ \hspace*{-0.05cm}
			\sum_{\ba_i \in \zosets(\NI_k)} \hspace*{-0.3cm}	 \frac{\bw_k \T(\ba_i,\bX_i) \Sigma^* (\ba_i, \bX_i, k) \bw_k (\ba_i,\bX_i)}{e^* (\ba_i \cond \bX_i, k)}  
			\\
			& \hspace*{5cm}
		+ \bigg\{ \hspace*{-0.05cm}
			\sum_{\ba_i \in \zosets(\NI_k) } \hspace*{-0.3cm} \bw_k \T (\ba_i,\bX_i)  \OR^* (\ba_i, \bX_i, k) - \uT_k^*
			\bigg\}	^2 \, \bigg| \, \type_i = k
		\bigg] \ .
		\nonumber
\end{align*}
Moreover, $\varphi(\bT^*)$ is also the efficient influence function of $\bT^*$ in model $\modelive= \big\{ P_{O,\type} \in \modeliv \cond e^*\text{ is known} \big\}$.
\end{lemma}
The proof is presented in Section \ref{sec:prooflemma2}.

We consider the case of known $p_k^*$s and derive the efficient influence function and the semiparametric efficiency bound for $\oT^* \in \Paraset$, which is the extension of Theorem \ref{thm:EIFVC} in the main paper. The result is formally presented in Theorem \ref{thm:EIF-known}.
\begin{theorem}[Semiparametric Efficiency Bound of $\oT^* \in \Paraset$ under known $p_k^*$] 		\label{thm:EIF-known}
	Let $\oT^* \in \Paraset$ be the parameter defined in \eqref{eq-31003-1} in the main paper. Suppose that the conditions in Assumption \ref{Assump:VC} in the main paper hold. If $p_k^*$s are known, the efficient influence function of $\oT^*$ in model $\modeliv$ (and in model $\modelive$) is 
	\begin{align*}
		 \eifuv ( \oT^* )
		 &
		 =  \sum_{k=1}^\NT  v_k(p_k^*) \varphi_k ( \uT_k^* )
	\end{align*}
	where $\eifv_k$ is defined in Lemma \ref{lmm:EIFbasic}. Moreover, the semiparametric efficiency bound of $\oT^*$ in model $\modeliv$ is 
	\begin{align*}
		\VAR \big\{  \eifuv (\oT^*) \big\} 
		= \sum_{k=1}^\NT v_k(p_k^*)^2 \text{SEB}_k (\uT_k^*)
	\end{align*}
	where $\text{SEB}_k$ is defined in Lemma \ref{lmm:EIFbasic}.
\end{theorem}
The proof is presented in Section \ref{sec:appendix3-TA1}. Compared to the result of Theorem \ref{thm:EIFVC} in the main paper, the semiparametric efficiency bound under known $p_k^*$ is smaller than or equal to the semiparametric efficiency bound under unknown $p_k^*$. That is, we gain efficiency because of the knowledge of $p_k^*$.

We show that the equation \eqref{eq-34001} can be used to construct a doubly robust estimator. Specifically, for each cluster type $k$, let $\widehat{\uT}_k(e' , \OR')$ be the solution to an estimating equation constructed from \eqref{eq-34001} by using some $e'$ and $\OR'$, i.e.,
\begin{align*}
	\widehat{\uT}_k(e' , \OR')
	= \frac{1}{\NC_k} \sum_{i: \type_i = k} \sum_{\ba_i \in \zosets(\NI_k) } \bw_k \T ( \ba_i,\bX_i )  \Bigg[   \frac{\ind(\bA_i = \ba_i)}{e'(\ba_i \cond \bX_i, k)} \Big\{ \bY_i - \OR' (\ba_i, \bX_i, k) \Big\}  + \OR' (\ba_i, \bX_i, k) \Bigg] \ .
\end{align*}
We define an estimator of $\oT^*$ based on $\widehat{\uT}_k(e' , \OR') $ as 
\begin{align}										\label{eq-31004}
	\widehat{\oT} (e' , \OR') 
	= \widehat{\bv}\T \widehat{\bT}	(e' , \OR')
	= \sum_{k=1}^\NT \widehat{v}_k \widehat{\uT}_k(e' , \OR') \quad , \quad
	 \widehat{\bT}(e',\OR') = \big( \widehat{\uT}_1(e' ,\OR' ) , \ldots ,  \widehat{\uT}_\NT(e' ,\OR') \big)\T 
		 \end{align}
		 where $\widehat{\bv} = \big( \widehat{v}_1,\ldots, \widehat{v}_\NT \big)\T$ is an unbiased estimator of $\bv(\bp^*)$ constructed based on the variable $\type_i$. Theorem \ref{thm:DRVC} shows that $\widehat{\oT}(e' , \OR')$ is unbiased even if the propensity score or the outcome model, but not both, is mis-specified.
\begin{theorem}[Double Robustness of $\widehat{\oT}(e',\OR')$] 					\label{thm:DRVC}
	Suppose that the conditions in Assumption \ref{Assump:VC} hold and $\widehat{\bv}$ is an unbiased estimator of $\bv(\bp^*)$. Then, $\widehat{\oT}$ is doubly robust in the sense that $\widehat{\oT} (e',  \OR ') $ is an unbiased estimator of $\oT$, i.e., $\EXP \big\{ \widehat{\oT} (e',  \OR ') \big\} = \oT^*$, if either $e' =e^*$ or $\OR' = \OR^*$.
\end{theorem}
The proof is presented in Section \ref{sec:ThmS2}. We believe Theorem \ref{thm:DRVC} can be used as a basis to construct robust machine learning and cross-fitting estimators \citep{Victor2018} under partial interference where half of the clusters are used to non-parametrically estimate $e$ and $\OR$ and the other half is used to estimate $\oT$; see Section \ref{sec:NPOR} in th main paper. 

Finally, we conclude the section by briefly discussing estimators of direct and indirect effects under our framework. 
Specifically, after some algebra, the efficient influence functions of direct and indirect effects are
\begin{align*}
	&
	\eifuv \big(  \tDE (\alpha) \big)
		= \sum_{k=1}^\NT \ind(\type_i=k) 
		\Big\{  \psi_k ( 1 , e^* , \OR^* ,  \alpha ) - \psi_k ( 0 , e^* , \OR^* ,  \alpha )  \Big\}
		- \tDE (\alpha)
	\\
	\nonumber
	&
	  \eifuv \big( \tIE (\alpha , \alpha') \big)
	  =  \sum_{k=1}^\NT \ind(\type_i=k) 
	 \Big\{  \psi_k ( 0 , e^* , \OR^* ,  \alpha ) - \psi_k ( 0 , e^* , \OR^* ,  \alpha' )  \Big\}
	  - \tIE (\alpha, \alpha')
	\end{align*}
	where
	\begin{align*}
	& \psi_k ( a , e , \OR , \alpha ) 
	= \frac{1}{\NI_k} \sum_{j=1}^{\NI_k} 
		\sum_{ \substack{ \ba_i \in \zosets(\NI_k) \\ a_{ij} = a} } \Bigg[ \frac{ \ind ( \bA_i = \ba_i ) }{ e(\ba_i \cond \bX_i , k)  } \Big\{ Y_{ij} - \indOR_j( \ba_i , \bX_i , k ) \Big\} 
		+ 
	 \indOR_j ( \ba_i , \bX_i, k ) \Bigg] \pi ( \ba_\eij \con \alpha )
	 \ .	
	 \nonumber
\end{align*}
Here $\indOR_j$ is $j$th component of $\OR$, i.e., the outcome regression of individual $j$. Also, by following \eqref{eq-31004}, the efficient influence functions above lead to the following doubly robust estimators of $\tDE(\alpha)$ and $\tIE(\alpha, \alpha')$.
\begin{align*}
&
	\widehat{\oT}^{\DE}(\alpha \con e' , \OR')
	= \sum_{k=1}^\NT \widehat{p}_k \cdot \widehat{\uT}_k^{\DE}(\alpha \con e' , \OR')  
	, \
	\widehat{\oT}^{\IE}(\alpha, \alpha' \con e' , \OR' )
	= \sum_{k=1}^\NT \widehat{p}_k \cdot \widehat{\uT}_k^{\IE}(\alpha, \alpha' \con e' , \OR')  
\end{align*} 
where $e'$ and $\OR'$ are pre-specified functions of the propensity score and the outcome regression, respectively, $\widehat{\uT}_k^{\DE}(\alpha \con e' , \OR')  
	 = \NC_k^{-1} \sum_{i: \type_i = k} 
	 \big\{ \psi_k ( 1 , e' , \OR' ,  \alpha ) - \psi_k ( 0 , e' , \OR' ,  \alpha ) \big\}$, 
	 and $\widehat{\uT}_k^{\IE}(\alpha, \alpha' \con e' , \OR')  
	 = \NC_k^{-1} \sum_{i: \type_i = k} \big\{ \psi_k ( 0 , e' , \OR' ,  \alpha ) - \psi_k ( 0 , e' , \OR' ,  \alpha' ) \big\}$. Moreover, $e'$ and $\OR'$ can be replaced with estimates that satisfy Theorem \ref{thm:ParaPopEst} in the main paper. As a consequence, we obtain $\widehat{\oT}^\DE(\alpha)$ and $\widehat{\oT}^\IE(\alpha,\alpha')$ in \eqref{eq-32005} in the main paper.

\subsection{Details of Section \ref{sec:Mestimation-general} in the Main Paper}							\label{sec:LocEffsupp}

To prove Theorem \ref{thm:ParaPopEst}, we present Lemma \ref{lmm:ParaEst} and the regularity conditions (R1)--(R4).
\begin{lemma}[M-Estimators of $\bT^*$] 	\label{lmm:ParaEst}
Suppose Assumption \ref{Assump:VC} and the following regularity conditions (R1)--(R4) hold for the estimating equation $\ee (\bT, \beta) = \big( \ee_\uT \T( \bT,\paraT) , \ee_\beta \T (\paraT) \big)\T$: (R1) $(\bT, \paraT)$ lies in an open set $\Omega_0$, which is a subset of a compact, finite dimensional Euclidean space. Also, $\paraPS$ and $\paraOR$ are variationally independent; (R2) $\ee(\bT,\paraT)$ is twice continuously differentiable in $(\bT,\paraT)$; (R3) There is a unique value $(\bT^*,\paraT^\dagger)$ where (i) $E \big\{ \ee( \bT^*, \paraT^\dagger) \big\} = 0 $, (ii) $E \big\{ \big\| \ee( \bT^*, \paraT^\dagger) \big\|_2^2 \big\} < \infty$, and (iii) $E \big\{	\partial \ee( \bT^*, \paraT^\dagger) / \partial ( \bT, \paraT) 	\big\} $ exists and is non-singular; (R4) Every element of the second order partial derivatives $ \partial^2 \ee( \bT, \paraT) / \{ \partial ( \bT, \paraT) \partial ( \bT, \paraT)\T \} $ are dominated by a fixed integrable function for all $( \bT, \paraT)$ in a neighborhood of $( \bT^*, \paraT^\dagger)$. 	
Let $\paraT^\dagger$ be the probability limit of $\widehat{\paraT}$. Then, under model $\modelie \cup \modelig$, we have
	\begin{align*}
		\sqrt{\NC} \big( \widehat{\bT} - \bT^* \big) 
		= 
		\frac{1}{\sqrt{\NC}}\sum_{i=1}^\NC \varphi^{\rm Par} (\bT^* , \paraT^\dagger) 
		+ o_P(1)
	\end{align*}
	where $
		\varphi^{\rm Par} \big( \bT^* , \paraT^\dagger \big)
		= \big[
			\varphi_1^{\rm Par} \big( \uT_1^* , \paraT^\dagger \big) \ , \ \ldots \ , \
			\varphi_\NT^{\rm Par} \big( \uT_\NT^* , \paraT^\dagger \big)
		\big] \T$ and
	\begin{align*}
	\varphi_k^{\rm Par} \big( \uT_k^* , \paraT^\dagger \big)= 	\frac{1}{p_k^*} \bigg[ \ee_{\uT,k} (\uT_k^*, \paraT^\dagger) 
		- 
		\EXP  \bigg\{	 \frac{\partial \ee_{\uT ,k} (\uT_k^* , \paraT^\dagger)   }{\partial \paraT \T } \bigg\}
		\bigg[ \EXP  \bigg\{	\frac{ \ee_\beta(\paraT^\dagger)  }{\partial \paraT \T } \bigg\} \bigg]^{-1}
		\ee_\beta(\paraT^\dagger) \bigg] \ .
	\end{align*}
		\end{lemma}	
	The proof is presented in Section \ref{sec:prooflmm41}.

\subsection{Details of Adaptive Estimation in Section \ref{sec:EffNonpar} in the Main Paper}									\label{sec:detail5}

This section investigates two properties related to adaptive estimation under partial interference, adaption to the knowledge about the propensity score and adaption to the knowledge about the inference pattern/covariance structure. We remark that without interference, the aforementioned works on the augmented inverse probability-weighted estimator for the average treatment effect showed that the estimator adapts to the knowledge about the propensity score and the variance.

First, suppose the propensity score $e^*$ is known as in a two-stage randomized experiment from \citet{HH2008}. In adaptive estimation, we want to understand whether having this knowledge can lead to more efficient estimation of $\oT^*$. 
Unfortunately, but in alignment with the results without interference, Theorem \ref{thm:NoGainFromPS} shows that the estimator from Theorem \ref{thm:ParaPopEst} in the main paper that does not use this information still achieves the semiparametric efficiency bound of $\oT^*$ when the propensity score is known.
\begin{theorem}[Adaptation to Known Propensity Score]						\label{thm:NoGainFromPS}
	Suppose conditions in Theorem \ref{thm:ParaPopEst} in the main paper hold. Let $\widehat{\oT}$ be the estimator in Theorem \ref{thm:ParaPopEst} in the main paper where $e$ and $g$ are estimated from $\modelie \cap \modelig$. Then, the asymptotic variance of $\widehat{\oT}$ achieves the semiparametric efficiency bound of $\oT^*$ under $\modelive = \big\{ P_{O,\type} \in \modeliv \cond e^*\text{ known} \big\}$.
\end{theorem}
The proof is presented in Section \ref{sec:proofS3}. In words, the efficient estimators in Theorem \ref{thm:ParaPopEst} in the main paper that does not use the knowledge that the propensity score is known can adapt and still achieve the best possible variance regardless of the knowledge of $e^*$.

Second, without interference, a somewhat under-emphasized fact about the augmented inverse probability-weighted estimator of the average treatment effect is that the estimator remains efficient  irrespective of the investigator's knowledge about the true variance of the outcome, i.e. the estimator adapts. We ask a similar type of question under partial interference, specifically whether having certain a priori knowledge about the interference pattern/exposure mapping \citep{Aronow2017} would affect efficiencies of the proposed estimators. Unlike the case without interference, we show that in a setting where the true model has no interference, but the investigator, out of caution or lack of awareness, uses the estimators in \eqref{eq-32005} that take interference into account, the estimators are consistent, but no longer efficient; in short, the estimators do not adapt to the underlying true interference pattern. 

Formally, consider an ``interference-free'' model space $\modelii  \subseteq \modeliv$ which is equal to
\begin{align*}
	\modelii
	= \big\{ P_{\bO,\type} \in \modeliv \, \big| \, &
	Y_{ij} \cond ( \bA_i = \ba_i , \bX_i = \bx , \type_i = k) 
	=
	Y_{ij} \cond ( A_{ij} = a_{ij} , \bX_i = \bx , \type_i = k) 
	\nonumber
	\\
	& \text{almost surely for all } \ba_i \in \zosets(\NI_k) , \ \bx \in \mathcal{X}(k) , \ k=1,\ldots,\NT
	\big\} \ .
\end{align*}
Next, we introduce $\widetilde{\mathcal{M}}_{\rm NoInt}$, which is defined as
		\begin{align*}
			\widetilde{\mathcal{M}}_{\rm NoInt}
			=
			\big\{
				P_{\bO,\type} \in \modeliv
				\, \big| \, & 
				\EXP\big( Y_{ij} \cond \bA_i , \bX_i  , \type_i = k \big) 
				= 
				\EXP\big( Y_{ij} \cond \bA_i' , \bX_i , \type_i = k \big) \ ,  \\
				& 
				\text{where }
				\bA_i , \bA_i' \in \zosets ( \NI_k)\text{ such that } A_{ij} = A_{ij}' \ , \
				 j=1,\ldots,\NI_k \ , \\
				&
				\bX_i \in \mathcal{X}(k) \ , \ 
				k = 1, \ldots , \NT 
			\big\} \ .
		\end{align*}
		$\widetilde{\mathcal{M}}_{\rm NoInt}$ is a collection of models without first-order (i.e., in mean) interference. Trivially, $\widetilde{\mathcal{M}}_{\rm NoInt}$ includes $\mathcal{M}_{\rm NoInt}$, the collection of ``truly'' no interference models because of the nested relationship imposed by the definition of each set. 

Lastly, we define the ``interference-free'' outcome model $\modeliig \subseteq \modelig \cap \modelii$ which is equal to
\begin{align*}
	\modeliig 
	= \big\{   P_{\bO,\type} \in \modelig \cap \modelii  	\, \big| \, 
&
	\indPOR_j (\ba_i , \bx , k \con \paraOR ) = \indPOR_j (\ba_i' , \bx , k \con \paraOR ) 		
	\text{ for all } j =1,\ldots, \NI_k, \  \nonumber 
	\\ 
	& \ba_i , \ba_i' \in \zosets(\NI_k) \text{ with } a_{ij} = a_{ij}' , \ \bx \in \mathcal{X}(k) , \ k =1,\ldots,\NT \big\}  \ .
\end{align*}
Under model $\modeliig$ and the consistency condition (A1), the outcome regression of unit $j$ in cluster $i$ does not depend on the treatment status of $i$'s peers so that $\indOR_j^* ( \ba_i , \bx , k)$ is a function only of unit $j$'s treatment assignment $ a_{ij}$. Furthermore, we introduce a function $\indIOR_j(a_{ij}, \bx_i, k)$ which is the same as $\indOR_j^*(\ba_i , \bx_i, k)$, but the former emphasizes the lack of dependence on $\ba_\eij$. Therefore, under model $\modeliig$, we obtain the following equivalence at the true parameter $\paraOR$ for all $\ba_\eij \in \zosets(\NI_k-1)$.
		\begin{align}											\label{eq:nointequiv}
			\indIOR_j(a_{ij}, \bx_i, k) = \indPIOR_j(a_{ij}, \bx_i, k \con \paraOR^*) = \indOR_j^* (\ba_i, \bx_i, k) = \indPOR_j (\ba_i, \bx_i, k \con \paraOR^* ) \ .
		\end{align}

We study the behavior of the direct effect and the indirect effect across the counterfactual parameters under model $\modelii$. Let the average treatment effect $\oT^\ATE$ where $\oT^\ATE = \sum_{k=1}^\NT p_k^*  \uT_k^\ATE$ and $
	\uT_k^\ATE
	=
	\NI_k^{-1} \sum_{j=1}^{\NI_k} \EXP  \big\{ Y_{ij}(a_{ij} = 1) - Y_{ij}(a_{ij} = 0) \cond \type_i = k \big\}$. Here $Y_{ij}(a_{ij}=a)$ is the potential outcome of unit $j$ in cluster $i$ when the unit's treatment status is $a \in \{0,1\}$. Lemma \ref{lmm:NoInterference} shows that $\oT^\DE(\alpha)$ and $\oT^\IE(\alpha , \alpha')$ defined in the main paper are the same as $\oT^\ATE$ and $0$, respectively, in model $\widetilde{\mathcal{M}}_{\rm NoInt}$.
	\begin{lemma}						\label{lmm:NoInterference}
		Suppose that the true model belongs to $\widetilde{\mathcal{M}}_{\rm NoInt}$. Then, $\oT^\DE(\alpha) = \oT^\ATE$ and $\oT^\IE(\alpha , \alpha')=0$ for all $\alpha , \alpha' \in (0,1)$.
	\end{lemma}
	The proof is presented in Section \ref{sec:appendix3-LA1}. Note that the above Lemma also holds after replacing $\widetilde{\mathcal{M}}_{\rm NoInt}$ with ${\mathcal{M}}_{\rm NoInt}$ because of the nested relationships. Importantly, this implies that we can use the proposed locally efficient estimators of the direct and indirect effects in Section \ref{sec:Mestimation-general}. Theorem \ref{thm:RobustATE} shows that while these estimators remain consistent, they do not adapt to the no-interference structure and are no longer efficient.
\begin{theorem}[Non-Adaptation to Exposure Mapping]				\label{thm:RobustATE}
Suppose conditions in Theorem \ref{thm:ParaPopEst} hold. Let $e$ and $g$ be estimated from $\modelie \cap \modelig$. Then, for all $\alpha, \alpha' \in (0,1)$, $\widehat{\oT}^{\DE}( \alpha)$ and $ \widehat{\oT}^{\IE}(  \alpha, \alpha')$ are consistent for $\oT^\ATE$ and $0$, respectively, under $ \modelie \cap \modeliig$. Also, unless the outcome and the propensity score models satisfy invariance conditions in Assumption \ref{assp:1} below, $\widehat{\oT}^{\DE}( \alpha) $ does not achieve the semiparametric efficiency bound for $\oT^\ATE$ under $\modelii$. 
\end{theorem}
The proof is presented in Section \ref{sec:B10}. 
Theorem \ref{thm:RobustATE} shows that if the investigator uses estimators that account for interference, but the true data has no interference, the estimators are consistent, but generally inefficient; in short, the investigator pays a price in terms of efficiency. 

The additional invariance conditions are given as follows.
	\begin{assumption}[Invariance Condition]							\label{assp:1}
	Suppose the outcome regression and the propensity score satisfy the following invariance conditions, respectively.
	\begin{itemize}
			\item[(a)] (Propensity Score): For all $\alpha \in (0,1)$ and $\ba_\eij \in \zosets(\NI_k-1)$,  $\pi(\ba_\eij \con \alpha) \Sigma_{jj}^*(\ba_i, \bx, k) /  e^* (\ba_i \cond \bx,k) $ is identical  where $\Sigma_{jj}^*(\ba_i, \bx, k)$ is the $j$th diagonal element of $\Sigma^*(\ba_i, \bx, k)$.

		\item[(b)] (Outcome Regression): For all $j =1,\ldots, \NI_k$, $\by_\eij \in \R^{\NI_k-1}$, $\ba_i \in \zosets(\NI_k)$, $\bx_i \in \mathcal{X}(k)$, $k=1,\ldots,\NT$, the following identities hold under model $\modelii$.
\begin{align*}
	&
	\EXP \big( Y_{ij} \, \big| \,  \bY_\eij = \by_\eij  , \bA_i = \ba_i , \bX_i =  \bx_i, \type_i = k \big) \\
	& =	\EXP \big( Y_{ij} \, \big| \,  \bA_i = \ba_i , \bX_i =  \bx_i, \type_i = k \big)
	= \EXP \big( Y_{ij} \, \big| \, A_{ij} = a_{ij} , \bX_i =  \bx_i, \type_i = k \big) \ .
\end{align*}
	Note that the second identity is trivial under model $\modelii$.
	\end{itemize}
	\end{assumption}
	Condition (a) of Assumption \ref{assp:1} states that the propensity score is invariant to $\alpha$ and $\ba_i$; note that condition (b) holds under randomized experiment where ${\rm pr}(A_{ij}=1) = \alpha$ and $\Sigma_{jj}^*(a_i, x, k)$ does not depend on $a_i$. 
	
	Condition (b) of Assumption \ref{assp:1} roughly states that the peers' outcomes do not provide any information about one's own outcome. The first immediate result under condition (a) is that the conditional variance matrix $\Sigma^*(\ba_i, \bx_i, k) = \VAR(\bY_i \cond \bA_i = \ba_i , \bX_i = \bx_i , \type_i = k) $ is diagonal. Lemma \ref{lmm:diagSigma} formally states this.
	\begin{lemma}							\label{lmm:diagSigma}
		Let the $(i,j)$th entry of $\Sigma^*(\ba_i, \bx_i, k) $ be $\Sigma_{ij}^*(\ba_i, \bx_i, k) $. 	Suppose that condition (a) of Assumption \ref{assp:1} holds. Then,  $\Sigma^*(\ba_i, \bx_i, k)$ is diagonal and $\Sigma_{jj}^*(\ba_i, \bx_i, k)$ does not depend on $\ba_\eij$. That is,  $
			\Sigma^*(\ba_i, \bx_i, k) 
			=
			{\rm diag} \big[ \Sigma_{11}^*(a_{i1}, \bx_i, k) , \ldots , \Sigma_{\NI_k \NI_k}^*(a_{i\NI_k}, \bx_i, k) \big]$.
	\end{lemma}
	The proof is presented in Section \ref{sec:appendix3-LA2}.
	
	Next, we derive the efficient influence function  of $\oT^\ATE$ in model $\modelii$ under Assumption \ref{assp:1}. Lemma \ref{lmm:NIEIF} formally shows the result.
	\begin{lemma}						\label{lmm:NIEIF}
		Suppose that the conditions in Assumption \ref{Assump:VC} in the main paper and Assumption \ref{assp:1} hold.	Then, the efficient influence function of $\oT^\ATE$ in model $\modelii$, denoted by $\varphi ( \oT^\ATE) $, is 
		\begin{align*}
			&
			\varphi ( \oT^\ATE ) 
			 =
			\sum_{k=1}^\NT \ind(\type_i = k) \bigg[
		\frac{1}{\NI_k} \sum_{j=1}^{\NI_k} 
		\bigg[
		\Big\{ \ind ( A_{ij} = 1 ) - \ind ( A_{ij} = 0 ) \Big\}	
		\frac{Y_{ij} - \indIOR_j ( A_{ij} , \bX_i , k ) }{e_j^*(A_{ij} \cond \bX_i,k)}
		\\
		& \hspace*{6cm} + \Big\{ \indIOR_j(1,\bX_i,k) - \indIOR_j(0,\bX_i,k) \Big\}		\bigg]
	\bigg] - \oT^\ATE \ .
		\end{align*}
		where $\indIOR_j$ is defined in \eqref{eq:nointequiv} and $e_j^*$ is the conditional probability of $A_{ij}$ being assigned to a certain treatment indicator; i.e., 
		\begin{align*}
			e_j^*(a \cond \bX_i, k) = \sum_{\ba_i \in \zosets(\NI_k)} \ind(a_{ij} = a) e^*(\ba_i \cond \bX_i, k) 
			= {\rm pr}(A_{ij} = a \cond \bX_i, \type_i = k) \ .
		\end{align*}
		Therefore, the semiparametric efficiency bound of $\oT^\ATE$ in model $\modelii$ is
\begin{align*}
	\EXP \big\{ \varphi(\oT^\ATE)^2 \big\} 
	& = 
	\sum_{k=1}^\NT \frac{p_k^* }{\NI_k^2} \sum_{j=1}^{\NI_k} \EXP \bigg\{ \frac{\Sigma_{jj}^*(1, \bX_i, k)}{e_j^*(1 \cond \bX_i, k)} + \frac{\Sigma_{jj}^*(0, \bX_i, k)}{e_j^*(0 \cond \bX_i, k)} \, \bigg| \, \type_i = k  \bigg\} \\
	& 
	+ \EXP \bigg[ \bigg[ \sum_{k=1}^\NT \frac{ \ind(\type_i = k) }{\NI_k} \sum_{j=1}^{\NI_k} \Big\{ \indIOR_j(1,\bX_i,k) - \indIOR_j(0,\bX_i,k) \Big\} - \oT^\ATE \bigg]^2  \bigg]  
\end{align*}
where $\Sigma_{jj}^*(a_{ij}, \bx_i, k)$ is defined in Lemma \ref{lmm:diagSigma}.
	\end{lemma}
	The proof is presented in Section \ref{sec:appendix3-LA3}. Note that $\varphi(\oT^\ATE)$ is an extension of \citet{Hahn1998} to clustered data. Specifically, if all units are independent and identically distributed, this leads to $\NT=1$, $\NI_1=1$, and $j=1$. Therefore, the efficient influence function  of $\varphi ( \oT^\ATE ) $ and the semiparametric efficiency bound of $\oT^\ATE$ reduce to
	\begin{align*}
		\varphi ( \oT^\ATE ) 
			& =
			\frac{\ind(A_{i1} = 1)}{e_1^*(1 \cond \bX_i , 1) } \Big\{ Y_{i1} - \indIOR_1( 1 , \bX_i , 1 ) \Big\} 
			-
			\frac{\ind(A_{i1} = 0)}{e_1^*(0 \cond \bX_i , 1) } \Big\{ Y_{i1} - \indIOR_1( 0 , \bX_i , 1 ) \Big\} 
			\\
			& \hspace*{3cm}
			+ \Big\{ \indIOR_1(1,\bX_i,1) - \indIOR_1(0,\bX_i,1)  - \oT^\ATE \Big\} \ ,
			\\
			\EXP \big\{ \varphi(\oT^\ATE)^2 \big\} 
	& = 
	\EXP \bigg[ \frac{\Sigma_{jj}^*(1, \bX_i, 1)}{e_1^*(1 \cond \bX_i, 1)} + \frac{\Sigma_{jj}^*(0, \bX_i, 1)}{e_1^*(0 \cond \bX_i, 1)}
	+ \Big\{ \indIOR_1(1,\bX_i,1) - \indIOR_1(0,\bX_i,1) - \oT^\ATE \Big\}^2  \bigg] 
	\end{align*}
	which are equivalent to the results under no interference originally introduced in \citet{Hahn1998}.

Finally, we conduct a small simulation study to visually illustrate Theorem \ref{thm:RobustATE}. Suppose we only have one cluster type (i.e., $\NT=1$) and the cluster size is two (i.e., $\NI_1=2$). We assume the true model has no interference and has the following form.
\begin{align*}
& (\text{Model of }X) 
\ \ X_{ij} \stackrel{\text{i.i.d.}}{\sim} N(0,1)
\quad \quad (\text{Model of } A) \ \
A_{ij} \stackrel{\text{i.i.d.}}{\sim} \text{Ber}(p_A^*) \ , \ p_A^*  \in \big\{  0.3 , 0.5 , 0.7 \big\}\ , 
 \\
& (\text{Model of } Y) 
\ \ Y_{ij} = 1 +  3 A_{ij} + 2 X_{ij} + 0.5 X_\eij + \epsilon_{ij} \ , \quad \epsilon_{ij} \stackrel{\text{i.i.d.}}{\sim} N(0,1)  \ .
\end{align*}
Briefly, each unit has one pre-treatment covariate, following a standard normal distribution, and pre-treatment covariates are independent from each other. The treatment is completely randomized with probability ${\rm pr}(A_{ij} =1 ) = p_A^* \in \{0.3,0.5,0.7\}$. The outcome variable is generated from a regression model that has no interference between units, but depends on peers' covariate $X_{i(-j)}$
; our numerical results will be similar if we remove the peer's covariate in the outcome model. We generate $\NC = 10,000$ samples from the simulation model and compute $\widehat{\oT}^\DE(\alpha)$ and $\widehat{\oT}^\IE(\alpha,0.9)$ that allow for interference. We compute these two estimates for a range of policy parameter from $\alpha=0.01$ to $\alpha=0.99$. We repeat the simulation $1,000$ times.

Before we discuss our numerical results, we first study what is expected from our theory by calculating the theoretical variance of $\widehat{\oT}^\DE(\alpha)$ 
using Theorem \ref{thm:ParaPopEst} and the semiparametric efficiency bound of $\oT^\ATE$.
\begin{align*}
	\VAR\big\{ \widehat{\oT}^\DE(\alpha) \big\} 
	& 
	=
	\frac{1}{2} \bigg\{
		 \frac{ (1-\alpha)^2 }{(1-p_A^*)^2 }
		+
		\frac{ \alpha^2 + (1-\alpha)^2 }{p_A^*(1-p_A^*) }
		+
		\frac{ \alpha^2 }{p_A^{*2}}
	\bigg\} 
	\ , \
	\EXP \big\{ \varphi(\oT^\ATE)^2 \big\} 
	=
		\frac{1}{2p_A^*(1-p_A^*)}  \ .
\end{align*}
Notice that $\VAR\big\{ \widehat{\oT}^\DE(\alpha) \big\} $ is uniquely minimized at $\alpha=p_A^*$, which satisfies the invariance condition, and becomes the semiparametric efficiency bound of $\oT^\ATE$. Specifically, $\VAR\big\{ \widehat{\oT}^\DE(p_A^*) \big\} $ is minimized at $p_A^*=0.5$ 
and maximized at either $(\alpha, p_A^*) = (0.99,0.3)$ or $(\alpha, p_A^*) = (0.01,0.7)$.

\begin{figure}[!htb]
	\centering
	\includegraphics[width=0.9\textwidth]{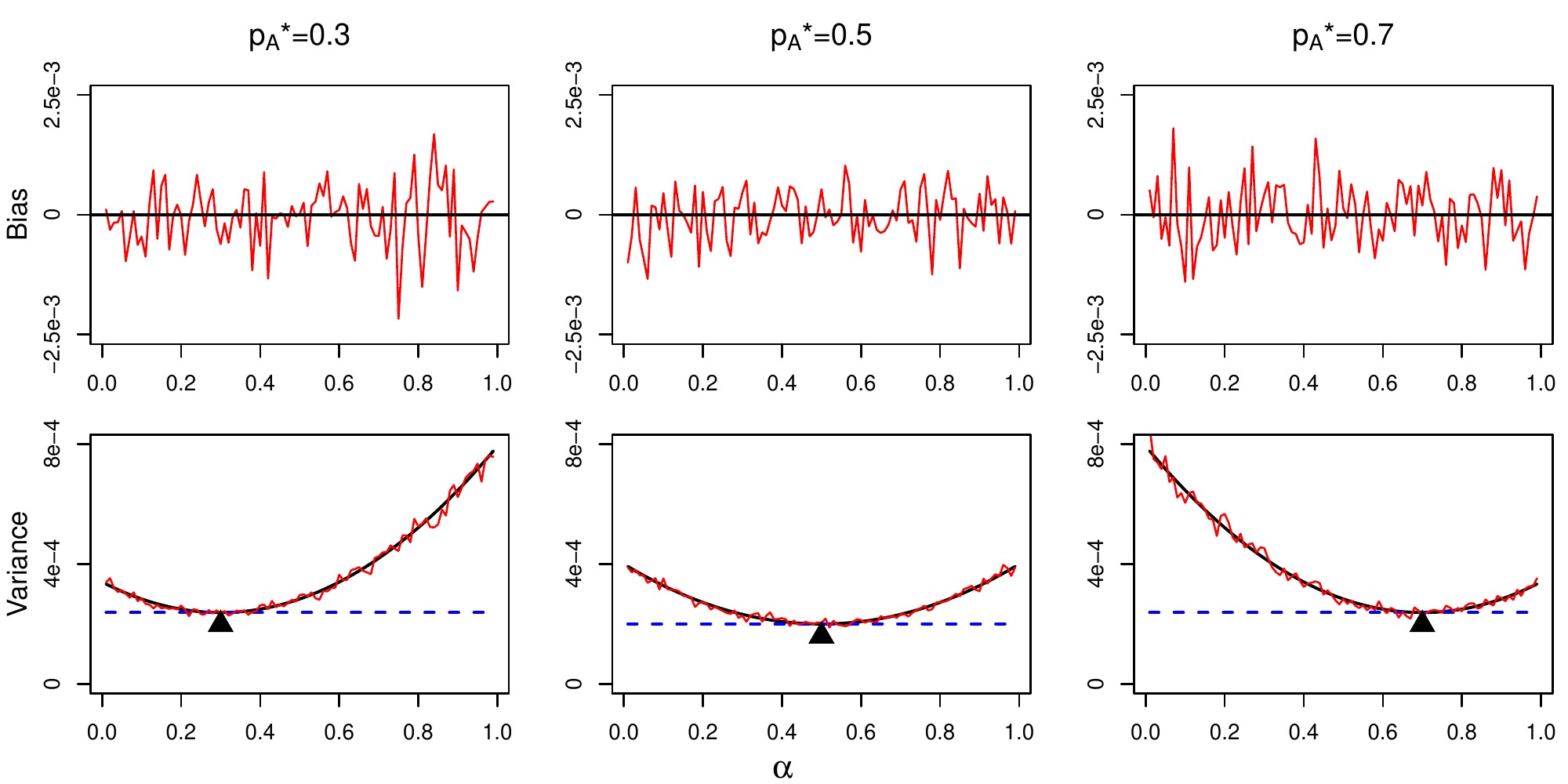}
   \caption{Graphical illustration of adaption under no interference. Left, middle, and right plots correspond to treatment assignment probabilities $p_A^*=0.3$, $0.5$, and $0.7$, respectively. Top plots show the empirical biases of direct effect estimates and bottom plots show the theoretical and empirical variances of direct effect estimates. The $x$ axis represents the policy parameter $\alpha$ and the $y$ axis represents either bias or variance. The black solid lines(\rule[0.5ex]{0.5cm}{1pt}) are the theoretical biases and variances of $\widehat{\oT}^\DE(\alpha)$. The red solid lines({\color{red}\rule[0.5ex]{0.5cm}{1pt}}) are the empirical biases and variances of $\widehat{\oT}^\DE(\alpha)$ from the simulation. The blue dashed lines({\color{blue}\rule[0.5ex]{0.1cm}{1pt}\hspace{0.1cm}\rule[0.5ex]{0.1cm}{1pt}\hspace{0.1cm}\rule[0.5ex]{0.1cm}{1pt}}) are the semiparametric efficiency bounds of $\oT^\ATE$. 
The filled triangle (\mytriangle{black}) is the policy parameter $\alpha$ that equals $p_A^*$. 
}
   \label{Fig:3}
\end{figure}

Next, Figure \ref{Fig:3} summarizes the numerical results from our simulation. We see that the empirical biases of $\widehat{\oT}^\DE(\alpha)$ 
are negligible and centered around zero, agreeing with the theory developed in Theorem \ref{thm:RobustATE}-(i); the empirical biases of $\widehat{\oT}^\IE(\alpha, 0.9)$, which are not reported, are also negligible. 
Also, the empirical variances of $\widehat{\oT}^\DE(\alpha)$ agrees with our theoretical discussion above where the variances are minimized when $\alpha = p_{A}^*$ and maximized when $\alpha$ is near the  ``edges'' of the plots. In short, the estimator is consistent, but is not always efficient when the true data has no interference. More generally, the results highlight that unlike non-interference settings, knowing the interference pattern may affect the efficiency of an estimator designed for interference and future work should be cognizant of this phenomena.

\subsection{Details of Section \ref{sec:Estimator} in the Main Paper}				\label{sec:DetailSec4}

Based on model in Section \ref{sec:GLMM}, we conduct a small simulation study to illustrate the theoretical results. Suppose we have two cluster types (i.e., $\NT=2$) and the cluster size of the two types is three and four (i.e., $\NI_1=3$ and $\NI_2=4$). We assume that the true data generating model has the following form.
\begin{align*}
	& 
	(\text{Model of } X) && \bX_{ij} = ( C_{i} , W_{ij} ) \T = ( C_{i} , W_{ij1} , W_{ij2}  ) \T  , \ 
	\\
	& &&
	C_{i} \stackrel{\text{i.i.d.}}{\sim} N(0,1)  , \
	W_{ij1} \stackrel{i.i.d}{\sim} \text{Ber}(0.5) , \ W_{ij2} \stackrel{\text{i.i.d.}}{\sim} N(0,1)  , \
	\\
	& &&
	C_i , W_{ij1}, \text{ and } W_{ij2} \text{ are independent}
	\\
		& 
	(\text{Model of } A) && 
		{\rm logit} 
		\big\{
			{\rm pr}(A_{ij} = 1 \cond \bX_i, L_i = k, b_i) 
		\big\}
		=
		\Big[ 1, \bW_{ij}\T,  \sum_{\ell \neq j} \bW_{i\ell}\T \Big] \paraPStype{k} + b_i
		\\
		& &&
		\paraPStype{1} = \big[ -1.25, 2, 0.3, 0.2, 0.1 \big]\T
		\ , \
		\paraPStype{2} = \big[ -1, 1.25, 0.2, 0.15, 0.1 \big]\T
		\ , \
		\\
		& &&
		b_i \stackrel{\text{i.i.d.}}{\sim} N(0, 0.25)
	 \\
	& 
	(\text{Model of } Y) && 
	Y_{ij} 
		=
		\Big[ 1, A_{ij}, \sum_{\ell \neq j} A_{i\ell} , A_{ij} C_i  , \Big( \sum_{\ell \neq j} A_{i\ell} \Big) C_i  , \bX_{ij}\T , \sum_{\ell \neq j} \bW_{i\ell} \T \Big]  \paraORtype{k}
		+ \xi_i + \epsilon_{ij}
		\\
		& &&
	\paraORtype{1} 
		=
		\big[
			2, 3, 0.8, 1, 0.5, 0.8, -1,0.5,-0.3,0.15
		\big] \T
		\ , \
		\\
		& &&
		\paraORtype{2}
		=
		\big[
			1,2, 0.4, 0.5, 0.3 ,0.6,-0.8,0.4,-0.2,0.1
		\big] \T
		\ , \
		\\
		& &&
		 \xi_i \stackrel{\text{i.i.d.}}{\sim} N(0, 0.1) \ , \ \epsilon_{ij} \stackrel{\text{i.i.d.}}{\sim} N(0,1)
\end{align*}
Each study unit has thee pre-treatment covariates, one binary ($W_{ij1}$) and two continuous ($C_i$ and $W_{ij2}$), where $W_{ij1}$ and $W_{ij2}$ are individual-level covariates and $C_i$ is a cluster-level covariate and the three covariates are independent of each other. Treatment is generated from a logistic mixed effects regression model.
The outcome is generated from a linear mixed effect that has interactions between covariates and treatment indicator. 
We generate $\NC =2,000$ samples and compute $\widehat{\oT}^\DE(0.4)$ and $\widehat{\oT}^\IE(0.8,0.2)$. 
The true values are $\oT^\DE(0.4)=2.75$ and $\oT^\IE(0.8,0.2)=0.9$, respectively. 

We consider two following specifications for the outcome regression estimation: (CO) the outcome regression is correctly specified; (MO) the outcome regression is mis-specified as the model below:
\begin{align*}
	Y_{ij} 
		=
		\Big[ 1, A_{ij}, \sum_{\ell \neq j} A_{i\ell} , \bZ_{ij}\T , \sum_{\ell \neq j} \bZ_{i\ell} \T \Big]  \paraORtype{k}
		+ \xi_i + \epsilon_{ij}; 
\end{align*}
Similarly, we consider two following specifications for the propensity score estimation: (CP) the propensity score is correctly specified; (MP) the propensity score is mis-specified as the model below:
\begin{align*}
		&
		{\rm logit} 
		\big\{
			{\rm pr}(A_{ij} = 1 \cond \bX_i, L_i = k, b_i) 
		\big\}
		=
		\Big[ 1, \bZ_{ij}\T,  \sum_{\ell \neq j} \bZ_{i\ell}\T \Big]  \paraPStype{k} + b_i;
		\end{align*}
Finally, we consider three specification for the cluster type: (CT) cluster type is correctly specified based on $\NI_i$ with $\NT=2$; (MT) cluster type is mis-specified where all clusters are considered to be generated from the same distribution with $\NT=1$; (OT) cluster type is over-specified based on the combination of $\NI_i$ and $\ind (C_i < 1.5)$ with $\NT=4$. 

Among possible combinations, we report the following six model specification scenarios. First, we report four cases when the outcome regression and propensity score are correctly specified and mis-specified, respectively, under correctly specified cluster type; these cases are denoted by (CO,CP,CT), (CO,MP,CT), (MO,CP,CT), and (MO,MP,CT), respectively. Second, we report two cases when cluster type is mis-specified and over-specified under correctly specified outcome regression and propensity score; these cases are denoted by (CO,CP,MT) and (CO,CP,OT), respectively. We repeat the simulation $1,000$ times. 


Table \ref{tab:sim1} reports the empirical bias, the empirical standard errors, and the coverage of 95\% confidence intervals based on the theory introduced in Section \ref{sec:GloEff} of the main paper. The empirical biases are negligible so long as either the outcome regression model or the propensity score is correctly specified. Also, the standard error is the smallest when both the outcome regression and the propensity score are correctly specified. When cluster types is over-specified, the estimates are consistent and asymptotically normal but inefficient. On the other hand, when cluster types are mis-specified, the estimates are inconsistent. Overall, despite the small simulation study, we believe the numerical results corroborate the theory presented in prior sections.

\begin{table}[!htp]
 \renewcommand{\arraystretch}{1.1} \centering
 \setlength{\tabcolsep}{2pt}
 \small
\begin{tabular}{|c|c|c|c|c|c|c|}
\hline
\multirow{2}{*}{Model specification} & \multicolumn{3}{c|}{$\DE(0.4)$}                          & \multicolumn{3}{c|}{$\IE(0.2,0.8)$}                      \\ \cline{2-7} 
                                     & Bias\,$(\times 10^{4})$ & SE\,$(\times 10^{2})$ & Coverage & Bias\,$(\times 10^{4})$ & SE\,$(\times 10^{2})$ & Coverage \\ \hline
(CO,CP,CT)                           & -3.88                  & 4.51                 & 0.957    & 6.83                   & 4.88                 & 0.959    \\ \hline
(CO,MP,CT)                           & -8.00                  & 4.78                 & 0.956    & -7.95                  & 5.37                 & 0.955    \\ \hline
(MO,CP,CT)                           & -16.47                 & 7.24                 & 0.961    & 6.48                   & 9.11                 & 0.957    \\ \hline
(MO,MP,CT)                           & 203.23                 & 7.99                 & 0.938    & 622.70                 & 10.55                & 0.883    \\ \hline
(MO,CP,MT)                           & -121.18                & 5.41                 & 0.951    & 224.87                 & 7.35                 & 0.956    \\ \hline
(MO,CP,OT)                           & -2.51                  & 4.88                 & 0.956    & 11.89                  & 5.91                 & 0.955    \\ \hline
\end{tabular}
\caption{Estimation results for $\DE(0.4)$ and $\IE(0.2,0.8)$.}
\label{tab:sim1}
\end{table}


Next, we present Algorithm \ref{al:CF} that shows the details of the cross-fitting procedure introduced in Section \ref{sec:NPOR} in the main paper.
\begin{algorithm}[!h]
\caption{Cross-fitting Procedure in Section \ref{sec:NPOR}} \label{al:CF}
\begin{tabbing}
   \qquad Let $\mathcal{D}_k $ be the set of indices that $\type_i=k$, i.e. $\mathcal{D}=\big\{ i \cond \type_i = k \big\}$.  \\
   \qquad Let $\mathcal{I}_{1,k}$ and $\mathcal{I}_{2,k}$ be randomly split two disjoint sets of $\mathcal{D}_k$.  \\
   \qquad Let $\mathcal{I}_1 = \bigcup_{k=1}^K \mathcal{I}_{1,k}$ and $\mathcal{I}_2 = \bigcup_{k=1}^K \mathcal{I}_{2,k}$, respectively. \\
   \enspace For each disjoint set $\mathcal{I}_{\ell}$ $(\ell=1,2)$: \\
   \qquad Estimate $g$ using data in $\mathcal{I}_\ell^c$. Let $\widetilde{\OR}_{(-\ell)}$ denote the estimated outcome regression. \\
   \qquad Estimate $e$ using data in $\mathcal{I}_\ell^c$. Let $\widetilde{e}_{(-\ell)}$ denote the estimated propensity score. \\
   \qquad Evaluate $\widetilde{\OR}_{(-\ell)}(\ba_i, \bX_i, k) $ and $\widetilde{e}_{(-\ell)}(\ba_i, \bX_i, k)$ for $\ba_i \in \zosets(\NI_k)$ and $i \in \mathcal{I}_\ell$.  \\\
\enspace Compute $\widetilde{\theta}_k$ and $\widetilde{\tau}$ in Section \ref{sec:NPOR} of the main paper. 
\end{tabbing}
\end{algorithm}

\subsection{Details of Section \ref{sec:application} in the Main Paper}									\label{sec:detailapplication}

We present the details of the data analysis in Section \ref{sec:application}. First, Table \ref{tab:Colombia} shows the exact distribution of treatment assignment in our analysis stratified by household size.				
\begin{table}[!htp]
		\renewcommand{\arraystretch}{1.1} \centering
		\small
\begin{tabular}{|c|c|c|c|c|c|c|c|c|}
\hline
\multicolumn{2}{|c|}{\multirow{2}{*}{}}                                                          & \multicolumn{6}{c|}{\begin{tabular}[c]{@{}c@{}}Number of Treated Children\\ in a Household\end{tabular}} & \multirow{2}{*}{Total} \\ \cline{3-8}
\multicolumn{2}{|c|}{}                                                                           & 0                 & 1                 & 2               & 3              & 4             & 5             &                        \\ \hline
\multirow{4}{*}{\begin{tabular}[c]{@{}c@{}}Number of Children\\ in a Household\end{tabular}} 
                                                                                             & 2 & 127               & 408               & 376             & -              & -             & -             & 911                    \\ \cline{2-9} 
                                                                                             & 3 & 3                 & 23                & 40              & 26             & -             & -             & 92                      \\ \cline{2-9} 
                                                                                             & 4 & 0                 & 0                 & 1               & 5              & 2             & -             & 8                      \\ \cline{2-9} 
                                                                                             & 5 & 0                 & 0                 & 0               & 0              & 1             & 0             & 1                      \\ \hline
\multicolumn{2}{|c|}{Total}                                                                      & 1748              & 2943              & 417             & 31             & 3             & 0             & 4790                   \\ \hline
\end{tabular}
\caption{Study Design of Conditional Cash Transfer Program in \citet{BO2011}. Each cell shows the number of households in the dataset with the total number of children in a household (row) and the number of treated children (column).}
\label{tab:Colombia}
\end{table}

We include the following methods and the corresponding R packages in our super learner library: linear regression via \texttt{glm}, lasso/elastic net via \texttt{glmnet} \citep{glmnet}, spline via \texttt{earth} \citep{earth} and \texttt{polspline} \citep{polspline}, generalized additive model via \texttt{gam} \citep{gam}, boosting  via \texttt{xgboost} \citep{xgboost} and \texttt{gbm} \citep{gbm}, random forest via \texttt{ranger} \citep{ranger}, and neural net  via 
\texttt{RSNNS} \citep{RSNNS}. 
Next we apply sample splitting \citep{Victor2018} by splitting the data into two folds and assigning one fold as the main sample and the other fold as the auxiliary sample. Third, we further split the main sample into the training and test sets. Using the training set, we obtain candidate estimates for $\OR$ from each method and we obtain an ensemble estimate $\widetilde{g}_{(-\ell)}$ by evaluating the performance on the test set. The ensemble estimate $\widetilde{g}_{(-\ell)}$ is evaluated at the auxiliary sample to construct $\widetilde{\oT}^\DE(\alpha)$ and $\widetilde{\oT}^\IE(\alpha,\alpha^*)$. 

\newpage

\section{Proof of the Lemmas and Theorems in the Main Paper}											\label{sec:appendix2}

\subsection{Notation}

To help guide the proof, we introduce all the notations used throughout the paper and the supplementary materials in a table. They are roughly listed in the order of appearance in the main paper.

		\begin{table}[!htp]
		\small
		\hspace*{-1cm}
		\renewcommand{\arraystretch}{1.1} \centering
		\begin{tabular}{|c|l|} \hline
		Notation & Definition \\ \hline\hline
		$\NT$ & Number of cluster types. \\ \hline
		$\NC$ & Number of clusters. \\ \hline
		$\NC_k$ & Number of clusters from cluster type $k$. \\ \hline
		$\NI_k$ & Size of cluster type $k$ (i.e., number of units in cluster type $k$).  \\ \hline
		$\zosets(t)$ & Collection of $t$-dimensional binary vectors (e.g., $\zosets(2) = \{(0,0),(0,1),(1,0), (1,1) \}$). \\ \hline
		$\mathcal{X}(k)$ & Finite dimensional support of the pre-treatment covariates from cluster type $k$. \\ \hline
		$Y_{ij}$ & Univariate outcome of unit $j$ in cluster $i$. \\ \hline
		$A_{ij}$ & Treatment indicator of unit $j$ in cluster $i$.  \\ \hline
		$\bX_{ij}$ & Pre-treatment covariate of unit $j$ in cluster $i$. \\ \hline 
		$\type_i$ & Cluster type variable. $\type_i \in \{ 1,\ldots , \NT\}$. \\ \hline
		$\bY_i$ & Vectorized outcomes of cluster $i$.  $\bY_i=(Y_{i1},\ldots,Y_{i\NI_k} )\T \in \R^{\NI_k}$ for $\type_i = k$. \\ \hline
		$\bA_i$ & Vectorized treatment indicator of cluster $i$. $\bA_i =(A_{i1},\ldots,A_{i\NI_k} )\T \in \zosets(\NI_k)$  for $\type_i = k$ \\ \hline
		$\bA_\eij$ & Vector of treatment indicators for all units in cluster $i$ except unit $j$. $\bA_\eij \in \zosets(\NI_k-1)$. \\ \hline
		$\bX_i$ & Vectorized pre-treatment vector of cluster $i$. $\bX_i=(\bX_{i1}\T,\ldots,\bX_{i\NI_k}\T )\T \in \mathcal{X}(k)$ for $\type_i = k$. \\ \hline
		$\bO_i$ & All the observed data from cluster $i$. $\bO_i=(\bY_i , \bA_i, \bX_i)$. \\ \hline
		$\ba_i , \ba$ & Realized value of $\bA_i$. \\ \hline
		$a_{ij}, \ba_\eij$ & Realized values of $A_{ij}$ and $\bA_\eij$, respectively. \\ \hline
		$Y_{ij} (\ba_i) $ & Potential outcome of unit $j$ in cluster $i$ under treatment vector $\ba_i \in \zosets(\NI_k)$ for $\type_i = k$. \\ \hline
		$\bY_i(\ba_i)$ & Vectorized potential outcomes of cluster $i$ under $\ba_i$. $\bY(\ba_i)=(Y_{i1}(\ba_i) , \ldots , Y_{i \NI_k}(\ba_i) )\T  \in \R^{\NI_k}$ for $\type_i = k$ \\ \hline
		$ \Parasetk_k$ & Cluster level parameter space; see Section \ref{sec:potout} in the main paper. \\ \hline
		$\bw_k(\ba , \bx )$ &  Weight vector associated with cluster level parameter at treatment vector $\ba \in \zosets(\NI_k)$,\\ 		
		$\bw_k(\ba, \bx \con \alpha_k , \alpha_k')$ &  pre-treatment covariate vector $\bx \in \mathcal{X}(k)$, and treatment allocation strategies $(\alpha, \alpha')$. \\ \hline 		
		$\uT_k $ & Cluster level parameter associated with cluster type $k$. \\ 
		$=\uT_k(\alpha_k , \alpha_k')$ & $\uT_k(\alpha_k ,\alpha_k') = \sum_{\ba} \bw_k\T(\ba \con \alpha_k , \alpha_k) \EXP \{ \bY_i(\ba) \cond \type_i = k \} \in \Parasetk_k$.  \\ \hline
		$\bT= \bT(\balpha,\balpha')$ & Vectorized cluster level parameter. $\bT(\balpha,\balpha') = (\uT_1(\alpha_1,\alpha_1'), \ldots , \uT_\NT(\alpha_\NT,\alpha_\NT'))\T$ where $\uT_k(\alpha_k , \alpha_k') \in \Parasetk_k$. \\ \hline
		$p_k$ & Cluster type probability. $p_k = \EXP\{ \ind(\type_i = k) \}$ \\ \hline
		$\bp$ & Vectorized cluster type probability. $\bp = (p_1, \ldots , p_\NT)\T$. \\ \hline
		$ \Paraset$ & Super-population level parameter space; see Section \ref{sec:potout} in the main paper. \\ \hline
		$v_k(p_k)$ & Weight function associated with super-population level parameter space. \\ \hline
		$\bv(\bp)$ & Vectorized $v_k(p_k)$. $\bv(\bp) = ( v_1(p_1) , \ldots , v_\NT (p_\NT))\T$.
		\\ \hline
		$\oT = \oT(\balpha, \balpha')$ & Super-population level parameter. $\oT(\balpha, \balpha') = \bv\T(\bp)\bT(\balpha,\balpha') = \sum_{k=1}^\NT v_k(p_k) \uT_k(\alpha_k , \alpha_k') \in \Paraset$.
		\\ \hline
		$\oT^\DE(\alpha)$ & Direct effect under $\alpha$-policy, i.e. $\tau^\DE(\alpha) = \EXP \big\{ \overline{Y}_i ( 1 \con \alpha)   - \overline{Y}_i ( 0 \con \alpha) \big\}$. \\ \hline
		$\oT^\IE(\alpha, \alpha')$ & Indirect effect under $(\alpha,\alpha')$-policy, i.e. $\tau^\IE(\alpha, \alpha') = \EXP \big\{ \overline{Y}_i ( 0 \con \alpha)   - \overline{Y}_i ( 0 \con \alpha') \big\}$. \\ \hline
	\end{tabular}
	\vspace*{-1cm}
	\end{table}
	
	\newpage

		\begin{table}[!htp]
		\small
		\hspace*{-1cm}
		\renewcommand{\arraystretch}{1.1} \centering
		\begin{tabular}{|c|l|} \hline
		Notation & Definition \\ \hline\hline
		$e(\ba \cond \bx , k)$ & Propensity score in cluster type $k$. $e(\ba \cond \bx , k ) = {\rm pr}(\bA_i = \ba \cond \bX_i = \bx , \type_i = k)$. \\ \hline
		$e_j(a \cond \bx , k)$ & Propensity score of unit $j$ from cluster type $k$. $e_j(a \cond \bx , k ) = {\rm pr}(A_{ij} = a \cond \bX_i = \bx , \type_i = k)$. \\ \hline
		$\OR(\ba, \bx, k)$ & Outcome regression in cluster type $k$. $\OR(\ba, \bx , k ) = \EXP(\bY_i \cond \bA_i = \ba , \bX_i = \bx , \type_i = k)$. \\ \hline
		$\indOR_j(\ba, \bx, k)$ & Outcome regression of unit $j$ from cluster type $k$. $\indOR_j(\ba, \bx , k ) = \EXP(Y_{ij} \cond \bA_i = \ba , \bX_i = \bx , \type_i = k)$. \\ \hline
		$\Sigma(\ba, \bx, k)$ & Conditional variance matrix of the outcome vector.  $\Sigma(\ba , \bx , k) = \VAR( \bY_i \cond \bA_i = \ba , \bX_i = \bx, \type_i = k )$. 	\\ \hline
		Superscript $^*$ & True value of parameters and functions. (e.g, $\bT^*$, $\oT^*$, $e^*$, $\OR^*$)
		\\ \hline
		$\varphi(\bT^*) , \ \varphi(\oT^*)$ & Efficient influence function of $\bT^*$ and $\oT^*$, respectively.
		\\ \hline
		\multirow{2}{*}{$\modeliv$} & Nonparametric model space satisfying equation \eqref{eq:NPmodel} in the main paper; \\ 
		& see Section \ref{sec:EffNonpar} in the main paper.  \\ \hline
		\multirow{2}{*}{$\modelie$} & Parametric submodel of $\modeliv$ with the correctly specified propensity score; \\
		& see Section \ref{sec:Mestimation-general} in the main paper. \\ \hline
		\multirow{2}{*}{$\modelig$} & Parametric submodel of $\modeliv$ with the correctly specified outcome regression; \\
		& see Section \ref{sec:Mestimation-general} in the main paper.  \\ \hline
		$\modelii$ & Nonparametric model space without interference; see Section \ref{sec:detail5}. \\ \hline
		$\modeliig$ & Parametric submodel of $\modelii \cap \modelig$; see Section \ref{sec:detail5}. \\ \hline
		$\paraPS$ & Propensity score parameter. \\ \hline
		$\paraOR$ & Outcome regression parameter. \\ \hline		
		$\paraT$ & Collection of the propensity score and the outcome regression parameter. $\paraT = (\paraPS \T , \paraOR \T)\T$
		\\ \hline
		$\Pe(\ba \cond \bx , k \con \paraPS )$ & Propensity score of cluster type $k$  parametrized by $\paraPS$. \\ \hline
		$\POR(\ba , \bx , k \con \paraOR )$ & Outcome regression of cluster type $k$ parametrized by $\paraPS$.
		\\ \hline
		$\ee$ & Entire estimating equation. \\ \hline 
		$\ee_p$ & Estimating equation to estimate $\bp$. $\ee_p = ( \ee_{p,1} , \ldots ,\ee_{p,\NT})\T $; see equation \eqref{proof:eep}. \\ \hline
		$\ee_\uT$ & Estimating equation to estimate $\bT$. $\ee_\uT = ( \ee_{\uT,1} , \ldots , \ee_{\uT,\NT})\T$; see Section \ref{sec:Mestimation-general} in the main paper. \\ \hline
		$\ee_\beta$ & Estimating equation to estimate $\paraT$. $\ee_\beta = ( \ee_e \T , \ee_\indOR\T)\T$; see Section \ref{sec:Mestimation-general} in the main paper. \\ \hline
		\multirow{2}{*}{$\widehat{\bp} , \widehat{\bT} , \widehat{\paraT}$}  & Solution to the estimation equation $0 = \sum_{i=1}^\NC \ee_\bp$, $0 = \sum_{i=1}^\NC \ee_\uT$, $0 = \sum_{i=1}^\NC \ee_\beta$, respectively;\\
		& see Section \ref{sec:Mestimation-general} in the main paper, \ref{sec:prooflmm41}, and \ref{sec:proofthm42}.
		\\ \hline
		$\ePe(\ba \cond \bx ,k)$ & Parametrically estimated propensity score. $\ePe(\ba \cond \bx ,k) = \Pe(\ba \cond \bx , k \con \estparaPS{})$.
		\\ \hline
		$\ePOR(\ba, \bx ,k)$ & Parametrically estimated outcome regression. $\ePOR(\ba, \bx ,k) = \POR(\ba, \bx , k \con \estparaOR{})$.
		\\ \hline
		$\paraT^\dagger$ & Probability limit of $\widehat{\paraT}$.
		\\ \hline
		$\varphi^{\rm Par} (\bT^* , \paraT^\dagger)$ & 
		Influence function of $\bT^*$ obtained from the M-estimation.  \\ \hline
		$\varphi^{\rm Par} (\oT^* , \paraT^\dagger)$ & 
		Influence function of $\oT^*$ obtained from the M-estimation.  
		\\ \hline
		\multirow{2}{*}{$\indIOR_j(a, \bx, k)$} & Outcome regression of unit $j$ from cluter type $k$ without interference. \\
		& $\indIOR_j(a, \bx, k) = \EXP(Y_{ij} \cond A_{ij} = a , \bX_i = \bx, \type_i = k)$. \\ \hline
	$\indPIOR_j(a, \bx, k \con \paraOR)$ & Outcome regression of unit $j$ from cluter type $k$ without interference parametrized by $\paraOR$. \\ \hline
	$\oT^\ATE$ & Average treatment effect under the absence of interference; see Section \ref{sec:detail5}. \\ \hline
	$\mathcal{T}$ & Tangent space for a model. \\ \hline
	$a \lesssim (\gtrsim) \, b$ & For some constant $C$ independent of $a$ and $b$,  $a \leq (\geq)\,  C \cdot  b$ holds. \\ \hline
	\end{tabular}
	\end{table}

\subsection{Proof of Theorem \ref{thm:EIFVC} in the Main Paper}

The proof is similar to the proof of Lemma \ref{lmm:EIFbasic}. The density of $(\bO_i , \type_i) = (\bY_i, \bA_i, \bX_i, \type_i)$ with respect to some $\sigma$-finite measure is 
\begin{align*}
	P^* ( \by , \ba , \bx , k) 
	& = \py^* (\by \cond \ba , \bx , k) e^*(\ba \cond \bx , k) \px^* ( \bx \cond k) p_k^*
\end{align*}
where $\py^*$ is the conditional density of $\bY_i$ given $(\bA_i, \bX_i, \type_i)$ and $\px^*$ is the conditional density of $\bX_i$ given $\type_i$. An asterisk in superscript of (conditional) density represents the true (conditional) density. A smooth regular parametric submodel parametrized by a possibly multi-dimensional parameter $\eta$ is
\begin{align*}
	P ( \by , \ba , \bx , k \con \eta) 
	& = \py (\by \cond \ba , \bx , k \con \eta) e(\ba \cond \bx , k \con \eta) \px ( \bx \cond k \con \eta) p_k(\eta)  
\end{align*}
where the smoothness and regularity conditions are given in Definition A.1 of the appendix in \citet{Newey1990}. We assume the density of the parametric submodel $P(\cdot \con \eta)$ equals the true density $P^*$ at $\eta=\eta^*$. The corresponding score function is
\begin{align}												\label{proof:1-002}
	s (\by,\ba,\bx, k \con \eta) 
	&=  s_Y( \by , \ba , \bx, k \con \eta ) + s_A( \ba , \bx , k \con \eta)  +  s_X ( \bx , k \con \eta) + s_\type( k \con \eta)  
\end{align}
where
\begin{align}											\label{proof:1-score}
	& s_Y(\by, \ba , \bx , k \con \eta) = \frac{\partial}{\partial \eta} \, \log \py ( \by \cond \ba , \bx , k \con \eta) \ ,
	&& s_A( \ba , \bx , k \con \eta) = \frac{\partial}{\partial \eta} \, \log e (  \ba \cond \bx , k \con \eta)\ , \\ 
	\nonumber
	& s_X(\bx , k \con \eta) = \frac{\partial}{\partial \eta} \, \log \px( \bx \cond k \con \eta)\ ,
	&& s_\type(k \con \eta) = \frac{\partial }{\partial \eta} \, p_k(\eta) \ .
\end{align}
The 1-dimensional tangent space for $1$-dimensional parameters is
\begin{align}							\label{proof:2-001}
	\nonumber 
	\mathcal{T} =
	\Big\{
		 S (\by, \ba, \bx, k)  \in \R \,
		      \Big| 	\,	      
			& S(\by , \ba , \bx , k) 	   
			= S_Y (\by , \ba, \bx, k)  + S_A (\ba, \bx, k)  + S_X ( \bx, k)  + S_\type (k)  \ , \
			\\
			\nonumber
		      \, &  \EXP \big\{ S_Y (\bY_i , \ba, \bx, k) \cond \bA_i = \ba, \bX_i =  \bx , \type_i = k \big\} = 0 \text{ for all } (\ba,  \bx ,  k) \ , \   \\
		      \nonumber
	    	\, &   \EXP \big\{ S_A (\bA_i , \bx , k ) \cond \bX_i = \bx , \type_i = k \big\} = 0  \text{ for all } ( \bx,  k ) \  , \  \\
	  \, &  	\EXP \big\{ S_X (\bX_i , k ) \cond \type_i = k \big\} = 0  \text{ for all } k \ , \   \EXP\big\{ S_\type (\type_i ) \big\} = 0
	    \Big\} \ .
\end{align}

The estimand $\oT^*$ is re-represented as $ \oT(\eta) = \sum_{k=1}^\NT v_k\big( p_k(\eta) \big) \uT_k(\eta)$ at parameter $\eta$ in the regular parametric submodel where $\uT_k(\eta)$ has the following functional form.
\begin{align}									\label{proof:1-taueta}
	\uT_k(\eta)
	& = \sum_{\ba \in \zosets(\NI_{k}) } \Bigg[ \iint  \big\{ \bw_k \T (\ba,\bx) \by  \big\} \py(\by \cond \ba , \bx , k \con \eta)  \px(\bx \cond k \con \eta) \, d \by \, d \bx \bigg] \ .
\end{align}
We find that $\oT(\eta^*)$ equals the true $\oT^*$. Thus, the derivative of $\oT$ evaluated at true $\eta^*$ is 
\begin{align}						\label{proof:2-002}
	\frac{\partial \oT(\eta^*)}{\partial \eta} 
	= 
	\sum_{k=1}^\NT v_k(p_k^*) \frac{\partial \uT_k ( \eta^* )  }{\partial \eta}
	+
	\sum_{k=1}^\NT \uT_k^* \frac{\partial v_k ( p_k^* ) }{\partial \eta} \ .
\end{align}
The conjectured efficient influence function  of $\oT^*$ is
\begin{align}						\label{proof:2-EIF}
	\varphi(\oT^*) 
	& = \sum_{k=1}^\NT v_k(p_k^*) \cdot  \varphi_k ( \uT_k^* )
		 + \sum_{k=1}^\NT \Big\{ \ind(\type_i=k) -p_k^* \Big\} \frac{\partial v_k(p_k^*)}{\partial p_k }   \uT_k^*  \ .
\end{align}

First, we show that $\oT(\eta)$ is a differantiable parameter, i.e.,
\begin{align}				\label{proof:2-003}
	\frac{\partial \oT (\eta^*) }{\partial \eta} 
	& = 
	\EXP
	\Big\{
		\varphi (\oT^*) \cdot s  (\bY_i, \bA_i, \bX_i, \type_i \con \eta^* )
	\Big\} \\
	\nonumber	
	& =
	\sum_{k=1}^\NT  \EXP\Big\{ v_k(p_k^*) \cdot  \varphi_k ( \uT_k^* ) \cdot s  (\bY_i, \bA_i, \bX_i, \type_i \con \eta^* ) \Big\}  \\
	& \hspace*{2cm} + \sum_{k=1}^\NT  \EXP\bigg[ \Big\{ \ind(\type_i=k) -p_k^* \Big\} \frac{\partial v_k(p_k^*)}{\partial p_k }   \uT_k^*  \cdot s  (\bY_i, \bA_i, \bX_i, \type_i \con \eta^* ) \bigg] \ .
	\nonumber	
\end{align}
From the identity \eqref{proof:1-007}, we obtain an equivalence between the first pieces of \eqref{proof:2-002} and \eqref{proof:2-003}.
\begin{align}                                  \label{proof:2-004}
	\sum_{k=1}^\NT  \EXP\Big\{ v_k(p_k^*) \cdot  \varphi_k ( \uT_k^* ) \cdot s  (\bY_i, \bA_i, \bX_i, \type_i \con \eta^* ) \Big\}
	= 
	\sum_{k=1}^\NT v_k(p_k^*) \cdot \frac{\partial \uT_k ( \eta^* )  }{\partial \eta} \ .
\end{align}
For the equivalence between the second pieces of \eqref{proof:2-002} and \eqref{proof:2-003}, we find
\begin{align}                                   \label{proof:2-005}
	\nonumber 
	 &  \sum_{k=1}^\NT \EXP  \bigg[
		  \Big\{ \ind(\type_i=k) - p_k^* \Big\} \frac{\partial v_k ( p_k^* ) }{\partial p_k }  \uT_k^* \cdot s (\bY_i , \bA_i , \bX_i , \type_i \con \eta^* ) 
	\bigg]
	\\
	\nonumber & = \sum_{k=1}^\NT p_k^*  \frac{\partial v_k (p_k^*) }{\partial p_k } \uT_k^*  \cdot \EXP\Big\{ s (\bY_i , \bA_i , \bX_i , k \con \eta^* )  \, \Big| \, \type_i = k \Big\}  
	-  \bigg\{ \sum_{k=1}^\NT p_k^*  \frac{\partial v_k(p_k^*) }{\partial p_k }  \uT_k^*  \bigg\} \cdot \EXP \Big\{ s (\bY_i , \bA_i , \bX_i , \type_i \con \eta^* ) \Big\} \\
	 & = \sum_{k=1}^\NT p_k^*  \frac{\partial v_k(p_k^*) }{\partial p_k }  \uT_k^*   \cdot s_\type ( k \con \eta^* )  
	= \sum_{k=1}^\NT p_k^*  \frac{\partial v_k(p_k^*) }{\partial p_k }  \uT_k^*  \cdot \frac{\partial p_k(\eta^*)}{\partial \eta} \frac{1}{p_k^*} 
	= \sum_{k=1}^\NT \uT_k^*  \frac{\partial v_k ( p_k^* ) }{\partial \eta} 
	 \ .
\end{align}
The first identity is based on the total law of expectation. The second identity is from the definition of $s$ in \eqref{proof:1-002} and the property of score functions. The third identity is from the definition of $s_\type$. The fourth identity is straightforward from the chain rule. Combining \eqref{proof:2-004} and \eqref{proof:2-005}, we arrive at \eqref{proof:2-003}, i.e., $\oT(\eta)$ is a differantiable parameter. 

Next we show $\varphi(\oT^*) $ belongs to $\mathcal{T}$ in \eqref{proof:2-001} by showing that each piece of $\varphi(\oT^*)$ satisfies the conditions  imposed on $\mathcal{T}$ in \eqref{proof:2-001}. Since the first piece of $\varphi(\oT^*) $ in \eqref{proof:2-EIF} is a linear combination of $\varphi_k(\uT_k^*)$s, the first piece of $\varphi(\oT^*) $ in \eqref{proof:2-EIF} also satisfies the same conditions and belongs to $\mathcal{T}$ in \eqref{proof:2-001}. To show that the second piece of $\varphi(\oT^*) $ in \eqref{proof:2-EIF} also belongs to $\mathcal{T}$ in \eqref{proof:2-001}, we check the mean-zero condition on $S_\type$.
\begin{align*}
	\EXP \bigg[
	\sum_{k=1}^\NT \Big\{ \ind(\type_i=k) -p_k^* \Big\} \frac{\partial v_k(p_k^*)}{\partial p_k }   \uT_k^* \bigg] 
	= \sum_{k=1}^\NT (p_k^* - p_k^*) \frac{\partial v_k(p_k^*)}{\partial p_k }   \uT_k^* = 0 
	 \ .
\end{align*}
Therefore, $\varphi(\oT^*) \in \mathcal{T} $ and, thus, $\varphi(\oT^*)$ is the efficient influence function  of $\oT^*$ by \cite{Newey1990}.

The semiparametric efficiency bound is the variance of $\varphi(\oT^*)$ which is equivalent to the expectation of the square of $\varphi(\oT^*)$. Therefore,
\begin{align}                                  \label{proof:2-006}
	\VAR \big\{ \varphi(\oT^*) \big\}
	\nonumber
	 = &
	\EXP \bigg[
		\bigg\{	\sum_{k=1}^\NT v_k(p_k^*) \varphi_k(\uT_k^*) \bigg\}^2
	\bigg]
	+
	\EXP \bigg[
		\bigg[
		\sum_{k=1}^\NT \Big\{ \ind(\type_i=k) -p_k^* \Big\} \frac{\partial v_k(p_k^*)}{\partial p_k }   \uT_k^*  \bigg]^2
	\bigg]
	\\
	& +
	2 \EXP \bigg[
	\bigg\{	\sum_{k=1}^\NT v_k(p_k^*) \varphi_k(\uT_k^*) \bigg\}
		\bigg[ \sum_{k=1}^\NT \Big\{ \ind(\type_i=k) -p_k^* \Big\} \frac{\partial v_k(p_k^*)}{\partial p_k }   \uT_k^*  \bigg]	\bigg]	
		 \ .
\end{align}

We study each term in \eqref{proof:2-006}. The first term is straightforward from the proof of Lemma \ref{lmm:EIFbasic}
\begin{align*}
	\EXP \bigg[
		\bigg\{	\sum_{k=1}^\NT v_k(p_k^*) \varphi_k(\uT_k^*) \bigg\}^2
	\bigg]
	=
	\bv \T (\bp^*) \EXP \big\{	 \varphi(\bT^*)\varphi \T (\bT^*) \big\} \bv (\bp^*) 
	=
	\sum_{k=1}^\NT v_k(p_k^*)^2 \text{SEB}_k (\uT_k^*)
	 \ . 
\end{align*}
The second term is represented as follows
\begin{align*}
	\nonumber
	 &
	\EXP \bigg[ 
		\bigg[
		\sum_{k=1}^\NT  \Big\{ \ind(\type_i=k) -p_k^* \Big\} \frac{\partial v_k(p_k^*)}{\partial p_k }   \uT_k^*  \bigg]^2
	\bigg]
	=  
	\sum_{k=1}^\NT p_k^* \bigg\{ \frac{\partial v_k(p_k^*) }{\partial p_k }  \uT_k^*  \bigg\}^2 - \bigg\{ 
			\sum_{k=1}^\NT p_k^* \frac{\partial v_k(p_k^*) }{\partial p_k }  \uT_k^*
		\bigg\}^2
		 \ .
\end{align*}
The last term is zero by observing the following
\begin{align*}
	\nonumber
	\EXP \bigg[
		\bigg\{	\sum_{k=1}^\NT & v_k(p_k^*) \varphi_k(\uT_k^*) \bigg\}
		\bigg[ \sum_{k=1}^\NT \Big\{ \ind(\type_i=k) - p_k^* \Big\} \frac{\partial v_k(p_k^*) }{\partial p_k } \uT_k^*  \bigg]
		 \bigg] 
		 \\
		\nonumber		 
		& = \EXP \bigg[
		\EXP \bigg\{	\sum_{k=1}^\NT v_k(p_k^*) \varphi_k(\uT_k^*) \, \bigg| \, \type_i \bigg\}
		\bigg[ \sum_{k=1}^\NT \Big\{ \ind(\type_i=k) - p_k^* \Big\} \frac{\partial v_k(p_k^*) }{\partial p_k } \uT_k^*  \bigg]
		 \bigg] 
		 \\
		 & = \EXP \bigg[
		0 \cdot \bigg[ \sum_{k=1}^\NT \Big\{ \ind(\type_i=k) - p_k^* \Big\} \frac{\partial v_k(p_k^*) }{\partial p_k } \uT_k^*  \bigg] \bigg] 
		 = 0
		  \ .
\end{align*}
Combining the result above, we get the explicit form of $\EXP \big\{  \varphi (\oT^*)^2  \big\}$.
\begin{align*}
	\VAR \big\{  \varphi(\oT^*)  \big\}
	& =
	\sum_{k=1}^\NT v_k(p_k^*)^2 \text{SEB}_k (\uT_k^*) 
	+
	\sum_{k=1}^\NT p_k^* \bigg\{ \frac{\partial v_k(p_k^*) }{\partial p_k }  \uT_k^*  \bigg\}^2 - \bigg\{ 
			\sum_{k=1}^\NT p_k^* \frac{\partial v_k(p_k^*) }{\partial p_k }  \uT_k^*
		\bigg\}^2
		 \ .
\end{align*}
This concludes the proof of the theorem.

	\subsection{Proof of Theorem \ref{thm:ParaPopEst} in the Main Paper}					\label{sec:proofthm42}

	We define the estimating equation of $\bp$.
	\begin{align}										\label{proof:eep}
		\ee_p(\bp) = \big( \ee_{p,1}(p_1) , \ldots , \ee_{p,\NT}(p_\NT) \big)\T
	\ ,\
		\ee_{p,k} (p_k) =	\ind(\type_i = k ) - p_k 
	\ , \ k = 1, \ldots , \NT
	 \ .
	\end{align}
	Therefore, $\widehat{\bp} = (\widehat{p}_1,\ldots,\widehat{p}_\NT)\T$ is the solution to the estimating equation $0 = \NC^{-1} \sum_{i=1}^\NC \ee_p(\widehat{\bp})$. The entire estimating equation is defined by $\ee ( \bp , \bT , \paraT ) 	=	\big(	\ee_p\T(\bp) , \ee_\uT\T(\bT , \paraT) ,\ee_\beta \T ( \paraT ) \big)\T$. It is straightforward to check that the regularity conditions on $\ee(\bT, \paraT)$ assumed in Lemma \ref{lmm:ParaEst} implies the regularity conditions on $\ee(\bp, \bT, \paraT)$. Hence, Theorem 5.41 of \citet{Vaart1998} gives the asymptotic result
	\begin{align}						\label{proof:7-000}
		\sqrt{\NC} \begin{bmatrix}
			\widehat{\bp} - \bp^* \\
			\widehat{\bT}  - \bT^* \\
			\widehat{\paraT} - \paraT^\dagger
		\end{bmatrix}
		= 
		- \frac{1}{\sqrt{\NC}} \bigg[  \underbrace{ \EXP \bigg\{ \frac{\partial \ee \big( \bp^* ,  \bT ^* , \paraT^\dagger \big) }{\partial \big( \bp, \bT  , \paraT \big)\T }  \bigg\}}_{(A)} \bigg]^{-1} \sum_{i=1}^\NC \ee \big( \bp^*, \bT^* , \paraT^\dagger \big) + o_P(1)
		 \ .
 	\end{align}
		Note that the expectation of the Jacobian matrix $(A)$ is 
	\begin{align*}
		(A) = 
		\EXP \bigg\{ \frac{\partial \ee \big( \bp^*, \bT ^* , \paraT^\dagger \big) }{\partial \big(\bp,  \bT  , \paraT \big)\T }  \bigg\} 
		= \begin{bmatrix}
			- I_\NT 
			& 0
			& 0 
			\\
			0 
			& 
			- \text{diag}(\bp^*) 
			& \displaystyle{ \EXP  \bigg\{	  \frac{\partial \ee_\uT \big(\bT^* , \paraT^\dagger \big)  }{\partial  \paraT \T }   \bigg\} } 
			\\[0.7cm]
			0 &
			0 & 
			\displaystyle{ \EXP  \bigg\{	\frac{\partial \ee_{\beta} (\paraT^\dagger)  }{\partial  \paraT \T }  \bigg\} } 
		\end{bmatrix}
	\end{align*}
	where $I_\NT$ is $\NT \times \NT$ identity matrix and $\text{diag}(\bp^*) = {\rm diag} \big[ p_1^*,\ldots,p_\NT^* \big]$. Therefore, we find 
	\begin{align*}
		(A)^{-1}		
		&
		=
		\bigg[ \EXP \bigg\{ \frac{\partial \ee \big( \bp^*, \bT ^* , \paraT^\dagger \big) }{\partial \big( \bp, \bT  , \paraT \big)\T }  \bigg\}  \bigg]^{-1} 
		\\
		&
		= \begin{bmatrix}
			-I_\NT 
			& 0 & 0 \\
			0 & 
			\displaystyle{ -  {\rm diag} (1/\bp^*) }
			& 
			\displaystyle{ {\rm diag} (1/\bp^*) \EXP  \bigg\{	  \frac{\partial \ee_\uT \big(\bT^* , \paraT^\dagger \big)  }{\partial  \paraT \T }   \bigg\}	\bigg[ \EXP  \bigg\{	\frac{\partial \ee_{\beta} (\paraT^\dagger)  }{\partial  \paraT \T }  \bigg\} \bigg]^{-1}		} \\[0.7cm]
			0 & 
			0 & \displaystyle{ \bigg[ \EXP  \bigg\{	\frac{\partial \ee_{\beta} (\paraT^\dagger)  }{\partial  \paraT \T }  \bigg\} \bigg]^{-1}}
		\end{bmatrix}
	\end{align*}
	where $\text{diag}(1/\bp^*) = {\rm diag} \big[ 1/p_1^*,\ldots,1/p_\NT^* \big]$. Replacing $(A)^{-1}$ in \eqref{proof:7-000} with the form above, we get the linear expansion of $(\widehat{\bp} , \widehat{\bT})$.
	\begin{align}					\label{proof:7-001}
		\sqrt{\NC} 
		\begin{bmatrix}
			\widehat{\bp} - \bp^* \\
			\widehat{\bT} - \bT^*
		\end{bmatrix}
		= 
		\frac{1}{\sqrt{\NC}} \sum_{i=1}^\NC \begin{bmatrix}
			\ee_p(\bp^*) \\
			\varphi^{\rm Par} (\bT^* , \paraT^\dagger)
		\end{bmatrix}
		+ o_P(1)
		 \ .
	\end{align}
	
	We consider a continuously differentiable function $h: \R^\NT \otimes \R^\NT \rightarrow \R$ with $h(\bp, \bT) = \bv \T (\bp) \bT$. Note that $h(\widehat{\bp} , \widehat{\bT} ) = \widehat{\oT}$ and $h(\bp^* , \bT^* ) = \oT^*$, respectively. Therefore, the standard delta method gives the asymptotic linear expansion of $\widehat{\oT} = h(\widehat{\bp} , \widehat{\bT} )$ at $\oT^* = h(\bp^* , \bT^* ) $.
	\begin{align}					\label{proof:7-002}
		\sqrt{\NC} \big(	\widehat{\oT} - \oT^* \big) 
		=
		\sqrt{\NC} \big\{ h(\widehat{\bp} , \widehat{\bT} ) - h(\bp^* , \bT^* ) \big\}
		= 
		\sqrt{\NC} \cdot 
		\big\{ \nabla  h  ( \bp^* , \bT^* ) \big\} \T
		\begin{bmatrix}
			\widehat{\bp} - \bp^* \\
			\widehat{\bT} - \bT^*
		\end{bmatrix}
		+ o_P(1)
	\end{align}
	where
	\begin{align*}
		\nabla  h  ( \bp , \bT )
		=
		\frac{\partial h  ( \bp , \bT )}{ \partial (\bp, \bT) }
		= 
		\bigg[
			\frac{\partial v_1(p_1) }{\partial p_1} \uT_1 , \ldots , \frac{\partial v_\NT(p_\NT) }{\partial p_\NT} \uT_\NT
			, v_1(p_1) , \ldots , v_\NT(p_\NT)
		\bigg] \T
		 \ .
	\end{align*}
	Combining \eqref{proof:7-001} and \eqref{proof:7-002}, we obtain
	\begin{align*}
		\sqrt{\NC} \big(	\widehat{\oT} - \oT^* \big) 
		& =
		\frac{1}{\sqrt{\NC}} \sum_{i=1}^\NC 
			\sum_{k=1}^{\NT} 
			\bigg\{
			\frac{\partial v_k(p_k^*) }{\partial p_k }  \uT_k^* \cdot \ee_{p,k}(p_k^*)
			+ v_k(p_k^*)  \varphi_k^{\rm Par}( \uT_k^* , \paraT^\dagger)
			\bigg\} + o_P(1)
		\\
		& =
		\frac{1}{\sqrt{\NC}} \sum_{i=1}^\NC 
		\underbrace{
		\sum_{k=1}^\NT \bigg[ v_k(p_k^*) \varphi_k^{\rm Par}( \uT_k^* , \paraT^\dagger)
			+ \Big\{ \ind(\type_i=k) -p_k^* \Big\} \frac{\partial v_k(p_k^*) }{\partial p_k }  \uT_k^* \bigg]}_{\varphi^{\rm Par}( \oT^* , \paraT^\dagger) } + o_P(1)
			 \ .
	\end{align*}
	This concludes the proof of the asymptotic normality of $\widehat{\oT}$.
	
	To prove the local efficiency of $\widehat{\oT}$, it suffices to show that $\varphi^{\rm Par}( \oT^* , \paraT^\dagger)$ is equivalent to $ \eifuv ( \oT^* )$ presented in Theorem \ref{thm:EIFVC} in the main paper under model $\modelie \cap \modelig$. This is obtained if $\varphi_k^{\rm Par} (\uT_k^*, \paraT^\dagger)$ is the same as $ \varphi_k ( \uT_k^*  ) $  where	
	\begin{align}						\label{proof:7-EIF}
	\varphi_k^{\rm Par} \big( \uT_k^* , \paraT^\dagger \big)
	 = 
	 \frac{1}{p_k^*} \bigg[ \ee_{\uT,k} (\uT_k^*, \paraT^\dagger) 
		- 
		\underbrace{ \EXP  \bigg\{	 \frac{\partial \ee_{\uT ,k} (\uT_k^* , \paraT^\dagger)   }{\partial \paraT \T } \bigg\} }_{(B)}
		\bigg[ \EXP  \bigg\{	\frac{ \ee_\beta(\paraT^\dagger)  }{\partial \paraT \T } \bigg\} \bigg]^{-1}
		\ee_\beta(\paraT^\dagger) \bigg] 
		 \ .
	\end{align}
	Since $\varphi_k ( \uT_k^*  )  = \ee_{\uT,k} (\uT_k^*, \paraT^\dagger)/ p_k^*$, it suffices to show that $(B)$ is zero. Note that $\paraPS^\dagger$ and $\paraOR^\dagger$ are the true parameters under model $\modelie \cap \modelig$; i.e., $\Pe (\ba \cond \bx , k \con \paraPS^\dagger) = \Pe (\ba \cond \bx , k \con \paraPS^*) = e^*(\ba \cond \bx , k)$ and $\OR^{\rm Par}(\ba , \bx , k \con \paraOR^\dagger) = \OR^{\rm Par}(\ba , \bx , k \con \paraOR^*) = \OR^*( \ba , \bx , k)$. Under model $\modelie \cap \modelig$, the derivative $\partial \ee_{\uT,k} / \partial\paraPS$ is
\begin{align*}
	&
	\frac{\partial \ee_{\uT,k} \big(\uT_k , \paraT \big)  }{\partial  \paraPS } 
	= 
	\ind( \type_i = k )
	\sum_{\ba_i \in \zosets(\NI_k) } \frac{\ind(\bA_i = \ba_i) \nabla_{\paraPS} \Pe (\ba_i \cond \bX_i, k \con \paraPS )}{ \Pe (\ba_i \cond \bX_i, k \con \paraPS )^2 } \bw_k \T ( \ba_i,\bX_i) \Big\{ \bY_i - \POR (\ba_i, \bX_i, k \con \paraOR ) \Big\} 
\end{align*}
where $\nabla_{\paraPS}  \Pe (\ba_i \cond \bX_i, k \con \paraPS )$ is the column vector of the partial derivative of $\Pe (\cdot \con \paraPS)$ with respect to $\paraPS$. The expectation of $\partial \ee_{\uT,k} / \partial\paraPS$ is $0$:
\begin{align}										\label{proof:7-003}
	\nonumber \EXP\bigg\{  \frac{\partial \ee_{\uT,k} \big(\uT_k^* , \paraT^\dagger \big)  }{\partial  \paraPS }  \bigg\} 
	& = 
	p_k^* \EXP \bigg\{
	\sum_{\ba_i \in \zosets(\NI_k) } \frac{ \nabla_{\paraPS} \Pe (\ba_i \cond \bX_i, k \con \paraPS^\dagger )}{ \Pe (\ba_i \cond \bX_i, k \con \paraPS^\dagger ) } \bw_k \T ( \ba_i,\bX_i)
	\\
	\nonumber
	& \hspace*{1cm} \times \underbrace{ \EXP \Big\{ \bY_i - \POR (\ba_i, \bX_i, k \con \paraOR^* ) \, \Big| \, \bA_i = \ba_i, \bX_i, \type_i = k \Big\} }_{=0} \, \bigg| \, \type_i = k
	\bigg\} \\
	& = 0  \ .
\end{align}
Next we find that the derivative $\partial \ee_{\uT,k} / \partial\paraOR$ is
\begin{align*}
	\frac{\partial \ee_{\uT,k} \big(\uT_k , \paraT \big)  }{\partial  \paraOR }  
	& =
	\ind(\type_i = k) \bigg\{
	- \sum_{\ba_i \in \zosets(\NI_k) } \frac{\ind(\bA_i = \ba_i) }{ \Pe (\ba_i \cond \bX_i, k \con \paraPS )} \bw_k \T ( \ba_i,\bX_i) \nabla_{\paraOR} \OR^{\rm Par} (\ba_i, \bX_i, k \con \paraOR ) 
	\\
	& \hspace*{4cm}
	+  \sum_{\ba \in \zosets(\NI_k) } \bw_k \T (\ba _i,\bX_i) \nabla_{\paraOR} \OR^{\rm Par} (\ba_i, \bX_i, k \con \paraOR ) \bigg\}\end{align*}
where $\nabla_{\paraOR} \POR (\ba_i , \bX_i, k \con \paraOR )$ is the column vector of the partial derivative of $\POR (\cdot \con \paraOR)$ with respect to $\paraOR$. The expectation of $\partial \ee_{\uT, k} /\partial \paraOR$ is
\begin{align}										\label{proof:7-004}
	\nonumber 
	& \EXP \bigg\{  \frac{\partial \ee_{\uT,k} \big(\bT^* , \paraT^\dagger \big)  }{\partial  \paraOR }  \bigg\}  
	\\
	\nonumber & =
	p_k^* \EXP \bigg\{
		- \sum_{\ba_i \in \zosets(\NI_k) } \underbrace{ \frac{ \EXP\big\{  \ind(\bA_i = \ba_i) \, \big| \, \bX_i , \type_i = k \big\} }{ \Pe (\ba_i \cond \bX_i, k \con \paraPS^* )} }_{=1}\bw_k \T ( \ba_i,\bX_i) \nabla_{\paraOR} \POR (\ba_i, \bX_i, k \con \paraOR^\dagger ) \\
		\nonumber
	& \hspace*{5cm} +
	\sum_{\ba_i \in \zosets(\NI_k) } \bw_k \T ( \ba_i,\bX_i) \nabla_{\paraOR} \POR (\ba_i, \bX_i, k \con \paraOR^\dagger ) \, \bigg| \, \type_i = k
	\bigg\} \\
	& = 0  \ .
\end{align}
Combining \eqref{proof:7-003} and \eqref{proof:7-004}, $(B)$ in \eqref{proof:7-EIF} is zero. Consequently, $\varphi_k^{\rm Par} (\uT_k^*, \paraT^\dagger) = \varphi_k(\uT_k^*)$ and $\varphi^{\rm Par} (\oT^* , \paraT^\dagger) = \varphi(\oT^*)$, respectively. This concludes the proof of the local efficiency of $\widehat{\oT}$.

\subsection{Proof of Corollary \ref{thm:Liu} in the Main Paper}

We only prove the case for the direct effect but the indirect effect can be proven in a similar manner. 
The estimator for the direct effect presented in \eqref{eq-32005} in the main paper is reduced to
\begin{align}	
	\widehat{\oT}^{\DE}(\alpha )  
	& = \sum_{k=1}^K \widehat{p}_k \cdot \widehat{\uT}^\DE_k( \alpha ) 
	= 
	\frac{1}{\NC} \sum_{i=1}^\NC \sum_{k=1}^\NT \ind \big( L_i = k \big) \Big\{
	\psi_k ( 1 , \ePe , \ePOR , \alpha ) - \psi_k ( 0 , \ePe , \ePOR , \alpha )
	\Big\}
	\label{disc:1-001}
\end{align}
where $\psi_k$ has the form of 
	\begin{align}							\label{disc:1-003}
	\nonumber 
	\psi_k ( a , \ePe , \ePOR , \alpha )
	& = \frac{1}{\NI_k} \sum_{j=1}^{\NI_k} \bigg[ 
	\sum_{ \substack{ \ba_i \in \zosets(\NI_k) \\ a_{ij} = a} } \frac{ \ind ( \bA_i = \ba_i ) }{ \ePe (\ba_i \cond \bX_i , k)  } \Big\{ Y_{ij} - \indePOR_j( \ba_i , \bX_i , k ) \Big\} \pi ( \ba_\eij \con \alpha) \\[-0.4cm]
	& \hspace*{4cm} + 
	\sum_{ \substack{ \ba_i \in \zosets(\NI_k) \\ a_{ij} = a} } \indePOR_j ( \ba_i , \bX_i, k ) \pi ( \ba_\eij \con \alpha )
	\bigg]
	 \ .
\end{align}

Next, we review the bias corrected doubly robust (DR$\cdot$BC) estimator proposed in \citet{Liu2019}. Adopting their notations, the DR$\cdot$BC estimator for the direct effect is defined by
\begin{align}						\label{disc:1-004}
	\widehat{\DE}^{\rm DR \cdot BC}(\alpha) 
	= \frac{1}{k} \sum_{i=1}^k \Big\{  \widehat{Y}_i^{\rm DR \cdot BC}(1 , \alpha)  - \widehat{Y}_i^{\rm DR \cdot BC}(0 , \alpha)   \Big\}
\end{align}
where
\begin{align}						\label{disc:1-005}
	\widehat{Y}_i^{\rm DR\cdot BC}(a, \alpha)
	= 
	\frac{1}{N_i} \sum_{j=1}^{N_i} \bigg[
		& \frac{ \ind ( A_{ij} = a ) }{ f(\ba_i \cond \bX_i \con \widehat{\gamma} )  } \Big\{ Y_{ij} - m_{ij}( \bA_i , \bX_i \con \widehat{\paraT} ) \Big\} \pi ( \bA_\eij \con \alpha) 
		\\
		\nonumber
		& \hspace*{1cm}
	+
	\sum_{ \ba_\eij  \in \zosets(N_i-1)  } m_{ij} ( a,  \ba_\eij , \bX_i \con \widehat{\paraT} ) \pi ( \ba_\eij \con \alpha )	
	\bigg]
	 \ .
\end{align}
We translate their notations into our notations as follows. First, $k$ and $N_i$ in \eqref{disc:1-004} and \eqref{disc:1-005} are the number of observed cluster and the cluster size of $i$th cluster which are written as $\NC$ and $\NI_1$ in our notation under unique cluster type assumption, respectively. Second, $f(\ba \cond \bx \con \widehat{\gamma})$ and $m_{ij}(\ba , \bx \con \widehat{\paraT})$ in \eqref{disc:1-005} are the estimated propensity score at the parameter $\widehat{\gamma}$ and the estimated outcome regression of $Y_{ij}$ at the parameter $\widehat{\paraT}$, which correspond to $\ePe(\ba \cond \bx , 1)$ and $\indePOR_j(\ba , \bx, 1)$, respectively, in our notation under unique cluster type assumption. Because of the assumptions (i) and (ii), we obtain $f(\ba \cond \bx \con \widehat{\gamma})=\ePe(\ba \cond \bx , 1)$ and $m_{ij}(\ba , \bx \con \widehat{\paraT}) = \indePOR_j(\ba , \bx, 1)$. Therefore, \eqref{disc:1-003} and \eqref{disc:1-005} are the same. Furthermore, \eqref{disc:1-001} and \eqref{disc:1-004} are the same. The DR$\cdot$BC estimator $\widehat{\DE}^{\rm DR \cdot BC}(\alpha)$ proposed in \citet{Liu2019} has the property introduced in Theorem \ref{thm:ParaPopEst} in the main paper under model $\modelie \cap \modelig$. That is, $\widehat{\DE}^{\rm DR \cdot BC}(\alpha)$ is locally efficient  under model $\modelie \cap \modelig$.

\subsection{Proof of Corollary \ref{thm:ParaRand} in the Main Paper}							\label{sec:proofGLMM}

Before proving Corollary, we introduce new notations for brevity. We denote the parameters associated with the propensity score in cluster type $k$ as $\nparaPStype{k} = \big( \paraPStype{k}\T , \lambda_k \big)\T $ and the parameters associated with the outcome regression in cluster type $k$ as $\nparaORtype{k} = \big( \paraORtype{k}\T, \eta_k , \rho_k)\T $. We define the collection of parameters for all the propensity scores and outcome regressions as $\nparaPS = (\nparaPStype{1}\T,\ldots,\nparaPStype{\NT}\T)\T$ and  $\nparaOR = (\nparaORtype{1}\T,\ldots,\nparaORtype{\NT}\T)\T$, respectively. Since the propensity score and the outcome regression in cluster type $k$ do not depend on parameters of other cluster types, we replace $\paraOR$ and $\paraPS$ in $\indPOR_j(\cdot \con \paraOR)$ and $\Pe_j(\cdot \con \paraPS)$ presented in \eqref{eq-50002} and \eqref{eq-50003} in the main paper with $\paraORtype{k}$ and $\paraPStype{k}$, respectively.

Throughout the proof, we use tensor notations to denote second-order derivatives. Specifically, for a matrix $D=[D_{ij}] \in \R^{r_1\times r_2} \ (i=1,\ldots r_1, j=1,\ldots,r_2)$ and a vector $\bm{d} = (d_1,\ldots,d_{r_3}) \in \R^{r_3}$, we let $\bm{d}^{\otimes 2} = \bm{d} \bm{d}\T$ (i.e., outer product), $D \otimes \bm{d}$ be an order-three tensor where each element is $D_{ij} d_{k}$ ($i=1,\ldots r_1$, $j=1,\ldots r_2$, $k=1,\ldots r_3$), and $\bm{d}^{\otimes 3} = (\bm{d} \bm{d}\T) \otimes \bm{d}$.

We consider the estimating equation
	\begin{align*}
		\ee( \bp , \bT , \nparaPS , \nparaOR)
	=
	\Big[
		\ee_1 \T (p_1,\uT_1,\nparaPStype{1}, \nparaORtype{1} ) 
		\ , \ \ldots \ , \
		\ee_\NT \T  (p_\NT,\uT_\NT,\nparaPStype{\NT}, \nparaORtype{\NT} )
		\Big] \T
	\end{align*}
	where $\ee_k (p_k,\uT_k,\nparaPStype{k}, \nparaORtype{k} ) 
		=
		\big[
			\ee_{p,k}  (p_k)  \ , \
			\ee_{\uT,k} (\uT_k ,\nparaPStype{k},\nparaORtype{k} )  \ , \
			\ee_{e,k}\T (\nparaPStype{k}) \ , \
			\ee_{\indOR,k}\T (\nparaORtype{k}) 
		\big]\T$ and
	\begin{align}				
		\label{proof:8-EE1}
		  \ee_{p,k} (p_k) 
		 &  =
		 \ind(\type_i = k) - p_k \\
		\label{proof:8-EE2}		 
		 \ee_{\uT,k} (\uT_k, \nparaPStype{k} , \nparaORtype{k})
	& =
	\ind( \type_i = k)
		\bigg[ \frac{\bw_k \T ( \bA_i,\bX_i) \big\{ \bY_i - \POR (\bA_i, \bX_i, k \con \paraORtype{k} ) \big\} }{ \Pe( \nparaPStype{k} )} 
		\\ \nonumber
		& \hspace*{3.5cm} + \sum_{\ba_i \in \zosets(\NI_k) } \bw_k \T ( \ba_i,\bX_i) \POR (\ba_i, \bX_i, k \con \paraORtype{k} ) - {\uT}_k \bigg]  
		\\
		\label{proof:8-EE3}
	\ee_{e,k}  \big( \nparaPStype{k} \big)
	& = \ind(\type_i = k) \cdot \frac{ \nabla_{\nparaPStype{k}} \Pe (\nparaPStype{k}) }{\Pe (\nparaPStype{k}) } \\
	\label{proof:8-EE4}
	\ee_{\indOR,k}  \big( \nparaORtype{k} \big)
	& =
	\ind(\type_i = k) \cdot \nabla_{\nparaORtype{k}} \ell_{\indOR,k} (\nparaORtype{k})   \ .
	\end{align}
Here $e^{\rm Par}( \nparaPStype{k} ) $ and $\ell_{\indOR,k}(\nparaORtype{k})$ are defined in \eqref{proof:8-001} and \eqref{proof:8-LogL}, respectively. Also, \eqref{proof:8-EE3} and \eqref{proof:8-EE4} are the score functions of the propensity score and the outcome regression, respectively. 

First, we study the explicit form of \eqref{proof:8-EE3}. The conditional individual propensity score given the random effect $b_i = b$, $P(A_{ij} = a_{ij} \cond \bX_{ij} = \bx_{ij}, \type_i = k , b_i = b \con \paraPStype{k} ) $,  simplifies to $\Pe_j(a_{ij} \cond b \con \paraPStype{k})$. The conditional group propensity score given the random effect $P (\bA_i = \ba_i \cond \bX_i = \bx_i , \type_i = k , b_i = b \con \paraPStype{k} ) $ simplifies to $ \Pe(b \con \paraPStype{k})$. Hence, the group propensity score $\Pe (\ba_i \cond \bx_i , k \con \nparaPStype{k} )$, which we denote as $\Pe (\nparaPStype{k})$, has the form
\begin{align}												\label{proof:8-001}
	e^{\rm Par}( \nparaPStype{k} ) 
	& = \int \Pe (b  \con \paraPStype{k}) \phi(b \con \lambda_k) \, db 
	= \int \bigg\{ \prod_{j=1}^{\NI_k} \Pe_j (a_{ij} \cond b \con \paraPStype{k} ) \bigg\} \phi(b \con \lambda_k) \, db 
\end{align}
where $\phi(\cdot \con \lambda_k)$ is the probability density function of the normal distribution $N(0, \lambda_k^{-1})$. The explicit form of $\nabla_{\nparaPStype{k}} \Pe ( \nparaPStype{k})  = \big( \nabla_{\paraPStype{k}}\T \Pe( \nparaPStype{k}) , \nabla_{\lambda_k} \Pe( \nparaPStype{k}) \big) \T$ in \eqref{proof:8-EE3} is
\begin{align*}
	\nabla_{\paraPStype{k}} \Pe( \nparaPStype{k}) 
	& = 
	\int  e^{\rm Par}( b \con \paraPStype{k}) {R} (b \con \paraPStype{k}) \phi(b \con \lambda_k) \, db
	\\
	\nabla_{\lambda_k} \Pe( \nparaPStype{k}) 
	&  = 
	\int \frac{1-\lambda_k b^2}{2\lambda_k}  
	e^{\rm Par}( b \con \paraPStype{k}) \phi(b \con \lambda_k) \, db
\end{align*}
where
\begin{align*}
	& {R}(b \con \paraPStype{k})
	=
		\sum_{j=1}^{\NI_k} \big\{	\ind(a_{ij} = 1) - e_j^{\rm Par}(1 \cond b \con \paraPStype{k} ) 	\big\} \begin{bmatrix}
			1 \\
			\bx_{ij} 
		\end{bmatrix}
		\ .
\end{align*}
The derivative and integration are exchangeable because of the dominated convergence theorem.

Next, we study \eqref{proof:8-EE4}. The conditional density of $\by_i $ given $(\bA_i = \ba_i , \bX_i = \bx_i , \type_i = k)$ at the parameter $\nparaORtype{k}$ is the multivariate normal distribution $N\big( \POR(\ba_i , \bx_i , k \con \paraORtype{k}) , S_k  \big)$ Note that the $j$th component of $\POR$ is represented as
\begin{align*}
 	\indPOR_j(\ba_i , \bx_i , k \con \paraORtype{k} ) = {h}_j\T(\ba_i, \bx_i , k) \paraORtype{k} 
 	\ , \
 	{h}_j(\ba_i, \bx_i, k) = \begin{bmatrix}
 		1 , a_{ij} , \sum_{j'\neq j} a_{ij'} , \bx_{ij}\T , \sum_{j' \neq j} \bx_{ij'}\T
	\end{bmatrix}\T
	\ .
\end{align*} 
We define a matrix $H$ which has $\NI_k$ columns and the $j$th column is ${h}_j$. Then, $\POR$ can be written as
\begin{align*}
	\OR(\ba_i , \bx_i , k) 
	= 
	H\T(\ba_i , \bx_i , k) \paraORtype{k}
	\quad , \quad
	H(\ba_i, \bx_i, k) = 
	\begin{bmatrix}
		{h}_1 (\ba_i, \bx_i, k) & \vline & \ldots & \vline & {h}_{\NI_k}  (\ba_i, \bx_i, k)
	\end{bmatrix}
	\ .
\end{align*}											
Therefore, the log-likelihood of the outcome regression is
\begin{align}									\label{proof:8-LogL}
	\ell_{\indOR,k} ( \nparaORtype{k} ) = -\frac{1}{2} \log \text{det}(2\pi S_k ) - \frac{1}{2}  \bep\T (\nparaORtype{k})  S_k^{-1} \bep (\nparaORtype{k}) 
\end{align}
where $\bep( \nparaORtype{k} ) = \by_i -  \OR(\ba_i , \bx_i, k \con \paraORtype{k} ) =  \by_i -  H\T(\ba_i , \bx_i , k) \paraORtype{k}$. From the Sherman-Morrison formula, the invserve of $S_k$ and the determinant of $S_k$ are
\begin{align*}
	S_k^{-1} = \eta_k I - \frac{\rho_k \eta_k^2}{1+\NI_k \rho_k \eta_k} {1}_{\NI_k}{1}_{\NI_k}\T  \quad , \quad 
	\text{det}(S_k) = \eta_k^{-\NI_k } (1 + \NI_k \rho_k \eta_k)
\end{align*}
where $I \equiv I_{\NI_k}$ is an $\NI_k$-dimensional identity matrix and ${1}_{\NI_k}$ is an $\NI_k$-dimensional vector of ones. The non-zero first derivatives of $\bep (\nparaORtype{k})$, $S_k^{-1}$, and $\log \text{det}(S_k)$ with respect to $\paraORtype{k}$, $\eta_k$, and $\rho_k$ are
\begin{align*}
	& \frac{\partial \bep( \nparaORtype{k} ) }{\partial \paraORtype{k}\T} 
	= - H (\ba_i , \bx_i , k) 
	\\
	& \frac{\partial S_k^{-1} }{\partial \eta_k} 
	= I - \frac{\rho_k \eta_k(2+ \NI_k \rho_k \eta_k)}{(1+ \NI_k \rho_k \eta_k)^2} {1}_{\NI_k}{1}_{\NI_k}\T \ , \
	&& \frac{\partial S_k^{-1} }{\partial \rho_k} 
	= - \frac{\eta_k^2}{(1+\NI_k \rho_k \eta_k)^2} {1}_{\NI_k}{1}_{\NI_k}\T
	\\
	& \frac{\partial \log \text{det}(S_k) }{\partial \eta_k} 
	=
	- \frac{\NI_k}{\eta_k} + \frac{\NI_k \rho_k}{1+ \NI_k \rho_k \eta_k} \ , \
	&& \frac{\partial \log \text{det}(S_k) }{\partial \rho_k} 
	=
	\frac{ \NI_k \eta_k}{1+\NI_k \rho_k \eta_k} \ .
\end{align*}

The derivative of $\ell_{\indOR,k}$ with respect to $\nparaORtype{k}$ is $\nabla_{\nparaORtype{k}} \ell_{\indOR,k} (\nparaORtype{k} )   = \big( \nabla_{\paraORtype{k}}\T \ell_{\indOR,k} (\nparaORtype{k} ) , \nabla_{\eta_k} \ell_{\indOR,k} (\nparaORtype{k} ) ,\\  \nabla_{\rho_k} \ell_{\indOR,k} (\nparaORtype{k} ) \big) \T$ where each component is given below.
\begin{align*}
	& \nabla_{\paraORtype{k}} \ell_{\indOR,k} (\nparaORtype{k} )
	= H (\ba_i , \bx_i , k)  S_k^{-1} \bep (\nparaORtype{k})  \\
	& \nabla_{\eta_k} \ell_{\indOR,k} (\nparaORtype{k} )
	= -\frac{1}{2} \bep \T (\nparaORtype{k}) \bep  (\nparaORtype{k}) + \frac{1}{2} \frac{\rho_k \eta_k (2 + \NI_k \rho_k \eta_k) }{(1 + \NI_k \rho_k \eta_k)^2} \big\{ \bep\T(\nparaORtype{k} ) {1}_{\NI_k} \big\}^2 + \frac{\NI_k}{2 \eta_k} - \frac{\NI_k \rho_k}{2(1+ \NI_k \rho_k \eta_k)} \\
	& \nabla_{\rho_k} \ell_{\indOR,k} (\nparaORtype{k} )
	= \frac{\eta_k^2}{2(1+\NI_k \rho_k \eta_k)^2}  \big\{ \bep\T(\nparaORtype{k} ) {1}_{\NI_k} \big\}^2 - \frac{ \NI_k \eta_k}{2(1+ \NI_k \rho_k \eta_k)} \ .
\end{align*}

Next, we show that $\ee$ satisfies the regularity conditions (R1)-(R4) of Lemma \ref{lmm:ParaEst}. Condition (R1) is trivially held as long as the parameters of interest (i.e., $( \bp, \bT, \nparaPS, \nparaOR)$) are restricted in such parameter space. 

Condition (R2) also can be shown via brute force derivation of the second order partial derivatives of estimating equations. Note that the derivatives of $\ee_k$ with respect to $(p_{k'},\uT_{k'},\nparaPStype{k'}, \nparaORtype{k'} )$ is zero if $k\neq k'$. Therefore, it suffices to derive the second order partial derivatives of $\ee_k$ with respect to $(p_{k},\uT_{k},\nparaPStype{k}, \nparaORtype{k} )$. 

The derivatives of $\ee_{p,k}$ are $\nabla_{p_k} {\ee_{p,k}} (p_k) = -1$ and 0 for other derivatives. Therefore, all second order partial derivatives of $\ee_{p,k}$ is zero. 

The non-zero first order partial derivatives of $\ee_{\uT,k}$ are 
\begin{align*}
	& \nabla_{\uT_k} \ee_{\uT,k} (\uT_k, \nparaPStype{k} , \nparaORtype{k})  = 
	- \ind (\type_i = k) \\
	& \nabla_{\paraORtype{k} } \ee_{\uT,k} (\uT_k, \nparaPStype{k} , \nparaORtype{k})
	= 
	\ind( \type_i=k )
		\sum_{\ba_i \in \zosets(\NI_k) }  \bigg\{- \frac{\ind(\bA_i = \ba_i)}{ \Pe( \nparaPStype{k} ) } + 1 \bigg\} \Big\{ H (\ba_i , \bX_i , k) \bw_k(\ba_i,\bX_i)  \Big\} \\
	& \nabla_{\nparaPStype{k} } \ee_{\uT,k}  (\uT_k, \nparaPStype{k} , \nparaORtype{k})
	= 
	- \ind( \type_i=k)
		\frac{\bw_k \T (\bA_i,\bX_i) \bep(\nparaORtype{k}) }{ \Pe(  \nparaPStype{k} ) ^2 } \Big\{ \nabla_{\nparaPStype{k}} \Pe(  \nparaPStype{k} )  \Big\}  
		\ .
\end{align*}
The non-zero second order partial derivatives of $\ee_{\uT,k}$ are 
\begin{align}									\label{proof:8-2ndder2}
	\nonumber
	&   \nabla_{\paraORtype{k}} \nabla_{\nparaPStype{k} }\T \ee_{\uT,k}  (\uT_k, \nparaPStype{k} , \nparaORtype{k}) 
	\\
	\nonumber
	& =
	\ind( \type_i=k)
		\sum_{\ba_i \in \zosets(\NI_k) } \frac{ H (\bA_i , \bX_i , k) \bw_k(\bA_i,\bX_i)  }{ \Pe(  \nparaPStype{k} ) ^2 }  \Big\{ \nabla_{\nparaPStype{k}} \T \Pe(  \nparaPStype{k} )  \Big\} \\
		\nonumber
	 & \nabla_{\nparaPStype{k} }^2 \ee_{\uT,k}  (\uT_k, \nparaPStype{k} , \nparaORtype{k})
	\\
	& =
	- \ind( \type_i=k)
	 \frac{\bw_k \T (\bA_i,\bX_i)  \bep(\nparaORtype{k}) }{ \Pe( \nparaPStype{k} ) ^3 } \Big[ \Pe( \nparaPStype{k} )  \Big\{  \nabla_{\nparaPStype{k}}^2 \Pe(  \nparaPStype{k} ) \Big\}  - 2 \Big\{	\nabla_{\nparaPStype{k}} \Pe( \nparaPStype{k} ) \Big\}^{\otimes 2}
		\Big] \ .
\end{align}
Here $\nabla_{\nparaPStype{k}}^2 \Pe ( \nparaPStype{k} ) $ is a $(2\times 2)$-block matrix with the components
\begin{align*}	
	\nabla_{\nparaPStype{k}}^2  \Pe ( \nparaPStype{k} ) 
	= \begin{bmatrix}
		\nabla_{\paraPStype{k}}^2  \Pe ( \nparaPStype{k} ) 
		& \nabla_{\lambda_k} \nabla_{\paraPStype{k}}\T  \Pe ( \nparaPStype{k} ) 
		\\
		\nabla_{\lambda_k} \nabla_{\paraPStype{k}}  \Pe ( \nparaPStype{k} ) 
		&
		\nabla_{\lambda_k}^2  \Pe ( \nparaPStype{k} ) 
	\end{bmatrix} \ .
\end{align*}
The explicit forms of the components can be obtained by using the interchangeability of integration and differentiation.
\begin{align*}
	& \nabla_{\paraPStype{k}} ^2  \Pe ( \nparaPStype{k} ) 
	\\
	& = 
	\int e^{\rm Par}( b \con \paraPStype{k}) 
	\bigg\{
		{R} ^{\otimes 2}  (b \con \paraPStype{k})- \sum_{j=1}^{\NI_k} e_j^{\rm Par}(1 \cond b \con \paraPStype{k} ) e_j^{\rm Par}(0 \cond b \con \paraPStype{k} ) \begin{bmatrix}
			1 \\
			\bx_{ij}
		\end{bmatrix}^{\otimes 2}
	\bigg\} \phi(b \con \lambda_k) \, db \\
	& \nabla_{\paraPStype{k}}  \nabla_{\lambda_k} \Pe ( \nparaPStype{k} ) 
	= 
	\int \frac{1-\lambda_k b^2}{2\lambda_k} 
	e^{\rm Par}( b \con \paraPStype{k}) {R}  (b \con \paraPStype{k})
	\phi(b \con \lambda_k) \, db \\
	& \nabla_{\lambda_k}^2  \Pe ( \nparaPStype{k} ) 
	= 
	\int 
	\frac{\lambda_k^2 b^4 - 2 \lambda_k b^2 - 1}{4 \lambda_k^2} 
	e^{\rm Par}( b \con \paraPStype{k})
	\phi(b \con \lambda_k) \, db  \ .
\end{align*}

The non-zero first order partial derivative of $\ee_{e,k}$ is
\begin{align*}
		& \nabla_{\nparaPStype{k}} \ee_{e,k}(\nparaPStype{k} )
		=  
		\ind(\type_i=k) \bigg[
		\frac{ \nabla_{\nparaPStype{k}}^2  \Pe ( \nparaPStype{k} )  }{\Pe (\nparaPStype{k})  }	
		-
		\frac{ \big\{ \nabla_{\nparaPStype{k}} \Pe ( \nparaPStype{k} ) \big\}^{\otimes 2} }{\Pe (\nparaPStype{k})^2  } \bigg]
		 \ .
\end{align*}

The non-zero second partial order partial derivatives of $\ee_{e,k}$ is
\begin{align}										\label{proof:8-PS2Der}
	& \nabla_{\nparaPStype{k}}^2 \ee_{e,k}(\nparaPStype{k} )
	\\
	&
	=
	\ind(\type_i=k) \bigg[
		\frac{ \nabla_{\nparaPStype{k}}^3 \Pe(\nparaPStype{k}) }{\Pe(\nparaPStype{k})}
		-
		 \frac{ 2 \big\{ \nabla_{\nparaPStype{k}}^2 \Pe(\nparaPStype{k}) \big\} \otimes \big\{ \nabla_{\nparaPStype{k}} \Pe ( \nparaPStype{k} ) \big\} }{ \Pe(\nparaPStype{k})^2 }
		 +
		 \frac{2 \big\{ \nabla_{\nparaPStype{k}} \Pe ( \nparaPStype{k} ) \big\}^{\otimes 3} }{ \Pe(\nparaPStype{k})^3 }
	\bigg] \nonumber \ .
\end{align}
Note that $\nabla_{\nparaPStype{k}}^3 \Pe(\nparaPStype{k})$ is order-three tensor containing sub-tensors $\nabla_{\paraPStype{k}}^3 \Pe(\nparaPStype{k})$, $\nabla_{\paraPStype{k}}^2 \nabla_{\lambda_k} \Pe(\nparaPStype{k})$, $\nabla_{\paraPStype{k}} \nabla_{\lambda_k}^2 \Pe(\nparaPStype{k})$, and $\nabla_{\lambda_k}^3 \Pe(\nparaPStype{k})$, which are of the form
\begin{align}											\label{proof:8-PS3}
	\nonumber
	& \nabla_{\paraPStype{k}}^3 \Pe( \nparaPStype{k} )  \\
	\nonumber
	&
	 = 
	\int \Pe(b  \con \paraPStype{k})
	\bigg[ {R}^{\otimes 3} (b \con \paraPStype{k}) 
	- 3
	 \bigg\{ \sum_{j=1}^{\NI_k} \Pe_j ( 1 \cond b \con \paraPStype{k} ) \Pe_j (0 \cond b \con \paraPStype{k} )  \begin{bmatrix}
			1 \\[-0.25cm]
			\bx_{ij}
		\end{bmatrix}^{\otimes 2} \bigg\} \otimes {R} (b \con \paraPStype{k}) \\
	 \nonumber
	& \hspace*{2cm} - \sum_{j=1}^{\NI_k} \Pe_j (1 \cond b \con \paraPStype{k} )  \Pe_j (0 \cond  b \con \paraPStype{k} ) \big\{  1-2 \Pe_j (1 \cond b \con \paraPStype{k} )  \big\}   \begin{bmatrix}
			1 \\[-0.25cm]
			\bx_{ij}
		\end{bmatrix}^{\otimes 3}
	\bigg] \phi(b \con \lambda_k) \, db \\
	\nonumber
	& \nabla_{\paraPStype{k}}^2 \nabla_{\lambda_k} \Pe( \nparaPStype{k} ) \\
	\nonumber
	&
	= 
	\int \frac{1-\lambda_k b^2}{2\lambda_k}  
	e^{\rm Par}(b  \con \paraPStype{k}) 
	\bigg[
	{R}^{\otimes 2} (b \con \paraPStype{k})  
	- \sum_{j=1}^{\NI_k}  \Pe_j (1 \cond b \con \paraPStype{k} )  \Pe_j (0 \cond b \con \paraPStype{k} )  \begin{bmatrix}
			1 \\[-0.25cm]
			\bx_{ij}
		\end{bmatrix}^{\otimes 2}
	\bigg] 	\phi(b \con \lambda_k) \, db \\
	\nonumber
	& \nabla_{\paraPStype{k}} \nabla_{\lambda_k}^2  \Pe( \nparaPStype{k} ) 
	=
	\int \frac{\lambda_k^2 b^4 - 2 \lambda_k b^2 - 1}{4 \lambda_k^2}  
	e^{\rm Par}( b \con \paraPStype{k}) {R}(b \con \paraPStype{k}) 
	\phi(b \con \lambda_k) \, db \\
	& \nabla_{\lambda_k}^3 \Pe( \nparaPStype{k} ) 
	=
	\int 
	\frac{-\lambda_k^3 b^6 + 3 \lambda_k^2 b^4 + 3 \lambda_k b^2 + 3}{8 \lambda_k^3} e^{\rm Par}( b \con \paraPStype{k}) \phi(b \con \lambda_k)  \, db 
	 \ .
\end{align}

The  non-zero first order partial derivative of $\ee_{\indOR,k}$ consists of components which are the product  of the second derivative of $\ell_{\indOR,k}$ and $\ind(\type_i=k)$ where 
  \begin{align*}
  	& \nabla_{\paraORtype{k}}^2 \ell_{\indOR,k} ( \nparaORtype{k}  )
  	= - H (\ba_i , \bx_i, k) S_k^{-1} H\T(\ba_i , \bx_i, k) \\
	& \nabla_{\paraORtype{k}} \nabla_{\eta_k} \ell_{\indOR,k} ( \nparaORtype{k}  )
	= H(\ba_i , \bx_i , k) \bep(\nparaORtype{k}) - \frac{\rho_k\eta_k(2 + \NI_k \rho_k \eta_k)}{(1+\NI_k\rho_k\eta_k)^2}\big\{ \bep\T(\nparaORtype{k}) {1}_{\NI_k} \big\}   \big\{ H(\ba_i, \bx_i , k) {1}_{\NI_k} \big\} \\
	& \nabla_{\paraORtype{k}} \nabla_{\rho_k} \ell_{\indOR,k} ( \nparaORtype{k}  )
	= - \frac{\eta_k^2}{(1+\NI_k\rho_k\eta_k)^2}\big\{ \bep\T(\nparaORtype{k}) {1}_{\NI_k} \big\}   \big\{ H(\ba_i, \bx_i , k) {1}_{\NI_k} \big\}  \\
	& \nabla_{\eta_k}^2  \ell_{\indOR,k} ( \nparaORtype{k}  )
	=  \frac{\rho_k }{(1+\NI_k \rho_k \eta_k)^3} \big\{ \bep\T(\nparaORtype{k}) {1}_{\NI_k} \big\} ^2	
	- \frac{\NI_k}{2 \eta_k^2} + \frac{\NI_k^2 \rho_k ^2}{2(1+\NI_k \rho_k \eta_k)^2} \\
	& \nabla_{\eta_k} \nabla_{\rho_k}  \ell_{\indOR,k} ( \nparaORtype{k}  ) 
	=  \frac{\eta_k }{(1+\NI_k \rho_k \eta_k)^3}\big\{ \bep\T(\nparaORtype{k}) {1}_{\NI_k} \big\} ^2	
	- \frac{\NI_k}{2(1+\NI_k \rho_k \eta_k)^2}  
	\\
	& \nabla_{\rho_k}^2 \ell_{\indOR,k} ( \nparaORtype{k}  )
	= - \frac{\NI_k \eta_k^3}{(1 + \NI_k \rho_k \eta_k)^3} \big\{ \bep\T(\nparaORtype{k}) {1}_{\NI_k} \big\} ^2	
	 + \frac{\NI_k^2\eta_k^2}{2(1+ \NI_k \rho_k \eta_k )^2 }
	  \ .
  \end{align*}
  The non-zero second order partial derivatives of $\ee_{\indOR,k}$ are the product of the non-zero third derivative of $\ell_{\indOR,k} ( \nparaORtype{k}  )$ and $\ind(\type_i=k)$ where 
  \begin{align*}								
	\nonumber
	& \nabla_{\paraORtype{k}}^2 \nabla_{\eta_k}  \ell_{\indOR,k} ( \nparaORtype{k}  )
	= 
	- H(\ba_i, \bx_i , k) \bigg\{ I - \frac{\rho_k \eta_k(2+\NI_k \rho_k \eta_k)}{(1+\NI_k \rho_k \eta_k)^2} {1}_{\NI_k}{1}_{\NI_k}\T \bigg\} H\T(\ba_i, \bx_i , k)
	\\
	\nonumber
	& \nabla_{\paraORtype{k}}^2 \nabla_{\rho_k}  \ell_{\indOR,k} ( \nparaORtype{k}  )
	= 
	\frac{\eta_k^2}{(1+\NI_k \rho_k \eta_k)^2} \big\{ H(\ba_i, \bx_i , k) {1}_{\NI_k} \big\} ^{\otimes 2} 
	\\
	& \nabla_{\paraORtype{k}} \nabla_{\eta_k}^2 \ell_{\indOR,k} ( \nparaORtype{k}  )
	= - \frac{2\rho_k}{(1+\NI_k\rho_k\eta_k)^3}\big\{ \bep\T(\nparaORtype{k}) {1}_{\NI_k} \big\}   \big\{ H(\ba_i, \bx_i , k) {1}_{\NI_k} \big\} 
	\end{align*}
	\begin{align}							\label{proof:8-2ndder4}
	\nonumber
	& \nabla_{\paraORtype{k}} \nabla_{\eta_k} \nabla_{\rho_k} \ell_{\indOR,k} ( \nparaORtype{k}  )
	= - \frac{2\eta_k}{(1+\NI_k\rho_k\eta_k)^3}\big\{ \bep\T(\nparaORtype{k}) {1}_{\NI_k} \big\}   \big\{ H(\ba_i, \bx_i , k) {1}_{\NI_k} \big\} \\
	\nonumber
	& \nabla_{\paraORtype{k}} \nabla_{\rho_k}^2 \ell_{\indOR,k} ( \nparaORtype{k}  )
	= \frac{2 \NI_k \eta_k^3}{(1+\NI_k\rho_k\eta_k)^3}\big\{ \bep\T(\nparaORtype{k}) {1}_{\NI_k} \big\}   \big\{ H(\ba_i, \bx_i , k) {1}_{\NI_k} \big\}
	\\
	\nonumber
	& \nabla_{\eta_k}^3 \ell_{\indOR,k} ( \nparaORtype{k}  )
	=  - \frac{3 \NI_k \rho_k^2 }{(1+\NI_k \rho_k \eta_k)^4} \big\{ \bep\T(\nparaORtype{k}) {1}_{\NI_k} \big\} ^2	
	+ \frac{\NI_k}{\eta_k^3} - \frac{\NI_k^3 \rho_k ^3}{(1+\NI_k \rho_k \eta_k)^3} \\
	\nonumber
	& \nabla_{\eta_k}^2 \nabla_{\rho_k} \ell_{\indOR,k} ( \nparaORtype{k}  )
	=  - \frac{2 \NI_k \rho_k \eta_k - 1 }{(1+\NI_k \rho_k \eta_k)^4} \big\{ \bep\T(\nparaORtype{k}) {1}_{\NI_k} \big\} ^2	
	+ \frac{\NI_k^2 \rho_k}{(1+\NI_k \rho_k \eta_k)^3} \\
	\nonumber
	& \nabla_{\eta_k} \nabla_{\rho_k}^2 \ell_{\indOR,k} ( \nparaORtype{k}  )
	=  - \frac{3 \NI_k \eta_k^2}{(1+\NI_k \rho_k \eta_k)^4} \big\{ \bep\T(\nparaORtype{k}) {1}_{\NI_k} \big\} ^2	
	+ \frac{\NI_k^2 \eta_k}{(1+\NI_k \rho_k \eta_k)^3} \\
	& \nabla_{\rho_k}^3 \ell_{\indOR,k} ( \nparaORtype{k}  )
	=  \frac{3 \NI_k^2 \eta_k^4}{(1+\NI_k \rho_k \eta_k)^4} \big\{ \bep\T(\nparaORtype{k}) {1}_{\NI_k} \big\} ^2	
	- \frac{\NI_k^3 \eta_k^3}{(1+\NI_k \rho_k \eta_k)^3} 
	 \ .
  \end{align}

Consequently, all second order derivatives of $\ee$ are twice continuously differentiable with respect to $(\bp, \bT, \nparaOR , \nparaPS)$.

To show condition (R3), we see that at the true parameter, $\EXP\big\{ \ee (\bp^*, \bT^*, \nparaOR^* , \nparaPS^*) \big\} = 0$. We first study the expectation of the derivative of $\nabla_{\nparaPStype{k}} \ee_{e,k}$ and $\nabla_{\nparaORtype{k}}  \ee_{\indOR,k}$
\begin{align*}
	& \EXP \big\{ \nabla_{\nparaPStype{k}} \ee_{e,k}(\nparaPStype{k}) \big\} 
	= p_k^* \EXP \big\{ \nabla_{\nparaPStype{k}} \ee_{e,k}(\nparaPStype{k}) \, \big| \, \type_i = k \big\} 
	\ , \
	\\
	& \EXP \big\{ \nabla_{\nparaORtype{k}} \ee_{\indOR,k}(\nparaORtype{k}) \big\} 
	= p_k^* \EXP \big\{ \nabla_{\nparaORtype{k}} \ee_{\indOR,k}(\nparaORtype{k}) \, \big| \, \type_i = k \big\}
	 \ .
\end{align*}
The negative conditional expectations of the matrices in the above become the Fisher information matrices at $\nparaPStype{k} = \nparaPStype{k}^*$ and $\nparaORtype{k} = \nparaORtype{k}^*$, respectively. The Fisher information matrix of the propensity score is assumed to be invertible. Also, the Fisher information matrix of the outcome regression is invertible through the following argument.
\begin{align}					\label{proof:8-Fisher}
	- & \EXP \big\{ \nabla_{\nparaORtype{k}} \ee_{\indOR,k}(\nparaORtype{k}^*) \, \big| \, \type_i = k \big\} \\
	\nonumber
	& = - \begin{bmatrix}
		- \EXP \big\{ H (\bA_i , \bX_i, k) \big( S_k^*\big) ^{-1} H\T(\bA_i , \bX_i, k)  \cond \type_i = k \big\} & 0 & 0 \\
		0 & B_{11}(\eta_k^*, \rho_k^*) & B_{12}(\eta_k^*, \rho_k^*) \\
		0 & B_{12}(\eta_k^*, \rho_k^*) & B_{22}(\eta_k^*, \rho_k^*)
	\end{bmatrix} \ .
\end{align}
Here $S_k^*$ is the true variance matrix $S_k(\eta_k^* , \rho_k^*)$ and 
\begin{align*}
	B_{11}(\eta_k, \rho_k) 
	& =
	\frac{1}{2\eta_k^2(1+\NI_k\rho_k\eta_k)^2} \Big\{
		2\NI_k\rho_k\eta_k -\NI_k(1+\NI_k\rho_k\eta_k)^2 + \NI_k^2\rho_k^2\eta_k^2
	\Big\} \\
	B_{12}(\eta_k, \rho_k) 
	& =
	\frac{\NI_k\eta_k^2}{2\eta_k^2(1+\NI_k\rho_k\eta_k)^2} 
	\quad , \quad 
	B_{22}(\eta_k, \rho_k) 
	=
	- \frac{\NI_k^2\eta_k^4}{2\eta_k^2(1+\NI_k\rho_k\eta_k)^2} 	
	 \ .
\end{align*}
Since the columns of $H$ do not degenerate, the first leading diagonal element is invertible. The determinant of the $[2,3]\times [2,3]$-block matrix of \eqref{proof:8-Fisher} is non-zero for all $\nparaORtype{k}$ in the parameter space because $\NI_k > 1$, i.e., 
\begin{align*}
	\text{det} 
		\begin{bmatrix}
			B_{11}(\eta_k, \rho_k) & B_{12}(\eta_k, \rho_k) \\
			B_{12}(\eta_k, \rho_k) & B_{22}(\eta_k, \rho_k) 
		\end{bmatrix}
	& =
	- \frac{\NI_k^2\eta_k^4}{4\eta_k^4 (1 + \NI_k \rho_k \eta_k)^4 }
	\Big\{
		2\NI_k\rho_k\eta_k -\NI_k(1+\NI_k\rho_k\eta_k)^2 + \NI_k^2\rho_k^2\eta_k^2 +1
	\Big\} \\
	& = \frac{\NI_k^2\eta_k^4}{4\eta_k^4 (1 + \NI_k \rho_k \eta_k)^4 }
	(\NI_k-1)(\NI_k \rho_k \eta_k + 1)^2
	\neq 0
	 \ .
\end{align*}
Therefore, the matrix presented in \eqref{proof:8-Fisher} is invertible.

Now, we discuss (R3)-(i). The uniqueness is guaranteed if the parameters are globally identifiable. First, $p_k$ and $\uT_k$ are uniquely defined by the form of the estimating equation. Second, $\nparaPStype{k}$ is assumed to be identifiable. Lastly, $\nparaORtype{k}$ is identifiable if the Fisher information of $\nparaORtype{k}$ is invertible because the outcome regression, which is a normal distribution, belongs to the exponential family. In \eqref{proof:8-Fisher}, the Fisher information is shown to be invertible, so $\nparaORtype{k}$ is identifiable.

Next, we discuss (R3)-(ii). It suffices to show $\EXP \big\{ \big\| \ee_k (p_k^*, \uT_k^*, \nparaPStype{k}^* , \nparaORtype{k}^* ) \big\|_2^2 \big\}$ is finite for all $k$ where 
\begin{align}							\label{proof:8-ee3}
&
	\big\| \ee_k (p_k^*, \uT_k^*, \nparaPStype{k}^* , \nparaORtype{k}^* ) \big\|_2^2
	\\
	\nonumber
	&
	= 
	\ee_{p,k} (p_k^*)^2 + \ee_{\uT,k}(\uT_k^*,  \nparaPStype{k}^* , \nparaORtype{k}^*) ^2 +
	\ee_{e,k}\T (\nparaPStype{k}^*) \ee_{e,k} (\nparaPStype{k}^*)  + \ee_{\indOR,k}\T(\nparaORtype{k}^*) \ee_{\indOR,k}(\nparaORtype{k}^*) \ .
\end{align}
The expectation of the first term $\ee_{p,k}^2 $ in \eqref{proof:8-ee3} is finite
\begin{align*}
	\EXP \big\{ \ee_{p,k} (p_k^*)^2 \big\} 
	= \EXP \big[ \big\{ \ind(\type_i = k ) - p_k^* \big\}^2 \big] 
	= p_k^*(1-p_k^*) < \infty \ .
\end{align*}
The expectation of the second term $\ee_{\uT,k}^2 $ in \eqref{proof:8-ee3} is finite
\begin{align*}
	\EXP \big\{ \ee_{\uT,k} (\uT_k^*, \nparaORtype{k}^* , \nparaPStype{k}^*) ^2\big\}
	=
	p_k^* \EXP \big\{ \ee_{\uT,k} (\uT_k^*, \nparaORtype{k}^* , \nparaPStype{k}^*)^2 \cond \type_i = k \big\}
	= p_k^{*3} \cdot \text{SEB}_k < \infty 
\end{align*}
because $\text{SEB}_k$ defined in Lemma \ref{lmm:EIFbasic} is finite. The third term $\ee_{e,k} \T \ee_{e,k}$ in \eqref{proof:8-ee3} can be decomposed into 
\begin{align}							\label{proof:8-PSbound}
	\nonumber
	\ee_{e,k} \T (\nparaPStype{k}) \ee_{e,k}(\nparaPStype{k}) 
	& =
	\frac{\ind( \type_i = k )}{ \Pe( \nparaPStype{k})^2 } \Big[ \big\{ \nabla_{\paraPStype{k}}  \Pe (\nparaPStype{k}) \big\}\T \big\{ \nabla_{\paraPStype{k}} \Pe (\nparaPStype{k}) \big\} 
	+ \big\{ \nabla_{\lambda_k} \Pe (\nparaPStype{k}) \big\}^2 \Big] \\
	& \leq \frac{\ind( \type_i = k )}{c^2 }
	\Big[
	\big\| \nabla_{\paraPStype{k}} \Pe (\nparaPStype{k}) \big\|_2^2
	+ 
	\big\{ \nabla_{\lambda_{k}} \Pe (\nparaPStype{k}) \big\}^2
	\Big] \ .
\end{align}
The inequality is from condition (A3) in Assumption \ref{Assump:VC} in the main paper. Note that $\big\| \nabla_{\paraPStype{k}} \Pe (\nparaPStype{k}) \big\|_2$ is bounded by the following quantity 
\begin{align*}	
	\big\| \nabla_{\paraPStype{k}} \Pe (\nparaPStype{k})  \big\|_2
	& = 
	\bigg\| \int  \Pe ( b \con \paraPStype{k}) {R} (b \con \paraPStype{k}) \phi(b \con \lambda_k) \, db \bigg\|_2 \\
	& \leq \int \Pe (b \con \paraPStype{k} ) \big\| {R} (b \con \paraPStype{k}) \big\|_2 \phi(b \con \lambda_k) \, db 
	\leq \int  \big\| {R} (b \con \paraPStype{k}) \big\|_2 \phi(b \con \lambda_k) \, db 
\end{align*}
where the first inequality is from the Jensen's inequality and the second inequality is from the boundedness of the probability $e^{\rm Par}(b  \con \paraPStype{k}) \leq 1$. $\big\| {R} (b \con \paraPStype{k}) \big\|_2$ is further bounded by
\begin{align*}
	\big\| {R} (b \con \paraPStype{k}) \big\|_2
	& \leq 
	\sum_{j=1}^{\NI_k} \big| \ind(a_{ij} = 1) - \Pe_j (1 \cond b \con \paraPStype{k} ) 	\big| \Big\| \big[ 1 , \bx_{ij}\T \big]\T  \Big\|_2  
	\\
	&
	\leq 2 \sum_{j=1}^{\NI_k}  \Big\| \big[ 1 , \bx_{ij}\T \big]\T \Big\|_2
	\leq 2 \NI_k \Big\| \big[ 1 , \bx_i\T \big]\T \Big\|_2
	 \ .
\end{align*}
Therefore, we see that the expectation of the first term in \eqref{proof:8-PSbound}  is bounded by the moment related to the covariate as follows
\begin{align}										\label{proof:8-004}
	\EXP \Big\{ \ind (\type_i = k) \big\| \nabla_{\paraPStype{k}} \Pe (\nparaPStype{k}) \big\|_2^2  \Big\}
	& = 
	p_k^* \EXP \Big\{ \big\| \nabla_{\paraPStype{k}} \Pe( \nparaPStype{k}) \big\|_2^2  \, \Big| \, \type_i = k \Big\} \nonumber \\
	& \leq
	4 p_k^* \NI_k^2 \Big\{  1 +  \EXP \big( \big\|\bX_i \big\|_2^2 \, \big| \, \type_i = k \big) \Big\} \ .
\end{align}
Since $\EXP \big( \| \bX_i \|_2^2 \cond \type_i=k \big)$ is finite by assumption, \eqref{proof:8-004} is also finite. Therefore, we find that the expectation of the second term in \eqref{proof:8-PSbound}  is bounded by the moment related to the covariate as follows. We can find that $\nabla_{\lambda_k} \Pe ( \nparaPStype{k})$ is bounded above by $1/\lambda_k$
\begin{align*}
	\big| \nabla_{\lambda_k} \Pe (  \nparaPStype{k}) \big|
	&  
	= 
	\frac{1}{2\lambda_k} \bigg|
	\int \big( 1-\lambda_k b^2 \big)
	\Pe( b \con \paraPStype{k}) \phi(b \con \lambda_k) \, db \bigg| \\
	& 
	\leq
	\frac{1}{2\lambda_k} 
	\int \big| 1-\lambda_k b^2 \big|
	\Pe ( b \con \paraPStype{k}) \phi(b \con \lambda_k) \, db 
	\leq
	\frac{1}{2\lambda_k} 
	\int \big( 1 + \lambda_k b^2 \big) \phi(b \con \lambda_k) \, db 
	= \frac{1}{\lambda_k}  \ .
\end{align*}
The last equality is from the variance of $b_i \sim N(0,\lambda_k^{-1})$. Therefore,  the expectation of $\ind(\type_i=k) \big\{ \nabla_{\lambda_k} \Pe ( \nparaPStype{k}) \big\}^2$ is bounded above by $1/ \lambda_k^2$. This concludes that $\EXP \big\{ \ee_{e,k} \T (\nparaPStype{k}) \ee_{e,k}(\nparaPStype{k})  \big\}$ is finite. 

The fourth term $\ee_k \ee_k\T$ is $\ee_{\indOR,k}\ee_{\indOR,k}\T$  in \eqref{proof:8-ee3} and can be decomposed into
\begin{align}										\label{proof:8-005}
	& \ee_{\indOR, k}\T (\nparaORtype{k}) \ee_{\indOR,k} (\nparaORtype{k})
	\nonumber
	\\
	& = 
	\ind (\type_i = k)  \Big[
		\big\| \nabla_{\paraORtype{k}} \ell_{\indOR,k} (\nparaORtype{k}  ) \big\|_2^2 
		+  \big\{ \nabla_{\eta_k} \ell_{\indOR,k} (\nparaORtype{k}  ) \big\}^2 
		+  \big\{ \nabla_{\rho_k} \ell_{\indOR,k} (\nparaORtype{k}  ) \big\}^2 \Big] \ .
\end{align}
Note that the expectation of the first term in \eqref{proof:8-005} is
\begin{align}									\label{proof:8-006}
	&
	\nonumber
	\EXP \Big\{ \big \| \nabla_{\paraORtype{k}} \ell_{\indOR,k} (\nparaORtype{k}  )  \big \|_2^2 \, \Big| \, \type_i = k \Big\}
	\\
	\nonumber
	& = 
	\EXP \Big\{\bep\T (\nparaORtype{k}) S_k^{-1}  H\T (\bA_i , \bX_i , k)  H (\bA_i , \bX_i , k)  S_k^{-1} \bep (\nparaORtype{k}) \, \Big| \,  \type_i = k  \Big\} \\
	& = \EXP \Big[  \text{tr} \Big\{   H (\bA_i , \bX_i , k)  S_k^{-1}  H\T (\bA_i , \bX_i , k) \Big\}  \, \Big| \,  \type_i = k    \Big]
	 \ .
\end{align}
The identity is straightforward by switching the order of expectation and trace. 
We find that 
\begin{align*}
	\text{tr} \Big\{ & H (\ba_i , \bx_i , k)    S_k^{-1}  H\T (\ba_i , \bx_i , k) \Big\} \\
	& = 
	\text{tr} \bigg[ \eta_k  H (\ba_i , \bx_i , k) H\T (\ba_i , \bx_i , k)  
	- \frac{\rho_k \eta_k^2}{1+\NI_k\rho_k\eta_k} \big\{ H (\ba_i , \bx_i , k) {1}_{\NI_k} \big\}  \big\{ H (\ba_i , \bx_i , k) {1}_{\NI_k}  \big\}\T \bigg] \\
	& =
	\eta_k \sum_{j=1}^\NT  h_j\T  (\ba_i , \bx_i , k) h_j  (\ba_i , \bx_i , k)
	- \frac{\rho_k \eta_k^2}{1+\NI_k\rho_k\eta_k} \bigg\| \sum_{j=1}^{\NI_k} h_j(\ba_i , \bx_i , k) \bigg\|_2^2 \ .
\end{align*}
Note that the second equality uses $\text{tr}(d d \T) = d\T d$. Therefore, $\EXP \big\{ \big \| \nabla_{\paraORtype{k}} \ell_{\indOR,k} (\nparaORtype{k}^*  )  \big  \|_2^2 \, \big| \,  \type_i = k \big\}$ is also finite if $\EXP\big\{ h_j\T  (\bA_i , \bX_i , k) h_j  (\bA_i , \bX_i , k) \cond \type_i = k \big\}$ is finite for all $j$. Since $h_j  ( \ba_i , \bx_i , k) = \big[ 1 , a_{ij} , \sum a_\eij  , \bx_{ij}\T , \sum \bx_\eij \T \big] \T$ and $\EXP \big( \| \bX_i \|_2^2 \cond \type_i=k \big)$ is finite by assumption, \eqref{proof:8-006} is also finite.

The expectations of the last two components of \eqref{proof:8-005} are equivalent to $p_k^*\EXP \big[ \big\{ \nabla_{\eta_k} \ell_{\indOR,k} (\nparaORtype{k}^*  ) \big\}^2  \cond \type_i = k \big]$ and $p_k^* \EXP \big[  \big\{ \nabla_{\rho_k} \ell_{\indOR,k} (\nparaORtype{k}^*  ) \big\}^2  \cond \type_i = k \big]$, respectively. These quantities are finite because the maximum order of the quantities are the fourth power of $\bep(\nparaORtype{k}^*)$ which follows a multivariate normal distribution. Since the expectations of the four components in \eqref{proof:8-ee3} are finite, $\EXP \big\{ \big\| \ee_k (p_k^*, \uT_k^*, \nparaPStype{k}^* , \nparaORtype{k}^* ) \big\|_2^2 \big\}$ is finite as well. 

Lastly, we discuss (R3)-(iii). We can get the derivative of $\ee$ with respect to $(\bp, \bT, \nparaPS, \nparaOR)$, the parameters associated with cluster type $k$, as follows
\begin{align*}
	\frac{\partial \ee (\bp, \bT , \nparaPS , \nparaOR) }{\partial (\bp, \bT, \nparaPS, \nparaOR)\T}
	= \begin{bmatrix}
		\displaystyle{\frac{\partial \ee_1 (p_1, \uT_1, \nparaPStype{1} , \nparaORtype{1} )}{\partial (p_1, \uT_1, \nparaPStype{1} , \nparaORtype{1} ) \T }} & 0 & \ldots & 0  \\
		0 & \displaystyle{ \frac{\partial \ee_2 (p_2, \uT_2, \nparaPStype{2} , \nparaORtype{2} )}{\partial (p_2, \uT_2, \nparaPStype{2} , \nparaORtype{2} ) \T } } & \ldots & 0 \\
		\vdots & \vdots & \ddots & \vdots \\
		0 & 0 & \ldots & \displaystyle{ \frac{\partial \ee_\NT (p_\NT, \uT_\NT, \nparaPStype{\NT} , \nparaORtype{\NT} )}{\partial (p_\NT, \uT_\NT, \nparaPStype{\NT} , \nparaORtype{\NT} ) \T } }
	\end{bmatrix}
\end{align*}
where 
\begin{align}											\label{proof:8-007}
	\frac{\partial \ee_k (p_k, \uT_k, \nparaPStype{k} , \nparaORtype{k} )}{\partial (p_k, \uT_k, \nparaPStype{k} , \nparaORtype{k} ) \T }
	=
	\begin{bmatrix}
		\nabla_{p_k} \ee_{p,k} &  0 &  0 &  0 \\ 
		0 & \nabla_{\uT_k} \ee_{\uT, k} &  \nabla_{\nparaPStype{k}} \ee_{\uT, k}  &  \nabla_{\nparaORtype{k}} \ee_{\uT, k}  \\
		0 & 0 & \nabla_{\nparaPStype{k}} \ee_{e, k} & 0 \\
		0 & 0 & 0 & \nabla_{\nparaORtype{k}} \ee_{\indOR, k}
	\end{bmatrix}
	 \ .
\end{align}
To show the invertibility of $\EXP \big\{\partial \ee (\bp, \bT , \nparaPS , \nparaOR) / \partial (\bp, \bT, \nparaPS, \nparaOR)\T \big\} $, it suffices to show that the diagonal entries in  \eqref{proof:8-007} are invertible at true parameter values. The first diagonal entry is $\nabla_{p_k} \ee_{p,k} (p_k^*)=-1$, so it is invertible. The second diagonal entry has expectation $\EXP \big\{ \nabla_{\bT_k} \ee_{\uT, k} (\uT_k^*, \nparaPStype{k}^*, \nparaORtype{k}^*)  \big\} = - p_k^* \neq 0 $ so it is invertible. The expectation of third entry is invertible because it is the negative Fisher information matrix. The expectation of the last entry is shown to be invertible based on the form in \eqref{proof:8-Fisher}.

To show condition (R4), we consider the neighborhood of $(p_k^*, \uT_k^*,\nparaPStype{k}^*,\nparaORtype{k}^*)$ defined by
\begin{align*}
	\mathcal{N}_{k,r}
	\equiv \Big\{ ( p_k, \uT_k , \nparaPStype{k} , \nparaORtype{k} ) \, \Big| \, \Big\| ( p_k, \uT_k , \nparaPStype{k} , \nparaORtype{k} ) - ( p_k^*, \uT_k^* , \nparaPStype{k}^* , \nparaORtype{k}^* ) \Big\|_2 < r \Big\} \ .
\end{align*}
for positive constant $r>0$. Note that any neighborhood of $(\bp^*, \bT^*, \nparaPS^*, \nparaOR^*)$ is included in the Cartesian product $\prod_{k=1}^\NT \mathcal{N}_{k,r}$ for some $r>0$. Therefore, it suffices to show that every element of the second order partial derivatives of $\ee_k $ having entries in \eqref{proof:8-EE1}-\eqref{proof:8-EE4} is bounded by a fixed intergrable function for cluster type $k$ parameters that belong to $\mathcal{N}_{k,r}$.

First, we show $\| \bep(\paraORtype{k})\|_2^2$ is bounded by a integrable function. Note that 
\begin{align*}
	\| \bep(\paraORtype{k}) \|_2^2 
	& = \| \by_i - \POR(\ba_i, \bx_i, k \con \paraORtype{k}^* ) + \POR(\ba_i, \bx_i, k \con \paraORtype{k}^* ) - \POR(\ba_i, \bx_i, k \con \paraORtype{k}) \|_2^2 \\
	& \leq 2  \| \by_i - \POR(\ba_i, \bx_i, k \con \paraORtype{k}^* ) \|_2^2 + 2  \| \POR(\ba_i, \bx_i, k \con \paraORtype{k}^* ) - \OR(\ba_i, \bx_i, k \con \paraORtype{k}) \|_2^2 \\
	& \leq 2  \|\bep(\nparaORtype{k}^*) \|_2^2 + 2  \| H\T(\ba_i, \bx_i, k) (\paraORtype{k}^* - \paraORtype{k} )   \|_2^2  \ .
\end{align*}
Since $ \|\bep(\nparaORtype{k}^*) \|_2^2$ is the residual sum of squares, it is an  integrable function whose value is finite for all $(\eta_k, \rho_k) \in \mathcal{N}_r$. Also, $ \| H\T(\ba_i, \bx_i, k) (\paraORtype{k}^* - \paraORtype{k} )   \|_2^2$ is upper bounded by $r^2 \{ \NI_k + \NI_k^3 + \NI_k \|\bx_i \|_2^2\} $.
\begin{align*}
	&
	 \| H\T(\ba_i, \bx_i, k) (\paraORtype{k}^* - \paraORtype{k} )   \|_2
	 \\
	 & \leq  \|H\T(\ba_i, \bx_i, k) \|_2  \| \paraORtype{k}^* - \paraORtype{k} \|_2 
	 \\
	 & \leq r \|H\T(\ba_i, \bx_i, k) \|_F
	 \\
	 & = r \sqrt{
	 \NI_k 
	  	+ \sum_{j=1}^{\NI_k} a_{ij}^2 
	 	+ \sum_{j=1}^{\NI_k} \bigg( \sum_{j' \neq j} a_{ij'} \bigg)^2
	 	+ \sum_{j=1}^{\NI_k} \|\bx_{ij}\|_2^2 
	 	+ \sum_{j=1}^{\NI_k} \bigg\| \sum_{j' \neq j} \bx_{ij'} \bigg\|_2^2 } 
	 \\
	 & \leq r \sqrt{ \NI_k+ \NI_k^3 + \NI_k \big\|\bx_i \big\|_2^2 }
\end{align*}
where $\| D \|_F$ is the Frobenius norm of a matrix $D$. First inequality uses the property of the induced matrix 2-norm. Second inequality is based on the relationship between induced matrix 2-norm and the Frobenius norm. Third identity is based on the definition of the Frobenius norm and the last inequality is straightforward from the definition of $\bx_i$ and the boundedness of $a_{ij}$s. Since $\EXP\big\{  \|\bX_i\|_2^2 \, \big| \, \type_i = k \big\}$ is finite, $\| \bep(\nparaORtype{k})\|_2^2$ is bounded by a fixed integrable function. 

Next we observe that the $r$th order partial derivatives of $\Pe(\nparaPStype{k})$ is bounded by $r$th degree polynomials of the covariates. For example, as in \eqref{proof:8-PS3}, we can show that each component of $\nabla_{\nparaPStype{k}}^3 e(\nparaPStype{k})$ is bounded by third order polynomial $c_3 \cdot \big\| \bx_i \big\|_2^3 + c_2 \cdot \big\| \bx_i \big\|_2^2 + c_1 \cdot \big\| \bx_i \big\|_2 + c_0$ where $c_0$, $c_1$, $c_2$, $c_3$s are generic constants. Similarly, the first and the second order derivatives of $\Pe(\nparaPStype{k})$ are bounded by the the first and the second order polynomials in $\|\bx_i\|_2$, respctively.

Lastly, $\Pe(\nparaPStype{k})$ is bounded between $[c , 1-c]$ for some constant $c$ if the parameters belong to $\mathcal{N}_r$. All second derivatives of $\ee_k $ have the forms in \eqref{proof:8-2ndder2}, \eqref{proof:8-PS2Der}, \eqref{proof:8-2ndder4}. Note that these functions are bounded by a function of the form $C_3 \cdot \big\| \bx_i \big\|_2^3 + C_2 \cdot \big\| \bx_i \big\|_2^2 + C_1 \cdot \big\| \bx_i \big\|_2 + C_0$  for all $(p_k, \uT_k, \nparaPStype{k}, \nparaORtype{k}) \in \mathcal{N}_{k,r}$ where $C_0$, $C_1$, $C_2$, $C_3$ are generic constants because of the previously established results regarding the boundedness of $	\| \bep(\nparaORtype{k}) \|_2^2 $, the derivatives of the propensity scores, and the propensity score. Therefore, since $\EXP \big\{ \| \bX_i \|_2^3 \cond \type_i = k \big\}$ is bounded, all elements of the second order derivatives of $\ee_k$ are bounded by a fixed integrable function for every $(p_k , \uT_k, \nparaPStype{k}, \nparaORtype{k}) \in \mathcal{N}_{k,r}$. 

This shows that $\ee$ satisfies the regularity conditions (R1)-(R4) of Lemma \ref{lmm:ParaEst}. It is straightforward to show the results  regarding $\widehat{\oT}$ in Corollary \ref{thm:ParaRand} in the main paper by following the proof of Theorem \ref{thm:ParaPopEst} in the main paper.

\subsection{Proof of Corollary \ref{thm:MLOR} in the Main Paper}						\label{sec:proof:MLOR}

We decompose $\widetilde{ \oT } - \oT^*$ as $\big( \widetilde{\oT} - \overline{\delta} \big) + \big( \overline{\delta} - \oT^* \big)$ where $\overline{\delta} = \sum_{i=1}^\NC \delta_i /\NC $ with the uncentered efficient influence function $\delta_i = \varphi(\oT^*) + \oT^*$. That is, $\overline{\delta} - \oT^*$ is the average of the efficient influence function in Theorem \ref{thm:EIFVC} in the main paper. Thus, it suffices to show $\NC^{1/2} \big( \widehat{\oT} - \overline{\delta} \big) = o_P(1)$. 
	
	Without loss of generality, let $i$ be the cluster in type $t$, i.e. $\type_i=t$. Then, $\delta_i$ is written as
	\begin{align*}
		\delta_i
		& =
		\frac{v_t(p_t^*) \big\{ \phi_t (\bO_i, e^*, \OR^*) -\uT_t^* \big\} }{p_t^*} + v_t'(p_t^*) \uT_t^* - \sum_{k=1}^\NT p_k^* v_k'(p_k^*) \uT_k^* + \oT^*
	\end{align*}
	where $v_k'(p_k) = \partial v_k(p_k)/\partial p_k$. Thus, we find
	\begin{align*}
		\overline{\delta}
		& =
		\frac{1}{\NC} \sum_{k=1}^\NT \sum_{i : \type_i = k} \frac{v_k(p_k^*) \big\{ \phi_k (\bO_i, e^*, \OR^*) -\uT_k^* \big\} }{p_k^*} 
		+
		\frac{1}{\NC} \sum_{k=1}^\NT \sum_{i: \type_i = k} v_k'(p_k^*) \uT_k^* -  \sum_{k=1}^\NT p_k^* v_k'(p_k^*) \uT_k^* + \oT^*
		\\
		& = 
		\sum_{k=1}^\NT \frac{\widehat{p}_k v_k(p_k^*)}{p_k^*} \underbrace{ \frac{1}{\NC_k} \sum_{i:\type_i = k} \phi_k(\bO_i,e^*,\OR^*) }_{\overline{\phi}_k(e^*, \OR^*)}
		+
		\sum_{k=1}^\NT \frac{( p_k ^* - \widehat{p}_k) v_k(p_k^*) \uT_k^*}{p_k^*}
		-
		\sum_{k=1}^\NT (p_k^* - \widehat{p}_k) v_k'(p_k^*) \uT_k^* 
	\end{align*}
	where  $\oT^* = \sum_{k=1}^\NT v_k (p_k^*) \uT_k^*$ is in the second equality. With some algebra, we find $abc - xyz = (a-x)(b+y)(c+z)/4 + (a+x)(b-y)(c+z)/4 + (ab+xy)(c-z)/2$. As a consequence, $\widetilde{\oT} - \overline{\delta}$ is
	\begin{align*}
		\nonumber
		\widetilde{\oT} - \overline{\delta}
		&
		=
		\sum_{k=1}^\NT
		\frac{p_k^* v_k(\widehat{p}_k) \widetilde{\uT}_k
			-
			\widehat{p}_k v_k(p^*) \overline{\phi}_k(e^*, \OR^*)}{p_k^*} 
		-
		\sum_{k=1}^\NT \frac{( p_k ^* - \widehat{p}_k) v_k(p_k^*) \uT_k^*}{p_k^*}
		+
		\sum_{k=1}^\NT (p_k^* - \widehat{p}_k) v_k'(p_k^*) \uT_k^* 
		\\
		\nonumber
		&
		= 
		\sum_{k=1}^\NT
		\frac{1}{p_k^*}
		\bigg[
			\frac{1}{4} \big( p_k^* - \widehat{p}_k \big) \Big\{ v_k(\widehat{p}_k) + v_k(p_k^*) \Big\}
			\Big\{ \widetilde{\uT}_k + \overline{\phi}_k(e^*, \OR^*) \Big\}
			\\
			\nonumber
			& \hspace*{2cm}
			+ \frac{1}{4} \big( p_k^* + \widehat{p}_k \big) \Big\{ v_k(\widehat{p}_k) - v_k(p_k^*) \Big\}
			\Big\{  \widetilde{\uT}_k + \overline{\phi}_k(e^*, \OR^*) \Big\}
			\\
			\nonumber
			& \hspace*{2cm}
			+ \frac{1}{2} \Big\{ p_k^*  v_k(\widehat{p}_k) + \widehat{p}_k v_k(p_k^*) \Big\}
			\Big\{ \widetilde{\uT}_k - \overline{\phi}_k(e^*, \OR^*) \Big\}
			\\
			& \hspace*{2cm}
			- ( p_k ^* - \widehat{p}_k) v_k(p_k^*) \uT_k^* + p_k^*  (p_k^* - \widehat{p}_k) v_k'(p_k^*) \uT_k^* 
		\bigg] \ .
	\end{align*}
	We study elementary terms in the above expression. From the central limit theorem, we have $\widehat{p}_k - p_k^* = O_P(\NC^{-1/2})$ and, combined with the delta method, this implies $v_k(\widehat{p}_k) - v_k (p_k^*) = v_k'(p_k^*) \big( \widehat{p}_k - p_k^* \big) + R_{1,\NC}$ where $R_{1,\NC} = o_P(\NC^{-1/2})$. Hence, $v_k(\widehat{p}_k) + v_k(p_k^*) = v_k(\widehat{p}_k) - v_k(p_k^*) + 2 v_k(p_k^*) = 2 v_k(p_k^*) + O_P(\NC^{-1/2})$.
	
	Next we establish the result about $\widetilde{\uT}_k - \overline{\phi}_k(e^*, \OR^*) $ which is further decomposed as follows.
	\begin{align}								\label{proof-NPOR1-1}
		\nonumber
		\widetilde{\uT}_k - \overline{\phi}_k(e^*, \OR^*)
		& 
		=
		\frac{1}{2} \sum_{\ell =1}^2 \frac{2}{\NC_k} \sum_{i \in \mathcal{I}_\ell } \Big\{ \overbrace{ \phi_k (\bO_i, \widetilde{e}_{(-\ell)}, \widetilde{\OR}_{(-\ell)} ) - \phi_k( \bO_i, e^*, \OR^*) } ^{D_i} \Big\}
		\\
		&
		=
		\frac{1}{2} \sum_{\ell =1}^2 \bigg[ \underbrace{ \frac{2}{\NC_k} \sum_{i \in \mathcal{I}_\ell} D_i
		-
		\EXP \big\{ D_i \cond \mathcal{I}_\ell^c \big\} }_{B_1} + \underbrace{ \EXP \big\{ D_i \cond \mathcal{I}_\ell^c \big\} }_{B_2} \bigg]
		 \ .
	\end{align}
	We show $B_1$ and $B_2$ are $o_P(\NC^{-1/2})$ under the assumption in the theorem.	First, the conditional expectation of $B_1^2$ given $\mathcal{I}_\ell^c$ is upper bounded by
	\begin{align}									\label{proof-NPOR1-2}
		&
		\EXP \big( B_1^2 \cond \mathcal{I}_\ell^c \big)
		=
		\frac{2}{\NC_k} \VAR \big( D_i \cond \mathcal{I}_\ell^c \big)
		\leq
		\frac{1}{\NC } \frac{2\NC }{\NC_k} \EXP \big( D_i^2 \cond \mathcal{I}_\ell^c \big) 
		=
		\frac{2/p_k^* + o_P(1) }{\NC } \EXP \big( D_i^2 \cond \mathcal{I}_\ell^c \big) \ .
	\end{align}
	Here, $\EXP \big( D_i^2 \cond \mathcal{I}_\ell^c \big) $ is represented as
	\begin{align*}
		&
		\EXP \big( D_i^2 \cond \mathcal{I}_\ell^c \big) 
		\\
		&
		=
		\int \ind(\type_i = k) \bigg[
			\frac{
				\big\{ e^*(\bA_i \cond \bX_i, k) - \widetilde{e}_{(-\ell)}(\bA_i \cond \bX_i, k)  \big\}
				\bw_k\T(\bA_i, \bX_i) \bY_i }{e^*(\bA_i \cond \bX_i, k) \widetilde{e}_{(-\ell)}(\bA_i \cond \bX_i, k) } 
				\\
				& 
				\hspace*{1cm}
				+
				\frac{
				\big\{ \widetilde{e}_{(-\ell)}(\bA_i \cond \bX_i, k) - e^*(\bA_i \cond \bX_i, k)  \big\}
				\bw_k\T(\bA_i, \bX_i)
				\big\{ \OR^* \big(\ba_i, \bX_i, k \big) +  \widetilde{\OR}_{(-\ell)} \big( \ba_i, \bX_i, k \big) \big\}
			}{2 e^*(\bA_i \cond \bX_i, k) \widetilde{e}_{(-\ell)}(\bA_i \cond \bX_i, k) } 
			\\
				& 
				\hspace*{1cm}
				+
				\frac{
				\big\{ \widetilde{e}_{(-\ell)}(\bA_i \cond \bX_i, k) + e^*(\bA_i \cond \bX_i, k)  \big\}
				\bw_k\T(\bA_i, \bX_i)
				\big\{ \OR^* \big(\ba_i, \bX_i, k \big) -  \widetilde{\OR}_{(-\ell)} \big( \ba_i, \bX_i, k \big) \big\}
			}{2 e^*(\bA_i \cond \bX_i, k) \widetilde{e}_{(-\ell)}(\bA_i \cond \bX_i, k) } 
			\\
			&
			\hspace*{1cm}
			+
		\sum_{\ba_i \in \zosets(\NI_k)}
		\Big[
			\bw_k\T \big( \ba_i, \bX_i \big) \Big\{ \widetilde{\OR}_{(-\ell)} \big( \ba_i, \bX_i, k \big) - \OR^* \big(\ba_i, \bX_i, k \big)
		\Big\} \Big]
		\bigg]^2
		\, dP(\bO_i) 
		\\
		&
		\lesssim
		p_k^* \int 
			\sum_{\ba_i \in \zosets(\NI_k)}
			\frac{
				\big\{ e^*(\ba_i \cond \bX_i, k) - \widetilde{e}_{(-\ell)}(\ba_i \cond \bX_i, k)  \big\}^2
				\bw_k\T(\ba_i, \bX_i) 
				\big[
				\big\{  \OR^* \big( \ba_i, \bX_i, k \big) \big\}^{\otimes 2} + \Sigma^*(\ba_i, \bX_i, k) \big] \bw_k(\ba_i, \bX_i)  }{e^*(\ba_i \cond \bX_i, k) \{ \widetilde{e}_{(-\ell)}(\ba_i \cond \bX_i, k) \}^2 } 
				\\
				&
				\hspace*{1cm}
				+
				\sum_{\ba_i \in \zosets(\NI_k)}
				\frac{
				\big\{ \widetilde{e}_{(-\ell)}(\ba_i \cond \bX_i, k) - e^*(\ba_i \cond \bX_i, k)  \big\}^2 \big[
				\bw_k\T(\ba_i, \bX_i)
				\big\{ \OR^* \big(\ba_i, \bX_i, k \big) +  \widetilde{\OR}_{(-\ell)} \big( \ba_i, \bX_i, k \big) \big\} \big]^2
			}{4 e^*(\ba_i \cond \bX_i, k) \big\{ \widetilde{e}_{(-\ell)}(\ba_i \cond \bX_i, k)  \big\}^2 } 
			\\
				&
				\hspace*{1cm}
				+
				\sum_{\ba_i \in \zosets(\NI_k)}
				\frac{
				\big\{ \widetilde{e}_{(-\ell)}(\ba_i \cond \bX_i, k) + e^*(\ba_i \cond \bX_i, k)  \big\}^2 \big[
				\bw_k\T(\ba_i, \bX_i)
				\big\{ \OR^* \big(\ba_i, \bX_i, k \big) -  \widetilde{\OR}_{(-\ell)} \big( \ba_i, \bX_i, k \big) \big\} \big]^2
			}{4 e^*(\ba_i \cond \bX_i, k) \big\{ \widetilde{e}_{(-\ell)}(\ba_i \cond \bX_i, k)  \big\}^2 } 
			\\
			& 
			\hspace*{1cm}
			+
		\bigg[ \sum_{\ba_i \in \zosets(\NI_k)}
			\bw_k\T \big( \ba_i, \bX_i \big) \Big\{ \widetilde{\OR}_{(-\ell)} \big( \ba_i, \bX_i, k \big) - \OR^* \big(\ba_i, \bX_i, k \big)
		\Big\} 
		\bigg]^2
		\, dP \big( \bO_i \cond \type_i = k \big)
			\ .
	\end{align*}
	The equality holds from straightforward math and the inequality holds from the total law of expectation and $(a+b+c+d)^2 \lesssim a^2 + b^2 + c^2 + d^2$. 
	Using the Cauchy-Schwartz inequality, the property of the matrix 2-norm, and the boundedness of the quantities from the assumption, we find the upper bounds for each term in the right hand side. For example, the first term is upper-bounded as follows.
	\begin{align*}
		&
		\sum_{\ba_i \in \zosets(\NI_k)}
		\frac{
				\big\{ e^*(\ba_i \cond \bX_i, k) - \widetilde{e}_{(-\ell)}(\ba_i \cond \bX_i, k)  \big\}^2
				\bw_k\T(\ba_i, \bX_i)
				\big[
				\big\{  \OR^* \big( \ba_i, \bX_i, k \big) \big\}^{\otimes 2} + \Sigma^*(\ba_i, \bX_i, k) \big] \bw_k(\ba_i, \bX_i)  }{e^*(\ba_i \cond \bX_i, k) \{ \widetilde{e}_{(-\ell)}(\ba_i \cond \bX_i, k) \}^2 } 
				\\
				&
				\leq
				\sum_{\ba_i \in \zosets(\NI_k)}
				\frac{
				\big\| e^*(\ba_i \cond \bX_i, k) - \widetilde{e}_{(-\ell)}(\ba_i \cond \bX_i, k)  \big\|_2^2
				\big\|
				\bw_k(\ba_i, \bX_i) \big\|_2^2
				\big\|
				\big\{  \OR^* \big( \ba_i, \bX_i, k \big) \big\}^{\otimes 2} + \Sigma^*(\ba_i, \bX_i, k) \big\|_2 }{e^*(\ba_i \cond \bX_i, k) \{ \widetilde{e}_{(-\ell)}(\ba_i \cond \bX_i, k) \}^2 } 
				\\
				&
				\lesssim 
				\sum_{\ba_i \in \zosets(\NI_k)}
				\big\| e^*(\ba_i \cond \bX_i, k) - \widetilde{e}_{(-\ell)}(\ba_i \cond \bX_i, k)  \big\|_2^2 \ .
	\end{align*}
	Similarly, the other terms are upper-bounded by $\sum_{\ba_i \in \zosets(\NI_k)}
				\big\| e^*(\ba_i \cond \bX_i, k) - \widetilde{e}_{(-\ell)}(\ba_i \cond \bX_i, k)  \big\|_2^2$ or $\sum_{\ba_i \in \zosets(\NI_k)}
				\big\| \OR^* \big( \ba_i, \bX_i, k \big) - \widetilde{\OR}_{(-\ell)} \big( \ba_i, \bX_i, k \big) \big\|_2^2$.
	As a consequence, $\EXP \big( D_i^2 \cond \mathcal{I}_\ell^c \big) $ is upper-bounded as follows.	
	\begin{align}							\label{proof-NPOR1-2-1}
		\EXP \big( D_i^2 \cond \mathcal{I}_\ell^c \big) 
		&
		\lesssim
		\int \sum_{\ba_i \in \zosets(\NI_k)}
		\big\| e^*(\ba_i \cond \bX_i, k) - \widetilde{e}_{(-\ell)}(\ba_i \cond \bX_i, k)  \big\|_2^2
		\, dP \big( \bO_i \cond \type_i = k \big)		
		\\
		& \hspace*{1cm}
		+
		\int \sum_{\ba_i \in \zosets(\NI_k)}
		\big\| \widetilde{\OR}_{(-\ell)} \big( \ba_i, \bX_i, k \big) - \OR^* \big(\ba_i, \bX_i, k \big)
		\big\|_2^2
		\, dP \big( \bO_i \cond \type_i = k \big)
		\nonumber
		\\
		&
		=
		O_P(r_{e,\NC}^2) + O_P( r_{g,\NC}^2)
		\nonumber
	\end{align}
	where the right hand side is $o_P(1)$. From \eqref{proof-NPOR1-2}, we find $\EXP \big(  B_1^2 \cond \mathcal{I}_\ell^c \big) = o_P(\NC^{-1})$ and $B_1=o_P(\NC^{-1/2})$ from Lemma 6.1 of \citet{Victor2018}. 
	
	Second, the conditional expectation of $D_i$ given $\mathcal{I}_\ell^c$ is zero as follows.
		\begin{align*}
		&
		\EXP \big( D_i \cond \mathcal{I}_\ell^c \big) 
		\\
		&
		=
		\int \ind(\type_i = k) \bigg[
			\frac{
				\big\{ e^*(\bA_i \cond \bX_i, k) - \widetilde{e}_{(-\ell)}(\bA_i \cond \bX_i, k)  \big\}
				\bw_k\T(\bA_i, \bX_i) \bY_i }{e^*(\bA_i \cond \bX_i, k) \widetilde{e}_{(-\ell)}(\bA_i \cond \bX_i, k) } 
				\\
				& 
				\hspace*{1cm}
				+
				\frac{
				\big\{ \widetilde{e}_{(-\ell)}(\bA_i \cond \bX_i, k) - e^*(\bA_i \cond \bX_i, k)  \big\}
				\bw_k\T(\bA_i, \bX_i)
				\big\{ \OR^* \big(\ba_i, \bX_i, k \big) +  \widetilde{\OR}_{(-\ell)} \big( \ba_i, \bX_i, k \big) \big\}
			}{2 e^*(\bA_i \cond \bX_i, k) \widetilde{e}_{(-\ell)}(\bA_i \cond \bX_i, k) } 
			\\
				& 
				\hspace*{1cm}
				+
				\frac{
				\big\{ \widetilde{e}_{(-\ell)}(\bA_i \cond \bX_i, k) + e^*(\bA_i \cond \bX_i, k)  \big\}
				\bw_k\T(\bA_i, \bX_i)
				\big\{ \OR^* \big(\ba_i, \bX_i, k \big) -  \widetilde{\OR}_{(-\ell)} \big( \ba_i, \bX_i, k \big) \big\}
			}{2 e^*(\bA_i \cond \bX_i, k) \widetilde{e}_{(-\ell)}(\bA_i \cond \bX_i, k) } 
			\\
			&
			\hspace*{1cm}
			+
		\sum_{\ba_i \in \zosets(\NI_k)}
		\Big[
			\bw_k\T \big( \ba_i, \bX_i \big) \Big\{ \widetilde{\OR}_{(-\ell)} \big( \ba_i, \bX_i, k \big) - \OR^* \big(\ba_i, \bX_i, k \big)
		\Big\} \Big]
		\bigg]
		\, dP(\bO_i) 
		\\
		&
		=
		p_k^* \int 
		\sum_{\ba_i \in \zosets(\NI_k)}
		\frac{1}{\widetilde{e}_{(-\ell)}(\ba_i \cond \bX_i, k) } 
		 \big\{e^*(\ba_i \cond \bX_i, k) - \widetilde{e}_{(-\ell)} (\ba_i \cond \bX_i, k) \big\} 
		 \\
		 &
		 \hspace*{4cm} \times 
		 \bw_k\T(\ba_i, \bX_i) \big\{ \OR^*\big( \ba_i, \bX_i, k \big) - \widetilde{\OR}_{(-\ell)} \big( \ba_i, \bX_i, k \big) \big\}
		\, dP \big( \bO_i \cond \type_i = k \big)
		\\
		&
		\lesssim
		\sum_{\ba_i \in \zosets(\NI_k)} 
		\int 
		 \big\| e^*(\ba_i \cond \bX_i, k) - \widetilde{e}_{(-\ell)} (\ba_i \cond \bX_i, k) \big\|_2
		 \big\| \OR^*\big( \ba_i, \bX_i, k \big) - \widetilde{\OR}_{(-\ell)} \big( \ba_i, \bX_i, k \big) \big\|_2
		\, dP \big( \bO_i \cond \type_i = k \big)
		\\
		&
		\leq
		\sum_{\ba_i \in \zosets(\NI_k)} 
		\bigg[
		\int 
		 \big\| e^*(\ba_i \cond \bX_i, k) - \widetilde{e}_{(-\ell)} (\ba_i \cond \bX_i, k) \big\|_2^2 \, dP \big( \bO_i \cond \type_i = k \big)
		 \bigg]^{1/2}
		 \\
		 & \hspace*{3cm}
		 \times
		 \bigg[ \int
		 \big\| \OR^*\big( \ba_i, \bX_i, k \big) - \widetilde{\OR}_{(-\ell)} \big( \ba_i, \bX_i, k \big) \big\|_2^2
		\, dP \big( \bO_i \cond \type_i = k \big)
		\bigg]^{1/2}
	\\
	&
	=
	O_P(r_{e,\NC} r_{g,\NC})
	=
	o_P(\NC^{-1/2}) \ .
	\end{align*}
	The first equality holds from the Cauchy-Schwartz inequality and the boundedness of $\widetilde{e}_{(-\ell)}$ and $w_k$. The second inequality holds from the H\"older inequality. The convergence rate is from the assumption. This concludes $B_2 = \EXP \big( D_i \cond \mathcal{I}_\ell^c \big) = o_P(\NC^{-1/2})$. Consequently, we have $R_{2,\NC} : = \widetilde{\uT}_k - \overline{\phi}_k(e^*,\OR^*)  = o_P(\NC^{-1/2})$ from \eqref{proof-NPOR1-1}.
	
	We find that $\overline{\phi}_k(e^*,\OR^*)$ is the empirical mean of the uncentered efficient influence function for $\uT_k$. Therefore, $\overline{\phi}_k(e^*,\OR^*) = \uT_k^* + O_P(\NC^{-1/2})$ from the law of large number. Moreover, we observe $ \widetilde{\uT}_k + \overline{\phi}_k(e^*,\OR^*)  = \widetilde{\uT}_k - \overline{\phi}_k(e^*,\OR^*) + 2 \overline{\phi}_k(e^*,\OR^*) = 2\uT_k^* +  O_P(\NC^{-1/2}) + R_{2,\NC}$ and $\widetilde{\uT}_k - \uT_k^* = O_P(\NC^{-1/2})$.
	
	Combining the established results, we find $\NC^{1/2} \big(	\widetilde{\oT} - \overline{\delta} \big) = o_P(1)$ as follows.	
	\begin{align}						\label{proof-NPOR1-5}
		\sqrt{\NC} \Big(	\widetilde{\oT} - \overline{\delta} \Big)
		&
		= 
		\sum_{k=1}^\NT
		\frac{\sqrt{\NC}}{p_k} 
		\bigg[
			\frac{1}{4} \big( p_k^* - \widehat{p}_k \big) \Big\{ v_k(\widehat{p}_k) + v_k(p_k^*) \Big\}
			\Big\{ \widetilde{\uT}_k + \overline{\phi}_k(e^*,\OR^*)  \Big\}
			\nonumber
			\\
			\nonumber
			& \hspace*{2cm}
			+ \frac{1}{4} \big( p_k^* + \widehat{p}_k \big) \Big\{ v_k(\widehat{p}_k) - v_k(p_k^*) \Big\}
			\Big\{  \widetilde{\uT}_k + \overline{\phi}_k(e^*,\OR^*)  \Big\}
			\\
			\nonumber
			& \hspace*{2cm}
			+ \frac{1}{2} \Big\{ p_k^*  v_k(\widehat{p}_k) + \widehat{p}_k v_k(p_k^*) \Big\}
			\Big\{ \widetilde{\uT}_k - \overline{\phi}_k(e^*,\OR^*)  \Big\}
			\\
			\nonumber
			& \hspace*{2cm}
			- ( p_k ^* - \widehat{p}_k) v_k(p_k^*) \uT_k^* + p_k^*  (p_k^* - \widehat{p}_k) v_k'(p_k^*) \uT_k^* 
		\bigg]
		\\
		\nonumber
		&
		= 
		\sum_{k=1}^\NT
		\frac{\sqrt{\NC}}{p_k^*} 
		\bigg[
			\frac{1}{4} 
			\big( p_k^* - \widehat{p}_k \big) 
			 \Big\{ v_k(\widehat{p}_k) + v_k(p_k^*) \Big\}
			\Big\{  \widetilde{\uT}_k + \overline{\phi}_k(e^*,\OR^*)  \Big\}
			\\
			\nonumber
			& \hspace*{2cm}
			+ \frac{1}{4} \big( p_k^* + \widehat{p}_k \big) 
			\Big\{ v_k'(p_k^*) \big( \widehat{p}_k - p_k^* \big) + R_{1,\NC} \Big\}
			\Big\{  \widetilde{\uT}_k + \overline{\phi}_k(e^*,\OR^*)  \Big\}
			\\
			\nonumber
			& \hspace*{2cm}
			+  \frac{1}{2} \Big\{ p_k^*  v_k(\widehat{p}_k) + \widehat{p}_k v_k(p_k^*) \Big\}
			R_{2,\NC}
			\\
			\nonumber
			& \hspace*{2cm}
			- ( p_k ^* - \widehat{p}_k) v_k(p_k^*) \uT_k^* + p_k^*  (p_k^* - \widehat{p}_k) v_k'(p_k^*) \uT_k^* 
		\bigg]
		\\
		& = 
		o_P(1) \ .
	\end{align}
	This concludes $\NC^{1/2} \big( \widetilde{\oT} - \oT^* \big)$ weakly converges to $N(0, \VAR\{ \varphi(\oT^*) \} )$.
	
	To complete the proof, we claim $\widetilde{\sigma}^2$ is consistent for $\sigma^2 = \VAR \big\{ \varphi(\oT^*) \big\} $. We define $\overline{\sigma}^2$ where
	\begin{align*}
		\overline{\sigma}^2
		=
		\frac{1}{\NC} \sum_{i=1}^\NC \varphi(\oT^*)^2
		=
		\frac{1}{\NC} \sum_{i=1}^\NC \bigg[
			\sum_{k=1}^\NT v_k(p_k^*) \varphi_k(\uT_k^*) 
			+
			\sum_{k=1}^\NT \Big\{ \ind (\type_i = k) - p_k^* \Big\} v_k'(p_k^*) \uT_k^*
		\bigg]^2 \ .
	\end{align*}
	It is trivial that $\overline{\sigma}^2 - \sigma^2 = o_P(1)$ from the law of large numbers, so it suffices to show $\widetilde{\sigma}^2 - \overline{\sigma}^2 = o_P(1)$. The difference between $\widetilde{\sigma}^2$ and $\overline{\sigma}^2$ is
	\begin{align*}
		\widetilde{\sigma}^2 - \overline{\sigma}^2
		& =
		\frac{1}{\NC} \sum_{\ell=1}^2 \sum_{i \in \mathcal{I}_\ell} 
		\bigg[ \sum_{k=1}^\NT \frac{v_k(\widehat{p}_k) }{\widehat{p}_k} \Big\{  \phi_k(\bO_i, \widetilde{e}_{(-\ell)} ,\widetilde{\OR}_{(-\ell)} ) - \widetilde{\uT}_k \Big\} + \sum_{k=1}^\NT\Big\{  \ind  (\type_i = k) - \widehat{p}_k \Big\} v_k' (\widehat{p}_k) \widetilde{\uT}_k \bigg]^2
		\\
		&
		\hspace*{2cm}
		-
		\bigg[
		\underbrace{ 
			\sum_{k=1}^\NT \frac{v_k(p_k^*)}{p_k^*}  \Big\{  \phi_k(\bO_i, e^* , \OR^* ) - \uT_k^* \Big\} 
			+
			\sum_{k=1}^\NT \Big\{ \ind (\type_i = k) - p_k^* \Big\} v_k'(p_k^*) \uT_k^*}_{\varphi_i ( \oT^*)}
		\bigg]^2
		\bigg]
		\\
		& =
		\frac{1}{\NC} \sum_{\ell=1}^2 \sum_{i \in \mathcal{I}_\ell} F_i \big\{ F_i + 2 \varphi_i ( \oT^*) \big\}
	\end{align*}
	where
	\begin{align*}
		F_i	&
		:=
		\sum_{k=1}^\NT \frac{v_k(\widehat{p}_k) }{\widehat{p}_k} \Big\{  \phi_k(\bO_i, \widetilde{e}_{(-\ell)} ,\widetilde{\OR}_{(-\ell)} ) - \widetilde{\uT}_k \Big\} + \sum_{k=1}^\NT\Big\{  \ind  (\type_i = k) - \widehat{p}_k \Big\} v_k' (\widehat{p}_k) \widetilde{\uT}_k \\
		& \hspace*{2cm}
		-
			\sum_{k=1}^\NT \frac{v_k(p_k^*)}{p_k^*}  \Big\{  \phi_k(\bO_i, e^* , \OR^* ) - \uT_k^* \Big\} 
			-
			\sum_{k=1}^\NT \Big\{ \ind (\type_i = k) - p_k^* \Big\} v_k'(p_k^*) \uT_k^* \ .
	\end{align*}
	From the H\"older's inequality, we find
	\begin{align}							\label{proof-NPOR1-4-0}
		\big| \widetilde{\sigma}^2 - \overline{\sigma}^2 \big| 
		\leq
		\bigg[ \frac{1}{\NC} \sum_{\ell=1}^2 \sum_{i \in \mathcal{I}_\ell} F_i^2 \bigg]
		+
		2
		\bigg[ \frac{1}{\NC} \sum_{\ell=1}^2 \sum_{i \in \mathcal{I}_\ell} F_i^2 \bigg]^{1/2}
		\bigg[ \frac{1}{\NC} \sum_{i =1}^\NC \varphi_i(\oT^*) ^2 \bigg] \ .
	\end{align}
	Since $ \NC^{-1} \sum_{i =1}^\NC \varphi_i(\oT^*) ^2 = \overline{\sigma}^2 = O_P(1)$, we have $\widetilde{\sigma}^2 - \overline{\sigma}^2 = o_P(1)$ if $ \NC^{-1} \sum_{\ell=1}^2 \sum_{i \in \mathcal{I}_\ell} F_i^2 = o_P(1)$, and it is sufficient to show $ (2/\NC) \sum_{i \in \mathcal{I}_\ell} F_i^2 = o_P(1)$. With some algebra, we find an upper bound of $(2/\NC) \sum_{i \in \mathcal{I}_\ell} F_i^2$ as follows.
	\begin{align}								\label{proof-NPOR1-4}
		&
		\frac{2}{\NC} \sum_{i \in \mathcal{I}_\ell} F_i^2
		\\
		 \nonumber
		&
		\lesssim
		\frac{2}{\NC} \sum_{i \in \mathcal{I}_\ell} 
		\sum_{k=1}^\NT
		\bigg[
			\bigg\{ \frac{v_k(\widehat{p}_k)}{\widehat{p}_k} \bigg\}^2
			 \Big\{  \phi_k(\bO_i, \widetilde{e}_{(-\ell)} ,\widetilde{\OR}_{(-\ell)} ) -  \phi_k(\bO_i, e^* , \OR^* )  \Big\}^2
			 \bigg]
			 \\
			 \nonumber
			 & \hspace*{1cm} + 
		\frac{2}{\NC} \sum_{i \in \mathcal{I}_\ell} 
		\sum_{k=1}^\NT
		\bigg[
			 \bigg\{ \frac{v_k(\widehat{p}_k)}{\widehat{p}_k} - \frac{v_k(p_k^*)}{p_k^*} \bigg\}^2 \phi_k(\bO_i, e^* , \OR^* ) ^2
		\bigg]
		\\
		\nonumber
		& \hspace*{1cm} 
		+ \sum_{k=1}^\NT \bigg[ \Big\{ v_k' (\widehat{p}_k)\widetilde{\uT}_k - v_k' (p_k^*) \uT_k^* \Big\}^2
		+
		 \bigg\{
			\frac{ v_k(p_k^*) \uT_k^* }{p_k^*} - \frac{ v_k(\widehat{p}_k) \widetilde{\uT}_k }{ \widehat{p}_k} \bigg\}^2
			+
		\Big\{
			v_k' (\widehat{p}_k) \widehat{p}_k \widetilde{\uT}_k - v_k' (p_k^*) p_k^* \uT_k^*
		\Big\}^2
		\bigg] \ .
	\end{align}
	We show that each term of the upper bound in the above display is $o_P(1)$ in the rest of the proof.
	
	The first term of the upper bound in \eqref{proof-NPOR1-4} is $o_P(1)$ as follows.
	\begin{align*}
		&
	\frac{2}{\NC} \sum_{i \in \mathcal{I}_\ell} 
		\sum_{k=1}^\NT
			\bigg\{ \frac{v_k(\widehat{p}_k)}{\widehat{p}_k} \bigg\}^2
			 \Big\{  \phi_k(\bO_i, \widetilde{e}_{(-\ell)} , \widetilde{\OR}_{(-\ell)} ) -  \phi_k(\bO_i, e^* , \OR^* )  \Big\}^2
			 \\
			 & = 
			\sum_{k=1}^\NT \widehat{p}_k \bigg\{ \frac{v_k(\widehat{p}_k)}{\widehat{p}_k} \bigg\}^2 \frac{2}{\NC_k}\sum_{i \in \mathcal{I}_\ell} \Big\{  \phi_k(\bO_i, \widetilde{e}_{(-\ell)} ,\widetilde{\OR}_{(-\ell)} ) -  \phi_k(\bO_i, e^* , \OR^* )  \Big\}^2
			\\
			 & = 
			\sum_{k=1}^\NT \underbrace{ \widehat{p}_k \bigg\{ \frac{v_k(\widehat{p}_k)}{\widehat{p}_k} \bigg\}^2 }_{O_P(1)} \underbrace{ \frac{2}{\NC_k}\sum_{i \in \mathcal{I}_\ell, i : \type_i= k} D_i^2 }_{o_P(1)}
			= o_P(1) \ .
	\end{align*}
	The second equality holds from the definition of $D_i^2$ in \eqref{proof-NPOR1-1}. The third equality holds from the law of large numbers as follows.
	\begin{align*}
		\frac{2}{\NC_k}\sum_{i \in \mathcal{I}_\ell} D_i^2
		=
		\EXP \big( D_i^2 \cond \mathcal{I}_k^c \big) + o_P(1) 
		=
		o_P(1)
	\end{align*}
	where $\EXP \big( D_i^2 \cond \mathcal{I}_k^c \big) = o_P(1)$ is established in \eqref{proof-NPOR1-2-1}. 

	The second term of the upper bound in \eqref{proof-NPOR1-4} is $o_P(1)$ as follows.
	\begin{align*}
		&
		\frac{2}{\NC} \sum_{i \in \mathcal{I}_\ell} 
		\sum_{k=1}^\NT
		\bigg[
			 \bigg\{ \frac{v_k(\widehat{p}_k)}{\widehat{p}_k} - \frac{v_k(p_k^*)}{p_k^*} \bigg\}^2 \phi_k(\bO_i, e^* , \OR^* ) ^2
		\bigg]
		\\
		& =
		\sum_{k=1}^\NT \underbrace{ \widehat{p}_k \bigg\{ \frac{v_k(\widehat{p}_k)}{\widehat{p}_k} - \frac{v_k(p_k^*)}{p_k^*} \bigg\}^2 }_{ o_P(1) } \underbrace{ \frac{2}{\NC_k}\sum_{i \in \mathcal{I}_\ell}  \phi_k(\bO_i, e^* , \OR^* ) ^2 }_{O_P(1)} 
		= o_P(1) \ .
	\end{align*}
	The second equality holds from the law of large numbers with the continuous mapping theorem and finite $\EXP \big\{ \phi_k(\bO_i, e^*, \OR^*)^2 \big\}$. More specifically,
	\begin{align*}
		&
		\frac{v_k(\widehat{p}_k) }{\widehat{p}_k} - \frac{v_k(p_k^*)}{p_k^*}
		=
		\frac{v_k(\widehat{p}_k) p_k^* - v_k(p_k^*) \widehat{p}_k}{p_k^* \widehat{p}_k}
		=
		\frac{v_k(p_k^*) p_k^* - v_k(p_k^*) p_k^* + o_P(1) }{(p_k^*)^2 + o_P(1)} = o_P(1)
		\ ,
		\\
		&
		\frac{2}{\NC_k}\sum_{i \in \mathcal{I}_\ell}  \phi_k(\bO_i, e^* , \OR^* ) ^2
		=
		\EXP \big\{ \phi_k(\bO_i, e^*, \OR^*)^2 \big\} + o_P(1)
		< \infty \ .
	\end{align*}
	
	Lastly, the third term of the upper bound in  \eqref{proof-NPOR1-4} is $o_P(1)$ from the law of large numbers, the continuous mapping theorem, and the consistency of $\widetilde{\uT}_k$ for $\uT_k^*$.
	\begin{align*}
		&
	 	\sum_{k=1}^\NT \bigg[ \Big\{ v_k' (\widehat{p}_k)\widetilde{\uT}_k - v_k' (p_k^*) \uT_k^* \Big\}^2
		+
		 \bigg\{
			\frac{ v_k(p_k^*) \uT_k^* }{p_k^*} - \frac{ v_k(\widehat{p}_k) \widetilde{\uT}_k }{ \widehat{p}_k} \bigg\}^2
			+
		\Big\{
			v_k' (\widehat{p}_k) \widehat{p}_k \widetilde{\uT}_k - v_k' (p_k^*) p_k^* \uT_k
		\Big\}^2
		\bigg]
		\\
		&
		=
		\sum_{k=1}^\NT \bigg[ \Big\{ v_k' (p_k^*) \uT_k^* - v_k' (p_k^*) \uT_k^* + o_P(1) \Big\}^2
		+
		 \bigg\{
			\frac{ v_k(p_k^*) \uT_k^* }{p_k^*} - \frac{ v_k(p_k^*) \uT_k^* + o_P(1) }{ p_k ^* + o_P(1) } \bigg\}^2
			\\
			& \hspace*{3cm}
			+
		\Big\{
			v_k' ( p_k ^* ) p_k^* \uT_k^* - v_k' (p_k^*) p_k^* \uT_k^* + o_P(1)
		\Big\}^2
		\bigg]
		\\
		&
		=
		o_P(1) \ . 
	\end{align*}
	This show the upper bound in \eqref{proof-NPOR1-4}  is $o_P(1)$. Thus, we have $\widetilde{\sigma}^2 = \overline{\sigma}^2 + o_P(1) $ from \eqref{proof-NPOR1-4-0}. As mentioned, this implies $\widetilde{\sigma}^2  = \sigma^2 + o_P(1)$ from the law of large numbers.

\subsection{Proof of Corollary \ref{thm:MLOR_KnownPS} in the Main Paper}											\label{sec:proof:MLOR_KnownPS}

The proof is similar to the proof of Corollary \ref{thm:MLOR} in Section \ref{sec:proof:MLOR}. We decompose $\widetilde{ \oT } - \oT^*$ as $\big( \widetilde{\oT} - \overline{\delta}' \big) + \big( \overline{\delta}' - \oT^* \big)$ where $\overline{\delta}' = \sum_{i=1}^\NC \delta_i' /\NC$ with
\begin{align*}
	\delta_i'
		& =
		\frac{v_t(p_t^*) \big\{ \phi_t (\bO_i, e^*, \OR') -\uT_t^* \big\} }{p_t^*} + v_t'(p_t^*) \uT_t^* - \sum_{k=1}^\NT p_k^* v_k'(p_k^*) \uT_k^* + \oT^* \ .
\end{align*}
From the central limit theorem, $\sqrt{\NC}\big( \overline{\delta}' - \oT^* \big)$ converges to $N(0, \sigma'^2)$ where $\sigma'^2 = \VAR(\delta_i')$. Thus, it suffices to show $\NC^{1/2} \big( \widehat{\oT} - \overline{\delta}' \big) = o_P(1)$ to show the asymptotic normality of $\widetilde{\oT}$. 

Similar to the previous math, we find
	\begin{align*}
		\overline{\delta}'
		& = 
		\sum_{k=1}^\NT \frac{\widehat{p}_k v_k(p_k^*)}{p_k^*} \underbrace{ \frac{1}{\NC_k} \sum_{i:\type_i = k} \phi_k(\bO_i,e^*,\OR') }_{\overline{\phi}_k(e^*, \OR')}
		+
		\sum_{k=1}^\NT \frac{( p_k ^* - \widehat{p}_k) v_k(p_k^*) \uT_k^*}{p_k^*}
		-
		\sum_{k=1}^\NT (p_k^* - \widehat{p}_k) v_k'(p_k^*) \uT_k^* \ ,
	\end{align*}
	and
		\begin{align*}
		\nonumber
		\widetilde{\oT} - \overline{\delta}'
		&
		= 
		\sum_{k=1}^\NT
		\frac{1}{p_k^*}
		\bigg[
			\frac{1}{4} \big( p_k^* - \widehat{p}_k \big) \Big\{ v_k(\widehat{p}_k) + v_k(p_k^*) \Big\}
			\Big\{ \widetilde{\uT}_k + \overline{\phi}_k(e^*, \OR') \Big\}
			\\
			\nonumber
			& \hspace*{2cm}
			+ \frac{1}{4} \big( p_k^* + \widehat{p}_k \big) \Big\{ v_k(\widehat{p}_k) - v_k(p_k^*) \Big\}
			\Big\{  \widetilde{\uT}_k + \overline{\phi}_k(e^*, \OR') \Big\}
			\\
			\nonumber
			& \hspace*{2cm}
			+ \frac{1}{2} \Big\{ p_k^*  v_k(\widehat{p}_k) + \widehat{p}_k v_k(p_k^*) \Big\}
			\Big\{ \widetilde{\uT}_k - \overline{\phi}_k(e^*, \OR') \Big\}
			\\
			& \hspace*{2cm}
			- ( p_k ^* - \widehat{p}_k) v_k(p_k^*) \uT_k^* + p_k^*  (p_k^* - \widehat{p}_k) v_k'(p_k^*) \uT_k^* 
		\bigg] \ .
	\end{align*}

	We study elementary terms in the above expression. From the central limit theorem, we have $\widehat{p}_k - p_k^* = O_P(\NC^{-1/2})$ and, combined with the delta method, this implies $v_k(\widehat{p}_k) - v_k (p_k^*) = v_k'(p_k^*) \big( \widehat{p}_k - p_k^* \big) + R_{1,\NC}$ where $R_{1,\NC} = o_P(\NC^{-1/2})$. Hence, $v_k(\widehat{p}_k) + v_k(p_k^*) = v_k(\widehat{p}_k) - v_k(p_k^*) + 2 v_k(p_k^*) = 2 v_k(p_k^*) + O_P(\NC^{-1/2})$. Lastly, $\widetilde{\uT}_k - \overline{\phi}_k(e^*, \OR') $ is further decomposed as follows.
	\begin{align}								\label{proof-NPOR2-1}
		\nonumber
		\widetilde{\uT}_k - \overline{\phi}_k(e^*, \OR')
		& 
		=
		\frac{1}{2} \sum_{\ell =1}^2 \frac{2}{\NC_k} \sum_{i \in \mathcal{I}_\ell } \Big\{ \overbrace{ \phi_k (\bO_i, e^*, \widetilde{\OR}_{(-\ell)} ) - \phi_k( \bO_i, e^*, \OR') } ^{D_i'} \Big\}
		\\
		&
		=
		\frac{1}{2} \sum_{\ell =1}^2 \bigg[ \underbrace{ \frac{2}{\NC_k} \sum_{i \in \mathcal{I}_\ell} D_i'
		-
		\EXP \big\{ D_i' \cond \mathcal{I}_\ell^c \big\} }_{B_1'} + \underbrace{ \EXP \big\{ D_i' \cond \mathcal{I}_\ell^c \big\} }_{B_2'} \bigg]
		 \ .
	\end{align}	
		We show $B_1'$ and $B_2'$ are $o_P(\NC^{-1/2})$ under the assumption in the theorem.	First, the conditional expectation of $B_1'^2$ given $\mathcal{I}_\ell^c$ is upper bounded by
	\begin{align}									\label{proof-NPOR2-2}
		&
		\EXP \big( B_1'^2 \cond \mathcal{I}_\ell^c \big)
		=
		\frac{2}{\NC_k} \VAR \big( D_i' \cond \mathcal{I}_\ell^c \big)
		\leq
		\frac{1}{\NC } \frac{2\NC }{\NC_k} \EXP \big( D_i'^2 \cond \mathcal{I}_\ell^c \big) 
		=
		\frac{2/p_k^* + o_P(1) }{\NC } \EXP \big( D_i'^2 \cond \mathcal{I}_\ell^c \big) \ .
	\end{align}
	Similar to \eqref{proof-NPOR1-2-1}, $\EXP \big( D_i'^2 \cond \mathcal{I}_\ell^c \big) $ is upper-bounded as
	\begin{align}						\label{proof-NPOR2-2-1}
& 		\EXP \big( D_i'^2 \cond \mathcal{I}_\ell^c \big) 
		\\
		&
		\lesssim 
		\int 
		\sum_{\ba_i \in \zosets(\NI_k)}
			\big\| \bw_k(\ba_i, \bX_i) \big\|_2^2 \big\| \OR' \big( \ba_i, \bX_i, k \big) - \widetilde{\OR}_{(-\ell)} \big( \ba_i, \bX_i, k \big) \big\|_2^2 \, dP \big( \bO_i \cond \type_i = k \big)
			\ ,
			\nonumber
	\end{align}
	and the right hand side is $O_P(r_{g,\NC}^2)$ implying $o_P(1)$. From \eqref{proof-NPOR2-2}, we find $\EXP \big(  B_1'^2 \cond \mathcal{I}_\ell^c \big) = o_P(\NC^{-1})$ and $B_1'=o_P(\NC^{-1/2})$ from Lemma 6.1 of \citet{Victor2018}. 
	
	Second, the conditional expectation of $D_i'$ given $\mathcal{I}_\ell^c$ is zero as follows.
		\begin{align*}
		\EXP \big( D_i' \cond \mathcal{I}_\ell^c \big) 
		&
		=
		\int \ind(\type_i = k) \bigg[
			\frac{\bw_k\T(\bA_i, \bX_i, k) \big\{ \OR'\big( \bA_i, \bX_i \big) - \widetilde{\OR}_{(-\ell)} \big( \bA_i, \bX_i, k\big) \big\} }{e^*(\bA_i \cond \bX_i) } 
			\\
			&
			\hspace*{0.5cm}
			+
		\sum_{\ba_i \in \zosets(\NI_k)}
		\Big[
			\bw_k\T \big( \ba_i, \bX_i \big) \Big\{ \widetilde{\OR}_{(-\ell)} \big( \ba_i, \bX_i, k \big) - \OR' \big(\ba_i, \bX_i, k \big)
		\Big\} \Big]
		\bigg]
		\, dP(\bO_i) 
		\\
		&
		=
		p_k^* \int 
			\sum_{\ba_i \in \zosets(\NI_k)}
			\bw_k\T(\ba_i, \bX_i) \big\{ \OR'\big( \ba_i, \bX_i, k \big) - \widetilde{\OR}_{(-\ell)} \big( \ba_i, \bX_i, k \big) \big\}
			\\
			& 
			\hspace*{0.5cm}
			+
			\sum_{\ba_i \in \zosets(\NI_k)}
			\bw_k\T \big( \ba_i, \bX_i \big) \Big\{ \widetilde{\OR}_{(-\ell)} \big( \ba_i, \bX_i, k \big) - \OR' \big(\ba_i, \bX_i, k \big)
		\Big\} 
		\, dP \big( \bO_i \cond \type_i = k \big)
		= 0 \ .
	\end{align*}
	The second equality holds from the total law of expectation. This concludes $B_2' = \EXP \big( D_i' \cond \mathcal{I}_\ell^c \big) = 0 = o_P(\NC^{-1/2})$. Consequently, we have $R_{2,\NC} : = \widetilde{\uT}_k - \overline{\phi}_k(e^*,\OR')  = o_P(\NC^{-1/2})$ from \eqref{proof-NPOR2-1}.

	We find that $\overline{\phi}_k(e^*,\OR') = \uT_k^* + O_P(\NC^{-1/2})$ from the law of large number. Moreover, we observe $ \widetilde{\uT}_k + \overline{\phi}_k(e^*,\OR')  = \widetilde{\uT}_k - \overline{\phi}_k(e^*,\OR') + 2 \overline{\phi}_k(e^*,\OR') = 2\uT_k^* +  O_P(\NC^{-1/2}) + R_{2,\NC}$ and $\widetilde{\uT}_k - \uT_k^* = O_P(\NC^{-1/2})$. Therefore, we find $\NC^{1/2} \big(	\widetilde{\oT} - \overline{\delta}' \big) = o_P(1)$ from the similar reason in \eqref{proof-NPOR1-5}. This concludes $\NC^{1/2} \big( \widetilde{\oT} - \oT^* \big)$ weakly converges to $N(0, \sigma'^2 )$ where $\sigma'^2 = \VAR(\delta_i')$.

	To complete the proof, we claim $\widetilde{\sigma}^2$ is consistent for $\sigma^2 = \VAR(\delta_i')$. We define $\overline{\sigma}'^2$ where
	\begin{align*}
	&
		\overline{\sigma}'^2
		=
		\frac{1}{\NC} \sum_{i=1}^\NC \delta_i'^2
		=
		\frac{1}{\NC} \sum_{i=1}^\NC \bigg[
			\sum_{k=1}^\NT v_k(p_k^*) \varphi_k'(\uT_k^*) 
			+
			\sum_{k=1}^\NT \Big\{ \ind (\type_i = k) - p_k^* \Big\} v_k'(p_k^*) \uT_k^*
		\bigg]^2 \ ,
		\\
		&
		\varphi_k'(\uT_k^*) 
		=
		\ind( \type_i = k) \frac{\phi_k (\bO_i, e^*, \OR') - \theta_k^*}{p_k^*}
	\end{align*}
	It is trivial that $\overline{\sigma}'^2 - \sigma'^2 = o_P(1)$ from the law of large numbers, so it suffices to show $\widetilde{\sigma}^2 - \overline{\sigma}'^2 = o_P(1)$. From analogous steps, the difference between $\widetilde{\sigma}^2$ and $\overline{\sigma}'^2$ is
	\begin{align*}
		\widetilde{\sigma}^2 - \overline{\sigma}'^2
		=
		\frac{1}{\NC} \sum_{\ell=1}^2 \sum_{i \in \mathcal{I}_\ell} F_i ' \big\{ F_i' + 2 \varphi_i' ( \oT^*) \big\}
	\end{align*}
	where
	\begin{align*}
		F_i'	&
		:=
		\sum_{k=1}^\NT \frac{v_k(\widehat{p}_k) }{\widehat{p}_k} \Big\{  \phi_k(\bO_i, e^* ,\widetilde{\OR}_{(-\ell)} ) - \widetilde{\uT}_k \Big\} + \sum_{k=1}^\NT\Big\{  \ind  (\type_i = k) - \widehat{p}_k \Big\} v_k' (\widehat{p}_k) \widetilde{\uT}_k \\
		& \hspace*{2cm}
		-
			\sum_{k=1}^\NT \frac{v_k(p_k^*)}{p_k^*}  \Big\{  \phi_k(\bO_i, e^* , \OR' ) - \uT_k^* \Big\} 
			-
			\sum_{k=1}^\NT \Big\{ \ind (\type_i = k) - p_k^* \Big\} v_k'(p_k^*) \uT_k^* \ .
	\end{align*}
	From the H\"older's inequality, we find
	\begin{align}							\label{proof-NPOR2-4-0}
		\big| \widetilde{\sigma}^2 - \overline{\sigma}'^2 \big| 
		\leq
		\bigg[ \frac{1}{\NC} \sum_{\ell=1}^2 \sum_{i \in \mathcal{I}_\ell} F_i'^2 \bigg]
		+
		2
		\bigg[ \frac{1}{\NC} \sum_{\ell=1}^2 \sum_{i \in \mathcal{I}_\ell} F_i'^2 \bigg]^{1/2}
		\bigg[ \frac{1}{\NC} \sum_{i =1}^\NC \varphi_i'(\oT^*) ^2 \bigg] \ .
	\end{align}
	Since $ \NC^{-1} \sum_{i =1}^\NC \varphi_i' (\oT^*) ^2 = \overline{\sigma}'^2 = O_P(1)$, we have $\widetilde{\sigma}^2 - \overline{\sigma}'^2 = o_P(1)$ if $ \NC^{-1} \sum_{\ell=1}^2 \sum_{i \in \mathcal{I}_\ell} F_i'^2 = o_P(1)$, and it is sufficient to show $ (2/\NC) \sum_{i \in \mathcal{I}_\ell} F_i'^2 = o_P(1)$. With some algebra, we find an upper bound of $(2/\NC) \sum_{i \in \mathcal{I}_\ell} F_i'^2$ as follows.
	\begin{align*}
		&
		\frac{2}{\NC} \sum_{i \in \mathcal{I}_\ell} F_i'^2
		\\
		 \nonumber
		&
		\lesssim
		\frac{2}{\NC} \sum_{i \in \mathcal{I}_\ell} 
		\sum_{k=1}^\NT
		\bigg[
			\bigg\{ \frac{v_k(\widehat{p}_k)}{\widehat{p}_k} \bigg\}^2
			 \Big\{  \phi_k(\bO_i, e^* ,\widetilde{\OR}_{(-\ell)} ) -  \phi_k(\bO_i, e^* , \OR' )  \Big\}^2
			 \bigg]
			 \\
			 \nonumber
			 & \hspace*{1cm} + 
		\frac{2}{\NC} \sum_{i \in \mathcal{I}_\ell} 
		\sum_{k=1}^\NT
		\bigg[
			 \bigg\{ \frac{v_k(\widehat{p}_k)}{\widehat{p}_k} - \frac{v_k(p_k^*)}{p_k^*} \bigg\}^2 \phi_k(\bO_i, e^* , \OR' ) ^2
		\bigg]
		\\
		\nonumber
		& \hspace*{1cm} 
		+ \sum_{k=1}^\NT \bigg[ \Big\{ v_k' (\widehat{p}_k)\widetilde{\uT}_k - v_k' (p_k^*) \uT_k^* \Big\}^2
		+
		 \bigg\{
			\frac{ v_k(p_k^*) \uT_k^* }{p_k^*} - \frac{ v_k(\widehat{p}_k) \widetilde{\uT}_k }{ \widehat{p}_k} \bigg\}^2
			+
		\Big\{
			v_k' (\widehat{p}_k) \widehat{p}_k \widetilde{\uT}_k - v_k' (p_k^*) p_k^* \uT_k^*
		\Big\}^2
		\bigg] \ .
	\end{align*}
	From similar manners in the proof under $\widetilde{e}_{(-\ell)}$ and $\widetilde{\OR}_{(-\ell)}$, we can show that each term of the upper bound in the above display is $o_P(1)$. Thus, we have $\widetilde{\sigma}^2 = \overline{\sigma}'^2 + o_P(1) $ from \eqref{proof-NPOR2-4-0} and $\widetilde{\sigma}^2  = \sigma'^2 + o_P(1)$. This concludes the proof.

\subsection{Proof of Corollary \ref{thm:Kernel} in the Main Paper}											\label{proof:NWkernel}

We follow the approach in Chapter 4.4 of \citet{LiRacine2007}. We first construct the kernel $\mathcal{K}_h$ for our setting. Let $\bX_{ij}^{(c)} \in \R^{p_1}$ and $\bX_\eij^{(c)} \in \R^{p_2}$ be continuous covariates of $\bX_{ij}$ and $\bX_\eij$, respectively, and let $\bX_{ij}^{(d)} \in \R^{q_1}$ and $\bX_{\eij}^{(d)} \in \R_{q_2}$ be discrete covariates of $\bX_{ij}$ and $\bX_\eij$, respectively. We denote $\bX_{i}^{(c)} \in \R^p$ and $\bX_i^{(d)} \in \R^q$ where $p = p_1+p_2$ and $q = q_1+q_2$. For continuous covariates, we define 
\begin{align*}
	&
	\mathcal{W}_{h_c} (\bx_i^{(c)} - \bX_i^{(c)} )
	=
	\frac{1}{h^p}
	\bigg\{
	\prod_{\ell=1}^{p_1}
	W 
	\bigg(
		\frac{\bx_{ij\ell}^{(c)} - \bX_{ij\ell}^{(c)}}{h_c}
	\bigg) \bigg\}
	\bigg\{
	\prod_{\ell=1}^{p_2}
	W 
	\bigg(
		\frac{\bx_{\eij\ell}^{(c)} - \bX_{\eij\ell}^{(c)}}{h_c}
	\bigg) \bigg\}
\end{align*}
where $W(\cdot)$ is a symmetric, nonnegative, univariate kernel function and $h_c \in (0,\infty)$ is the bandwidth for continuous variables. For discrete covariates, we define
\begin{align*}
	&
	\mathcal{L}_{h_d} ( \ba_\eij - \bA_\eij, \bx_{i}^{(d)} - \bX_{i}^{(d)}  )
	=
	\bigg[
	\prod_{\ell \neq j} h_d^{ \ind \{ \ba_{i\ell} \neq \bA_{i\ell} \} }
	\bigg]
	\bigg[
	\prod_{\ell=1}^{q_1} h_d^{  \ind \{ \bx_{ij\ell}^{(d)} \neq \bX_{ij\ell}^{(d)} \} }
	\bigg]
	\bigg[
	\prod_{\ell=1}^{q_2} h_d^{ \ind \{ \bx_{\eij\ell}^{(d)} \neq \bX_{\eij\ell}^{(d)} \} }
	\bigg]
\end{align*}
where $h_d \in [0,1]$ is the bandwidth for discrete variables. 
We define $\mathcal{K}_{h_c,h_d}$ as follows.
\begin{align*}
	&
	\mathcal{K}_{h_c,h_d}(a_{i(-j)} - \bA_\eij, x_{ij}-X_{ij}, \bx_{\eij} - X_{i(-j)}) 
	\\
	&
	=
	\mathcal{W}_{h_c} (\bx_i^{(c)} - \bX_i^{(c)})
	\mathcal{L}_{h_d} ( \ba_\eij - \bA_\eij, \bx_i^{(d)} - \bX_i^{(d)} )
	\ .
\end{align*}

We find that the indicator function $\ind (A_{ij} = a_{ij}, \type_i = k)$ is a special case of $L$ at $h_d=0$, so $\mathcal{L}_{h_d}$ can incorporate $A_{ij}$ and $\type_i$ as follows.
\begin{align*}
	&
	\mathcal{L}_{h_d} ( \ba_{i} - \bA_{i}, \bx_i^{(d)} - \bX_i^{(d)} , k - \type_i )	
	\\
	&
	=
	\ind (A_{ij} = a_{ij}, \type_i = k)
	\mathcal{L}_{h_d} ( \ba_\eij - \bA_\eij, \bx_{i}^{(d)} - \bX_{i}^{(d)})
	\ .
\end{align*}
Accordingly, $\mathcal{K}_{h_c,h_d}$ is adjusted as follows.
\begin{align*}
	&
	\mathcal{K}_{h_c,h_d}(\ba_i - \bA_i, x_{i}-X_{i}, k - \type_i ) 
	\\
	&
	=
	\mathcal{K}_{h_c,h_d}(a_{i(-j)} - \bA_\eij, x_{ij}-X_{ij}, \bx_{\eij} - X_{i(-j)}) 
	\ind (A_{ij} = a_{ij}, \type_i = k)
	\\
	&
	=
	\mathcal{W}_{h_c} (\bx_{i}^{(c)} - \bX_{i}^{(c)} )
	\mathcal{L}_{h_d} ( \ba_i - \bA_i, \bx_{i}^{(d)} - \bX_{i}^{(d)}, k - \type_i )
	\ .
\end{align*}
Then, $\widetilde{\OR}_{j,(-\ell)}^{\rm NW}$ is represented as
\begin{align*}
	&\widetilde{\OR}_{j,(-\ell)}^{\rm NW}(a_{ij}, \ba_{\eij}, x_{ij},\bx_{\eij},k) 
= \frac{\sum_{ij} Y_{ij} \mathcal{K}_{h_c,h_d} (\ba_i - \bA_i, \bx_i - \bX_i, k - \type_i) } {\sum_{ij}  \mathcal{K}_{h_c,h_d} (\ba_i - \bA_i, \bx_i - \bX_i, k - \type_i) }
\end{align*}
which has the same form as equation (4.20) of \citet{LiRacine2007}.

We assume the following conditions to show the convergence of $\widetilde{\OR}_{j,(-\ell)}^{\rm NW}$: 
(K1) two constants $\kappa_0 = \int W(t)^2 \, dt $ and $ \kappa_2 = \int t^2  W(t)^2 \, dt$ are finite;
(K2) the support of $(\bX_i^{(c)}$ is a compact set in an Euclidean space and the support of $(\bA_i, \bX_i^{(d)}, \type_i)$ has finite number of elements;
(K3) the second derivative of $\mu_k(a_{ij}, \ba_\eij, \bx_{ij}, \bx_\eij)$ with respect to $\bx_i^{(c)}$ is uniformly bounded and continuous; 
(K4) $\sigma_k(a_{ij}, \ba_\eij, \bx_{ij}, \bx_\eij)$ is uniformly bounded; 
(K5) the density of $(\bA_i, \bX_i, \type_i)$, denoted by $f$, and its derivative with respect to $\bx_i^{(c)}$ are bounded between $[f_L, f_U]$ where $0<f_L$ and $f_U<\infty$.

The quantities $B_{1s}(\ba_i, \bx_i, k)$, $B_{2s}(\ba_i, \bx_i, k)$ in page 137 of \citet{LiRacine2007} are represented as follows
\begin{align*}
	&
	B_{1s}(\ba_i, \bx_i, k)
	\\
	&
	=
	\frac{\kappa_2}{2} 
	\bigg[
		\bigg\{
		\frac{\partial^2 \mu_k(\ba_i, \bx_i, k) }{\partial (\bx_i^{(c)})^2} 
		\bigg\}_{(s,s)}
		+
		\frac{2}{f(\ba_i, \bx_i, k)} \bigg\{ \frac{\partial \mu_k(\ba_i, \bx_i, k)}{\partial \bx_i^{(c)}} \bigg\}_{(s)} \bigg\{ \frac{\partial f(\ba_i, \bx_i, k)}{\partial \bx_i^{(c)}} \bigg\}_{(s)}
	\bigg]
\ , \\
&
B_{2s}(\ba_i, \bx_i, k)
\\
&
	=
	\frac{1}{c_s-1} \sum_{ ( \ba_i', \bx_i'^{(d)}, k' ) } \ind_s\Big( (\ba_i, \bx_i^{(d)} , k) , (\ba_i', \bx_i'^{(d)} , k') \Big)
	\\[-0.5cm]
	&
	\hspace*{4cm} \times
	\big\{ \mu_k(\ba_i', \bx_i^{(c)}, \bx_i'^{(d)} , k') - \mu_k(\ba_i, \bx_i, k) \big\} \frac{f (\ba_i', \bx_i^{(c)}, \bx_i'^{(d)} , k') }{f(\ba_i, \bx_i, k)}
	\ .
\end{align*}
Here $\{ A \}_{s}$ is the $s$-th component of vector $A$, $c_s$ is the number of possible values for $s$th component of $ ( \ba_i', \bx_i'^{(d)}, k' )$, and 
\begin{align*}
	\ind_s
	(\bz, \bz')
	=
	\ind(\bz_s \neq bz_s') \prod_{j \neq s} \ind (\bz_j = \bz_j')
	\ .
\end{align*}
Therefore, under conditions (K1)-(K4), we find $| B_{1s} |$ and $|B_{2s}|$ are bounded above by $B< \infty$. Moreover, equation (4.21) and (4.22) implies that 
\begin{align*}
	\int \Big\{ 
		\widehat{g}_{j,(-\ell)}^{\rm NW} (\ba_i, \bx_i, k) 
		-
		\mu_k (\ba_i, \bx_i, k) 
	\Big\}^2
	\, 
	dP(\ba_i, \bx_i, k)
	=
	O_p (\eta_1 + \eta_2^2)
\end{align*}
where $\eta_1$ and $\eta_2$ in page 137 of \citet{LiRacine2007} are represented as $ \eta_1 = \NC^{-1} h_c^{-p}$, $	\eta_2 = p h_c^2 + q h_d$. Therefore, by choosing $h_c = O(\NC^{-1/(4+p)})$ and $h_d = h_c^2 = O(\NC^{-2/(4+p)})$, we find $O_p (\eta_1 + \eta_2^2) = O_P(\NC^{-4/(4+p)})$. Since the dimension of $\OR$ is upper bounded by a fixed integer, this implies 
\begin{align*}
	\int \Big\|
		\widehat{\OR}_{(-\ell)}^{\rm NW} (\ba_i, \bx_i, k) 
		-
		\OR^*(\ba_i, \bx_i, k)
	\Big\|_2^2
	\, 
	dP(\ba_i, \bx_i, k)
	=
	O_P(\NC^{-4/(4+p)})
\end{align*}
This concludes the proof.

%

\section{Proof of the Lemmas and Theorems in Section \ref{sec:appendix1}}							\label{sec:appendix3}

\subsection{Proof of Lemma \ref{lem-AE}}									\label{lem:embedding}

We find that $\overline{Y}_{ij} (a \con \alpha_k)$ at cluster type $k$ is defined by
\begin{align*}	
	\overline{Y}_{ij} (a \con \alpha_k) & = 
	\hspace*{-0.2cm} \sum_{ \ba_i \in \zosets(\NI_k)  } \hspace*{-0.2cm} Y_{ij} ( \ba_i ) \cdot  \ind (a_{ij} = a) \cdot \pi( \ba_{\eij} \con \alpha_k ), \
	\pi(\ba_{\eij} \con \alpha) =  \prod_{\ell \neq j} \alpha^{a_{i\ell }} (1-\alpha)^{1 - a_{i\ell}}
	\ .
\end{align*}
Hence, $\oT^F(\balpha, \balpha')$ can be written as
\begin{align}												\label{proof:0-001}
	\oT^F(\balpha, \balpha')
	& =
	\frac{1}{\NC} \sum_{i=1}^\NC \sum_{k=1}^\NT \ind(\type_i = k) \hspace*{-0.2cm} \sum_{\ba_i \in \zosets(\NI_k) } \bw_k\T(\ba_i \con \alpha_k , \alpha_k') \bY_i(\ba_i)
	\ .
\end{align}
Here $\bw_k(\ba \con \alpha_k , \alpha_k') \in \R^{\NI_k}$ is defined as
\begin{align*}
	\bw_k(\ba \con \alpha_k , \alpha_k')
	& =
	\sum_{j=1}^{\NI_k}
	u_j \cdot \Big[
	\ind(a_{ij} = 1) \big\{ C_{1jk} \pi( \ba_\eij \con \alpha_k)  + C_{3jk}  \pi( \ba_\eij \con \alpha_k') \big\} \\
	& \hspace*{3cm} +	\ind(a_{ij} = 0) \big\{ C_{2jk} \pi( \ba_\eij \con \alpha_k)  + C_{4jk}  \pi( \ba_\eij \con \alpha_k') \big\} \Big]
\end{align*} 
where $u_j \in \reals^{\NI_k}$ is the $j$th standard unit vector. Another representation of \eqref{proof:0-001} is given below.
\begin{align}												\label{proof:0-002}
	\oT^F(\balpha, \balpha')
	=
	\sum_{k=1}^\NT \frac{\NC_k}{\NC} \sum_{\ba_i \in \zosets(\NI_k) } \bw_k\T(\ba_i \con \alpha_k , \alpha_k') \frac{1}{\NC_k} \sum_{i: \type_i = k}	 \bY_i(\ba_i)
	\ .
\end{align}
We have $\NC_k \rightarrow \infty$ almost surely as $\NC \rightarrow \infty$ for all $k$. Specifically, by the law of large numbers, we get the following result as $\NC \rightarrow \infty$.
\begin{align}											\label{proof:0-003}
	\frac{\NC_k}{\NC} \rightarrow p_k^* \in (0,1) \ \text{ almost surely},
\end{align}
and this leads the following result by the law of large numbers as $\NC_k \rightarrow \infty$.
\begin{align}							\label{proof:0-004}
	\frac{1}{\NC_k} \sum_{i: \type_i = k}	 \bY_i(\ba_i) \rightarrow \EXP \big\{ \bY_i(\ba_i)  \, \big| \, \type_i = k \big\} \ \text{ in probability}.
\end{align}
Combining \eqref{proof:0-002}, \eqref{proof:0-003}, and \eqref{proof:0-004}, we obtain the following result from the continuous mapping theorem as $ \NC \rightarrow \infty$.
\begin{align}										\label{proof:0-005}
	\oT^F(\balpha, \balpha) \rightarrow 
	\sum_{k=1}^\NT p_k^* \underbrace{\sum_{\ba_i \in \zosets(\NI_k) } \bw_k\T (\ba_i \con \alpha_k , \alpha_k') \EXP \big\{ \bY_i (\ba_i) \, \big| \, \type_i = k \big\} }_{\uT_k(\alpha_k , \alpha_k')} \text{ in probability.}
\end{align}
Note that $\| \bw_k(\ba \con \alpha_k , \alpha_k') \|_2$ is finite for all $\ba \in \zosets(\NI_k)$ and $\alpha_k , \alpha_k' \in (0,1)$. Therefore, $\uT_k$ defined above belongs to $\Parasetk_k$. Furthermore, by taking $v_k(p_k) = p_k$ for all $k$, \eqref{proof:0-005} is of the form $\sum_{k=1}^\NT v_k(p_k^*) \uT_k(\alpha_k , \alpha_k') $ so it belongs to $\Paraset$.

We show that  $\oT^{\DE,F}(\alpha)$ and $\oT^{\IE,F}(\alpha , \alpha')$ take the form of $\oT^F(\balpha, \balpha')$. First, we unify all $\alpha_k$s to $\alpha$ and $\alpha_k'$s to $\alpha'$. If we take $C_{1jk} = 1/\NI_k$, $C_{2jk} = -1/\NI_k$, and $C_{3jk} = C_{4jk} = 0$, then the $j$th entry of $\bw_k ( \ba_i \con \alpha_k , \alpha_k')$ becomes $\big\{ \ind(a_{ij} = 1) - \ind(a_{ij} = 0) \big\} \pi(\ba_\eij \con \alpha) / \NI_k $. Furthermore,
\begin{align*}
	\bw_k ( \ba_i \con \alpha_k , \alpha_k') 
	=
	\frac{1}{\NI_k} \sum_{j=1}^{\NI_k}
	u_j
	 \cdot \big\{ \ind ( a_{ij}=  1 ) -  \ind ( a_{ij}=  0 ) \big\} \cdot	\pi( \ba_\eij \con \alpha)
	 \ .
\end{align*}
Therefore, $\bw_k ( \ba \con \alpha_k , \alpha_k') $ and corresponding $\oT^F(\balpha,\balpha')$ are equivalent to $\bw_k^\DE(\alpha)$ and $\oT^{\DE,F}(\alpha)$, respectively. Similarly, if we take $C_{1jk} = C_{3jk} = 0$, $C_{2jk} = 1/\NI_k$, and $C_{4jk} = -1/\NI_k$, it is straightforward to show that $\bw_k ( \ba \con \alpha_k , \alpha_k') $ and corresponding $\oT^F(\balpha,\balpha')$ are equivalent to $\bw_k^\IE(\alpha,\alpha')$ and $\oT^{\IE,F}(\alpha,\alpha')$, respectively.

\subsection{Proof of Lemma \ref{lmm:EIFbasic}}										\label{sec:prooflemma2}

	To show that $\eifuv(\bT^*)$ is the efficient influence function  of $\bT^*$ in model $\modeliv$ and $\modelive$, we follow the proof technique laid out in \citet{Newey1990} and \citet{Hahn1998}.

First, we consider the result under model $\modeliv$. The density of $(\bO_i , \type_i) = (\bY_i, \bA_i, \bX_i, \type_i)$ with respect to some $\sigma$-finite measure is 
\begin{align*}
	P^* ( \by , \ba , \bx , k) 
	& = \py^* (\by \cond \ba , \bx , k) e^*(\ba \cond \bx , k) \px^* ( \bx \cond k) p_k^*
\end{align*}
where $\py^*$ is the conditional density of $\bY_i$ given $(\bA_i, \bX_i, \type_i)$ and $\px^*$ is the conditional density of $\bX_i$ given $\type_i$. An asterisk in superscript of (conditional) density represents the true (conditional) density. A smooth regular parametric submodel parametrized by a possibly multi-dimensional parameter $\eta$ is
\begin{align}							\label{proof-submodel}
	P ( \by , \ba , \bx , k \con \eta) 
	& = \py (\by \cond \ba , \bx , k \con \eta) e(\ba \cond \bx , k \con \eta) \px ( \bx \cond k \con \eta) p_k(\eta)  
\end{align}
where the smoothness and regularity conditions are given in Definition A.1 of the appendix in \citet{Newey1990}. We assume the density of the parametric submodel $P(\cdot \con \eta)$ equals the true density $P^*$ at $\eta=\eta^*$. The corresponding score function is
\begin{align*}
	s (\by,\ba,\bx, k \con \eta) 
	&=  s_Y( \by , \ba , \bx, k \con \eta ) + s_A( \ba , \bx , k \con \eta)  +  s_X ( \bx , k \con \eta) + s_\type( k \con \eta)  
\end{align*}
where
\begin{align*}
	& s_Y(\by, \ba , \bx , k \con \eta) = \frac{\partial}{\partial \eta} \, \log \py ( \by \cond \ba , \bx , k \con \eta)\ ,
	&& s_A( \ba , \bx , k \con \eta) = \frac{\partial}{\partial \eta} \, \log e (  \ba \cond \bx , k \con \eta)\ , \\ 
	\nonumber
	& s_X(\bx , k \con \eta) = \frac{\partial}{\partial \eta} \, \log \px( \bx \cond k \con \eta) \ ,
	&& s_\type(k \con \eta) = \frac{\partial }{\partial \eta} \, p_k(\eta)
	\ .
\end{align*}
From the parametric submodel, we obtain the $\NT$-dimensional tangent space which is the mean closure of all $\NT$-dimensional linear combinations of scores, i.e.,
\begin{align}							\label{proof:1-003}
	\mathcal{T} =
	\Big\{
		 S (\by,  \ba,  \bx, k)    \, \Big| \, 
		 & S (\by, \ba, \bx, k) = \big(  S_1 (\by, \ba, \bx , k) , \ldots , S_\NT(\by, \ba, \bx , k) \big)\T \in \R^\NT \ , \ \\
		      \nonumber		      	      
		      & \text{For all } \by \in \R^{\NI_k} , \ba \in \zosets(\NI_k) , \bx \in \mathcal{X}(k), k, \ell =1, \ldots , \NT \ , \ \\
		      & \nonumber 
			S_\ell (\by , \ba , \bx , k) 	
			= S_{\ell Y}(\by , \ba, \bx, k)  + S_{\ell A}(\ba, \bx, k)  + S_{\ell X}( \bx, k)  + S_{\ell \type}(k), 
			\\
			\nonumber
		      &  \EXP \big\{ S_{\ell Y}(\bY_i , \ba, \bx, k) \cond \bA_i = \ba, \bX_i =  \bx , \type_i = k \big\} = 0, \\
		      \nonumber
	    	 &   \EXP \big\{ S_{\ell A} (\bA_i , \bx , k ) \cond \bX_i = \bx , \type_i = k \big\} = 0, \\
	    	\nonumber
	    	&   \EXP \big\{ S_{\ell X}(\bX_i , k ) \cond \type_i = k \big\} = 0, \quad
	    	\EXP\big\{ S_{\ell \type} (\type_i ) \big\} = 0
	    \Big\}
	    \ .
\end{align}

The estimand $\bT^*$ is re-represented as $\bT( \eta ) = \big( \uT_1(\eta) , \ldots , \uT_\NT (\eta) \big)\T$ at parameter $\eta$ in the regular parametric submodel and $\uT_k(\eta)$ has a following functional form as in \eqref{proof:1-taueta}.
\begin{align*}	
	\uT_k(\eta)
	& = \sum_{\ba \in \zosets(\NI_{k}) } \bigg[ \iint  \big\{ \bw_k \T (\ba,\bx) \by  \big\} \py(\by \cond \ba , \bx , k \con \eta)  \px(\bx \cond k \con \eta) \, d \by \, d \bx \bigg]
	\ .
\end{align*}
The sum and integral are interchangeable by Fubini's Theorem along with bounded dimension of $\bx$.  Note that $\uT_k(\eta^*)$ equals the true $\uT_k^*$. Therefore, the derivative of $\uT_k$ evaluated at true $\eta^*$ is 
\begin{align}                                  \label{proof:1-004}
	&
	\frac{\partial \uT_k ( \eta^* )  }{\partial \eta} 
	\\
	 &
	=     \sum_{\ba \in \zosets(\NI_{k}) }  \iint  \big\{ \bw_k \T (\ba , \bx)  \by \big\} s_Y (\by , \ba, \bx, k \con \eta^* ) \py^* (\by \cond \ba , \bx , k  )  \px^* (\bx \cond k ) \, d \by \, d \bx 
	\nonumber 
	\\
	\nonumber 
	& 
	\hspace*{2cm}
	+ \sum_{\ba \in \zosets(\NI_{k}) }  \iint \big\{ \bw_k \T (\ba , \bx)  \by \big\} \py^* (\by \cond \ba , \bx , k  )  s_X (\bx, k \con \eta^*) \px^* (\bx \cond k ) \, d \by \, d \bx
	\ .
\end{align}

The conjectured efficient influence function  of $\bT^*$ is $\varphi(\bT^*) = \big( \varphi_1(\uT_1^*) , \ldots , \varphi_\NT(\uT_\NT^*) \big) \T$ where $\varphi_k(\uT_k^*)$ is defined by
\begin{align*}                                 
	\varphi_k ( \uT_k^*  ) 
	= 
	\frac{\ind (\type_i = k)}{p_k^*} \bigg[
		\sum_{\ba_i \in \zosets( \NI_k ) } & \frac{\ind(\bA_i = \ba_i)}{e^*(\ba_i \cond \bX_i, k)}  
		\bw_k \T(\ba_i , \bX_i)  \Big\{ \bY_i - \OR^*(\ba_i, \bX_i, k) \Big\}  \\
		&
		+
		\sum_{\ba_i \in \zosets( \NI_k ) }
		 \bw_k \T (\ba_i, \bX_i)  \OR^*(\ba_i, \bX_i, k) - \uT_k^* \bigg]
		 \ .
\end{align*}
For brevity, we introduce $F_{1k}$ and $F_{2k}$ satisfying $\varphi_k(\uT_k^*) = F_{1k} ( \bY_i, \bA_i, \bX_i , \type_i) + F_{2k} (\bX_i, \type_i)$ as well as
\begin{align}                                 \label{proof:1-FG}
&
	F_{1k} ( \bY_i, \bA_i, \bX_i , \type_i) 
		= \frac{\ind(\type_i = k)}{p_k^*} 
	\bigg[
		\sum_{\ba_i \in \zosets( \NI_k ) } \frac{\ind(\bA_i = \ba_i)}{e^*(\ba_i \cond \bX_i, k)} 
		\bw_k \T(\ba_i, \bX_i)  \Big\{ \bY_i - \OR^*(\ba_i, \bX_i, k) \Big\}  
		\bigg]
	\nonumber		
		\\
	& 		
	F_{2k} (\bX_i, \type_i)
	= \frac{\ind(\type_i = k)}{p_k^*} 
	\bigg\{
		 \sum_{\ba_i \in \zosets( \NI_k ) } \bw_k \T (\ba_i, \bX_i)  \OR^*(\ba_i, \bX_i, k) - \uT_k^* \bigg\}
		 \ .
\end{align}

We first show that $\bT(\eta)$ is a differentiable parameter, i.e.,
\begin{align*}
	\frac{\partial \bT (\eta^*) }{\partial \eta} 
	= 
	\EXP
	\Big\{
		\varphi (\bT^*) \cdot s (\bY_i, \bA_i, \bX_i, \type_i \con \eta^* )
	\Big\} \ .
\end{align*}
To show this, consider the following entrywise derivatives for all $k$.
\begin{align*}
	\frac{\partial \uT_k (\eta^*) }{\partial \eta} 
	& = 
	\EXP
	\Big\{
		\varphi_k (\uT_k^*) \cdot s (\bY_i, \bA_i, \bX_i, \type_i \con \eta^* )
	\Big\}
	\\
	&
	= \EXP \Big[
	\big\{ F_{1k} (\bY_i, \bA_i, \bX_i, \type_i )  + F_{2k} (\bX_i, \type_i) \big\} \cdot s (\bY_i, \bA_i, \bX_i, \type_i \con \eta^* ) 
	\Big]
\end{align*}
where $\partial \uT_k (\eta^*) / \partial \eta$ has the form in \eqref{proof:1-004}.

The expectation of $F_{1k} \cdot s$ is 
\begin{align*}
    &  \EXP \Big\{ F_{1k} (\bY_i, \bA_i, \bX_i, \type_i ) \cdot s (\bY_i, \bA_i, \bX_i, \type_i \con \eta^* )  \Big\} \\
    & = \sum_{k'=1}^\NT p_{k'}^* \cdot  \EXP \Big\{ F_{1k} (\bY_i, \bA_i, \bX_i, k' ) \cdot s (\bY_i, \bA_i, \bX_i, k' \con \eta^* )  \, \Big| \, \type_i = k' \Big\}
		\\
	& = \sum_{\ba_i \in \zosets(\NI_k)} \EXP \bigg[ \frac{\ind(\bA_i = \ba_i)}{e^*(\ba_i \cond \bX_i, k)} 
		\bw_k \T(\ba_i,\bX_i)  \Big\{ \bY_i - \OR^*(\ba_i, \bX_i, k) \Big\} \cdot s(\bY_i, \bA_i, \bX_i, k \con \eta^*) \, \bigg| \, \type_i = k  \bigg] \\
		  & = \sum_{\ba_i \in \zosets(\NI_k)} \EXP \bigg[  \bw_k \T(\ba_i,\bX_i) \EXP \Big[
		 \Big\{ \bY_i - \OR^*(\ba_i, \bX_i, k) \Big\}  \times  \Big\{ s_Y(\bY_i, \ba_i, \bX_i, k \con \eta^*) \\
		 & \hspace*{1.5cm} + s_A(\ba_i, \bX_i, k \con \eta^*)  + s_X(\bX_i, k \con \eta^*) + s_\type(k \con \eta^*)\Big\}
		 \, \Big| \, \bA_i = \ba_i , \bX_i, \type_i = k \Big]
		 \, \bigg| \, \type_i = k  \bigg] \ .
\end{align*}
The first identity is straightforward from the law of total expectation. The second and the last identities are based on the definitions of $F_{1k}$ and $s$, respectively. We study the conditional expectations of the product of $\bY_i - \OR^*(\ba, \bX_i, k) $ and the score functions. The first product $ \big\{ \bY_i - \OR^*(\ba, \bX_i, k) \big\} \cdot s_Y$ has the following conditional expectation.
\begin{align*}
	\EXP \Big[ \big\{ \bY_i - \OR^*(\ba, \bX_i, k) \big\} &  \cdot s_Y(\bY_i, \ba, \bX_i, k \con \eta^*) \, \Big| \, \bA_i = \ba , \bX_i = \bx, \type_i = k \Big] \\
	& = \EXP \Big\{ \bY_i \cdot s_Y(\bY_i, \ba, \bX_i, k \con \eta^*) \, \Big| \, \bA_i = \ba , \bX_i = \bx, \type_i = k \Big\} \\
	& =  \int \by \cdot s_Y (\by , \ba, \bx, k \con \eta^* ) \py^*(\by \cond \ba , \bx , k  ) \, d \by
	\ .
\end{align*}
The first identity is based on the conditional mean zero property of $s_Y$. The second identity is the functional representation of conditional expectations. The second product $ \big\{ \bY_i - \OR^*(\ba, \bX_i, k) \big\} \cdot \big\{ s_A + s_X + s_\type \big\}$ has zero conditional expectation.
\begin{align*}
	 & \EXP \Big[ \big\{ \bY_i  - \OR^*(\ba, \bX_i, k) \big\}  \big\{ s_A( \ba, \bX_i, k \con \eta^*) + s_X(\bX_i, k \con \eta^*) + s_\type(k \con \eta^*)  \big\} \, \Big| \, \bA_i = \ba , \bX_i, \type_i = k \Big] \\
	& = \overbrace{\EXP \big\{  \bY_i - \OR^*(\ba, \bX_i, k) \, \big| \, \bA_i = \ba , \bX_i, \type_i = k \big\}}^{= \ 0} \big\{ s_A( \ba, \bX_i, k \con \eta^*) + s_X(\bX_i, k \con \eta^*) + s_\type(k \con \eta^*)  \big\} \\
	& = 0
	\ .
\end{align*}
Combining the results above, we obtain the following functional form of $F_{1k} \cdot s$.
\begin{align}                         \label{proof:1-005}
	 \EXP \Big\{ F_{1k} & (\bY_i,  \bA_i, \bX_i, \type_i ) \cdot s (\bY_i, \bA_i, \bX_i, \type_i \con \eta^* )  \Big\}  \\
	\nonumber
	 & =  \sum_{\ba \in \zosets(\NI_{k}) }  \iint \big\{  \bw_k \T (\ba,\bx)  \by \big\} \cdot s_Y (\by , \ba, \bx, k \con \eta^* ) \py^*(\by \cond \ba , \bx , k  )  \px^*(\bx \cond k ) \, d \by \, d \bx
	 \ .
\end{align}

Next we study the expectation of $F_{2k} \cdot s$.
\begin{align*}
	 \EXP  & \Big\{  F_{2k} (\bX_i, \type_i ) \cdot s (\bY_i, \bA_i, \bX_i, \type_i \con \eta^* )  \Big\}  \\
	 & = \sum_{k'=1}^\NT p_{k'}^* \cdot  \EXP \Big\{ F_{2k} (\bX_i, k' ) \cdot s (\bY_i, \bA_i, \bX_i, k' \con \eta^* )  \, \Big| \, \type_i = k' \Big\} \\
	 & =\sum_{\ba_i \in \zosets( \NI_k ) } \EXP \Big[ \bw_k \T (\ba_i,\bX_i) \OR^*(\ba_i, \bX_i, k)  \Big\{ s_Y(\bY_i, \bA_i, \bX_i, k \con \eta^*) 
	 \\
	 & \hspace*{4cm} 
	 + s_A(\bA_i, \bX_i, k \con \eta^*) + s_X(\bX_i, k \con \eta^*) \Big\} \, \Big| \, \type_i = k  \Big] \\
	 & \hspace*{2cm} + \underbrace{
	 \EXP \bigg\{ \sum_{\ba_i \in \zosets( \NI_k ) } \bw_k \T (\ba_i,\bX_i)  \OR^*(\ba_i, \bX_i, k) - \uT_k^* \, \bigg| \, \type_i = k  \bigg\}}_{= \ 0 } s_\type(k \con \eta^*) \\
	 & =\sum_{\ba_i \in \zosets( \NI_k ) } \EXP \Big[ \bw_k \T (\ba_i,\bX_i) \OR^*(\ba_i, \bX_i, k)  \Big\{ s_Y(\bY_i, \bA_i, \bX_i, k \con \eta^*) 
	 \\
	 & \hspace*{4cm} 
	 + s_A(\bA_i, \bX_i, k \con \eta^*) + s_X(\bX_i, k \con \eta^*) \Big\} \, \Big| \, \type_i = k  \Big] 
	 \ .
\end{align*}
The first identity is straightforward from the law of total expectation. The second identity is from the definition of $F_{2k}$ and $s$ along with the property of score functions. Finally, the last identity is from the definition of $\uT_k^*$. We study the conditional expectations of the product of $\bw\T(\ba,\bX_i) \OR^*(\ba, \bX_i, k)$ and the score functions. The first product $\bw\T(\ba,\bX_i) \OR^*(\ba, \bX_i, k) \cdot \{ s_Y+s_A \}$ has zero conditional expectation.
\begin{align*}
	&  \EXP \Big[ \bw\T(\ba,\bX_i) \OR^*(\ba, \bX_i, k) \cdot \big\{  s_Y(\bY_i, \bA_i, \bX_i, k \con \eta^*) + s_A( \bA_i, \bX_i, k \con \eta^*)\big\}  \, \Big| \, \type_i = k \Big] \\
	& = \EXP \Big[ \bw\T(\ba,\bX_i) \OR^*(\ba, \bX_i, k) \cdot \EXP \big\{    s_Y(\bY_i, \bA_i, \bX_i, k \con \eta^*) + s_A(\bA_i, \bX_i, k \con \eta^*) \, \big| \, \bX_i, \type_i = k \big\} \, \Big| \, \type_i = k \Big] \\
	& = 0
	\ .
\end{align*}
The first identity uses the law of total expectation and the second identity is from the property of score functions. The product $\bw\T(\ba,\bX_i) \OR^*(\ba, \bX_i, k) \cdot s_X$ has the following conditional expectation.
\begin{align*}
	&
	\EXP \Big\{ \bw\T(\ba,\bX_i) \OR^*(\ba, \bX_i, k)  \cdot  s_X(\bX_i, k \con \eta^*)   \, \Big| \, \type_i = k \Big\} 
	\\
	&
	=    
     \iint  \big\{ \bw_k \T (\ba,\bx) \by \big\} \py^*(\by \cond \ba , \bx , k  )  s_X (\bx, k \con \eta^*) \px^*(\bx \cond k ) \, d \by \, d \bx
     \ .
\end{align*}
Combining all the results above, we obtain the following functional form of $F_{2k} \cdot s$.
\begin{align}                         \label{proof:1-006}
	\EXP   \Big\{  F_{2k} & (\bX_i, \type_i ) \cdot s (\bY_i, \bA_i, \bX_i, \type_i \con \eta^* )  \Big\}  
	\nonumber
	\\
	& =
	\sum_{\ba \in \zosets(\NI_{k}) }  \iint \big\{ \bw_k \T (\ba,\bx)  \by \big\} \cdot \py^*(\by \cond \ba , \bx , k  )  s_X (\bx, k \con \eta^*) \px^*(\bx \cond k ) \, d \by \, d \bx
	\ .
\end{align}
This proves that $\uT_k(\eta)$ and $\bT(\eta)$ are differentiable parameters, i.e., 
\begin{align}                         \label{proof:1-007}
	& 
	\EXP
	\Big\{
		\varphi_k  (\uT_k^*)  \cdot s (\bY_i, \bA_i, \bX_i, \type_i \con \eta^* )
	\Big\} \\
	\nonumber 
	& = 
	\EXP \Big[ \big\{ F_{1k}  (\bY_i, \bA_i, \bX_i, \type_i )  + F_{2k} (\bX_i, \type_i) \big\} \cdot s (\bY_i, \bA_i, \bX_i, \type_i \con \eta^* ) 
	\Big] \\
	\nonumber 	
	& =
	\sum_{\ba \in \zosets(\NI_{k}) }  \iint  \big\{ \bw_k \T (\ba,\bx)  \by \big\} \cdot s_Y (\by , \ba, \bx, k \con \eta^* ) \py^*(\by \cond \ba , \bx , k  )  \px^*(\bx \cond k ) \, d \by \, d \bx \\
	\nonumber 	
	& \hspace*{1cm} + \sum_{\ba \in \zosets(\NI_{k}) }  \iint \big\{ \bw_k \T (\ba,\bx)  \by \big\} \cdot \py^*(\by \cond \ba , \bx , k  )  s_X (\bx, k \con \eta^*) \px^*(\bx \cond k ) \, d \by \, d \bx
	\\
	\nonumber 
	& = 
\frac{\partial \uT_k ( \eta^* )  }{\partial \eta} 
\end{align}
where the first identity is a direct consequence of \eqref{proof:1-005} and \eqref{proof:1-006} and the second identity holds from \eqref{proof:1-004}.

Next, we claim that $\varphi (\bT^*)$ belongs to $\mathcal{T}$ in \eqref{proof:1-003}. By showing that $\varphi_k (\uT_k^*) = F_{1k} ( \bY_i, \bA_i, \bX_i , \type_i) + F_{2k} (\bX_i, \type_i) $ satisfies the elementry-wise conditions on $S (\by, \ba, \bx, k)$. First, $F_{1k}$ satisfies the conditional mean zero condition given $\bA_i, \bX_i, \type_i$.
\begin{align*}
	\EXP \big\{ F_{1k} ( \bY_i& ,  \ba, \bx , k')  \, \big| \, \bA_i = \ba , \bX_i = \bx , \type_i = k' \big\} \\
		& = 
				\frac{\ind(k' = k)}{p_k^*}
	\frac{\bw_k \T(\ba.\bx) }{e^*(\ba \cond \bx, k)} 
	\underbrace{ \EXP \Big\{ \bY_i - \OR^*(\ba, \bx, k) \, \Big| \, \bA_i = \ba , \bX_i = \bx , \type_i = k \Big\} }_{= \ 0}
			 = 0 \ .
\end{align*}
Next, $F_{2k}$ satisfies the conditional mean zero condition given $\bX_i, \type_i$.
\begin{align*}
	&
	\EXP \big\{ F_{2k} ( \bX_i ,   k')  \, \big| \, \type_i = k' \big\}
	\\
	&
	=
	 \frac{\ind(k' = k)}{p_k^*} \bigg[ \underbrace{ \sum_{\ba_i \in \zosets( \NI_k ) } \EXP \big\{  \bw_k \T (\ba_i,\bX_i) \OR^*(\ba_i, \bX_i, k) \, \big| \type_i = k \big\} 
				- \uT_k^*}_{= \ 0} \bigg] = 0
				\ .
\end{align*}
This concludes that each $\varphi_k(\uT_k^*)$ satisfies the conditions on each component of $\mathcal{T}$. Therefore, $\varphi(\bT^*) \in \mathcal{T}$. Moreover, $\varphi(\bT^*) $ is the efficient influence function  of $\bT^*$ by \citet{Newey1990}.

Second, we consider the result under model $\modelive$. Due to the knowledge of $e^*$, the submodel in \eqref{proof-submodel} becomes 
\begin{align*}	
	P ( \by , \ba , \bx , k \con \eta) 
	& = \py (\by \cond \ba , \bx , k \con \eta) e^*(\ba \cond \bx , k ) \px ( \bx \cond k \con \eta) p_k(\eta)  \ .
\end{align*}
The corresponding score function is
\begin{align*}
	s_{e^*} (\by,\ba,\bx, k \con \eta) 
	&=  s_Y( \by , \ba , \bx, k \con \eta )  +  s_X ( \bx , k \con \eta) + s_\type( k \con \eta)  
\end{align*}
where $s_Y$, $s_X$, and $s_\type$ are the same as in the previous proof under $\modeliv$ case. Accordingly, the $\NT$-dimensional tangent space becomes
\begin{align*}
	\mathcal{T}_{e^*} =
	\Big\{
		 S (\by,  \ba,  \bx, k)    \, \Big| \, 
		 & S (\by, \ba, \bx, k) = \big(  S_1 (\by, \ba, \bx , k) , \ldots , S_\NT(\by, \ba, \bx , k) \big)\T \in \R^\NT \ , \ \\
		      \nonumber		      	      
		      & \text{For all } \by \in \R^{\NI_k} , \ba \in \zosets(\NI_k) , \bx \in \mathcal{X}(k), k, \ell =1, \ldots , \NT \ , \ \\
		      & \nonumber 
			S_\ell (\by , \ba , \bx , k) 	
			= S_{\ell Y}(\by , \ba, \bx, k) + S_{\ell X}( \bx, k)  + S_{\ell \type}(k), 
			\\
			\nonumber
		      &  \EXP \big\{ S_{\ell Y}(\bY_i , \ba, \bx, k) \cond \bA_i = \ba, \bX_i =  \bx , \type_i = k \big\} = 0, \\
	    	\nonumber
	    	&   \EXP \big\{ S_{\ell X}(\bX_i , k ) \cond \type_i = k \big\} = 0, \quad
	    	\EXP\big\{ S_{\ell \type} (\type_i ) \big\} = 0
	    \Big\} \ .
\end{align*}
By following the similar approach, we find that $\varphi(\bT^*)$ satisfies 
\begin{align}                        \label{proof:Adatp-DiffPara}
	\frac{\partial \bT (\eta^*) }{\partial \eta} 
	= 
	\EXP
	\Big\{
		\varphi (\bT^*) \cdot s_{e^*} \T (\bY_i, \bA_i, \bX_i, \type_i \con \eta^* )
	\Big\} 
\end{align}
and $\varphi(\bT^*) \in \mathcal{T}_{e^*}$. Therefore, $\varphi(\bT^*)$ is the efficient influence function of $\bT^*$ in model $\modelive$.

The semiparametric efficiency bound is the variance of the efficient influence function . Because of the zero mean property of $\varphi(\bT^*)$ and the orthogonality between each component of $\varphi (\bT^*)$, we obtain
\begin{align*}	
	\VAR \big\{ \varphi(\bT^*) \big\}
	= 
	\EXP \big\{ 	\varphi(\bT^*) \varphi \T (\bT^*)  \big\}
	=
	\text{diag} \Big[ 	 \EXP \big\{ \varphi_1 (\uT_1^*)^2 \big\} , \ldots , \EXP \big\{ \varphi_\NT (\uT_\NT^*)^2  \big\} \Big]
	\ .
\end{align*}
Each component is represented as 
\begin{align}                        		 \label{proof:1-008}
	&
	\EXP \big\{  \varphi_k (\uT_k^*)^2  \big\}
	\\
	\nonumber
	& =
	\EXP \Big[ \EXP \big\{  \varphi_k (\uT_k^*)^2   \, \big| \, \type_i \big\} \Big] \\ 
	\nonumber	
	& =
	p_k^*  \EXP \big\{  \varphi_k (\uT_k^*)^2 \, \big| \, \type_i = k \big\}
	\\
	\nonumber		
	& = 	
	p_k^* \EXP \Big\{  F_{1k}(\bY_i, \bA_i, \bX_i, k)^2 + F_{2k}(\bX_i, k)^2 + 2 F_{1k}(\bY_i, \bA_i, \bX_i, k) F_{2k}(\bX_i, k)  \, \Big| \, \type_i = k \Big\} 
	\ .
\end{align}
The first identity holds from the law of total expectation and the second identity is from the form of $\varphi_k$. The last identity is from the definition of $F_{1k}$ and $F_{2k}$ in \eqref{proof:1-FG}.

We study the explicit form of each piece in \eqref{proof:1-008}. First, the conditional mean of $F_{1k}^2$ is given by
\begin{align*}
	& \EXP  \Big\{  F_{1k}(\bY_i, \bA_i, \bX_i, k)^2 \, \Big| \, \type_i = k \Big\} 
	\\
	\nonumber
	& = \sum_{\ba_i \in \zosets(\NI_k) } \EXP \Big[ e^*(\ba_i \cond \bX_i, k )  \EXP \Big\{  F_{1k}(\bY_i, \ba_i, \bX_i, k)^2  \, \Big| \, \bA_i = \ba_i , \bX_i , \type_i = k \Big\}  \, \Big| \, \type_i = k \Big] \\
	\nonumber	
	& = \frac{1}{p_k^{*2}}\sum_{\ba_i \in \zosets(\NI_k) } 
	\EXP \bigg[  \frac{\bw_k\T(\ba_i,\bX_i)}{e^*(\ba_i \cond \bX_i, k) } 
	\EXP \Big[  \big\{ \bY_i - \OR^*(\ba_i, \bX_i, k) \big\} \\
	& \hspace*{4cm} \times  \big\{ \bY_i - \OR^*(\ba_i, \bX_i, k) \big\}\T
	\, \Big| \, \bA_i = \ba_i , \bX_i , \type_i = k \Big] \bw_k(\ba_i,\bX_i) \, \bigg| \, \type_i = k \bigg]
	\\	
	\nonumber
	& = \frac{1}{p_k^{*2}}\sum_{\ba_i \in \zosets(\NI_k) } 
	\EXP \bigg\{ 
	\frac{\bw_k\T(\ba_i,\bX_i)  \Sigma^*(\ba_i, \bX_i, k) \bw_k(\ba_i,\bX_i)}{e^*(\ba_i \cond \bX_i, k) }	
	 \, \bigg| \, \type_i = k \bigg\}
	 \ .
\end{align*}
Second, the conditional mean of $F_{2k}^2$ follows from straightforward algebra.
\begin{align*}
	&
	\EXP  \Big\{  F_{2k}(\bY_i,\bA_i, \bX_i, k)^2  \, \Big| \, \type_i = k \Big\} 
	\\
	&
	= \frac{1}{p_k^{*2}} \EXP\bigg[ \bigg\{
		 \sum_{\ba_i \in \zosets( \NI_k ) } \bw_k \T (\ba_i,\bX_i)  \OR^*(\ba_i, \bX_i, k) - \uT_k^* \bigg\}^2 \, \bigg| \, \type_i = k \bigg]
		 \ .
\end{align*}
Lastly, $F_{1k}$ and $F_{2k}$ are orthogonal, i.e.,
\begin{align*} 
	&
	\EXP \Big\{ F_{1k}(\bY_i, \bA_i, \bX_i, k) F_{2k}(\bX_i, k)  \, \Big| \, \type_i = k \Big\} \\
		\nonumber
	& =
	\EXP \Big[ \underbrace{ \EXP \Big\{ F_{1k}(\bY_i, \bA_i, \bX_i, k) \, \Big| \, \bA_i, \bX_i, \type_i = k \Big\} }_{= \ 0} F_{2k}(\bX_i, k)  \, \Big| \, \type_i = k \Big] = 0 
	\ .
\end{align*}
Again, the first identity is from the law of total expectation and the second identity is straightforward from the definition of $F_{1k}$. Combining the results above, we get the explicit form of $\EXP \big\{  \varphi_k (\uT_k^*)^2  \big\}$.
\begin{align*}
	\EXP \big\{  \varphi_k (\uT_k^*)^2  \big\}
	& =
		\frac{1}{p_k^*} \EXP \bigg[ 
			\sum_{\ba_i \in \zosets(\NI_k)} \frac{\bw_k \T(\ba_i,\bX_i) \Sigma^* (\ba_i, \bX_i, k) \bw_k (\ba_i,\bX_i)}{e^* (\ba_i \cond \bX_i, k)} 
			\\
			& \hspace*{2cm}
		+ \bigg\{
			\sum_{\ba_i \in \zosets(\NI_k) } \bw_k \T (\ba_i,\bX_i)  \OR^* (\ba_i, \bX_i, k) - \uT_k^*
			\bigg\}	^2 \, \bigg| \, \type_i = k
	\bigg] 
	\\
	& =\text{SEB}_k \big( \uT_k^* \big)
\end{align*}
where $\text{SEB}_k \big( \uT_k^* \big)$ is defined in the theorem.

\subsection{Proof of Theorem \ref{thm:EIF-known}}						\label{sec:appendix3-TA1}

We take a similar approach to the proof of Theorem \ref{thm:EIFVC}. Consider a smooth regular parametric submodel parametrized by a possibly multi-dimensional parameter $\eta$.
\begin{align*}
	P ( \by , \ba , \bx , k \con \eta) 
	& = \py (\by \cond \ba , \bx , k \con \eta) e(\ba \cond \bx , k \con \eta) \px ( \bx \cond k \con \eta) p_k^*
\end{align*}
where the smoothness and regularity conditions are given in Definition A.1 of the appendix in \citet{Newey1990}. Since $p_k^* = \EXP\{ \ind (\type_i = k)\}$ is known, it does not depend on $\eta$. We assume the density of the parametric submodel $P(\cdot \con \eta)$ equals the true density $P$ at $\eta=\eta^*$. The corresponding score function is
\begin{align*}
	s (\by,\ba,\bx, k \con \eta) 
	&=  s_Y( \by , \ba , \bx, k \con \eta ) + s_A( \ba , \bx , k \con \eta)  +  s_X ( \bx , k \con \eta) 
\end{align*}
where the score functions are defined in \eqref{proof:1-score}. 
From the parametric submodel, we obtain the $1$-dimensional tangent space for $1$-dimensional parameters which is the mean closure of all $1$-dimensional linear combinations of scores, that is, 
\begin{align}							\label{proof:3-002}
	\nonumber 
	\mathcal{T} =
	\Big\{
		 S (\by, \ba, \bx, k)  \,
		      \Big| 	\,	      
			& S(\by , \ba , \bx , k) 	   
			= S_Y (\by , \ba, \bx, k)  + S_A (\ba, \bx, k)  + S_X ( \bx, k)  \in \R  \ , \ 
			\\
			\nonumber
		      \, &  \EXP \big\{ S_Y (\bY_i , \ba, \bx, k) \cond \bA_i = \ba, \bX_i =  \bx , \type_i = k \big\} = 0 \ \text{ for all } ( \ba,  \bx ,  k)   \ , \ \\
		      \nonumber
	    	\, &   \EXP \big\{ S_A (\bA_i , \bx , k ) \cond \bX_i = \bx , \type_i = k \big\} = 0  \ \text{ for all } ( \bx, k)  \ , \   \\
	  \, &  	\EXP \big\{ S_X (\bX_i , k ) \cond \type_i = k \big\} = 0 
	   \ \text{ for all } k
	    \Big\}
	    \ .
\end{align}

The estimand $\oT^*$ can be represented as $ \oT(\eta) = \sum_{k=1}^\NT v_k\big( p_k(\eta) \big) \uT_k(\eta)$ at parameter $\eta$ in the regular parametric submodel where $\uT_k(\eta)$ has the functional form \eqref{proof:1-taueta}. Note that $\oT(\eta^*)$ equals the true $\oT^*$. Therefore, the derivative of $\oT$ evaluated at true $\eta^*$ is 
\begin{align*}
	\frac{\partial \oT(\eta^*)}{\partial \eta} 
	= 
	\sum_{k=1}^\NT v_k(p_k^*) \frac{\partial \uT_k ( \eta^* )  }{\partial \eta}\ .
\end{align*}
The conjectured efficient influence function  of $\oT^*$ is given as
\begin{align*}
	\varphi(\oT^*) 
	& = \sum_{k=1}^\NT v_k(p_k^*) \cdot  \varphi_k ( \uT_k^* )
	\ .
\end{align*}
First, we show that $\oT(\eta)$ is a differantiable parameter, which suffices to show
\begin{align*}
	\frac{\partial \oT (\eta^*) }{\partial \eta} 
	& = 
	\EXP
	\Big\{
		\varphi (\oT^*) \cdot s  (\bY_i, \bA_i, \bX_i, \type_i \con \eta^* )
	\Big\} 
	=
	\sum_{k=1}^\NT  \EXP\Big\{ v_k(p_k^*) \cdot  \varphi_k ( \uT_k^* ) \cdot s  (\bY_i, \bA_i, \bX_i, \type_i \con \eta^* ) \Big\} 
	\ .
\end{align*}
This is straightforward from the identity \eqref{proof:1-007} obtained in the proof of Lemma \ref{lmm:EIFbasic}.
\begin{align*}                             
	\sum_{k=1}^\NT  \EXP\Big\{ v_k(p_k^*) \cdot  \varphi_k ( \uT_k^* ) \cdot s  (\bY_i, \bA_i, \bX_i, \type_i \con \eta^* ) \Big\}
	= 
	\sum_{k=1}^\NT v_k(p_k^*) \cdot \frac{\partial \uT_k ( \eta^* )  }{\partial \eta}\ .
\end{align*}

We claim that $\varphi(\oT^*)$ belongs to $\mathcal{T}$ in \eqref{proof:3-002}, which suffices to show $\varphi_k(\uT_k^*) \in \mathcal{T}$. This is straightforward because $\varphi_k(\uT_k^*)$ satisfies the conditions imposed on $S_Y$ and $S_X$ in \eqref{proof:3-002}, which is shown in the proof of Lemma \ref{lmm:EIFbasic}. Therefore, $\varphi(\oT^*) $ is the efficient influence function  of $\oT^*$ under known $p_k^*$s and under $\modeliv$. 

We can show that $\varphi(\oT^*)$ is also the efficient influence function of $\oT^*$ under known $p_k^*$s and under $\modelive$ from similar manner. Specifically, the score function and the tangent space are updated as
\begin{align*}
	s_{e^*} (\by,\ba,\bx, k \con \eta) 
	=  	&
	s_Y( \by , \ba , \bx, k \con \eta )  +  s_X ( \bx , k \con \eta) 
\end{align*}
and	
\begin{align*}
\mathcal{T}_{e^*} =
	\Big\{
		 S (\by, \ba, \bx, k)  \,
		      \Big| 	\,	      
			& S(\by , \ba , \bx , k) 	   
			= S_Y (\by , \ba, \bx, k)  + S_X ( \bx, k)  \in \R  \ , \ 
			\\
			\nonumber
		      \, &  \EXP \big\{ S_Y (\bY_i , \ba, \bx, k) \cond \bA_i = \ba, \bX_i =  \bx , \type_i = k \big\} = 0 \ \text{ for all } ( \ba,  \bx ,  k)   \ , \ \\
	  \, &  	\EXP \big\{ S_X (\bX_i , k ) \cond \type_i = k \big\} = 0 
	   \ \text{ for all } k
	    \Big\}
	    \ .
\end{align*}
We can easily show 
\begin{align*}
	\frac{\partial \oT (\eta^*) }{\partial \eta} 
	& = 
	\EXP
	\Big\{
		\varphi (\oT^*) s_{e^*}  (\bY_i, \bA_i, \bX_i, \type_i \con \eta^* )
	\Big\} 
	=
	\sum_{k=1}^\NT  \EXP\Big\{ v_k(p_k^*) \varphi_k ( \uT_k^* ) s_{e^*}  (\bY_i, \bA_i, \bX_i, \type_i \con \eta^* ) \Big\} 
\end{align*}
and $\varphi(\oT^*) \in \mathcal{T}_{e^*}$. Therefore, $\varphi(\oT^*) $ is the efficient influence function  of $\oT^*$ under known $p_k^*$s and under $\modelive$.

\subsection{Proof of Theorem \ref{thm:DRVC}}				\label{sec:ThmS2}

Conditional on the observed cluster types, $(\type_1,\ldots,\type_\NC) = (\ell_1,\ldots,\ell_\NC) \in \{1,\ldots,\NT\}^\NC$, we observe that
\begin{align}									\label{proof:4-001}
	\EXP \big\{ \widehat{\uT}_k &  ( e' , \OR') \cond (\type_1 , \ldots , \type_\NC) =  (\ell_1 , \ldots , \ell_\NC) \big\} \\
	& = 
	\frac{1}{\NC_k} \sum_{i=1}^\NC \ind(\ell_i = k) \EXP \bigg[ \sum_{\ba_i \in \zosets(\NI_k) } \frac{\ind(\bA_i = \ba_i )}{e' (\ba_i \cond \bX_i, k)} \bw_k \T ( \ba_i,\bX_i) \Big\{ \bY_i - \OR'(\ba_i, \bX_i, k) \Big\} \nonumber \\
	& \hspace*{6cm}  + \sum_{\ba_i \in \zosets(\NI_k) } \hspace*{-0.2cm} \bw_k \T(\ba_i,\bX_i) \OR'(\ba_i, \bX_i, k) \, \bigg| \, \type_i = k \bigg]
	\nonumber 
	\ .
\end{align}
The summand is not empty because of condition (A5)  in Assumption \ref{Assump:VC} in the main paper. 

We study the conditional expectation in \eqref{proof:4-001} for each mis-specification scenario assumed in the theorem. First, suppose $e'$ is correctly specified but $\OR'$ is mis-specified; i.e., $\widehat{\uT}_k(e', \OR') = \widehat{\uT}_k(e^*, \OR')$ where $\OR^* \neq \OR'$. Then, the conditional expectation in \eqref{proof:4-001} is
\begin{align}					\label{proof:4-002}
	\nonumber 
	& \EXP  \bigg[ \sum_{\ba_i \in \zosets(\NI_k) }  \frac{\ind(\bA_i = \ba_i)}{e' (\ba_i \cond \bX_i, k)} \bw_k \T ( \ba_i,\bX_i) \Big\{ \bY_i - \OR'(\ba_i, \bX_i, k) \Big\} 
	\\
	\nonumber
	& \hspace*{6cm}
	+  \sum_{\ba_i \in \zosets(\NI_k) } \hspace*{-0.2cm} \bw_k \T(\ba_i,\bX_i) \OR'(\ba_i, \bX_i, k) \, \bigg| \, \type_i = k \bigg]
	\\
	\nonumber 
	& = 
	\sum_{\ba_i \in \zosets(\NI_k) } \bigg[ \EXP \Big[ \bw_k \T ( \ba_i,\bX_i) \EXP \Big\{ \bY_i - \OR'(\ba_i, \bX_i, k)   \, \Big| \, \bA_i = \ba_i , \bX_i , \type_i = k \Big\} \, \Big| \, \type_i = k \Big]
	\\ 
	& \hspace*{6cm} +  \EXP \Big\{ \bw_k \T ( \ba_i,\bX_i) \OR'(\ba_i, \bX_i, k) \, \Big| \, \type_i = k \Big\}  \bigg]
	\nonumber
	\\
	\nonumber 
	& = \sum_{\ba_i \in \zosets(\NI_k) }  \EXP \Big\{  \bw_k \T ( \ba_i,\bX_i)  \EXP \big( \bY_i  \, \big| \, \bA_i = \ba_i , \bX_i , \type_i = k \big)  \, \Big| \, \type_i = k 	\Big\}  
	\\
	\nonumber 
	& = \sum_{\ba_i \in \zosets(\NI_k) } \EXP \big\{ \bw_k \T ( \ba_i,\bX_i) \OR^*(\ba_i , \bX_i, k) \, \big| \, \type_i = k \big\} 
	\\
	& = \uT_k^*\ .
\end{align}
Here, the first identity is based on an assumption on $e'$ and $\OR'$  along with the law of total expectation. The second identity is based on the law of total expectation applied to $\OR'$ which cancels out with the second term. The rest of the identities are from the definition of $\OR^*$ and $\uT_k^*$.

Next, suppose $\OR'$ is correctly specified but $e'$ is mis-specified; i.e., $\widehat{\uT}_k(e', \OR') = \widehat{\uT}_k(e', \OR^*)$ where $e^* \neq e'$. Then, the conditional expectation in \eqref{proof:4-001} is
\begin{align}				\label{proof:4-003}
\nonumber 
	&   \EXP \bigg[ \sum_{\ba_i \in \zosets(\NI_k) } \frac{\ind(\bA_i = \ba_i)}{e' (\ba_i \cond \bX_i, k)} \bw_k \T ( \ba_i,\bX_i) \Big\{ \bY_i - \OR'(\ba_i, \bX_i, k) \Big\} \\
	\nonumber 
	&
	\hspace*{6cm} + \sum_{\ba_i \in \zosets(\NI_k) } \bw_k \T(\ba_i) \OR'(\ba_i, \bX_i, k) \, \bigg| \, \type_i = k \bigg]
	\\
	\nonumber 
	& = 
	\sum_{\ba_i \in \zosets(\NI_k ) } \EXP \bigg[ \bw_k  \T(\ba_i,\bX_i)  \frac{e^* (\ba_i \cond \bX_i , k )}{e'(\ba_i \cond \bX_i , k )} \EXP \Big\{ \bY_i - \OR^*(\ba_i, \bX_i, k)   \, \Big| \, \bA_i = \ba_i , \bX_i , \type_i = k \Big\} \, \bigg| \, \type_i = k \bigg]  
	\\
	\nonumber 
	& \hspace*{6cm}
	 + \sum_{\ba_i \in \zosets(\NI_k) } \EXP \Big\{ \bw_k  \T(\ba_i,\bX_i) \OR^*(\ba_i, \bX_i , k )\, \Big| \, \type_i = k  \Big\}  \\
	& =  \uT_k^*
	\ .
\end{align}
The first identity is based on an assumption on $e'$ and $\OR'$  along with the law of total expectations. The second identity is straightforward from the definition of $\OR^*$ and $\uT_k^*$. Also, if  one of $e'$ or $\OR'$ is correctly specified, we obtain 
\begin{align*}
	\EXP \big\{ \widehat{\uT}_k ( e' , \OR') \cond (\type_1 , \ldots , \type_\NC) =  (\ell_1 , \ldots , \ell_\NC) \big\} 
	= \frac{1}{\NC_k} \sum_{i=1}^\NC \ind(\ell_i = k) \uT_k^* 
	= \uT_k^*
	\ .
\end{align*}
The first identity is from \eqref{proof:4-002} and \eqref{proof:4-003} and the second identity is straightforward from the definition of $\NC_k = \sum_{i=1}^\NC \ind( \ell_i = k)$. This implies that
\begin{align}				\label{proof:4-004}
	\EXP \big\{ \widehat{\uT}_k ( e' , \OR') \cond \type_1 , \ldots , \type_\NC \big\}	= \uT_k^*
	\ .
\end{align}
As a result, $\widehat{\oT} ( e' , \OR' )$ is an unbiased estimator for $\oT^*$. 
\begin{align*}
	\EXP \big\{ \widehat{\oT} ( e' , \OR' ) \big\}
	& =
	 \EXP \Big[
	\EXP \big\{ \widehat{\oT}(e' , \OR') 
	\, \big| \, \type_1 , \ldots , \type_\NC
	\big\} \Big] \\
	& =
	\sum_{k=1}^\NC
	 \EXP \Big[
	 \widehat{v}_k
	\EXP \big\{ \widehat{\uT}_k(e' , \OR') 
	\, \big| \, \type_1 , \ldots , \type_\NC
	\big\} \Big] 
	=
	\sum_{k=1}^\NC
	 \EXP \big( \widehat{v}_k \big)\uT_k^* 
	= \sum_{k=1}^\NC v_k(p_k) \uT_k^* = \oT^*
	\ .
\end{align*}
Again, the first identity is from the law of total expectation and the second identity is from the definition of $\widehat{\oT}(e', \OR')$ and $\widehat{v}_k$ is a function only of $\type_1,\ldots,\type_\NC$. The third identity is from \eqref{proof:4-004} and the fourth identity is based on the unbiasedness assumption on $\EXP(\widehat{v}_k) = v_k(p_k^*)$. The last identity is straightforward from the definition of $\oT^*$.

\subsection{Proof of Lemma \ref{lmm:ParaEst}}						\label{sec:prooflmm41}

We follow the M-estimation theory laid out in \citet{Stenfanski2002} and Section 5 of \citet{Vaart1998} to prove the theorem.

	We first show that $\big( \bT^* , \paraT^\dagger \big)$ is the unique root of the population equation $\EXP \big\{ \ee \big(\bT^*,   \paraT^\dagger   \big) \big\} = 0$ under model $\modelie \cup \modelig$. It is trivial that $\EXP\big\{ \ee_e(\paraPS^\dagger ) \big\}$, and $\EXP\big\{ \ee_\indOR(\paraOR^\dagger) \big\}$ are zero by the definition and the consistency result from M-estimation, so it suffices to show $\EXP \big\{ \ee_\uT \big( \bT^* , \paraT^\dagger \big) \big\}=0$.	Under model $\modelie$, $\paraPS^\dagger$ is equal to the true propensity score parameter $\paraPS^*$, so that $e^{\rm Par}( \ba \cond \bx , k \con \paraPS^\dagger) = e^{\rm Par}( \ba \cond \bx , k \con \paraPS^*) = e^*(\ba \cond \bx , k)$. Therefore, $\EXP \big\{ \ee_{\uT,k} \big( \uT_k^*, \paraT^\dagger \big) \big\}  = \EXP \big\{ \widehat{\uT}_k (e^* , \OR') \big\} - \uT_k^*$ where the form of $\widehat{\uT}_k(e^* , \OR')$ is presented in the main paper and $\OR'( \ba , \bx , k ) = \POR ( \ba , \bx , k \con \paraOR^\dagger )$ is  a possibly mis-specified outcome regression. Similarly, under model $\modelig$, $\paraOR^\dagger$ is equal to the true outcome regression parameter $\paraOR^*$, so that $\POR ( \ba , \bx , k \con \paraOR^\dagger) = \POR ( \ba , \bx , k \con \paraOR^*) = \OR^*(\ba , \bx , k)$. Therefore, $\EXP \big\{ \ee_{\uT,k} \big( \uT_k^*, \paraT^\dagger \big) \big\}  = \EXP \big\{ \widehat{\uT}_k (e' , \OR^*) \big\} - \uT_k^*$ where $e'( \ba \cond \bx , k ) = e^{\rm Par} ( \ba \cond \bx , k \con \paraPS^\dagger )$ is a possibly mis-specified propensity score. Thus, $\EXP \big\{ \ee_{\uT,k} \big( \uT_k^* , \paraT^\dagger   \big) \big\} = 0 $ under model $\modelie \cup \modelig$ from the intermediate results in \eqref{proof:4-001} and \eqref{proof:4-002} from the proof of Theorem \ref{thm:DRVC}. Since $\big( \bT^*, \paraT^\dagger \big)$ is the root of $\EXP \big\{ \ee \big( \bT^*, \paraT^\dagger \big) \big\} = 0$, Theorem 5.41 of \citet{Vaart1998} states that
	\begin{align}						\label{proof:6-000}
		\sqrt{\NC} \begin{bmatrix}
			\widehat{\bT}  - \bT^* \\
			\widehat{\paraT} - \paraT^\dagger
		\end{bmatrix}
		= 
		- \frac{1}{\sqrt{\NC}} \bigg[  \underbrace{ \EXP \bigg\{ \frac{\partial \ee \big( \bT ^* , \paraT^\dagger \big) }{\partial \big( \bT  , \paraT \big)\T }  \bigg\}}_{(A)} \bigg]^{-1} \sum_{i=1}^\NC \ee \big( \bT^* , \paraT^\dagger \big) + o_P(1)
		\ .
 	\end{align}
	Note that the expectation of the Jacobian matrix $(A)$ is 
	\begin{align*}
		(A) = 
		\EXP \bigg\{ \frac{\partial \ee \big( \bT ^* , \paraT^\dagger \big) }{\partial \big( \bT  , \paraT \big)\T }  \bigg\} 
		= \begin{bmatrix}
			- \text{diag}(\bp^*) 
			& \displaystyle{ \EXP  \bigg\{	  \frac{\partial \ee_\uT \big(\bT^* , \paraT^\dagger \big)  }{\partial  \paraT \T }   \bigg\} } 
			\\[0.7cm]
			0 & \displaystyle{ \EXP  \bigg\{	\frac{\partial \ee_{\beta} (\paraT^\dagger)  }{\partial  \paraT \T }  \bigg\} } 
		\end{bmatrix}
	\end{align*}
	where $\text{diag}(\bp^*) = {\rm diag} \big[ p_1^*,\ldots,p_\NT^* \big]$.	Therefore, we find 
	\begin{align*}
		(A)^{-1}		
		& =
		\bigg[ \EXP \bigg\{ \frac{\partial \ee \big( \bT ^* , \paraT^\dagger \big) }{\partial \big( \bT  , \paraT \big)\T }  \bigg\}  \bigg]^{-1} 
		\\
		&
		= \begin{bmatrix}
			\displaystyle{ -   \text{diag} \bigg( \frac{1}{\bp^*} \bigg) }
			& 
			\displaystyle{  \text{diag} \bigg( \frac{1}{\bp^*} \bigg) \EXP  \bigg\{	  \frac{\partial \ee_\uT \big(\bT^* , \paraT^\dagger \big)  }{\partial  \paraT \T }   \bigg\}	\bigg[ \EXP  \bigg\{	\frac{\partial \ee_{\beta} (\paraT^\dagger)  }{\partial  \paraT \T }  \bigg\} \bigg]^{-1}		} \\[0.7cm]
			0 & \displaystyle{ \bigg[ \EXP  \bigg\{	\frac{\partial \ee_{\beta} (\paraT^\dagger)  }{\partial  \paraT \T }  \bigg\} \bigg]^{-1}}
		\end{bmatrix}
	\end{align*}
	where $\text{diag}(1/\bp^*) = {\rm diag} \big[ 1/p_1^*,\ldots,1/p_\NT^* \big]$. Replacing $(A)^{-1}$ in \eqref{proof:6-000} with the form above, we get the linear expansion of $\NC^{1/2} \big( \widehat{\bT} -\bT^* \big)$
	\begin{align}					\label{proof:6-001}
	&
		\sqrt{\NC} \big( \widehat{\bT} -\bT^* \big)
		\\
		\nonumber
		&
		= 	\frac{1}{\sqrt{\NC}} \sum_{i=1}^\NC  \text{diag} \bigg( \frac{1}{\bp^*} \bigg)
		 \bigg[
			\ee_\uT (\bT^* , \paraT^\dagger)   - 
			\EXP  \bigg\{	  \frac{\partial \ee_\uT \big(\bT^* , \paraT^\dagger \big)  }{\partial \paraT \T }   \bigg\}	\bigg[ \EXP  \bigg\{	\frac{\partial \ee_{\beta} (\paraT^\dagger)  }{\partial  \paraT \T }  \bigg\} \bigg]^{-1}\hspace*{-0.2cm}	\ee_{\beta}(\paraT^\dagger)
		\bigg] + o_P(1)\ .
	\end{align}
	It is straightforward to check that the $k$th component of the influence function in \eqref{proof:6-001} is equivalent to $\varphi_k^{\rm Par} (\uT_k^*, \paraT^\dagger)$. Therefore, the influence function in \eqref{proof:6-001} is equivalent to $\varphi^{\rm Par} (\bT^*, \paraT^\dagger)$.

\subsection{Proof of Theorem \ref{thm:NoGainFromPS}}		\label{sec:proofS3}

It suffices to show that $\varphi(\oT^*)$ is the efficient influence function of $\oT*$ in model $\modelive$. The proof is similar to the proof of Theorem \ref{thm:EIFVC}. The density of $(\bO_i , \type_i) = (\bY_i, \bA_i, \bX_i, \type_i)$ with respect to some $\sigma$-finite measure is 
\begin{align*}
	P^* ( \by , \ba , \bx , k) 
	& = \py^* (\by \cond \ba , \bx , k) e^*(\ba \cond \bx , k) \px^* ( \bx \cond k) p_k^*
\end{align*}
where $\py^*$ is the conditional density of $\bY_i$ given $(\bA_i, \bX_i, \type_i)$ and $\px^*$ is the conditional density of $\bX_i$ given $\type_i$. An asterisk in superscript of (conditional) density represents the true (conditional) density. A smooth regular parametric submodel parametrized by a possibly multi-dimensional parameter $\eta$ is
\begin{align*}
	P ( \by , \ba , \bx , k \con \eta) 
	& = \py (\by \cond \ba , \bx , k \con \eta) e^*(\ba \cond \bx , k) \px ( \bx \cond k \con \eta) p_k(\eta)  
\end{align*}
where the smoothness and regularity conditions are given in Definition A.1 of the appendix in \citet{Newey1990}. We assume the density of the parametric submodel $P(\cdot \con \eta)$ equals the true density $P^*$ at $\eta=\eta^*$. The corresponding score function is
\begin{align*}		
	s_{e^*} (\by,\ba,\bx, k \con \eta) 
	&=  s_Y( \by , \ba , \bx, k \con \eta ) +  s_X ( \bx , k \con \eta) + s_\type( k \con \eta)  
\end{align*}
where $s_Y$, $s_X$, and $s_\type$ are defined in \eqref{proof:1-score}.

The 1-dimensional tangent space for $1$-dimensional parameters is
\begin{align}							\label{proof:Adapt-002}
	\nonumber 
	\mathcal{T}_{e^*} =
	\Big\{
		 S (\by, \ba, \bx, k)  \in \R \,
		      \Big| 	\,	      
			& S(\by , \ba , \bx , k) 	   
			= S_Y (\by , \ba, \bx, k)  + S_X ( \bx, k)  + S_\type (k)  \ , \
			\\
			\nonumber
		      \, &  \EXP \big\{ S_Y (\bY_i , \ba, \bx, k) \cond \bA_i = \ba, \bX_i =  \bx , \type_i = k \big\} = 0 \ \text{ for all } (\ba,  \bx ,  k)  \ , \   \\
	  \, &  	\EXP \big\{ S_X (\bX_i , k ) \cond \type_i = k \big\} = 0 \ \text{ for all } k \ , \   \EXP\big\{ S_\type (\type_i ) \big\} = 0
	    \Big\}
\end{align}

The estimand $\oT^*$ is re-represented as $ \oT(\eta) = \sum_{k=1}^\NT v_k\big( p_k(\eta) \big) \uT_k(\eta)$ at parameter $\eta$ in the regular parametric submodel where $\uT_k(\eta)$ has the following functional form \eqref{proof:1-taueta}. Note that $\oT(\eta^*)$ equals the true $\oT^*$. Therefore, the derivative of $\oT$ evaluated at true $\eta^*$ is 
\begin{align*}	
	\frac{\partial \oT(\eta^*)}{\partial \eta} 
	= 
	\sum_{k=1}^\NT v_k(p_k^*) \frac{\partial \uT_k ( \eta^* )  }{\partial \eta}
	+
	\sum_{k=1}^\NT \uT_k^* \frac{\partial v_k ( p_k^* ) }{\partial \eta} \ .
\end{align*}
The conjectured efficient influence function  of $\oT^*$ is
\begin{align*}
	\varphi(\oT^*) 
	& = \sum_{k=1}^\NT v_k(p_k^*) \cdot  \varphi_k ( \uT_k^* )
		 + \sum_{k=1}^\NT \Big\{ \ind(\type_i=k) -p_k^* \Big\} \frac{\partial v_k(p_k^*)}{\partial p_k }   \uT_k^* 
		  \ .
\end{align*}

First, we check that $\oT(\eta)$ is a differantiable parameter, i.e.
\begin{align}				\label{proof:Adapt-004}
	\frac{\partial \oT (\eta^*) }{\partial \eta} 
	& = 
	\EXP
	\Big\{
		\varphi (\oT^*) \cdot s_{e^*}  (\bY_i, \bA_i, \bX_i, \type_i \con \eta^* )
	\Big\} \\
	\nonumber	
	& =
	\sum_{k=1}^\NT  \EXP\Big\{ v_k(p_k^*) \cdot  \varphi_k ( \uT_k^* ) \cdot s  (\bY_i, \bA_i, \bX_i, \type_i \con \eta^* ) \Big\}  \\
	& \hspace*{2cm} + \sum_{k=1}^\NT  \EXP\bigg[ \Big\{ \ind(\type_i=k) -p_k^* \Big\} \frac{\partial v_k(p_k^*)}{\partial p_k }   \uT_k^*  \cdot s_{e^*}  (\bY_i, \bA_i, \bX_i, \type_i \con \eta^* ) \bigg]
	\nonumber	
	 \ .
\end{align}
From \eqref{proof:Adatp-DiffPara}, we find
\begin{align*} 
	\sum_{k=1}^\NT  \EXP\Big\{ v_k(p_k^*) \cdot  \varphi_k ( \uT_k^* ) \cdot s_{e^*}  (\bY_i, \bA_i, \bX_i, \type_i \con \eta^* ) \Big\}
	= 
	\sum_{k=1}^\NT v_k(p_k^*) \cdot \frac{\partial \uT_k ( \eta^* )  }{\partial \eta} 
	 \ .
\end{align*}
Following \eqref{proof:2-005}, we have
\begin{align*}
	\nonumber 
	 &  \sum_{k=1}^\NT \EXP  \bigg[
		  \Big\{ \ind(\type_i=k) - p_k^* \Big\} \frac{\partial v_k ( p_k^* ) }{\partial p_k }  \uT_k^* \cdot s_{e^*} (\bY_i , \bA_i , \bX_i , \type_i \con \eta^* ) 
	\bigg]
	= \sum_{k=1}^\NT \uT_k^*  \frac{\partial v_k ( p_k^* ) }{\partial \eta} 
	 \ .
\end{align*}
Therefore, this shows that \eqref{proof:Adapt-004} holds.

Next we show $\varphi(\oT^*) $ belongs to $\mathcal{T}_{e^*}$ in \eqref{proof:Adapt-002} which can be proven by the same way in the proof or Theorem \ref{thm:EIFVC}. This concludes that $\varphi(\oT^*)$ is the efficient influence function of $\oT*$ under model $\modelive$ and that $\widehat{\oT}$ achieves the semiparametric efficiency bound of $\oT^*$ under model $\modelive$.

\subsection{Proof of Lemma \ref{lmm:NoInterference}}
\label{sec:appendix3-LA1}

	We observe that
		\begin{align*}
			& \uT_k^\DE(\alpha) 
			\\
			& = 
			\sum_{ \ba_i \in \zosets(\NI_k) } 
				\big\{ \bw_k^\DE(\ba_i \con \alpha) \big\}\T \EXP \big\{ \bY_i( \ba_i  ) \cond \type_i = k \big\}  \\
			& =
			\frac{1}{\NI_k} \sum_{j=1}^{\NI_k} u_j\T  \sum_{ \ba_i \in \zosets(\NI_k) }  \big\{ \ind (a_{ij} = 1) - \ind ( a_{ij} = 0 ) \big\} \pi(\ba_{\eij} \con \alpha) \EXP \big\{ \OR^*(\ba_i , \bX_i , k) \cond \type_i = k \big\} \\
			& =
			\frac{1}{\NI_k} \sum_{j=1}^{\NI_k}  \sum_{ \ba_i \in \zosets(\NI_k) }  \big\{ \ind (a_{ij} = 1) - \ind ( a_{ij} = 0 ) \big\} \pi(\ba_{\eij} \con \alpha) \EXP \big\{ \indOR_j^* (\ba_i , \bX_i , k) \cond \type_i = k \big\}
		\end{align*}
		where $u_j$ is the $j$th standard $\NI_k$-dimensional unit vector. In the absence of interference, $\EXP \big\{ \indOR_j^* (\ba_i , \bX_i , k) \cond \type_i = k \big\}$ is the same as $\EXP \big\{ \indPIOR_j (a_{ij} , \bX_i , k \con \paraOR^*) \cond \type_i = k \big\}$ for all $\ba_\eij \in \zosets(\NI_k-1)$ based on \eqref{eq:nointequiv}. Furthermore, $\sum_{\ba_{\eij} \in \zosets(\NI_k-1)}\pi(\ba_\eij \con \alpha) = 1 $ for all $\alpha$. Therefore,
		\begin{align*}
			&
			\uT_k^\DE (\alpha) 
			\\
			& =
			\frac{1}{\NI_k} \sum_{j=1}^{\NI_k}  \sum_{ \ba_i \in \zosets(\NI_k) }  \big\{ \ind (a_{ij} = 1) - \ind ( a_{ij} = 0 ) \big\}
			\\
			& 
			\hspace*{2cm}
			\times \pi(\ba_{\eij} \con \alpha) \EXP \big\{ \indPIOR_j(a_{ij} , \bX_i , k \con \paraOR^* ) \cond \type_i = k \big\}
			\\
			& = 
			\frac{1}{\NI_k}  \sum_{j=1}^{\NI_k}  \Big[ \EXP \big\{ \indPIOR_j(1 , \bX_i , k \con \paraOR^*) - \indPIOR_j(0 , \bX_i , k \con \paraOR^*) \cond \type_i = k \big\} \Big] \sum_{ \ba_\eij  }  \pi(\ba_{\eij} \con \alpha) \\
			& = 
			\frac{1}{\NI_k} \sum_{j=1}^{\NI_k}  \Big[ \EXP \big\{ \indPIOR_j(1 , \bX_i , k \con \paraOR^*) \cond \type_i = k \big\} - \EXP \big\{ \indPIOR_j(0 , \bX_i , k \con \paraOR^*) \cond \type_i = k \big\} \Big]   \\
			& =	\uT_k^\ATE\ .
		\end{align*}
	Similarly, we find $\uT_k^\IE(\alpha , \alpha') = 0$ in the absence of interference.
	\begin{align*}
			&
			\uT_k^\IE(\alpha , \alpha')
			\\
			& = 
			\sum_{ \ba_i \in \zosets(\NI_k) } 
				\big\{ \bw_k^\IE (\ba_i \con \alpha , \alpha') \big\} \T \EXP \big\{ \bY_i( \ba_i ) \cond \type_i = k \big\} \\
			& =
			\frac{1}{\NI_k} \sum_{j=1}^{\NI_k} u_j\T  \sum_{ \ba_i \in \zosets(\NI_k) }  \ind (a_{ij} = 0) \big\{ \pi(\ba_{\eij} \con \alpha) - \pi(\ba_{\eij} \con \alpha') \big\} \EXP \big\{ \OR^* (\ba_i , \bX_i , k) \cond \type_i = k \big\} \\
			& =
			\frac{1}{\NI_k} \sum_{j=1}^{\NI_k}  \sum_{ \ba_i \in \zosets(\NI_k) } \ind (a_{ij} = 0) \big\{ \pi(\ba_{\eij} \con \alpha) - \pi(\ba_{\eij} \con \alpha') \big\} \EXP \big\{ \indOR_j^*(\ba_i , \bX_i , k) \cond \type_i = k \big\} \\
			& = 
			\frac{1}{\NI_k} \sum_{j=1}^{\NI_k} \EXP \big\{ \indPIOR_j(0 , \bX_i , k \con \paraOR^* ) \cond \type_i = k \big\}   \sum_{ \ba_\eij }  \big\{ \pi(\ba_{\eij} \con \alpha) - \pi(\ba_{\eij} \con \alpha') \big\} \\
			& = 
			0\ .
		\end{align*}

	\subsection{Proof of Theorem \ref{thm:RobustATE}}							\label{sec:B10}
	
		The result in (i) is straightforward by combining Theorem \ref{thm:ParaPopEst} in the main paper and Lemma \ref{lmm:NoInterference}. Theorem \ref{thm:ParaPopEst} in the main paper implies that $\widehat{\oT}^\DE(\alpha)$ and $\widehat{\oT}^\IE(\alpha, \alpha')$ are consistent estimators for $\oT^\DE(\alpha)$ and $\oT^\IE(\alpha, \alpha')$, respectively. In the absence of interference, Lemma \ref{lmm:NoInterference} shows that $\oT^\DE(\alpha) = \oT^\ATE$ and $\oT^\IE(\alpha,\alpha')=0$, respectively. Thus, we get the consistency of $\widehat{\oT}^\DE(\alpha)$ and $\widehat{\oT}^\IE(\alpha , \alpha')$ for $\uT^\ATE$ and $0$, respectively. 
	
	To claim the result in (ii), we derive the semiparametric efficiency bound under stated assumptions. The variance of $\varphi(\oT^\ATE)$ is already given in Lemma \ref{lmm:NIEIF}. That is, 
\begin{align*}
	\EXP \big\{ \varphi(\oT^\ATE)^2 \big\} 
	& = 
	\sum_{k=1}^\NT \frac{p_k^* }{\NI_k^2} \sum_{j=1}^{\NI_k} \EXP \bigg\{ \frac{\Sigma_{jj}^*(1, \bX_i, k)}{e_j^*(1 \cond \bX_i, k)} + \frac{\Sigma_{jj}^*(0, \bX_i, k)}{e_j^*(0 \cond \bX_i, k)} \, \bigg| \, \type_i = k  \bigg\} \\
	& 
	+ \EXP \bigg[ \bigg[ \sum_{k=1}^\NT \frac{ \ind(\type_i = k) }{\NI_k} \sum_{j=1}^{\NI_k} \Big\{ \indIOR_j(1,\bX_i,k) - \indIOR_j(0,\bX_i,k) \Big\} - \oT^\ATE \bigg]^2  \bigg] 
	 \ .
\end{align*}
The variance of $\varphi(\oT^\DE(\alpha))$ is given by 
\begin{align*}
	&
	\EXP \big\{ \varphi(\oT^\DE(\alpha)) ^2 \big\} 
	\\
	& =
	\sum_{k=1}^\NT \frac{p_k^*}{\NI_k^2} \sum_{j=1}^{\NI_k} \sum_{ \ba_i \in \zosets(\NI_k) } \EXP \bigg\{ \frac{ \pi(\ba_{\eij} \con \alpha)^2 \Sigma_{jj}^* (a_{ij}, \bX_i, k) }{e^* (\ba_i \cond \bX_i, k)} \, \bigg| \, \type_i = k  \bigg\} \\
	& \hspace*{1cm} 
	+ \EXP \bigg[ \bigg[ \sum_{k=1}^\NT \frac{ \ind(\type_i = k) }{\NI_k} \sum_{j=1}^{\NI_k} \Big\{ \indIOR_j(1,\bX_i,k) - \indIOR_j(0,\bX_i,k) \Big\} - \oT^\ATE \bigg]^2  \bigg] 
	 \ .
\end{align*}
The second equality is straightforward by observing $\{ \bw_k^\DE(\ba_i \con \alpha) \big\}\T = u_j\T   \big\{ \ind (a_{ij} = 1) - \ind ( a_{ij} = 0 ) \big\} \pi(\ba_{\eij} \con \alpha)$ where $u_j$ is the $j$th standard $\NI_k$-dimensional unit vector and following the results established in Lemma \ref{lmm:NoInterference}, i.e., $\Sigma^*(\ba_i, \bx_i, k) = \text{diag} \big[ \Sigma_{11}^*(a_{i1}, \bx_i, k) , \ldots , \Sigma_{\NI_k \NI_k}^*(a_{i\NI_k}, \bx_i, k) \big] $ and $\oT^\DE(\alpha) = \oT^\ATE$. Therefore, the gap between $\EXP \big\{ \varphi(\oT^\DE(\alpha) )^2 \big\} $ and $\EXP \big\{ \varphi(\oT^\ATE) ^2 \big\} $ is 
\begin{align}												\label{proof:12-000}
	& \EXP \big\{ \varphi(\oT^\DE(\alpha) )^2 \big\} - \EXP \big\{ \varphi(\oT^\ATE) ^2 \big\} 
	\\
	& = \sum_{k=1}^\NT \frac{p_k^*}{\NI_k^2}  \sum_{j=1}^{\NI_k}  \EXP \bigg[
	\sum_{ \ba_i \in \zosets(\NI_k) } \frac{ \pi(\ba_{\eij} \con \alpha)^2 \Sigma_{jj}^* (a_{ij}, \bX_i, k) }{e^*(\ba_i \cond \bX_i, k)} 	
	-  
	\sum_{a_{ij} = 0}^1
	\frac{\Sigma_{jj}^*(a_{ij}, \bX_i, k)}{e_j^*(a_{ij} \cond \bX_i, k)}\, \bigg| \, \type_i = k   \bigg]
	\nonumber
	\ .
\end{align}

First term of the conditional expectation in \eqref{proof:12-000} is lower bounded by the second term.
\begin{align*}
	&  \sum_{ \ba_i \in \zosets(\NI_k) } \frac{ \pi(\ba_{\eij} \con \alpha)^2 \Sigma_{jj}^* (a_{ij}, \bx_i, k) }{e^*(\ba_i \cond \bx_i, k)} 
	\\
	& =
	\sum_{ \ba_i : a_{ij} = 1 } \frac{ \pi(\ba_{\eij} \con \alpha)^2}{e^*(\ba_i \cond \bx_i, k)  \Sigma_{jj}^{*-1} (a_{ij}, \bx_i, k) }
	+ \sum_{ \ba_i : a_{ij} = 0 } \frac{ \pi(\ba_{\eij} \con \alpha)^2}{e(\ba_i \cond \bx_i, k)  \Sigma_{jj}^{*-1} (a_{ij}, \bx_i, k) }
	\\
	& \geq
	 \frac{ \big\{ \sum_{ \ba_i : a_{ij} = 1 } \pi(\ba_{\eij} \con \alpha) \big\} ^2}{ \big\{ \sum_{ \ba_i : a_{ij} = 1 } e^*(\ba_i \cond \bx_i, k)  \Sigma_{jj}^{*-1} (a_{ij} , \bx_i, k) \big\} }
	+ \frac{ \big\{ \sum_{ \ba_i : a_{ij} = 0 } \pi(\ba_{\eij} \con \alpha) \big\} ^2}{ \big\{ \sum_{ \ba_i : a_{ij} = 0 } e^*(\ba_i \cond \bx_i, k)  \Sigma_{jj}^{*-1} (a_{ij} , \bx_i, k) \big\} }
	 \\
	& = 
	 \frac{ 1 }{ e_j^*(1 \cond \bx_i, k)  \Sigma_{jj}^{*-1} (1, \bx_i, k)  }
	+ \frac{ 1 }{ e_j^*(0, \cond \bx_i, k) \Sigma_{jj}^{*-1} (0, \bx_i, k)  }
	 \\
	& = \frac{\Sigma_{jj}^*(1, \bX_i, k)}{e_j^*(1 \cond \bX_i, k)} + \frac{\Sigma_{jj}^*(0, \bX_i, k)}{e_j^*(0 \cond \bX_i, k)} 
	\ .
\end{align*}
The inequality in the third line is based on the Bergstr{\"o}m's inequality. Specifically, if $c_i \in \R$ and $d_i > 0$ for $i=1,\cdots,n$, we can use the Cauchy-Schwarz inequality to obtain
\begin{align*}
	\sum_{i=1}^n \frac{c_i^2}{d_i} \cdot \left( \sum_{i=1}^n d_i	\right) 
	=
	\sum_{i=1}^n \left( \frac{c_i}{\sqrt{d_i}} \right)^2 \cdot \left( \sum_{i=1}^n  \sqrt{d_i} 
	\right)^2 
	 \geq 
	\left(
		\sum_{i=1}^n \frac{c_i}{\sqrt{d_i}} \cdot \sqrt{d_i}
	\right)^2 
	=
	 \left(
		\sum_{i=1}^n c_i 
	\right)^2
	\ .
\end{align*}
As a result, we get
\begin{align*}
	\frac{c_1^2}{d_1} + \cdots + \frac{c_n^2}{d_n} \geq \frac{(c_1+ \cdots + c_n)^2}{d_1 + \cdots + d_n} \ . 
\end{align*}
The equality is only attained when $c_1/d_1= \ldots = c_n/d_n$. Replacing $c_i$ with $\pi(\ba_\eij \con \alpha)$ and $d_i$ with $e^*(\ba_i \cond \bx_i, k) \Sigma_{jj}^{*-1}(a_{ij}, \bx_i, k)$, the inequality in the third line and the conditions for the equality are proven. As a result, the gap $ \EXP \big\{ \varphi ( \oT^\DE(\alpha ) ) ^2 \big\} - \EXP \big\{ \varphi (\oT^\ATE) ^2 \big\}$ in \eqref{proof:12-000} is non-negative and becomes zero if and only if $\pi(\ba_\eij \con \alpha)/\{ e^*(\ba_i \cond \bx_i,k) \Sigma_{jj}^{*-1}(a_{ij}, \bx_i, k) \} = \pi(\ba_\eij \con \alpha) \Sigma_{jj}^*(a_{ij}, \bx_i, k) /  e^*(\ba_i \cond \bx_i,k) $ are identical for all $\alpha \in (0,1)$ and $\ba_\eij \in \zosets(\NI_k-1)$.

\subsection{Proof of Lemma \ref{lmm:diagSigma}}
\label{sec:appendix3-LA2}
	
		We denote $\EXP ( \bY_\eij \cond \bA_i = \ba_i , \bX_i = \bx_i, \type_i = k) = \OR_{(-j)}^*(\ba_i, \bx_i, k)$ for brevity. To show that $\Sigma^*$ is diagonal, it suffices to show $\text{Cov} ( Y_{ij} , \bY_\eij \cond \bA_i = \ba_i , \bX_i, \type_i = k)$ is zero. We see that
		\begin{align*}
			\text{Cov} & ( Y_{ij} , \bY_\eij  \cond \bA_i = \ba_i , \bX_i = \bx_i, \type_i = k) \\
			& =
			\EXP \bigg[
				\Big\{ Y_{ij} - \indOR_j^*(\ba_i, \bx_i, k)\Big\} 
				\Big\{ \bY_\eij - \OR_{(-j)}^*(\ba_i, \bx_i, k) \Big\}
				 \, \bigg| \, \bA_i = \ba_i, \bX_i= \bx_i, \type_i = k \bigg] \\
			& = 
			\EXP \bigg[
				\EXP \Big\{ Y_{ij} - \indOR_j^*(\ba_i, \bx_i, k) \, \Big| \, \bY_\eij ,  \bA_i = \ba_i, \bX_i= \bx_i, \type_i = k \Big\}  \\
				& \hspace*{4cm} \times
				\Big\{ \bY_\eij - \OR_{(-j)}^*(\ba_i, \bx_i, k) \Big\} 
				\, \bigg| \, \bA_i = \ba_i, \bX_i= \bx_i, \type_i = k \bigg] \\
			& = 
			\EXP \bigg[
				\Big\{
				\EXP ( Y_{ij} |  \bA_i = \ba_i, \bX_i= \bx_i, \type_i = k )  - \indOR_j^*(\ba_i, \bx_i, k)  \Big\} \\
				& \hspace*{4cm} \times
				\Big\{ \bY_\eij - \OR_{(-j)}^*(\ba_i, \bx_i, k) \Big\} 
				\, \bigg| \, \bA_i = \ba_i, \bX_i= \bx_i, \type_i = k \bigg] 	\\
				& =0\ .
		\end{align*}
		The first identity is from the definition of the conditional covariance. The second identity is from the law of total expectation. The third identity holds from Assumption \ref{assp:1}.	
		
		Next, we show that each diagonal element is a function of its own treatment indicator only. In model $\modelii$, the conditional distribution of $Y_{ij}$ given $(\bA_i, \bX_i, \type_i)$ is the same as the conditional distribution of $Y_{ij}$ given $(A_{ij}, \bX_i, \type_i)$. Therefore, we obtain $\EXP(Y_{ij}^2 \cond \bA_i = \ba_i, \bX_i= \bx_i, \type_i = k) = \EXP(Y_{ij}^2 \cond A_{ij} = a_{ij} , \bX_i= \bx_i, \type_i = k) $ and the identity presented in \eqref{eq:nointequiv}. Hence, the conditional variance of $Y_{ij}$ given $(\bA_i, \bX_i, \type_i)$ does not depend on $\bA_\eij$.
		\begin{align*}
			\VAR (Y_{ij} \cond \bA_i = \ba_i , \bX_i = \bx_i, \type_i = k)
			& = 
			\EXP \big( Y_{ij}^2 \, \big| \, \bA_i = \ba_i, \bX_i= \bx_i, \type_i = k \big)
			- \indOR_j^*(\ba_i, \bx_i, k) ^2  
			\\
			& = 
			\EXP \big( Y_{ij}^2 \, \big| \, A_{ij} = a_{ij}, \bX_i= \bx_i, \type_i = k \big)
			- \indOR_j^{\rm NoInt} (a_{ij}, \bx_i, k) ^2 \\
			& = 
			\VAR (Y_{ij} \cond A_{ij} = a_{ij} , \bX_i = \bx_i, \type_i = k)
			\ .
		\end{align*}
		Thus, each diagonal element of $\Sigma^*(\ba_i, \bx_i, k)$ can be represented as $\Sigma_{jj}^*(a_{ij}, \bx_i, k)$. 
	
	\subsection{Proof of Lemma \ref{lmm:NIEIF}}
	\label{sec:appendix3-LA3}
	
					We follow the proof of Theorem \ref{thm:EIFVC} in the main paper. Consider a smooth regular parametric submodel parametrized by a possibly multi-dimensional parameter $\eta$ is
\begin{align*}
	P ( \by , \ba , \bx , k \con \eta) 
	& = \pyi (\by \cond \ba , \bx , k \con \eta) e(\ba \cond \bx , k \con \eta) \px ( \bx \cond k \con \eta) p_k(\eta)
	\ .
\end{align*}
We assume the density of parametric submodel $P(\cdot \con \eta)$ equals the true density $P$ at $\eta=\eta^*$. The corresponding score function is
\begin{align*}
	s (\by,\ba,\bx,k \con \eta) 
	&=  s_Y ( \by , \ba , \bx, k \con \eta ) + s_A( \ba , \bx , k \con \eta)  +  s_X ( \bx , k \con \eta)  + s_\type ( k \con \eta)   
\end{align*}
where the score functions are defined in \eqref{proof:1-score}.

We find that $\modelii$ imposes the following identity for the conditional density.
\begin{align*}
	P_{Y,j}^* ( y_{ij} \cond \ba_i , \bx_i , k) 
	& =	
	\int P_Y^* (\by_i \cond \ba_i , \bx_i , k) \, d\by_\eij 
	\\
	&= 
	\int P_Y^* (\by_i \cond \ba_i' , \bx_i , k) \, d\by_\eij 
	=
	P_{Y,j}^* ( y_{ij} \cond \ba_i' , \bx_i , k) 
\end{align*}
where $P_{Y,j}^*$ is the conditional density of $Y_{ij}$ given $(\bA_i, \bX_i, \type_i)$ and the $j$th element of $\ba_i$ and $\ba_i'$ are the same; i.e., $a_{ij}=a_{ij}'$. Therefore, considering the parametric submodel and its derivative, we obtain the following restriction based on $\modelii$. 
\begin{align}									\label{proof:11-NoInt}
	\nonumber
	&
	\int s_Y(\by_i, \ba_i, \bx_i , k \con \eta) P_Y(\by_i \cond \ba_i , \bx_i , k \con \eta ) \, d \by_\eij 
	\\
	&
	= \int s_Y(\by_i, \ba_i' , \bx_i , k \con \eta) P_Y(\by_i \cond \ba_i' , \bx_i , k \con \eta ) \, d \by_\eij 
\end{align}
where $a_{ij} = a_{ij}'$. At $\eta=\eta^*$, we observe that
\begin{align}								\label{proof:11-000}
	\nonumber
	\EXP  \Big\{ Y_{ij} s_Y(\bY_i, \ba_i,  & \bX_i, k \con \eta^*) \, \Big| \, \bA_i = \ba_i , \bX_i = \bx_i, \type_i = k \Big\}
	\\
	\nonumber
	& = \iint
	y_{ij} s_Y(\by_i, \ba_i, \bx_i , k \con \eta^*) P_Y^* (\by_i \cond \ba_i , \bx_i , k ) \, d \by_\eij  \, dy_{ij}
	\\
	\nonumber
	& = \iint y_{ij}  s_Y(\by_i, \ba_i' , \bx_i , k \con \eta^*) P_Y^* (\by_i \cond \ba_i' , \bx_i , k ) \, d \by_\eij  \, d y_{ij}
	\\
	& = \EXP \Big\{ Y_{ij} s_Y(\bY_i, \ba_i', \bX_i, k \con \eta^*) \, \Big| \, \bA_i = \ba_i' , \bX_i = \bx_i, \type_i = k \Big\}
\end{align}
where $a_{ij} = a_{ij}'$. This implies that $\EXP  \big\{ Y_{ij} s_Y(\bY_i, \ba_i, \bX_i, k \con \eta^*) \, \big| \, \bA_i = \ba_i , \bX_i = \bx_i, \type_i = k \big\}$ does not depend on $\ba_\eij$. We emphasize the independence from $\ba_\eij$ by denoting the expectation as $\EXP  \big\{ Y_{ij} s_Y(\bY_i, \ba_{i(j=a)}, \bX_i, k \con \eta^*) \, \big| \, \bA_i = \ba_{i(j=a)} , \bX_i = \bx_i, \type_i = k \big\}$ where $\ba_{i(j=a)}$ is any treatment vector where the $j$th component $a_{ij}$ is equal to $a \in \{0,1\}$.

From the parametric submodel, we obtain the $1$-dimensional tangent space for $1$-dimensional parameters which is the mean closure of all $1$-dimensional linear combinations of scores, that is, 
\begin{align}							\label{proof:11-001}
	\nonumber 
	\mathcal{T} =
	\Big\{
		 S (\by, \ba, \bx, k)  \,
		      \Big| 	\,	      
			& S(\by , \ba , \bx , k) 	   
			= S_Y (\by , \ba, \bx, k)  + S_A (\ba, \bx, k)  + S_X ( \bx, k) + S_\type (k)   \in \R \ , \ 
			\\
			\nonumber
		      \, &  \EXP \big\{ S_Y (\bY_i , \ba, \bx, k) \cond \bA_i = \ba, \bX_i =  \bx , \type_i = k \big\} = 0  \
		      \text{ for all } (\ba,\bx,k) \ ,
		      \\
		      \nonumber
		      \, &  \int S_Y(\by_i, \ba_i, \bx_i , k) P_Y^* (\by_i \cond \ba_i , \bx_i , k ) \, d \by_\eij \\
		      \nonumber
		      \, & \hspace*{1cm}
				= \int S_Y(\by_i, \ba_i', \bx_i , k) P_Y^* (\by_i \cond \ba_i' , \bx_i , k ) \, d \by_\eij \text{ where } a_{ij} = a_{ij}',
				\\
		      \nonumber
	    	\, &   \EXP \big\{ S_A (\bA_i , \bx , k ) \cond \bX_i = \bx , \type_i = k \big\} = 0  
	    	\text{ for all } (\bx,k) \ ,
	    	\\
	  \, &  	\EXP \big\{ S_X (\bX_i , k ) \cond \type_i = k \big\} = 0  \ , \  ^\forall k  \text{ for all } k \ ,
	  \EXP\big\{ S_\type (\type_i ) \big\} = 0
	    \Big\}
	    \ .
\end{align}

	The ATE can be represented as $\oT^\ATE(\eta) = \sum_{k=1}^\NT p_k (\eta) \uT^\ATE_k(\eta) $ in the regular parametric submodel at parameter $\eta$ where 
	\begin{align*}
	& \uT_k^\ATE (\eta) 
	\\
	\nonumber
	& =
	\frac{1}{\NI_k}
	\sum_{j=1}^{\NI_k} \iint y_{ij}  \Big\{ \pyi (\by_i \cond \ba_{i(j=1)} , \bx_i , k \con \eta)  - \pyi (\by_i \cond \ba_{i(j=0)} , \bx_i , k \con \eta)  \Big\} \\
	\nonumber
	& \hspace*{5cm} \times  \px(\bx_i \cond k \con \eta) \, d \by_i d \bx_i
	  \ .
\end{align*}
The derivative of $\uT^\ATE_k(\eta)$ is
\begin{align}												\label{proof:11-003}
	\frac{\partial \oT^\ATE(\eta^*)  }{\partial \eta} 
	& = \underbrace{\sum_{k=1}^\NT p_k \frac{\partial \uT^\ATE_k(\eta^*) }{\partial \eta}}_{\equiv \ Q_1(\eta^*)}
	+ \underbrace{\sum_{k=1}^\NT \uT^\ATE_k \frac{\partial p_k(\eta^*)}{\partial \eta}}_{\equiv \ Q_2(\eta^*)}
\end{align}
where
\begin{align}                                  \label{proof:11-004}
	\frac{\partial \uT^\ATE_k(\eta) }{\partial \eta} 
	&
	=   \frac{1}{\NI_k} \sum_{j=1}^{\NI_k} \iint y_{ij} \Big\{ s_Y (\by_i , \ba_{i(j=1)}, \bx_i, k \con \eta) \pyi ( \by_i \cond \ba_{i(j=1)} , \bx_i, k \con \eta ) 	
	\\
	\nonumber & \hspace*{0.5cm} -
	s_Y (\by_i , \ba_{i(j=0)}, \bx_i, k \con \eta) \pyi ( \by_i \cond \ba_{i(j=0)} , \bx_i, k \con \eta ) 	\Big\}
	\px(\bx_i \cond k \con \eta) \, d\by_i d\bx_i \\
	\nonumber
	& + \frac{1}{\NI_k} \sum_{j=1}^{\NI_k} \iint y_{ij}  \Big\{ \pyi (\by_i \cond \ba_{i(j=1)} , \bx_i , k \con \eta)  - \pyi (\by_i \cond \ba_{i(j=0)} , \bx_i , k \con \eta)  \Big\} \\
	\nonumber 
	& \hspace*{0.5cm} \times s_X (\bx_i, k \con \eta) \px(\bx_i \cond k \con \eta ) \, d \by_i d \bx_i
	  \ .
\end{align}
	
	The conjectured efficient influence function  of $\oT^\ATE$ is
\begin{align*}
			\varphi ( \oT^\ATE ) 
			& =
			\sum_{k=1}^\NT \ind(\type_i = k) \bigg[
		\frac{1}{\NI_k} \sum_{j=1}^{\NI_k} \bigg[ \Big\{ \ind ( A_{ij} = 1 ) - \ind ( A_{ij} = 0 ) \Big\}	
		\frac{Y_{ij} - \indIOR_j ( A_{ij} , \bX_i , k ) }{e_j^*(A_{ij} \cond \bX_i,k)}
		\bigg]
		\\
		& \hspace*{2cm} + \frac{1}{\NI_k} \sum_{j=1}^{\NI_k} \Big\{ \indIOR_j(1,\bX_i,k) - \indIOR_j(0,\bX_i,k) \Big\}
	\bigg] - \oT^\ATE
	\\
	&
	= 
	\sum_{k=1}^\NT F_{1k} ( \bY_i, \bA_i, \bX_i , \type_i) 
	+ \sum_{k=1}^\NT  F_{2k} (\bX_i, \type_i)
	+ \sum_{k=1}^\NT F_{3k}(\type_i)
	\end{align*}
	where
\begin{align*}
	\nonumber 
	F_{1k} ( \bY_i, \bA_i, \bX_i , \type_i) 
	& = 
		\frac{\ind(\type_i = k)}{\NI_k} \sum_{j=1}^{\NI_k} \bigg[ \Big\{ \ind ( A_{ij} = 1 ) - \ind ( A_{ij} = 0 ) \Big\}	
		\frac{Y_{ij} - \indIOR_j ( A_{ij} , \bX_i , k ) }{e_j^*(A_{ij} \cond \bX_i,k)} \bigg] \\
		\nonumber
	F_{2k} (\bX_i, \type_i)
	& = \frac{\ind(\type_i = k)}{\NI_k}
	\bigg[
		 \sum_{j =1 }^{\NI_k} \Big\{ \indIOR_j(1,\bX_i,k) - \indIOR_j(0,\bX_i,k) - \uT_k^\ATE \Big\}  \bigg] \\
		 F_{3k}(\type_i) & =  \ind(\type_i=k) \big( \uT_k^\ATE - \oT^\ATE \big)  \ .
\end{align*}

		We first show that $\oT^\ATE(\eta)$ is a differentiable parameter, i.e.,
	\begin{align}                            \label{proof:11-Cond1}
		\frac{\partial \oT^\ATE(\eta^*)}{\partial \eta} 
		= Q_1(\eta^*) + Q_2(\eta^*)
		& =
		\EXP 
		\Big\{
			\varphi ( \oT^\ATE ) \cdot s(\bY_i, \bA_i, \bX_i , \type_i \con \eta^* )
		\Big\}   \ .
	\end{align}
	The elementary terms in $\varphi ( \oT^\ATE) \cdot s ( \bY_i , \bA_i , \bX_i , \type_i \con \eta^* )$ are represented below 
\begin{align}                                 \label{proof:11-006}
	 \varphi ( \oT^\ATE) \cdot & s ( \bY_i , \bA_i , \bX_i , \type_i \con \eta^* )
     \\
      & 
      \nonumber
      = 
      \sum_{k=1}^\NT F_{1k}(\bY_i, \bA_i, \bX_i, \type_i) s_Y(\bY_i, \bA_i, \bX_i, \type_i \con \eta^*) \\
      \nonumber
      & +
      \sum_{k=1}^\NT F_{1k}(\bY_i, \bA_i, \bX_i, \type_i) \big\{ s_A(\bA_i, \bX_i, \type_i \con \eta^*) + s_X(\bX_i, \type_i \con \eta^*) + s_\type(\type_i \con \eta^*) \big\}
      \nonumber
      \\      
     & +  \sum_{k=1}^\NT F_{2k}( \bX_i, \type_i)  \big\{  s_Y(\bY_i, \bA_i, \bX_i, \type_i \con \eta^*) + s_A(\bA_i, \bX_i, \type_i \con \eta^*) + s_\type(\type_i \con \eta^*) \big\}  
     \nonumber
     \\
      & +
      \sum_{k=1}^\NT F_{2k}(\bX_i, \type_i) s_X(\bX_i, \type_i \con \eta^*) 
      \nonumber
      \\ 
      & + \sum_{k=1}^\NT F_{3k}(\type_i)   \big\{  s_Y(\bY_i, \bA_i, \bX_i, \type_i \con \eta^*) + s_A(\bA_i, \bX_i, \type_i \con \eta^*) + s_X(\bX_i, \type_i \con \eta^*) \big\}  
      \nonumber
      \\
     & + \sum_{k=1}^\NT F_{3k}(\type_i) s_\type(\type_i \con \eta^*)
      \nonumber
        \ .
\end{align}

We study the expectation of each piece in \eqref{proof:11-006}. The expectation of the first piece in \eqref{proof:11-006} is represented as
\begin{align}					\label{proof:11-008}
	 & \EXP \Big\{  F_{1k}(\bY_i, \bA_i, \bX_i, \type_i) s_Y(\bY_i, \bA_i, \bX_i, \type_i \con \eta^*) \Big\} 
	\\
	\nonumber 
	& = \frac{p_k^*}{\NI_k}  \EXP\bigg[ 
	\sum_{j=1}^{\NI_k} 
	\bigg[ \EXP \Big\{ Y_{ij} s_Y(\bY_i, \ba_{i(j=1)}, \bX_i, k \con \eta^*) \, \Big| \, \bA_i = \ba_{i(j=1)} , \bX_i, \type_i = k \Big\} 
	\\
	\nonumber
	& \hspace*{2cm}
	- \EXP \Big\{ Y_{ij} s_Y(\bY_i, \ba_{i(j=0)}, \bX_i, k \con \eta^*) \, \Big| \, \bA_i = \ba_{i(j=0)} , \bX_i, \type_i = k \Big\}  \bigg]
	 \, \bigg| \, \type_i = k  \bigg] 
	 \\
	\nonumber	
	& = 
	\frac{p_k^*}{\NI_k} \sum_{j=1}^{\NI_k} \iint y_{ij} \Big\{ s_Y (\by_i , \ba_{i(j=1)}, \bx_i, k \con \eta^* ) \pyi^*  ( \by_i \cond \ba_{i(j=1)} , \bx_i, k) 	
	\\
	\nonumber
	& \hspace*{2cm} -
	s_Y (\by_i , \ba_{i(j=0)}, \bx_i, k \con \eta^* ) \pyi^*  ( \by_i \cond \ba_{i(j=0)} , \bx_i, k) 	\Big\}
	\px^*(\bx_i \cond k ) \, d\by_i d\bx_i 
	  \ .
\end{align}
The first equality holds from 
the law of total expectations and the definitions of  
\eqref{proof:11-000} and $e_j^*$. The last equality holds from the definition of conditional expectations.  The above quantity contains the first term of $\partial \uT_k^\ATE(\eta^*) / \partial \eta$ in \eqref{proof:11-004}.

The expectation of the second piece in \eqref{proof:11-006} is zero from the law of total expectation.
\begin{align}									\label{proof:11-008-2}
	\EXP \Big[ & F_{1k}(\bY_i, \bA_i, \bX_i, \type_i) \big\{ s_A(\bA_i, \bX_i, \type_i \con \eta^*) + s_X(\bX_i, \type_i \con \eta^*) + s_\type(\type_i \con \eta^*) \big\} \Big]
	\\
	\nonumber
	& =
	p_k^*  \EXP\bigg[ \sum_{\ba_i \in \zosets(\NI_k)}  e^*(\ba_i, \bX_i, k) \underbrace{ \EXP \Big\{ F_{1k}(\bY_i, \ba_i, \bX_i, k) \, \Big| \, \bA_i = \ba_i , \bX_i, \type_i = k \Big\} }_{=0}\\
	& \hspace*{3cm} 
		\times \Big\{ s_A(\bA_i, \bX_i, k  \con \eta^*) + s_X(\bX_i, k \con \eta^*) + s_\type(k \con \eta^*) \Big\} 
	\, \bigg| \, \type_i = k  \bigg] 
	= 0
	\nonumber
	  \ .
\end{align}

The expectation of the third piece in \eqref{proof:11-006} is also zero from the law of total expectation.
\begin{align}								\label{proof:11-008-3}
	\EXP \Big[ &  F_{2k}(\bX_i, \type_i) \big\{  s_Y(\bY_i, \bA_i, \bX_i, \type_i \con \eta^*) + s_A(\bA_i, \bX_i, \type_i \con \eta^*) + s_\type(\type_i \con \eta^*) \big\} \Big]
	\\
	\nonumber
	& =
	p_k^*  \EXP\bigg[ F_{2k}(\bX_i, k) \underbrace{ \EXP \Big\{  s_Y(\bY_i, \bA_i, \bX_i, k \con \eta^*) + s_A(\bA_i, \bX_i, k \con \eta^*) \, \Big| \, \bX_i , \type_i = k \Big\} }_{=0} \, \bigg| \, \type_i = k \bigg]	\\
	\nonumber
	& \hspace*{2cm} + p_k^* \underbrace{ \EXP\Big\{ F_{2k}(\bX_i, k) \, \Big| \, \type_i = k   \Big\} }_{=0} s_\type(k \con \eta^*)
	\\
	& = 0
	\nonumber	
	  \ .
\end{align}

The expectation of the fourth piece  in \eqref{proof:11-006} is represented as
\begin{align}								\label{proof:11-009}
	\EXP & \Big\{ F_{2k}(\bX_i, \type_i) s_X(\bX_i, \type_i \con \eta^*)  \Big\}
	\\
	\nonumber	
	& = \frac{p_k^* }{\NI_k} \sum_{j =1 }^{\NI_k} 
	\bigg[
		\Big\{ \indIOR_j(1,\bX_i,k) - \indIOR_j(0,\bX_i,k) - \uT_k^\ATE \Big\} s_X(\bX_i, k \con \eta^*)  \, \bigg| \, \type_i = k  \bigg] \\
	& = 
	 \frac{p_k^* }{\NI_k} \sum_{j=1}^{\NI_k}
	 \iint
	 y_{ij} \big\{ \py^* (\by_i \cond \ba_{i(j=1)} , \bx_i, k) - \py^* (\by_i \cond \ba_{i(j=0)} , \bx_i, k) \big\} \nonumber
	 \\
	 \nonumber
	 & \hspace*{4cm}
	 \times s_X(\bx_i, k \con \eta^*) \px^*  ( \bx_i \cond k)
	 \, d\by_i d\bx_i
	 \nonumber
	   \ .
\end{align}
The above quantity contains the second term of $\partial \uT_k^\ATE(\eta^*) / \partial \eta$ in \eqref{proof:11-004}.

The expectation of the fifth piece  in \eqref{proof:11-006} is zero because
\begin{align}									\label{proof:11-008-5}
	& 
	\EXP \Big[ F_{3k}(\type_i) \big\{  s_Y(\bY_i, \bA_i, \bX_i, \type_i \con \eta^*) + s_A(\bA_i, \bX_i, \type_i \con \eta^*) + S_X(\bX_i, \type_i \con \eta^*) \big\}   \Big]
	\\
	& =
	p_k ^* F_{3k}(k)  \underbrace{ \EXP \Big\{  s_Y(\bY_i, \bA_i, \bX_i, k \con \eta^*) + s_A(\bA_i, \bX_i, k \con \eta^*) + S_X(\bX_i, k \con \eta^*)  \, \Big| \, \type_i = k \Big\} }_{=0}
	= 0
\nonumber		
  \ .
\end{align}

The expectation of the last piece in \eqref{proof:11-006} is equal to the summand of $Q_2(\eta^*)$ in \eqref{proof:11-003}
\begin{align}							\label{proof:11-010}
	\sum_{k=1}^\NT \EXP \Big\{ F_{3k}(\type_i) s_\type(\type_i \con \eta^*) \Big\}
	& = 
	\sum_{k=1}^\NT p_k^* \uT_k^\ATE s_\type(k \con \eta^*)  - \oT^\ATE \EXP\big\{ s_\type(\type_i \con \eta^*) \big\} \\
	& = 
	p_k^* \uT_k^\ATE  \frac{1}{p_k(\eta^*)} \frac{\partial p_k(\eta^*) }{\partial \eta} 
	=   \uT_k^\ATE \frac{\partial p_k(\eta^*) }{\partial \eta}
	\nonumber	
\end{align}
where $\EXP\big\{ s_\type(\type_i \con \eta^*) \big\}=0$. Replacing each term in \eqref{proof:11-006} with the intermediate results in \eqref{proof:11-008}-\eqref{proof:11-010}, we show that the identity in \eqref{proof:11-Cond1} holds.
\begin{align*}
	\nonumber
				&  
	\EXP 
		\Big\{
\varphi ( \oT^\ATE ) \cdot  s(\bY_i, \bA_i, \bX_i , \type_i \con \eta^* )
		\Big\} \\
	\nonumber
	& = 
	\sum_{k=1}^\NT 
	p_k^* \bigg[ 
		\frac{1}{\NI_k} \sum_{j=1}^{\NI_k} \iint y_{ij} \Big\{ s_Y (\by_i , \ba_{i(j=1)}, \bx_i, k \con \eta^* ) \pyi^* ( \by_i \cond \ba_{i(j=1)} , \bx_i, k) 	
	\\
	\nonumber
	& \hspace*{2cm} -
	s_Y (\by_i , \ba_{i(j=0)}, \bx_i, k \con \eta^* ) \pyi^*  ( \by_i \cond \ba_{i(j=0)} , \bx_i, k) 	\Big\}
	\px^* (\bx_i \cond k ) \, d\by_i d\bx_i \\
	\nonumber
	& \hspace*{0.5cm}  + \frac{1}{\NI_k} \sum_{j=1}^{\NI_k}
	 \iint
	 y_{ij} \big\{ \py^* (\by_i \cond \ba_{i(j=1)} , \bx_i, k) - \py^* (\by_i \cond \ba_{i(j=0)} , \bx_i, k) \big\} \\
	\nonumber
	& \hspace*{2cm} \times s_X(\bx_i, k \con \eta^*) \px^*  ( \bx_i \cond k)
	 \, d\by_i d\bx_i \bigg] 
	  + \sum_{k=1}^\NT 
	\uT_k^\ATE \frac{\partial p_k(\eta_k^*)}{\partial \eta} \\
	& = Q_1(\eta^*) + Q_2(\eta^*) 
	= \frac{\partial \oT^\ATE(\eta^*)}{\partial \eta} 
	  \ .
\end{align*}
This shows that $\oT^\ATE(\eta)$ is a differentiable parameter.

Next, we claim that $\varphi(\oT^\ATE) $ belongs to $\mathcal{T}$ in \eqref{proof:11-001}. First, we show that $F_{1k}$ satisfies the conditions related to $S_Y$. Note that the conditional expectation of $F_{1k}$ given $\bA_i, \bX_i, \type_i$ is
\begin{align*}
	&
	\EXP \big\{ F_{1k} ( \bY_i ,  \ba_i, \bx_i , k')  \, \big| \, \bA_i = \ba_i , \bX_i = \bx_i , \type_i = k' \big\} \\
	& = 
	\ind(k' = k)
	\bigg[
		\frac{1}{\NI_k} \sum_{j=1}^{\NI_k} \bigg[ \frac{\ind ( a_{ij} = 1 )}{e_j^*(1 \cond \bX_i,k) } \Big\{ \EXP( Y_{ij} \cond \bA_i = \ba_i , \bX_i = \bx_i, \type_i = k)- \indOR_j^{\rm NoInt} ( 1 , \bX_i , k ) \Big\} \\
		\nonumber
	& \hspace*{2cm}	- 
		\frac{\ind ( a_{ij} = 0 )}{e_j^*(0 \cond \bX_i,k) } \Big\{ \EXP( Y_{ij} \cond \bA_i = \ba_i , \bX_i = \bx_i, \type_i = k) - \indOR_j^{\rm NoInt} ( 0 , \bX_i , k ) \Big\} 
		\bigg] \bigg]
		\\
		& = 0
		  \ .
\end{align*}
This shows $F_{1k}$ satisfies first condition imposed on $S_Y$. To show that $F_{1k}$ satisfies second condition imposed on $S_Y$, we first observe two intermediate results. We consider the integral of the product of $y_{ij'}$ and $P_Y^*(\by_i \cond \ba_i, \bx_i, k)$ with respect to $\by_\eij$. If $j=j'$, we get
\begin{align}								\label{proof:11-101}
	\int y_{ij} P_Y^*(\by_i \cond \ba_i , \bx_i , k) \, d \by_\eij 
	=
	y_{ij} P_{Y,j}^* (y_{ij} \cond \ba_i , \bx_i, k)
	= 
	y_{ij} P_{Y,j}^* (y_{ij} \cond a_{ij} , \bx_i, k)
	  \ .
\end{align}
The first identity is from the definition of conditional density and the second identity is from \eqref{proof:11-NoInt}. If $j\neq j'$, we get
\begin{align}							\label{proof:11-102}
	& \int y_{ij'} P_Y^* (\by_i \cond \ba_i , \bx_i , k) \, d \by_\eij 
	\\
\nonumber	
	& =
	\EXP( Y_{ij'} \cond Y_{ij} = y_{ij} , \bA_i = \ba_i, \bX_i = \bx_i , \type_i = k)   P_{Y,j}^* (y_{ij} \cond \ba_i , \bx_i, k)
	\\
	\nonumber
	& = \indIOR_{j'} (a_{ij'} , \bx_i, k) P_{Y,j}^* (y_{ij} \cond a_{ij} , \bx_i, k)
	  \ .
\end{align}
The first identity is from the basic decomposition of the density of $\bY_i$ and the second identity is based on Assumption \ref{assp:1} and \eqref{proof:11-NoInt}. Using \eqref{proof:11-101} and \eqref{proof:11-102}, each summand of $F_{1k}$ has the following integral with respect to $\by_\eij$ after being multiplied by $P_Y^*$.
\begin{align*}
	\nonumber
	\int & \bigg[ \big\{ \ind( a_{ij'}=1)  - \ind( a_{ij'}=0)  \big\} \frac{y_{ij'} - \indIOR_{j'} ( a_{ij'} , \bx_i , k )}{e_{j'}^*(a_{ij'} \cond \bx_i, k) }
	\bigg] P_Y^*(\by_i \cond \ba_i , \bx_i , k) \, d \by_\eij 
	\\
	& =
	\begin{cases}
			\big\{ \ind( a_{ij}=1)  - \ind( a_{ij}=0)  \big\} \frac{y_{ij} - \indIOR_{j} ( a_{ij} , \bx_i , k )}{e_{j}^*(a_{ij} \cond \bx_i, k) } P_{Y,j}^* (y_{ij} \cond a_{ij} , \bx_i, k)  & \text{ if } j = j' \\[0.5cm]
		\big\{ \ind( a_{ij'}=1)  - \ind( a_{ij'}=0)  \big\} \frac{\indIOR_{j'} ( a_{ij'} , \bx_i , k ) - \indIOR_{j'} ( a_{ij'} , \bx_i , k )}{e_{j'}^*(a_{ij'} \cond \bx_i, k) } = 0 & \text{ if } j \neq j' 
	\end{cases}
\end{align*}
Note that each summand is zero unless the index is omitted in $\by_\eij$. Hence, we observe that
\begin{align*}
	\int & F_{1k} ( \by_i, \ba_i, \bx_i , k') P_Y^* (\by_i \cond \ba_i , \bx_i , k') \, d \by_\eij
	\\
	& =
	\frac{\ind(k'= k) }{\NI_k} \sum_{j'=1}^{\NI_k}\int 
	\bigg[ \frac{\ind( a_{ij'}=1) }{e_{j'}^*(1 \cond \bx_i, k) } \Big\{ y_{ij'} - \indIOR_{j'} ( 1 , \bx_i , k ) \Big\}  \\
	& \hspace*{3cm}
	- 
	\frac{\ind( a_{ij'}=0) }{e_{j'}^*(0 \cond \bx_i, k) } \Big\{ y_{ij'} - \indIOR_{j'} ( 0 , \bx_i , k ) \Big\} 
	\bigg] P_Y^* (\by_i \cond \ba_i , \bx_i , k) \, d \by_\eij 
	\\
	& = 	
	\frac{\ind(k'= k) }{\NI_k} \bigg[
	\frac{\ind( a_{ij}=1) }{e_{j}^*(1 \cond \bx_i, k) } \Big\{ y_{ij} - \indIOR_{j} ( 1 , \bx_i , k ) \Big\} P_{Y,j}^*  (y_{ij} \cond a_{ij} , \bx_i, k) \\
	& \hspace*{3cm} -	
			\frac{\ind( a_{ij}=0) }{e_{j}^*(0 \cond \bx_i, k) } \Big\{ y_{ij} - \indIOR_{j} ( 0 , \bx_i , k ) \Big\} P_{Y,j}^* (y_{ij} \cond a_{ij} , \bx_i, k)
			\bigg] \\
			& =
			\int F_{1k} ( \by_i, \ba_i', \bx_i , k') P_Y^* (\by_i \cond \ba_i' , \bx_i , k') \, d \by_\eij
\end{align*}
where $a_{ij} = a_{ij}'$. This shows $F_{1k}$ satisfies both conditions imposed on $S_Y$ in $\mathcal{T}$ presented in \eqref{proof:11-001}. Next, $F_{2k}$ satisfies the mean zero condition given $\bX_i, \type_i$
	\begin{align*}
	&
	\EXP \big\{ F_{2k} ( \bX_i ,   k')  \, \big| \, \type_i = k' \big\}
	\\
	\nonumber
	& = \frac{\ind(k' = k)}{\NI_k}
	 \sum_{j =1 }^{\NI_k} 
	\EXP
		\Big\{ \indIOR_j(1,\bX_i,k) - \indIOR_j(0,\bX_i,k) - \uT_k^\ATE \, \Big| \, \type_i = k \Big\}  = 0
		  \ .
	\end{align*}
Therefore, $\sum_{k=1}^\NT F_{2k}$ satisfies the condition imposed on $S_X$ in \eqref{proof:11-001}. Lastly, $\sum_{k=1}^\NT F_{3k}$ satisfies the mean zero condition
	\begin{align*}
		\sum_{k=1}^\NT \EXP \big\{ F_{3k}(\type_i) \big\} & =  \sum_{k=1}^\NT p_k^* \big( \uT_k^\ATE - \oT^\ATE \big) = 0
		  \ .
	\end{align*}
	Therefore, $\sum_{k=1}^\NT F_{3k}$ satisfies the condition imposed on $S_\type$ in \eqref{proof:11-001}. Combining the above results, we have $\varphi(\oT^\ATE) \in \mathcal{T}$ in \eqref{proof:11-001}.
	
	The semiparametric efficiency bound is the expectation of the squared efficient influence function . Therefore, the result can be shown by following the proof of Lemma \ref{lmm:EIFbasic}.

\newpage
\bibliographystyle{apa}
\bibliography{EN.bib}

\end{document}